\documentclass[
11pt, 
english, 
singlespacing, 
headsepline, 
]{MastersDoctoralThesis} 

\usepackage[utf8]{inputenc} 
\usepackage[LGR,T1]{fontenc} 

\usepackage{mathpazo} 

\RequirePackage{preamble}

\usepackage[style=numeric,natbib=true]{biblatex} 

\usepackage[autostyle=true]{csquotes} 

\thesistitle{Predicates of the 3D Apollonius Diagram} 
\supervisor{Menelaos \textsc{Karavelas}} 
\examiner{} 
\degree{Doctor of Philosophy} 
\author{ \textsc{Manos Kamarianakis}} 
\addresses{} 

\subject{Mathematics} 
\university{\href{http://www.uoc.gr}{University Of Crete}} 
\department{\href{http://math.uoc.gr}{Department of Mathematics \& Applied Mathematics}} 
\group{Faculty of Sciences and Engineering} 
\faculty{} 

\AtBeginDocument{
\hypersetup{pdftitle=\ttitle} 
\hypersetup{pdfauthor=\authorname} 
\hypersetup{pdfkeywords=\keywordnames} 
}

\usepackage{hhline}

\begin{document}

\frontmatter 

\pagestyle{plain} 


\begin{titlepage}
\begin{center}
\vspace*{-4cm}
\vspace*{.06\textheight}
{\scshape\LARGE \univname\par}
{\scshape\Large \groupname\par}\vspace{0.2cm} 
{\scshape\Large \deptname}\vspace{1.3cm} 

\textsc{\Large Doctoral Thesis}\\[0.5cm] 

\HRule \\[0.4cm] 
{\huge  \ttitle\par}\vspace{0.4cm} 
\HRule \\[1.0cm] 

\begin{minipage}[t]{0.4\textwidth}
\begin{center} \Large
\emph{Author}\\
\href{http://www.tem.uoc.gr/~manosk}{\authorname} 
\end{center}
\end{minipage}\\[1cm]

 
\vspace{2cm}

\large \textit{A thesis submitted in fulfillment of the requirements\\ for the degree of \degreename}\\[1.0cm] 

\vspace{3cm}
\includegraphics[width=0.25\textwidth]{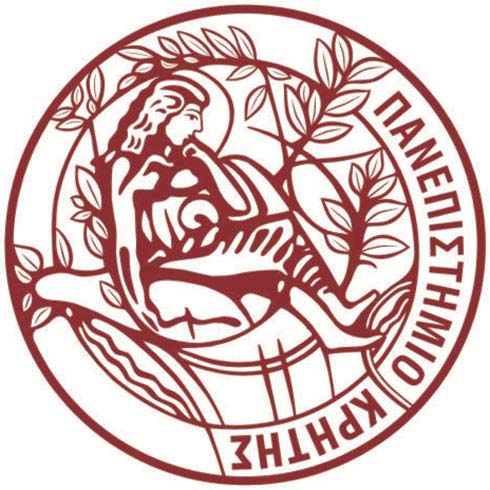}

\vfill

{\Large Heraklion, December 2018}

\end{center}
\end{titlepage}

\cleardoublepage


\HRule \par\vspace{0.4cm} 
{\Huge \emph{Thesis Committee}\par}\vspace{0.4cm} 
\HRule \\[1.5cm] 

{\huge \textbf{PhD advisory committee}}
{
\begin{itemize}
\item Ioannis Emiris, Professor, Department of Informatics \& Telecommunications, National and Kapodistrian University of Athens
\item Menelaos Karavelas, Associate Professor, Department of Mathematics \& Applied Mathematics, University of Crete, (\emph{Advisor})
\item Athanasios Pheidas, Professor, Department of Mathematics \& Applied Mathematics, University of Crete
\end{itemize}}

\vspace{1cm}
{\huge \textbf{Dissertation committee}}
{
\begin{itemize}
\item Michael Lambrou, Professor, Department of Mathematics \& Applied Mathematics, University of Crete
\item Leonidas Palios, Professor, Department of Computer Science and Engineering, University of Ioannina 
\item Michael Plexousakis, Assistant Professor, Department of Mathematics \& Applied Mathematics, University of Crete
\item Nikolaos G. Tzanakis, Professor, Department of Mathematics \& Applied Mathematics, University of Crete
\end{itemize}}

\cleardoublepage

\vspace*{0.2\textheight}

\noindent\enquote{\itshape A single idea, if it is right, saves us the labor of an infinity of experiences.}\bigbreak

\hfill Jacques Maritain


\begin{abstract}
\addchaptertocentry{\abstractname} 
  In this thesis we study one of the fundamental predicates required for
  the construction of the 3D Apollonius diagram (also known as the 3D
  Additively Weighted Voronoi diagram), namely the \conflict predicate:
  given five sites $S_i, S_j,S_k,S_l,S_m$ that define an edge
  $\edge{ijklm}$ in the 3D Apollonius diagram, and a sixth query site
  $S_q$, the predicate determines the portion of $\edge{ijklm}$
  that will disappear in the Apollonius diagram of the six sites due to
  the insertion of $S_q$.

  Our focus is on the algorithmic analysis of the predicate with the aim
  to minimize its algebraic degree. We decompose the main predicate into
  sub-predicates, which are then evaluated with the aid of
  additional primitive operations. We show that the maximum algebraic degree
  required to answer any of the sub-predicates and primitives, and, thus,
  our main predicate is 10 in non-degenerate configurations when the 
  trisector is of Hausdorff dimension 1. We also prove that all subpredicates developed can be 
  evaluated using 10 or 8-degree demanding operations for degenerate input 
  for these trisector types, depending on whether they require the 
  evaluation of an intermediate \insphere predicate or not. 

  Among the tools we use is the 3D inversion transformation 
  and the so-called qualitative symbolic perturbation scheme.
  Most of our analysis is carried out in the inverted space, 
  which is where our geometric observations and analysis is captured 
  in algebraic terms.

  \vfill


\end{abstract}


\begin{greekabstract}
\addchaptertocentry{Greek Abstract} 

\textgreek{ Στην εργασία αυτή μελετάμε ένα από τα κεντρικότερα 
κατηγορήματα το οποίο απαιτείται για την κατασκευή του 3-Διάστατου 
Απολλώνιου Διαγράμματος (γνωστό και ως \textlatin{Voronoi} διάγραμμα βεβαρυμένων σημείων), το επονομαζόμενο κατηγόρημα \conflict: 
δεδομένων 5 σφαιρών $S_i, S_j,S_k,S_l,S_m$ τα οποία ορίζουν μια 
ακμή $\edge{ijklm}$ στο Απολλώνιο διάγραμμα και μίας έκτης 
σφαίρας $S_q$, το κατηγόρημα αποφαίνεται ποιο υποσύνολο της 
$\edge{ijklm}$ θα πάψει να υπάρχει ως ακμή στο στο Απολλώνιο διάγραμμα
των 6 σφαιρών. 

Το κύριο μέλημά μας είναι η αλγοριθμική ανάλυση του κατηγορήματος αυτού, έχοντας ως βασικό στόχο την ελαχιστοποίηση του αλγεβρικού του βαθμού. 
Αρχικά αποσυνθέτουμε το βασικό κατηγόρημα σε υποκατηγορήματα, τα οποία 
βασίζονται με τη σειρά τους σε πιο βασικούς γεωμετρικοαλγεβρικούς ελέγχους. 
Αποδεικνύουμε ότι το σύνολο των υποκατηγορημάτων και άρα και το κεντρικό κατηγόρημα απαιτεί υπολογισμούς αλγεβρικού βαθμού το πολύ 10 για να απαντηθεί για μη εκφυλισμένες εισόδους. Επίσης, για του ίδιου τύπου τριχοτόμους, αποδεικνύουμε ότι όλα τα υποκατηγορήματα που έχουμε 
σχεδιάσει απαιτούν μέγιστο αλγεβρικό βαθμό 10 ή 8, εάν απαιτούν τον
ενδιάμεσο υπολογισμό του κατηγορήματος \textlatin{InSphere} ή όχι.

Ανάμεσα στα εργαλεία που χρησιμοποιούμε είναι ο μετασχηματισμός 
αντιστροφής και η συμβολική διαταραχή. Το 
μεγαλύτερο μέρος της ανάλυσής μας γίνεται στον αντεστραμμένο χώρο, 
όπου οι γεωμετρικές μας παρατηρήσεις μεταφράζονται σε αλγεβρικούς 
όρους. }

\vfill


\textgreek{\textbf{Λέξεις και φράσεις κλειδιά}}.
\textgreek{Υπολογιστική Γεωμετρία, αλγεβρικοί υπολογισμοί, γεω\-με\-τρικά κατηγορήματα, Ευκλείδειο Απολλώνιο Διάγραμμα, κατηγόρημα \textlatin{EdgeConflict}, συμβολική διαταραχή.} 
 
\end{greekabstract}


\begin{acknowledgements}
\addchaptertocentry{\acknowledgementname} 

The author was fully supported by Onassis Foundation via a scholarship.
\end{acknowledgements}


\tableofcontents 

\listoffigures 

\listoftables 

\dedicatory{Dedicated to those who believed in me\\ and, above all, Thanases} 


\mainmatter 

\pagestyle{thesis} 



\chapter{Preliminaries}

\section{Introduction}

  Voronoi diagrams have been among the most studied 
  structures in computational geometry since their 
  inception \cite{green1978computing, brown1979voronoi, 
  aurenhammer2000voronoi, guibas1990randomized}, due to their numerous
  applications, including motion planning and collision detection,
  communication networks, graphics, and growth of microorganisms 
  in biology.

  Despite being a central topic in research for many 
  years, generalized Voronoi diagrams, and especially the Voronoi 
  diagram of spheres (also known as the 3D Apollonius diagram) 
  have not been explored sufficiently \cite{kim2006region}; this is also pointed out by Aurenhammer et al. \cite{aurenhammer2013voronoi}. 
  Moreover, due to 
  recent scientific discoveries in biology and chemistry, 3D Apollonius 
  diagrams are becoming increasingly important for representing 
  and analysing the molecular 3D structure and surface \cite{joonghyun2005computation} or 
  the structure of the protein \cite{kim2005protein}.

  The methods used to calculate the Apollonius diagram usually rely 
  on the construction of a different diagram altogether. 
  Some methods include the intersection of cones \cite{aurenhammer1987power} with the lifted power diagram and lower envelope calculations \cite{WillThesis, will1998fast, hanniel2008computing}. 
  Boissonnat et al. use the convex hull to describe its construction 
  \cite{boissonnat2003combinatorial, boissonnat2005convex}. 
  Aurenhammer's lifting method has also been implemented for two dimensions
  \cite{anton2002exact}. 
  Karavelas and Yvinec \cite{karavelas2002dynamic} create the 2­d Apollonius diagram from its dual, using the predicates developed in \cite{Emiris2006Predicates}. In \cite{Emiris2006Predicates}, it is also reported  that the Apollonius diagram can be obtained as a 
  concrete case of the abstract Voronoi diagrams of Klein et al. 
  \cite{klein1992randomized}.

  Kim et al. made a major research contribution in the 
  domain of the Voronoi diagrams of spheres. Their work provides many new algorithms related to the Voronoi diagrams including the computation of three-dimensional Voronoi diagrams 
  \cite{kim2010calculating,kim2005voronoi}, Euclidean Voronoi diagram of 3D balls and its
  computation via tracing edges \cite{kim2005euclidean} and the Euclidean Voronoi diagrams of 3D spheres and applications to protein structure analysis \cite{kim2005protein,mgos}.


  Hanniel and Elber \cite{hanniel2008computing} provide an 
  algorithm for computation of the Voronoi diagrams for planes, 
  spheres and cylinders in $\RR^3$. Their algorithm relies on 
  computing the lower envelope of the bisector surfaces similar 
  to the algorithm of Will \cite{will1998fast}. However, none of 
  the current research efforts provide an exact method 
  for computing the Apollonius diagram (or its dual Delaunay graph) 
  of spheres. It is also true that very few works study the 
  \emph{exact} Voronoi diagram (or its dual
  Delaunay graph) for curved objects and relative predicates or invariants, such 
  as \cite{Emiris2006Predicates} for circles, \cite{emiris2009exact} for convex pseudo-circles
  and \cite{castro2015invariants} for $n$-dimensional spheres.

  Various  algorithms and implementations that construct 
  the Apollonius diagram of spheres (or its dual graph) can be found
  in the current bibliography \cite{mgos,voroplusplus,dupuis2004voro3d,olechnovivc2010voroprot,cazals2010revisiting,cazals2006revisiting}. 
  These approximative algorithms 
  however are not exact as numerical approximation of 
  roots of high degree polynomials is
  required throught their execution. 
  The propagation of the errors can be critical, 
  especially if the final output involves approximate intermediary 
  computations or when degenerate cases arise. Regarding degeneracies, 
  most of these implementations do not handle them separately or efficiently 
  if not at all.

  In this thesis, we are inspired by the the approach presented by 
  Emiris and Karavelas in \cite{Emiris2006Predicates} for the 
  exact evaluation of the 2D Apollonius diagram. 
  In order to extend their work for the Apollonius diagram for 3D spheres, we 
  develop equivalent predicates as the ones presented in their paper 
  for the 2D case. Our ultimate goal is to analyze the most degree demanding
  predicate, called the \conflict predicate: 
  given five sites $S_i, S_j,S_k,S_l,S_m$ that define a finite edge
  $\edge{ijklm}$ in the 3D Apollonius diagram, and a sixth query site
  $S_q$, the predicate determines the portion of $\edge{ijklm}$
  that will disappear in the Apollonius diagram of the six sites due to
  the insertion of $S_q$.

  A description of how these predicates could be used in the scope of a 
  randomized incremental algorithm that constructs the Apollonius diagram 
  is provided in \cite[Section~3]{Emiris2006Predicates} and 
  \cite{karavelas2002dynamic} for the 2D case. However, these 
  procedure could easily be adapted for the 3D case, with minor 
  modifications. The idea behind the algorithm is that one could 
  construct the Voronoi diagram $\mathcal{VD}(\Sigma)$ of a set 
  $\Sigma$ by initially constructing the Voronoi diagram 
  $\mathcal{VD}(\Sigma')$ of a subset 
  $\Sigma'=\{S_i, S_j,S_k,S_l,S_m\}$ of $\Sigma$. 
  Then, for a site $S_q\in\Sigma\backslash\Sigma'$, update 
  $\mathcal{VD}(\Sigma')$ to $\mathcal{VD}(\Sigma'\cup\{S_q\})$ and 
  set $\Sigma'\leftarrow \Sigma'\cup\{S_q\}$. If this process is 
  repeated until $\Sigma\backslash\Sigma'$ is empty, \ie, $Sigma'$ 
  becomes $\Sigma$, then we will have obtain $\mathcal{VD}(\Sigma)$. 
  The difficult task of updating the Voronoi diagram requires, among 
  other operations, to examine if a portion of a valid Voronoi edge
  $\edge{ijklm}$ in $\mathcal{VD}(\Sigma')$ will disappear in 
  $\Sigma'\leftarrow \Sigma'\cup\{S_q\}$. The last task is accomplished
  via the call of the \conflict predicate and it is evident that 
  multiple calls of the predicate are demanded even after the
  insertion of a single site $S_q$. It is therefore crucial that 
  an efficient design of this predicate would greatly impact the 
  overall efficiency of the incremental algorithm that
  constructs $\mathcal{VD}(\Sigma')$. 

  Below, we provide a list of the major predicates that we need 
  to incrementally construct the 3D Apollonius diagram,
  similarly with the 2D case (cf. \cite{Emiris2006Predicates,karavelas2002dynamic}). The indices of the input sites in the 
  following predicates are indicative. 
  \begin{itemize}
  \item {\sc{}NearestNeighbor}$(S_a)$, which computes the sphere of 
  $\Sigma'$ closest to $S_a$. This predicates involve finding the 
  minimum of $d(S_q,S_n)$ for all $S_n\in\Sigma'$, which is a 4 
  degree-demanding operation. 
  \item {\sc{}IsHidden}$(S_a,S_b)$, which computes if $S_a$ lies 
  inside $S_b$ and therefore the Voronoi region of $S_a$ lies inside 
  the region of $S_b$. This predicate requires operations of degree 2 
  as it corresponds to the sign of 
  $d(S_a,S_b)+2r_a=\sqrt{(x_a-x_b)^2+(y_a-y_b)^2+(z_a-z_b)^2}+r_a-r_b$.
  \item $\text{\conflict}(S_i,S_j,S_k,S_l,S_m,S_q)$, 
  which is the main predicate studied in this thesis and is fully 
  analyzed for non-degenerate inputs in the case where 
  the trisector $\tri{ijk}$ is either hyperbolic or elliptic (see 
  Sections~\ref{sec:non_degenerate_hyperbolic} and ~\ref{sec:non_degenerate_elliptic}). The subpredicates required to 
  answer the \conflict predicate for both of these trisector types,
  except the \insphere (also referred to as \vconflict in the 
  bibliography), are also studied for degenerate inputs in Chapter~\ref{sec:degenerate_hyperbolic}.
  \item $\text{\infrconflict}(S_i,S_j,S_k,S_l,S_q)$, 
  which computes the portion of a semi-finite edge $e$ that no longer 
  remains in the Voronoi diagram of the first four sites, after the 
  insertion of the fifth. This edge $e$ lies on the trisector 
  $\tri{ijk}$ and is bounded on the ``left'' by the Apollonius 
  vertex $v_{ijkl}$. This predicate can be called only in the 
  case where the trisector is infinite hence either of hyperbolic or parabolic type. It has been analyzed in Section~\ref{sec:the_edgeconflict_predicate_for_infinite_hyperbolic_edges} for
  non-degenerate configurations on hyperbolic trisectors.
  \item $\text{\inflconflict}(S_i,S_j,S_k,S_m,S_q)$, which is 
  the symmetric predicate of the previous one and 
  computes the portion of a semi-finite edge $e$ that no longer 
  remains in the Voronoi diagram of the first four sites, after the 
  insertion of the fifth.
  This edge $e$ lies on the trisector 
  $\tri{ijk}$ and is bounded on the ``right'' by the Apollonius 
  vertex $v_{ikjm}$. The predicates analysis for
  non-degenerate configurations on hyperbolic trisectors can
  be found in Section~\ref{sec:the_edgeconflict_predicate_for_infinite_hyperbolic_edges}. 
  \item {\sc{}InCone}$(S_a,S_b,S_c)$, which returns the relative 
  position of $S_c$ against the semi-cone defined by $S_a$ and $S_b$. 
  This predicate's detailed description and analysis can be found 
  in Sections~\ref{ssub:the_incone_and_tritype_predicates} and 
  \ref{sub:the_incone_predicate_analysis}. The predicate's analysis 
  for degenerate input is provided in Section~\ref{sub:the_perturbed_incone_predicate}. Via the \incone predicate, 
  we can determine the existence of disconnected components of the 
  skeleton of the 3D Apollonius diagram, e.g. when a smaller sphere
  is inserted inside the convex hull of two others.
  \item {\sc{}FaceConflict}$(S_a,S_q)$ which computes if there are 
  points of the Apollonius face of $S_a$ in $\mathcal{VD}(\Sigma)$ that 
  no longer are part of an Apollonius face in 
  $\mathcal{VD}(\Sigma\cup\{S_q\})$ due to the insertion of $S_q$; 
  we shall call that these points are \emph{in conflict} with $S_q$. 
  It holds that if $S_q$ is not hidden, then a subset of the 
  Apollonius face of it's nearest neighbor site, $NN(S_q)\in\Sigma$ 
  will be in conflict with $S_q$. If this subset contains points that 
  belong to Apollonius edges, then we can call the \conflict predicate 
  to detect the portion of the edge that is in conflict. However, it 
  does not hold that there are always points of Apollonius edges that 
  are in conflict, \eg, if a small sphere is inserted in the 
  convex hull of two larger spheres. There has not been any references 
  in the current bibliography on how the {\sc{}FaceConflict} predicate
  could be designed and/or implemented in the 3D Apollonius diagram. 
  It is however apparent that it will have to include calling the 
  \incone and/or \conflict as subpredicates.
  \end{itemize}
 
  In this thesis, we describe and analyze how all of the  
  predicates described above can be answered, except for 
  the {\sc{}FaceConflict}. 
  The development of these predicates along with the necessary subpredicates and primitives  was made taking into consideration the 
  modern shift of predicate design towards lower level algorithmic issues.
  Specifically, a critical 
  factor that influenced our design was our goal to minimize the 
  algebraic degree of the tested quantities (in terms of the 
  input parameters) during a predicate evaluation. Such a 
  minimization problem has become a main concern that 
  influences algorithm design especially in geometric predicates, where 
  zero tolerance in all intermediate computations is needed to obtain an 
  exact result \cite{devillers2002algebraic, geismann2001computing,berberich2002computational, wein2002high, wolpert2003jacobi} .

  Our main contribution in the research area is the development 
  of a list of subpredicates that were not analyzed, either 
  explicitly or implicitly, in the current bibliography and can 
  be used within the scope of an incremental algorithm that 
  constructs the 3D Apollonius diagram of a set of spheres.
  Our most outstanding result is the fact that all subpredicates 
  presented in this thesis along with the 
  \conflict predicate require at most 10-fold degree demanding 
  operation (with respect to the input quantities) in non-degenerate 
  configurations. 
  This is quite a low bound on the required degree
  since the equivalent \conflict predicate in the 2D Apollonius 
  diagram requires 6-fold operations 
  \cite{millman2007degeneracy} for non-degenerate inputs. Note that 
  our approach of resolving the \conflict predicate can also be 
  applied for the 2D case to yield similar algebraic degrees.

  Beside the implementation of these predicates under the
  no-degeneracies assumption, we also provide a way of
  resolving the predicates for degenerate inputs 
  using the so-called qualitative symbolical perturbation technique 
  (see Section~\ref{sub:qsp_introduction} below). 
  For every subpredicate appearing in the evaluation of the 
  \conflict predicate except for the \insphere predicate (also known as the \vconflict predicate), we prove that the maximum 
  algebraic degree required to evaluate it is 8, regardless of 
  the existence of degeneracies. Therefore,
  we have proven that the algebraic degree of deciding the \conflict predicate for degenerate inputs is $\max{d,8}$, where $d$ is the
  respective algebraic cost of \vconflict. Moreover, the tools presented
  in this thesis and especially the observations made in the 
  inverted space, suggest that the existing result $d=28$ 
  \cite{devillers2017qualitative}, provided by the author, can be as 
  low as 10. 

  \section{Introduction to Qualitative Symbolic Perturbation} 
  \label{sub:qsp_introduction}
  
    In the field of computational geometry, the first non-primitive
    predicates that were designed usually ignored degenerate configurations. 
    The sentences ``no four points are cocircular'' or 
    ``no four spheres are tagent to a common plane'' are common in 
    many papers; indeed, such assumptions can be found in almost all
    sections of Chapters~\ref{sec:non_degenerate_hyperbolic} and 
    ~\ref{sec:non_degenerate_elliptic}. The non-degeneracy hypothesis
    allows simpler design and analysis of algorithms as special 
    cases are not considered at all. 

    However, degenerate situations can not be ignored as they actually 
    occur in practice \cite{edelsbrunner1990simulation,yap1990symbolic,yap1990geometric}. Such degeneracies can be handled 
    in various ways, usually by applying a \emph{perturbation scheme}. 
    
    Before considering the evaluation of ``perturbed'' predicates, 
    let us first describe what happens during the call of a predicate. 
    Assume we have to evaluate the outcome of a 
    predicate $G(\textbf{x})$ for some input $\textbf{x}_0$. 
    In most such predicates, as the ones described in  
    Chapters~\ref{sec:non_degenerate_hyperbolic} and ~\ref{sec:non_degenerate_elliptic}, the outcome is decided by looking at 
    signs of various \emph{geometric subpredicates} for the same input. 
    Each of these subpredicates can be viewed as the sign of a polynomial 
    $P(\textbf{x})$, where $\textbf{x}\in\RR^{\mu}$
    includes the coordinates of all input sites. The original 
    algorithm that evaluates $G(\textbf{x})$ assumes that all these 
    signs are never 0,, \ie, $\textbf{x}_0$ is a no-degenerate 
    input. 

    Otherwise, a degenerate configuration $G(\textbf{x}_0)$ 
    will result in some $P(\textbf{x}_0)$ being 0. 
    We handle these cases by 
    exchanging the input $\textbf{x}_0$ with a proper function 
    $\pi(\textbf{x}_0,\epsilon)$; this is called a
    \emph{symbolic perturbation}. 
    The required properties of $\pi(\textbf{x}_0,\epsilon)$ is that 
    it has to be a continuous function of the parameter $\epsilon$, 
    $\pi(\textbf{x}_0,0)=\textbf{x}_0$ and that 
    $x=\pi(\textbf{x}_0,\epsilon)$ is not a degenerate input for 
    $G$ for sufficiently small positive values of $\epsilon$. 
    Under these assumptions, we can define the predicate 
    $G(\textbf{x}_0)$ as the limit of 
    $G(\textbf{x}_0,\epsilon)$ when $\epsilon\to 0^+$, denoted as 
    $G^{\epsilon}(\textbf{x}_0)$.

    In some works \cite{edelsbrunner1990simulation,seidel1998nature,
    emiris1995general,alliez2000removing,devillers2011perturbations,emiris1997efficient}, 
    the practical evaluation of the 
    perturbed predicate $G^{\epsilon}(\textbf{x}_0)$, involves expressing
    the sign of $P(\pi(\textbf{x}_0,\epsilon))$ as a polynomial 
    of $\epsilon$ and the coordinates of $\textbf{x}_0$. The evaluation 
    of the perturbed predicate $G^{\epsilon}(\textbf{x}_0)$ ultimately 
    amounts to determining the sign of the non-vanishing limit 
    $\lim_{\epsilon\to 0^+}P(\pi(\textbf{x}_0,\epsilon))$, for all 
    predicates $P$ such that $P(\textbf{x}_0)=0$. To decide the sign of 
    this complex limit, the polynomial $P(\pi(\textbf{x}_0,\epsilon))$ 
    is rewritten
    as a polynomial in $\epsilon$ and its monomials are ordered in terms
    of increasing degree. The first degree is actually $P(\textbf{x}_0)$ 
    while the rest are evaluated in increasing degrees in $\epsilon$ 
    until a non-vanishing coefficient is found. The sign of this 
    coefficient is then returned as the sign of the limit. 

    A different approach to practically evaluate 
    $G^{\epsilon}(\textbf{x}_0)$ is presented in 
    \cite{devillers2017qualitative} by Devillers, Karavelas and 
    Teillaud. Their approach, called \emph{Qualitative Symbolic 
    Perturbation}, is also the one we adopt in the
    analysis of Chapter~\ref{sec:degenerate_hyperbolic}. 
    In their paper, perturbed predicates are not resolved  
    algebraicly by considering non-vanishing terms as before. Instead, 
    the evaluation of the sign of $P(\pi(\textbf{x}_0,\epsilon))$ 
    when $\epsilon\to 0^+$ is considered in geometric terms. Such
    strategy can be followed as $P(\textbf{x}_0)$ itself expresses a 
    geometric property of the input configuration $\textbf{x}_0$; 
    we simply have to determine how this property changes as 
    the $\textbf{x}_0$ infinitesimally changes to 
    $\pi(\textbf{x}_0,\epsilon)$.

    As presented in \cite{devillers2017qualitative}, to 
    resolve a degenerate geometric predicate $P$ with input the spheres
    $S_n$, for $n\in I$, we first index the sites using the so-called
    \emph{max-weight ordering} that assigns a larger index to the site 
    with larger weight. To break ties between sites with the same 
    weights, we use the lexicographic ordering of their centers: 
    among two sites with the same weight, the site whose center is 
    smaller lexicographically than the other is assigned a smaller 
    max-weight order. If the site site $S_n$ has larger index than $S_m$,
    we will denote it by $n>m$.

    After the input sites are ordered, the predicate $P$ for a 
    degenerate input is resolved as follows. Initially, we symbolically 
    perturbe the site with the largest index \ie and consider the outcome
    of $P$ after the site is infinitesimally inflated, while the others
    remain can be considered as fixed. 
    If the configuration still remains degenerate, we enlarge the 
    second largest index and so on until the resulting configuration
    is non-degenerate, in which case the predicate is resolved. 
    Moreover, if more than one site has to be perturbed, each 
    subsequent site is enlarged by an amount that is smaller than the previous ones; the benefits 
    of this approach are presented in the respective paper 
    \cite{devillers2017qualitative}.

  This thesis is organised as follows. 
  In Chapters~\ref{sec:non_degenerate_hyperbolic} and 
  \ref{sec:non_degenerate_elliptic}, we present 
  an algorithm that decides the \conflict 
  predicate assuming a non-degenerate input and that the 
  supporting trisector of the Voronoi edge is \emph{hyperbolic} or 
  \emph{elliptic} respectively. In 
  Chapter~\ref{sec:degenerate_hyperbolic}, a description of the way
  we resolve degeneracies for all subpredicates is provided. Finally,
  in Section~\ref{sec:conclusion_and_future_progress}, we conclude 
  the thesis.

 \section{Basic Definitions} 
  \label{sub:definitions}
  
  Let $\mathcal{S}$ be a set of closed spheres $S_n$ (also referred as 
  \emph{sites}) 
  in $\EE^3$, with centers $C_n=(x_n,y_n,z_n)$ and radii $r_n$. 
  Define the Euclidean distance
  $\delta(p,S)$ between a point $p\in\EE^3$ and a sphere $S=\{C,r\}$
  as $\delta(p,S)=\|p-C\|-r$, where $\|\cdot\|$ stands for the
  Euclidean norm. The \emph{Apollonius diagram}
  is then defined as the subdivision of the space induced by 
  assigning each point $p\in\EE^3$ to its nearest neighbor with 
  respect to the distance function $\delta(\cdot,\cdot)$. 

  For each $i\neq j$, let 
  $H_{ij}=\{y\in\EE^3 :\delta(y,S_i)\leq \delta(y,S_j)\}$. Then 
  the (closed) \emph{Apollonius cell} $V_i$ of $S_i$ is defined to be 
  $V_i=\cap_{i\neq j} H_{ij}$. The set of points that belong to exactly 
  two Apollonius cells are called the \emph{Apollonius faces}, 
  whereas the 1-dimensional connected intersections of face closures 
  are called \emph{Apollonius edges}. Points that correspond to 
  intersections of Apollonius edges and therefore belong to more than
  three Apollonius cells are called \emph{Apollonius vertices}; 
  the\emph{ Apollonius diagram} $\mathcal{VD}(\mathcal{S})$ of 
  $\mathcal{S}$ is defined as the collection of the Apollonius cells, 
  faces, edges and vertices. 

  An Apollonius vertex $v$ is a point that is equidistant to 4 or more 
  sites. If we denote 4 of these sites by $S_i,S_j,S_k$ and $S_l$ and 
  let $r$ be the common distance of $v$ from each of them, then the 
  sphere centered at $v$ with radius $r$ must be externally tagent 
  to all four spheres. This cotagent sphere is called an
  \emph{external Apollonius sphere} of the sites $S_i,S_j,S_k$ and $S_l$.
  If $T_n$ denotes the point of tangency of the Apollonius
  sphere and $S_n$, for $n\in\{i,j,k,l\}$, then the tetrahedron 
  $T_iT_jT_kT_l$ can be
  either positively or negatively oriented, or even flat
  \cite{devillers2017qualitative}. If $T_iT_jT_kT_l$ is positively 
  oriented then the Apollonius vertex $v$ will be denoted as $v_{ijkl}$
  whereas if $T_iT_jT_kT_l$ is negatively oriented it will be denoted 
  as $v_{ikjl}$. Observe that a cyclic permutation of the indices 
  does not alter the Apollonius vertex choice. 

  The \emph{trisector} $\tri{ijk}$ of three different sites $S_i,S_j$
  and $S_k$ is the locus of points that are equidistant from  the
  three sites. In the absense of degeneracies its Hausdorff dimension
  is 1, and it is either (a branch of) a hyperbola, a line, an ellipse,
  a circle, or a parabola \cite{WillThesis}. We shall say that 
  the trisector is
  \begin{itemize}
  \item \emph{hyperbolic}, if it is a branch of a hyperbola or a line,
  \item \emph{elliptic}, if it is an ellipse or a circle, and 
  \item \emph{parabolic}, if it is a parabola.
  \end{itemize}

  In this thesis, we mainly focus on the cases of hyperbolic and 
  elliptic trisectors. The parabolic trisector can be viewed as a
  degenerate trisector as a symbolic perturbation of the sites 
  $S_i,S_j$ and $S_k$ would result in $\tri{ijk}$ becoming 
  hyperbolic or elliptic. Therefore, analysis of all predicates 
  on parabolic trisectors is reduced to the respective predicates 
  for either hyperbolic or elliptic trisectors.

  A finite Apollonius edge of a 3D Apollonius diagram is 
  denoted by $\edge{ijklm}$ if the edge lies on the trisector 
  $\tri{ijk}$ of the sites $S_i,S_j$ and $S_k$ different sites it's
  endpoints are $v_{ijkl}$ and $v_{ikjm}$.

  
 \section{Inversion} 
  \label{sub:inversion}
  The 3-dimensional \emph{inversion transformation} is a
  mapping from $\RR^3$ to $\RR^3$ that maps a point $z\in\RR^3$ to the
  point $W(z)=(z-z_0)/\|z-z_0\|^2$. The point $z_0$ is called the
  \emph{pole of inversion}.
  Inversion maps spheres that do not pass through the pole to
  spheres, spheres that pass through the pole to planes and 
  planes that pass through the pole to planes.
  
  In the Apollonius diagram context we call \zspace the space where
  the sites live. Since the Apollonius diagram does not change when we
  add to the radii of all spheres the same quantity, we will, most of
  the times, reduce the radii of the spheres  $S_i,S_j,S_k,S_l,S_m$ and $S_q$
  by the radius of a sphere $S_I$, where $I\in\{i,j,k,l,m,q\}$. 
  The new spheres have
  obviously the same centers, whereas their radii 
  become $r^\star_n=r_n-r_I$, $n\in\{i,j,k,l,m,q\}$. For
  convience, we call the image space of this radius-reducing
  transformation the $\mathcal{Z}^\star$-space. We may then apply
  inversion, with $C_I$ as the pole, to get a new set of spheres or
  planes; we call \wspace the space where the radius-reduced, inverted
  sites live.

  Note that we can safely assume that none of the 
  sites $S_i,S_j,S_k,S_l,S_m,S_q$ are contained inside another. Indeed, 
  in a Voronoi diagram $\mathcal{VD}(\mathcal{S})$, if 
  $S_a,S_b\in\mathcal{S}$ and $S_a\subset S_b$ then $S_a$ has an empty
  Voronoi cell, therefore the deletion of $S_a$ from $\mathcal{S}$ 
  does not alter the Voronoi diagram. 

  The non-inclusion assumption also implies that 
  the image of the sphere $S_n$ in \wspace,
  for $n\in\{i,j,k,l,m,n,q\}\backslash\{I\}$,
  is a sphere $\inv{S}_n$, centered at 
  $\inv{C}_n=(u_n,v_n,w_n)$ with radius $\rho_n$, where
  \begin{alignat}{4}
  u_n &= \dfrac{\inv{x}_n}{\inv{p}_n}, &\quad 
  v_n &= \dfrac{\inv{y}_n}{\inv{p}_n}, &\quad
  w_n &= \dfrac{\inv{z}_n}{\inv{p}_n}, &\quad
  \rho_n &= \dfrac{\inv{r}_n}{\inv{p}_n},\\
  \inv{x}_n &= x_n-x_I, &\quad
  \inv{y}_n &= y_n-y_I, &\quad
  \inv{z}_n &= z_n-z_I, &\quad
  \inv{r}_n &= r_n-r_I,
  \end{alignat}

  \noindent
  and $\inv{p}_n=(\inv{x}_n)^2+(\inv{y}_n)^2+(\inv{z}_n)^2-(\inv{r}_n)^2>0$.

  \textbf{Proof.} 
  Let's start with the equation of the sphere $S_n$ in \zspace:
  \begin{equation}
  (x-x_n)^2 + (y-y_n)^2 + (z-z_n)^2 = r_n^2  
  \end{equation}
  The image of $S_n$ in $\mathcal{Z^\star}$-space is the 
  sphere $S_n$ after its radius is reduced by $r_I$ and 
  therefore it's equation is:
  \begin{equation}
  (x-x_n)^2 + (y-y_n)^2 + (z-z_n)^2 = (r_n - r_I)^2 = (r_n^*)^2.
  \end{equation}
  This can be rewritten as 
  \begin{equation}\label{eq:inversion}
  (x-x_I - x_n^*)^2 + (y-y_I-y_n^*)^2 + (z-z_I-z_n^*)^2 = (r_n^*)^2.
  \end{equation}
  Now, we apply the inversion transformation using the point
  $(x_I,y_I,z_I)$ as the inversion pole and therefore a point 
  $(x,y,z)$ on $S_n$ is mapped to $(u,v,w)$ where
  \begin{equation}
  u = (x-x_I) / D,\ \ v = (y-y_I) / D,\ \ w = (z-z_I) / D,\ \ 
  \text{and } D = (x-x_I)^2 + (y-y_I)^2 + (z-z_I)^2.
  \end{equation}

  Let $E:=u^2 + v^2 + w^2 = 1/D$, it holds that 
  \begin{equation}
  x = uD+x_I, \  y = vD+y_I,\  \text{and }  z = wD+z_I.
  \end{equation}
  Substituting in \eqref{eq:inversion}, we get
  \begin{equation}
  (uD - x_n^*)^2 + (vD-y_n^*)^2 + (wD-z_n^*)^2 = (r_n^*)^2.
  \end{equation} 
  Multiplying by $E^2=1/D^2$ we get 
  \begin{equation}
  (u - Ex_n^*)^2 + (v-Ey_n^*)^2 + (w-Ez_n^*)^2 = (Er_n^*)^2.
  \end{equation} 
  Now we expand the squares and regroup terms:
  \begin{equation}
  u^2 + v^2 + w^2 - 2 u E x_n^* - 2 v E y_n^* - 2 w E z_n^* + E^2 [(x_n^*)^2 + (y_n^*)^2+(z_n^*)^2]= E^2 (r_n^*)^2.
  \end{equation}
  which can then be rewritten as (we use that $u^2 + v^2 + w^2 = E$)
  \begin{equation}
  E - 2 u E x_n^* - 2 v E y_n^* - 2 w E z_n^* + E^2 [(x_n^*)^2 + (y_n^*)^2+(z_n^*)^2]= E^2 (r_n^*)^2.
  \end{equation}. 
  Since $E$ is not zero, we get the equation
  \begin{equation}
  1 - 2 u x_n^* - 2 v y_n^* - 2 w z_n^* + E [(x_n^*)^2 + (y_n^*)^2+(z_n^*)^2-(r_n^*)^2]= 0 
  \end{equation}
  or equivalently 
  \begin{equation}
    1 - 2 u x_n^* - 2 v y_n^* - 2 w z_n^* + E p_n^*= 0.
  \end{equation}
  Since $p_n^*$ is strictrly positive due to the fact that $S_I$ is 
  not contained in $S_n$, we divide by $p_n^*$ and substitute 
  $E=u^2 + v^2 + w^2$:
  \begin{equation}
    u^2 - 2 u x_n^*/p_n^* +v^2 - 2 v y_n^*/p_n^* + w^2 - 2 w z_n^*/p_n^* = -1/p_n^*.
  \end{equation}
  After completing the squares, we get
  \begin{equation}\label{eq:inv_last}
  (u - x_n^*/p_n^*)^2 +(v-y_n^*/p_n^*)^2 +(w- z_n^*/p_n^*)^2 = 
  (x_n^*/p_n^*)^2+ (y_n^*/p_n^*)^2+ (z_n^*/p_n^*)^2-1/p_n^*.
  \end{equation}
  The right part of the equation can be rewritten as
  \begin{equation}
  (x_n^*/p_n^*)^2+ (y_n^*/p_n^*)^2+ (z_n^*/p_n^*)^2-1/p_n^*=
  \dfrac{(x_n^*)^2+ (y_n^*)^2+ (z_n^*)^2-p_n^*}{(p_n^*)^2} = 
  \dfrac{(r_n^*)^2}{(p_n^*)^2}.
  \end{equation}
  Therefore, \eqref{eq:inv_last} is written as
  \begin{equation}
  (u-u_n)^2+(v-v_n)^2+(z-z_n)^2=\rho_n^2,
  \end{equation}
  which is the equation of $\inv{S_n}$ in \wspace. \hfill $\square$

  We also define the quantities
  \begin{equation}
  \begin{gathered}
  D^{\pi\theta}_{\lambda\mu\nu} = 
  \begin{vmatrix}
  \pi_\lambda & \theta_\lambda & 1\\
  \pi_\mu & \theta_\mu & 1\\
  \pi_\nu & \theta_\nu & 1
  \end{vmatrix}, 
  \ \ \
  D^{\pi\theta\eta}_{\lambda\mu\nu} = 
  \begin{vmatrix}
  \pi_\lambda & \theta_\lambda & \eta_\lambda\\
  \pi_\mu & \theta_\mu & \eta_\mu\\
  \pi_\nu & \theta_\nu & \eta_\nu
  \end{vmatrix}, 
  \ \ \ 
  D^{\pi\theta\eta}_{\lambda\mu\nu\xi} = 
  \begin{vmatrix}
  \pi_\lambda & \theta_\lambda & \eta_\lambda & 1\\
  \pi_\mu & \theta_\mu & \eta_\mu & 1\\
  \pi_\nu & \theta_\nu & \eta_\nu & 1\\
  \pi_\xi & \theta_\xi & \eta_\xi & 1\\
  \end{vmatrix}, \\
  D^{\pi\theta\eta\zeta}_{\lambda\mu\nu\xi} = 
  \begin{vmatrix}
  \pi_\lambda & \theta_\lambda & \eta_\lambda & \zeta_\lambda\\
  \pi_\mu & \theta_\mu & \eta_\mu & \zeta_\mu\\
  \pi_\nu & \theta_\nu & \eta_\nu & \zeta_\nu\\
  \pi_\xi & \theta_\xi & \eta_\xi & \zeta_\xi\\
  \end{vmatrix},
  \end{gathered} 
  \end{equation}

  \noindent
  and
  \begin{equation}
  \begin{gathered}
  E^{\alpha\beta\gamma}_{\lambda\mu\nu} = 
  \begin{vmatrix}
  \inv{\alpha_\lambda} & \inv{\beta_\lambda} & \inv{\gamma_\lambda}\\
  \inv{\alpha_\mu} & \inv{\beta_\mu} & \inv{\gamma_\mu}\\
  \inv{\alpha_\nu} & \inv{\beta_\nu} & \inv{\gamma_\nu}
  \end{vmatrix}, \ \ \ 
  E^{\alpha\beta\gamma\delta}_{\lambda\mu\nu\xi} = 
  \begin{vmatrix}
  \inv{\alpha_\lambda} & \inv{\beta_\lambda} & \inv{\gamma_\lambda} & \inv{\delta_\lambda}\\
  \inv{\alpha_\mu} & \inv{\beta_\mu} & \inv{\gamma_\mu} & \inv{\delta_\mu}\\
  \inv{\alpha_\nu} & \inv{\beta_\nu} & \inv{\gamma_\nu} & \inv{\delta_\nu}\\
  \inv{\alpha_\xi} & \inv{\beta_\xi} & \inv{\gamma_\xi} & \inv{\delta_\xi}\\
  \end{vmatrix}, 
  \end{gathered}
  \end{equation}

  \noindent
  for $\pi,\theta,\eta,\zeta \in\{ x,y,z,r,u,v,w,\rho\}$, 
  $\alpha,\beta,\gamma,\delta \in\{x,y,z,r\}$ and 
  $\lambda,\mu,\nu,\xi\in\{i,j,k,l\}$.

  If it holds for all $\tau\in\{\lambda,\mu,\nu\}$  that 
  \begin{gather}
  \pi_\tau = \dfrac{\inv{\alpha}_\tau}{\inv{p}_\tau}, \ \ \
  \theta_\tau = \dfrac{\inv{\beta}_\tau}{\inv{p}_\tau},\ \ \
  \eta_\tau = \dfrac{\inv{\gamma}_\tau}{\inv{p}_\tau},
  \end{gather}

  \noindent
  for some $\pi,\theta,\eta \in \{u,v,w,\rho\}$ and 
  some $\alpha,\beta, \gamma \in\{x,y,z,r\}$, then it also holds that

  \begin{gather}
  D^{\pi\theta}_{\lambda\mu\nu} = \dfrac{1}{\inv{p_\lambda}\inv{p_\mu}
  \inv{p_\nu}}E^{\alpha\beta p}_{\lambda\mu\nu}, \ \ \
  D^{\pi\theta\eta}_{\lambda\mu\nu} = \dfrac{1}{\inv{p_\lambda}\inv{p_\mu}
  \inv{p_\nu}}E^{\alpha\beta\gamma}_{\lambda\mu\nu}.
  \end{gather}

  \section{Known tools} 
  \label{sec:known_tools}

  Let $\ov{v}=(v_1,v_2,v_3)$ and  $\ov{w}=(w_1,w_2,w_3)$ be two vectors
  of $\RR^3$. The inner product of these vectors is the quantity
  $\ov{v}\cdot\ov{w}:=v_1w_1+v_2w_2+v_3w_3$. It is known that the inner
  product of two vectors equals zero if and only if either vector is 
  the zero vector or the vectors are perpendicular. 

  The cross product of the vectors $\ov{v}$ and $\ov{w}$ is the vector
  $\ov{v}\times\ov{w}:=(v_2w_3-v_3w_2,-v_1w_3+v_3w_1,v_1w_2-v_1w_2)$. 
  It is known that the cross product of two vectors is the zero 
  vector if and only if either vector is the zero vector or the vectors 
  are parallel. 

  Given four points $K,L,M,N$ of $\RR^3$, the outcome of the  
  $\text{\orient}(K,L,M,N)$ predicate corresponds to the orientation 
  of the tetrahedron $KLMN$. Specifically, if the predicate returns 
  ``+'' or ``-'', then the tetrahedron $KLMN$ is positively or 
  negatively oriented, respectively. Otherwise, if the predicate returns
  ``0'', the tetrahedron is flat, \ie, the four points are coplanar.
  Note that the \orient predicate can be easily evaluated geometrically
  via the right-hand rule; if we orient our right hand with fingers 
  curled to follow the circular sequence $L,M,N$ then the 
  predicate returns ``+'' (resp., ``-'') iff our thumb points towards
  (resp., away from) $K$. Algebraicly, the \orient predicate can 
  be evaluated as

  \begin{equation}
  \text{\orient}(K,L,M,N)=
  \sgn\left(
  \begin{vmatrix}
  x_K & y_K & z_K & 1 \\
  x_L & y_L & z_L & 1 \\
  x_M & y_M & z_M & 1 \\
  x_N & y_N & z_N & 1
  \end{vmatrix}
  \right)=\sgn(D^{xyz}_{KLMN})
  \end{equation}
  \noindent
  where $P=(x_P,y_P,z_P)$, for $P\in\{K,L,M,N\}$.



\newpage


\chapter{Non-Degenerate Case Analysis for Hyperbolic Trisectors} 
 \label{sec:non_degenerate_hyperbolic}

  In this chapter, we introduce the subpredicates and the algorithm
  that are used to answer the \conflict predicate under the assumption 
  that the trisector 
  of the sites $S_i,S_j$ and $S_k$ is a branch of a hyperbola or a 
  straight line and that no degeneracies arise.
  The section is organized as follows. In Section~\ref{sub:orientation_hyperbolic_trisector} we provide a way of orienting 
  a \emph{hyperbolic} trisector and in Section~\ref{sub:voronoi_edges} 
  we make useful remarks regarding the Voronoi edges that 
  lie on such trisectors. In Section~\ref{sub:problem_outline}, we 
  provide the list of possible outcomes for the \conflict predicate 
  as well as the assumptions that are considered to hold 
  throughout this Chapter. 
  The description of several useful subpredicates takes
  place in Sections~\ref{sub:subpredicates}, whereas 
  in Section~\ref{sub:the_main_algorithm}, we provide the algorithm
  that combines them and decides the \conflict predicate. Lastly, 
  a detailed analysis for each of the subpredicates can be found in 
  Section~\ref{sec:algebraic_analysis} .

  \section{Orientation of a hyperbolic or linear trisector} 
  \label{sub:orientation_hyperbolic_trisector}
\begin{figure}[htbp]
   \centering
   \includegraphics[width=0.95\textwidth]{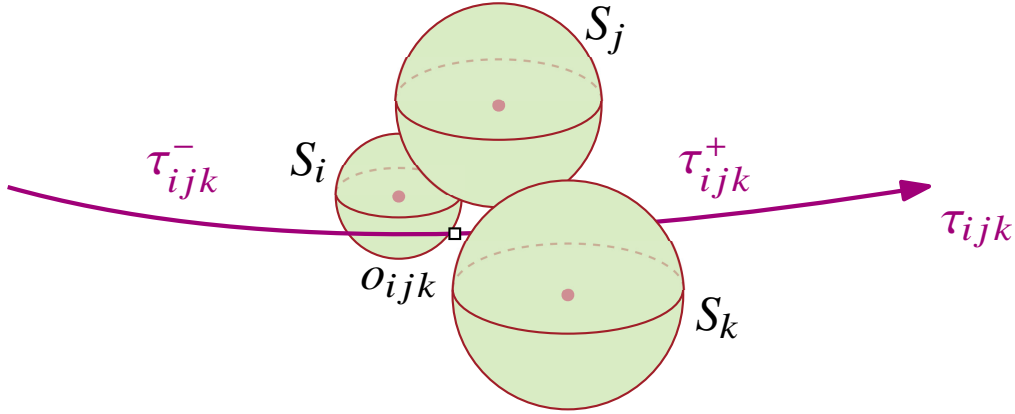}
   \caption[A hyperbolic trisector]{The case where the trisector $\tri{ijk}$ of the 
   spheres $S_i$, $S_j$ and $S_k$ is \emph{hyperbolic}. Notice the 
   orientation of $\tri{ijk}$ based on the ``right-hand rule''. 
   The point $o_{ijk}$ of the trisector, which is coplanar with the 
   centers $C_i$, $C_j$ and $C_k$, separates the $\tri{ijk}$ 
   into two semi-trisectors, $\tri{ijk}^-$ and $\tri{ijk}^+$.}
   \label{fig:hyperbolic_trisector}
  \end{figure}

  Under the assumption that the trisector $\tri{ijk}$ of the sites
  $S_i,S_j,S_k$ is a line or a hyperbola, the three centers
  $C_i,C_j,C_k$ cannot be collinear \cite{WillThesis}. 
  A natural way of orienting $\tri{ijk}$
  is accomplished via the well-known ``right-hand rule''; if we fold 
  our right hand to follow the centers $C_i,C_j$ and $C_k$ (in that order), 
  our thumb will be showing the positive ``end'' of $\tri{ijk}$ 
  (see Figure~\ref{fig:hyperbolic_trisector}).

  By orienting $\tri{ijk}$, we clearly define an ordering on the points of
  $\tri{ijk}$, which we denote by $\prec$.
  Let $\oo$ be the intersection of $\tri{ijk}$ and the plane 
  $\Pi_{ijk}$ going through the centers $C_i,C_j$ and $C_k$.
  We can now parametrize $\tri{ijk}$ as follows: if $\oo\prec p$
  then $\map{p}=\delta(p,S_i)-\delta(\oo,S_i)$; 
  otherwise $\map{p}=-(\delta(p,S_i)-\delta(\oo,S_i))$. 
  The function $\map{\cdot}$ is a 1-1 and onto mapping from
  $\tri{ijk}$ to $\RR$. Moreover, we define $\map{S}$, where 
  $S$ is an external tangent sphere to the sites $S_i,S_j$ and $S_k$,
  to be $\map{c}$, where $c\in\tri{ijk}$ is the center of $S$.

  We also use $\tri{ijk}^+$ (resp., $\tri{ijk}^-$) to denote 
  the positive (resp., negative) semi-trisector, \ie, the set of 
  points $p\in\tri{ijk}$ such that $\oo\prec p$ (resp., $p\prec\oo$).

 \section{Voronoi Edges on Hyperbolic Trisectors} 
  \label{sub:voronoi_edges}

  In order to better understand our initial problem, more insight 
  regarding the properties of a Voronoi diagram is required. 
  Let us look closer at 
  an edge $e_{ijklm}$ (we drop the subscript for convenience) 
  of $\mathcal{V}(\mathcal{S})$, 
  where $\mathcal{S}$ is a set of given sites that includes 
  $S_n$, for $n\in\{i,j,k,l,m\}$, and does not include $S_q$. 
  This edge $e$ lies on the trisector $\tri{ijk}$,  
  the locus of points that are equidistant to the sites
  $S_i,S_j$ and $S_k$. 
  In the scope of this chapter, 
  we assume that $\tri{ijk}$ is of Hausdorff dimension 1 and 
  is \emph{hyperbolic}, \ie, either (a branch of) a hyperbola or a line. 
  To ensure that the 
  spheres $S_i,S_j$ and $S_k$ meet this criteria, 
  the predicate \tritype$(S_i,S_j,S_k)$, described in 
  Section~\ref{ssub:the_incone_and_tritype_predicates} 
  and analyzed in Section~\ref{sub:the_tritype_predicate_analysis}, 
  must return ``hyperbolic''. 

  \begin{figure}[htbp]
  \centering
  \includegraphics[width=0.95\textwidth]{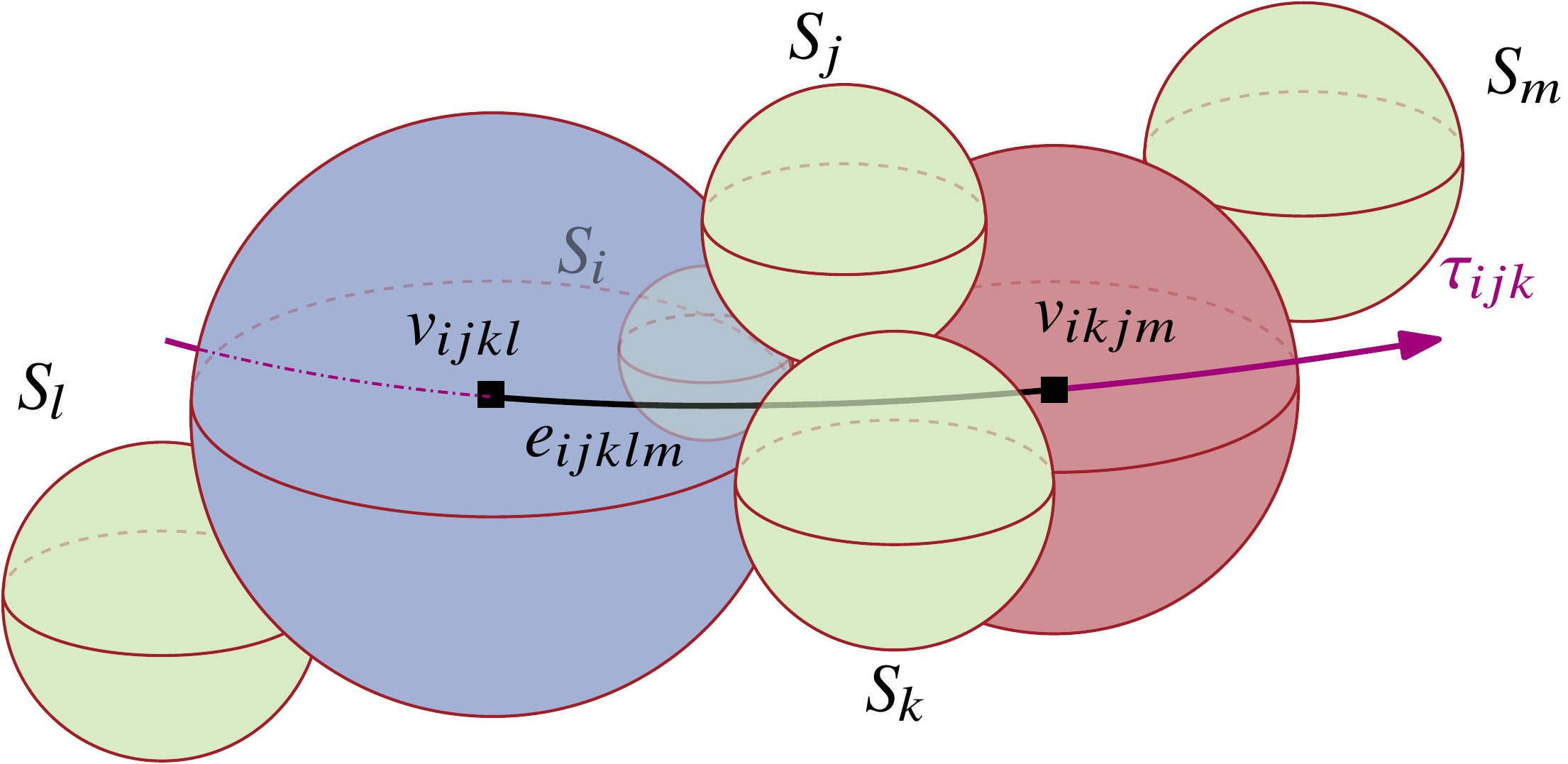}
  \caption[A finite edge on a hyperbolic trisector]{An edge $e_{ijklm}$
  that lies in a hyperbolic trisector $\tri{ijk}$. It's left 
  and right endpoints are  of the Apollonius vertices $v_{ijkl}$ 
  and $v_{ikjm}$, respectively. The blue and red sphere are 
  the Apollonius spheres centered at $v_{ijkl}$ 
  and $v_{ikjm}$, respectively.}
  \label{fig:edge_on_hyperbolic_trisector}
  \end{figure}

  We now focus on the edge $e$, \ie the open continuous subset of 
  $\tri{ijk}$ whose closure is bounded by the Voronoi vertices 
  $v_{ijkl}$ and $v_{ikjm}$. If $\tts{t}$ denotes the external 
  Apollonius sphere of the sites $S_i,S_j$ and $S_k$ that is 
  centered at $t$, the most crucial
  property of a point $t\in e$ is that $\tts{t}$ does not intersect 
  with any other site of $\mathcal{S}$. We call this the
  \emph{Empty Sphere Principle} since it is a property 
  that derives from the empty circle principle of a generic Voronoi 
  diagram and its basic properties.

  Notice that the original definition of the edge $e$ included the 
  fact that it was bounded by the Apollonius vertices $v_{ijkl}$ 
  and $v_{ikjm}$. 
  Using the Empty Sphere Principle, we can show that the former and
  latter vertice correspond to the left and right endpoint of the 
  edge, respectively (see Figure~\ref{fig:edge_on_hyperbolic_trisector}) ; this is equivalent to showing that 
  $v_{ijkl}\prec v_{ijkm}$ on the oriented trisector.

  We will now prove that the left endpoint is indeed 
  $v_{ijkl}$ and not $v_{ikjm}$. Assuming that $v_{ikjm}$ is the 
  left endpoint, we consider a point $t\in\tri{ijk}$
  such that initially $t\equiv v_{ikjm}$ and then move it 
  infinitesimally on the trisector towards its positive direction. 
  The  Apollonius sphere $\tts{t}$ was initially $\tts{v_{ikjm}}$ 
  and therefore tangent to $S_m$. Since $v_{ikjm}$
  was assumed to be the left endpoint of $e$, $t$ lies on $e$ after
  being moved and therefore the sphere $\tts{t}$ should not
  longer be tangent nor intersect $S_m$ due to the 
  Empty Sphere Principle. 
  However, the tetrahedron $T_iT_jT_kT_m$, where $T_n$ is the 
  tangency point of the spheres $\tts{v_{ikjm}}$ and $S_n$, 
  for $n\{i,j,k,l\}$, is negatively oriented by definition of the 
  Apollonius vertex $v_{ikjm}$ (see Section~\ref{sub:definitions}).
  Due to the tetrahedron's orientation and the movement of $t$ towards 
  the positive direction of $\tri{ijk}$, it must hold that 
  the sphere $\tts{t}$ contains $T_m$ and therefore intersects $S_m$, 
  yielding a contradiction. Consequently, we have proven that the left 
  point of $e$ is necessarily $v_{ijkl}$ and the 
  right endpoint is $v_{ikjm}$.


 \section{Problem Outline and Assertions} 
  \label{sub:problem_outline}
 
  For clarity reasons, we restate the \conflict predicate, highlighting 
  its input, output as well as the assertions we are making for the rest
  of this paper. 

  The \conflict predicate, one of the fundamental predicates required for
  the construction of the 3D Apollonius diagram (also known as the 3D
  Additively Weighted Voronoi diagram), takes as input five sites 
  $S_i, S_j,S_k,S_l$ and $S_m$ that define an edge $\edge{ijklm}$ in the 
  3D Apollonius diagram as well as a sixth query site $S_q$. The 
  predicate determines the portion of $\edge{ijklm}$ (we drop the 
  subscripts for convenience)
  that will disappear in the Apollonius diagram of the six sites due to
  the insertion of $S_q$ and therefore its output is one of the following  

  \begin{itemize}
  \item 
  \noconflict : no portion of $e$ is destroyed by the insertion of $S_q$ in the Apollonius diagram of the five sites.
  \item 
  \fullconflict : the entire edge $e$ is destroyed by the addition of $S_q$ in the Apollonius diagram of the five sites.
  \item 
  \leftvertex : a subsegment of $e$ adjacent to its origin vertex ($v_{ijkl}$) disappears in the Apollonius diagram of the six sites.
  \item 
  \rightvertex : is the symmetric case of the \leftvertex case; a subsegment of $e$ adjacent to the vertex $v_{ikjm}$ disappears in the Apollonius diagram of the six sites.
  \item 
  \verticesconflict : subsegments of $e$ adjacent to its two vertices disappear in the Apollonius diagram of the five sites.
  \item
  \middleconflict : a subsegment in the interior of $e$ disappears in the Apollonius diagram of the five sites.
  \end{itemize}

  In Section~\ref{sub:the_main_algorithm}, we prove that these are indeed
  the only possible answers to the studied predicate, under the assumption
  that no degeneracies occur. Specifically, all analysis presented in this section is done under the following two major assumptions: 
  \begin{itemize}
  \item The trisector $\tri{ijk}$ of the sites $S_i,S_j$ and $S_k$ is 
  ``hyperbolic'', \ie, it is either a branch of a hyperbola or a straight
  line. Therefore, the spheres must lie in \emph{convex position}; in other 
  words, there must exist two distinct planes commonly tangent to all three spheres.
  \item None of the subpredicates called during the algorithm presented 
  in Section~\ref{sub:the_main_algorithm} returns a degenerate answer. 
  Mainly, this is equivalent to the statement: \emph{All of the existing Apollonius vertices defined by the sites $S_i,S_j,S_k$ and $S_n$, for 
  $n\in\{l,m,q\}$, are distinct and the respective Apollonius spheres are all finite, \ie, they are not centered at infinity}. Such assertion 
  dictates that the edge $e$ is finite as none of its bounding vertices 
  $v_{ijkl}$ and $v_{ikjm}$ can lie at infinity.
  \end{itemize}

 

 \section{SubPredicates and Primitives} 
  \label{sub:subpredicates}

  In this section, we describe the various 
  subpredicates used throughout the evaluation of the \conflict predicate 
  via the main algorithm presented in 
  Section~\ref{sub:the_main_algorithm}. For convenience, only the input, output and specific geometric observations is provided in this section, whereas a detailed analysis along with an algebraic degree analysis of each subpredicate is found in Section~\ref{sec:algebraic_analysis}.

  \subsection{The \insphere predicate} 
  \label{sub:the_insphere_predicate}

 The $\text{\insphere}(S_i,S_j,S_k,S_a,S_b)$ predicate 
 returns $-,+$ or $0$ if and only if the sphere $S_b$ 
 intersects, does not intersect or is tangent to the 
 external Apollonius sphere of the sites $S_i,S_j,S_k$
 and $S_b$, centered at $v_{ijka}$. It is assumed that 
 $v_{ijka}$ exists and none of the first four inputed sites 
 are contained inside one another. In \cite{Iordanov}, it is 
 shown that the evaluation of the \insphere predicate requires
 operations of maximum algebraic degree 10, whereas in \cite{anton2011exact}
 an implicit \insphere predicate could be evaluated via the Delaunay graph, 
 using 6-fold degree operations (although it is not clear if 
 we could easily distinguish if we are testing against the Apollonius 
 sphere centered at $v_{ijka}$ or $v_{ikja}$). 

 Since degenerate configurations are beyond the scope of this chapter, 
 the \insphere tests evaluated during the main algorithm (see Section~\ref{sub:the_main_algorithm}) will always return $+$ or $-$. 
 We should also remark that, in bibliography, the \insphere predicate is also referred to 
 as the {\sc{}VertexConflict} predicate to reflect the fact that a negative 
 (resp., positive)
 outcome of $\text{\insphere}(S_i,S_j,S_k,S_a,S_b)$ amounts to the 
 Apollonius vertex $v_{ijka}$ in $\mathcal{VD}(\Sigma)$ vanishing (resp., remaining) in $\mathcal{VD}(\Sigma\cup\{S_b\})$, where $\Sigma$ contains 
 $S_i,S_j,S_k$ and $S_a$ but not $S_b$.

 \begin{lemma}
 The \insphere predicate can be evaluated by determining the sign
 of quantities of algebraic degree at most 10 (in the input 
 quantities).
 \end{lemma}


  \subsection{The \incone and {\sc{}Trisector} predicates} 
   \label{ssub:the_incone_and_tritype_predicates}

\begin{figure}[htbp]
    \centering
    \includegraphics[width=0.95\textwidth]{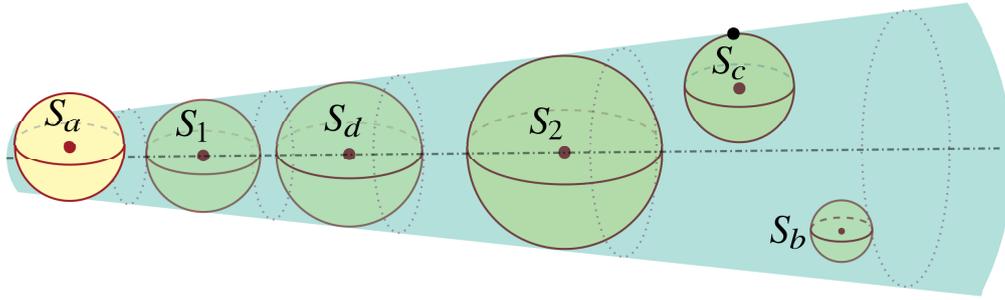}
    \caption[The Incone predicate possible outcomes]{Some of the possible locations of a sphere $S_n$ against 
    the cone $\cone$ defined by $S_1$ and $S_2$. The 
    $\text{\incone}(S_1,S_2;S_n)$ returns \outside, \inside, 
    \ptouch\xspace and \ctouch\xspace for $n=a,b,c$ and $d$ respectively.}
    \label{fig:03}
   \end{figure}

   Given three spheres $S_a,S_b$ and $S_c$, such that $S_a$ and $S_b$
   are not contained one inside the other, we want to determine the
   relative geometric position of $S_c$ with
   respect to the uniquely defined closed semi-cone $\cone(S_a,S_b)$ that is 
   tangent to both $S_a$ and $S_b$ and includes their centers 
   (see Figure~\ref{fig:03}). 
   We shall call this the $\text{\incone}(S_a,S_b;S_c)$ predicate. 

   In case the radii of $S_a$ and $S_b$ are equal, $\cone(S_a,S_b)$ 
   (we drop the parenthesis for convenience)
   degenerates into a cylinder without this having an impact to 
   the predicate. If $S^\circ_c$ is used to denote the open sphere 
   that corresponds to $S_c$, then all possible answers of the 
   predicate $\text{\incone}(S_a,S_b;S_c)$ are

   \begin{itemize}
   \item
   $\outside$ ,
      \text{if at least one point of $S_c$ is outside $\cone$}, 
   \item
   $\inside$,
     \text{if $S^\circ_c$ lies inside $\cone$ and 
     $S_c\cap \vartheta\cone=\emptyset$}, 
   \item
   $\ptouch$,
     \text{if $S^\circ_c$ lies inside $\cone$ and
     $S_c\cap \vartheta\cone$ is a point}, 
   \item
   $\ctouch$,
     \text{if $S^\circ_c$ lies inside $\cone$ and
     $S_c\cap \vartheta\cone$ is a circle.}
   \end{itemize}

   The last two answers are considered ``degenerate'' and therefore, we  
   may consider that whenever \incone is called during the algorithm 
   presented in Section~\ref{sub:the_main_algorithm}, it will either return 
   $\outside$ or $\inside$. 

   This predicate is basic tool used in various other sub-predicates 
   such as the \tritype, which returns the trisector type of a set 
   of three spheres. It is known (\cite{WillThesis}) that if the trisector 
   $\tri{abc}$ of $S_a,S_b,S_c$ has Hausdorff dimension 1, it can 
   either be a branch of a``hyperbola'', a ``line'', an 
   ``ellipse'', a ``circle'' or a ``parabola''; these are the 
   possible answers of the $\text{\tritype}(S_a,S_b,S_c)$ predicate.
   However, since the ``line'' and the ``circle'' type are sub-case of 
   the ``hyperbolic'' and ``elliptic'' trisector types respectively, 
   we can characterize a trisector as either ``hyperbolic'', ``elliptic'' 
   or ``parabolic''. 

   During the execution of the main algorithm of Section~\ref{sub:the_main_algorithm}, the $\text{\tritype}\allowbreak(S_i,S_j,S_k)$ has to be evaluated. 
   Being able to distinguish the type of the trisector $\tri{ijk}$ is essential since all the analysis presented in this section assumes that  
   $\tri{ijk}$ is hyperbolic.

   The analysis followed to determine the outcome of the \incone 
   or the \tritype predicate can be found in Sections~\ref{sub:the_incone_predicate_analysis} and \ref{sub:the_tritype_predicate_analysis} respectively, where the following lemma is proved.

  \begin{lemma}
  The \incone and \tritype predicates can be evaluated by determining 
  the sign of quantities of algebraic degree at most 4
  (in the input quantities).
  \end{lemma}

  \subsection{The {\sc{}Distance} predicate} 
   \label{ssub:the_distance_predicate}
   
   When the trisector $\tri{ijk}$ is a hyperbola or a line, there 
   exist two distinct planes, denoted by $\Pi^{-}_{ijk}$ and 
   $\Pi^{+}_{ijk}$, such that each one is commonly tangent 
   to the sites $S_i,S_j,S_k$ and leave their centers on the same 
   halfspace.

   Observe that $\Pi^{-}_{ijk}$ and
   $\Pi^{+}_{ijk}$ correspond to the two Apollonius spheres
   \emph{at infinity}, in the sense that they are centered at 
   \emph{infinity} and are cotangent to the spheres $S_i,S_j$ and 
   $S_k$. These planes are considered as oriented, and subdivide $\RR^3$
   into a positive and a negative halfspace, the positive being the
   halfspace containing the centers of the spheres. 

\begin{figure}[htbp]
    \centering
    \includegraphics[width=0.95\textwidth]{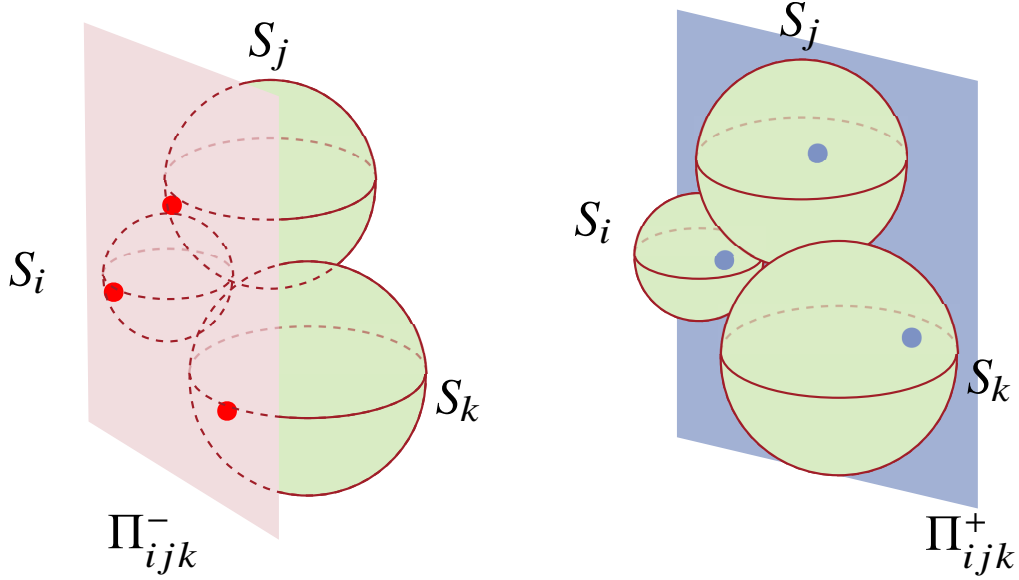}
    \caption[The planes commonly tagent to three spheres]{If the spheres $S_i, S_j$ and $S_k$ lie in a convex 
    position there exist two distinct planes, $\Pi_{ijk}^-$ and 
    $\Pi_{ijk}^+$, cotangent to all spheres. These planes 
    are considered as the Apollonius sphere of the sites 
    $S_i, S_j$ and $S_k$, centered at $p\in\tri{ijk}$, as $\map{p}$ 
    goes to $\pm\infty$ respectively.}
    \label{fig:04}
   \end{figure}

   If we consider a point $p$ that moves on the trisector $\tri{ijk}$ 
   such that $\map{p}$ goes to $-\infty$ or $+\infty$, the
   sphere $\tts{p}$ becomes the corresponding Apollonius sphere at
   infinity, \ie, the plane $\Pi^{-}_{ijk}$ or
   $\Pi^{+}_{ijk}$ (see Figure~\ref{fig:04}).

   Given the sites $S_i,S_j,S_k$ and $S_\alpha$, the
   \distance$(S_i,S_j,S_k,S_\alpha)$ predicate determines whether
   $S_\alpha$ intersects, is tangent to, or does not intersect the
   (closed) negative halfspaces delimited by the two planes $\Pi^{-}_{ijk}$
   and $\Pi^{+}_{ijk}$. The ``tangency'' case is considered as 
   degenerate and is beyond the scope of this paper.  
   This predicate is used in the evaluation of the \shadow predicate, 
   and is equal to
   $\text{\distance}(S_i,S_j,S_k,S_\alpha)=
    \big(\sgn(\delta(S_\alpha,\Pi^{-}_{ijk})),
    \sgn(\delta(S_\alpha,\Pi^{+}_{ijk}))
    \big)$,
   where $\delta(S,\Pi)=\delta(C,\Pi)-r$, and $\delta(C,\Pi)$
   denotes the signed Euclidean of $C$ from the plane $\Pi$ and $S$ is a sphere of radius $r$, centered at $C$. As for the
   \exist predicate, we reduce it to the computation
   of the signs of the two roots of a quadratic equation and prove the
   following lemma (see 
   Section~\ref{sub:the_distance_predicate_analysis}
   for this analysis).
   
   \begin{lemma}
   The \distance predicate can be evaluated by determining 
   the sign of quantities of algebraic degree at most 6 
   (in the input quantities).
   \end{lemma}

  \subsection{The {\sc{}Existence} predicate} 
   \label{ssub:the_exist_predicate}

   The next primitive operation we need for answering the 
   \conflict predicate is what we call the \exist predicate:
   given four sites $S_a,S_b,S_c$ and $S_n$, we would 
   like to determine the number of Apollonius spheres of the quadruple
   $S_a,S_b,S_c,S_n$.
   In general, given four sites there can be ``0'', ``1'', ``2'' or ``infinite'' Apollonius spheres 
   (cf. \cite{devillers2017qualitative}) including 
   the Apollonius sphere(s) at infinity.
   The $\text{\exist}(S_a,S_b,S_c,S_n)$ predicate only counts the 
   Apollonius spheres that are \emph{not} centered at infinity and since 
   degenerate configurations of the input sites are beyond the scope of this 
   paper, it is safe to assume that the outcome will always be ``0'',``1'' or ``2''. 
   It is also clear that in case of a 
   ``1'' outcome, the corresponding Apollonius center will either be 
   $v_{abcn}$ or $v_{acbn}$ but not both; the case where $v_{abcn}\equiv v_{acbn}$ is ruled out by our initial no-degeneracies assumption.

   The analysis of the \exist predicate can be found in 
   Section~\ref{sub:the_existence_predicate_analysis} where we 
   prove the following lemma.

   \begin{lemma}
   The \exist  predicate can be evaluated by determining 
   the sign of quantities of algebraic degree at most 8 
   (in the input quantities).
   \end{lemma}


  \subsection{The \shadow predicate} 
   \label{ssub:the_shadow_predicate}

   We now come to the second major subpredicate used by the \conflict
   predicate: the \shadow predicate.

\begin{figure}[htbp]
   \centering
   \includegraphics[width=0.95\textwidth]{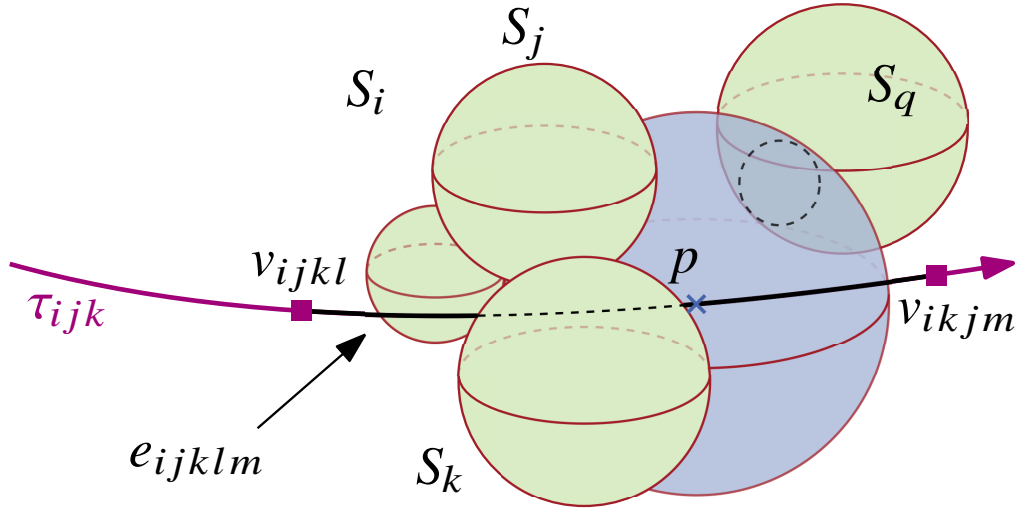}
   \caption[A finite edge $e_{ijklm}$ on a hyperbolic trisector]{A finite edge $e_{ijklm}$ of the Voronoi diagram 
   $\mathcal{VD(S)}$ of the set 
   $\mathcal{S}=\{S_n : n = i,j,k,l,m\}$ 
   is the locus of points $p\in\tri{ijk}$
   such that $v_{ijkl}\prec p \prec v_{ikjm}$. Any sphere 
   $\tts{p}$, centered at $p$ and cotangent to $S_i,S_j$ and $S_k$ 
   does not intersect any sphere of $\mathcal{S}$ 
   (\emph{Empty Sphere Principle}). However, after inserting $S_q$ 
   in the existing Voronoi diagram, $\tts{p}$ may intersect it. 
   All points $p\in\sh{S_q}$ will no longer exist on 
   in $\mathcal{VD(S}\cup\{ S_q \})$ as points on Voronoi edges.}
   \label{fig:02}
   \end{figure}

  Given three sites $S_i,S_j$ and $S_k$, we define the 
  \emph{shadow region} $\sh{S_\alpha}$ of a site $S_\alpha$, with 
  respect to the trisector $\tri{ijk}$, to be the locus of points $p$ 
  on $\tri{ijk}$ such that $\delta(C_\alpha,\tts{p})<r_\alpha$. The 
  shadow region of the sites $S_l, S_m$ and $S_q$ play an important 
  role when answering $\text{\conflict}(S_i,S_j,S_k,S_l,S_m,S_q)$ 
  (see Figure~\ref{fig:02} and Section~\ref{sub:the_main_algorithm}).

   The $\text{\shadow}(S_i,S_j,S_k,S_\alpha)$ predicate returns the type of
   $\sh{S_\alpha}$ seen as an interval, or union of intervals, in
   $\RR$. More precisely, the \shadow predicate returns the topological
   structure of the set $\map{\sh{S_\alpha}}=
   \map{\{p\in\tri{ijk}\mid{}\delta(C_\alpha,\tts{p})<r_\alpha\}}$, 
   which we denote by $SRT(S_\alpha)$. 
   
   Clearly, the boundary points of the closure of 
   $\text{\shadow}(S_i,S_j,S_k,S_\alpha)$ 
   are the points $p$ on $\tri{ijk}$ for which
   $\delta(C_\alpha,\tts{p})=r_\alpha$. These points are nothing
   but the centers of the Apollonius spheres of the four sites
   $S_i,S_j,S_k$ and $S_\alpha$, and, as such, there can only be 0, 1 or
   2 (assuming no degeneracies). 
   Since $\tri{ijk}$ is assumed to be hyperbolic, this immediately suggests that $SRT(S_\alpha)$ can have one of the 
   following 6 types:
   $\emptyset$, $(-\infty,\infty)=\RR$, $(-\infty,\phi)$, $(\chi,+\infty)$, 
   $(\chi,\phi)$, or $(-\infty,\phi)\cup(\chi,+\infty)$,
   where $\phi,\chi\neq\pm\infty$. 

\begin{figure}[htbp]
    \centering
    \includegraphics[width=0.95\textwidth]{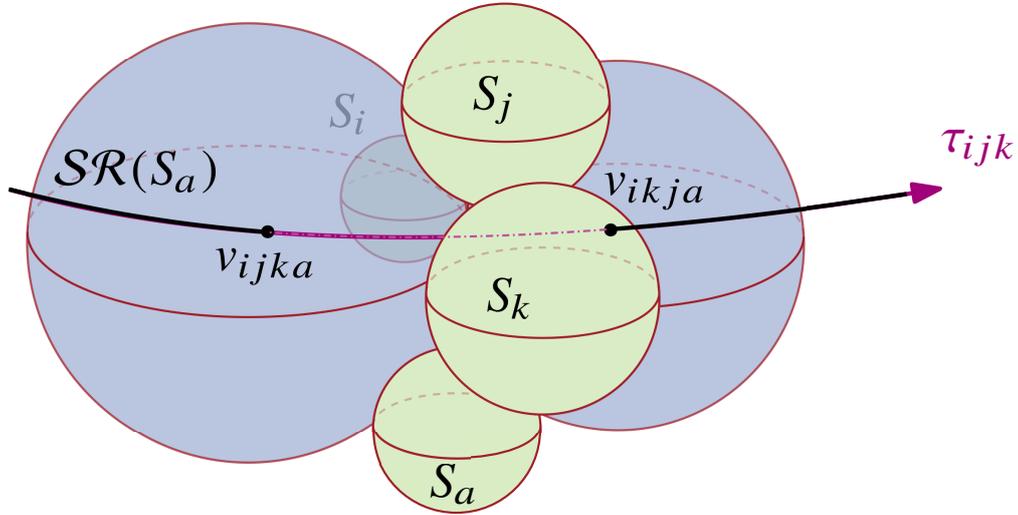}
    \caption[The shadow region of a sphere on a hyperbolic trisector]{Since there are two Apollonius spheres of the 
    sites $S_n$, for $n\in\{i,j,k,a\}$, centered at $v_{ijka}$ and 
    $v_{ikja}$, the $\sh{S_a}$ on $\tri{ijk}$ must have two endpoints. 
    In this specific configuration notice that, for every 
    point $p$ on the segments of $\tri{ijk}$ painted black, the 
    sphere $\tts{p}$ will intersect $S_a$. Therefore, the black 
    segments are indeed the shadow region $\sh{S_a}$ of the sphere 
    $S_a$ on the trisector $\tri{ijk}$.}
    \label{fig:05}
   \end{figure}

   For convenience, we will use the 
   $\sh{S_a}$ notation instead of $SRT(S_\alpha)$; for example, 
   the statement ``$\sh{S_a} = (-\infty,\phi)$'' will be 
   often used instead of ``$\sh{S_a}$'s type is $(-\infty,\phi)$'' or 
   ``$SRT(S_\alpha) = (-\infty,\phi)$'' (see Figure~\ref{fig:05} 
   for an example). This notation change further highlights the fact 
   that we are only interested in the topological structure of $\sh{S_a}$
   rather than the actual set itself.

   In Section~\ref{sub:the_shadowregion_predicate_analysis}, we 
   prove that the evaluation of the \shadow predicate only 
   requires the call of the respective \distance and 
   \exist predicate, yielding the following lemma.
   
   \begin{lemma}
   The \shadow  predicate can be evaluated by determining 
   the sign of quantities of algebraic degree at most 8 
   (in the input quantities).
   \end{lemma}

  \subsection{The {\sc{}Order} predicate} 
   \label{ssub:the_order_predicate}

   The most important sub-predicate used to evaluate the \conflict predicate is what we call the \order predicate.
   When $\text{\order}(S_i,S_j,S_k,S_a,S_b)$ is called, it returns 
   the order of appearance of any of the existing Apollonius vertices
   $v_{ijka},v_{ikja},v_{ijkb}$ and $v_{ikjb}$ on the oriented trisector 
   $\tri{ijk}$. 

   This sub-predicate is called during the main algorithm 
   that answers the $\text{\conflict}(\allowbreak S_i,\allowbreak S_j,S_k,S_l,S_m,S_q)$, for $(a,b)\in\{(l,q),(m,q)\}$, 
   only in the case that either $v_{ijkq}$, $v_{ikjq}$ or both exist. 
   Let us also recall that, in this chapter, the trisector 
   $\tri{ijk}$ is ``hyperbolic'' and that $\edge{ijklm}$ is a valid 
   finite Apollonius edge; the Apollonius vertices $v_{ijkl}$ and $v_{ikjm}$ both exist on (the oriented) $\tri{ijk}$ and $v_{ijkl} \prec v_{ikjm}$.

   In order to answer the \order predicate, we first call the 
   $\text{\shadow}(S_i,S_j,S_k,\allowbreak S_n)$ predicate, 
   for $n\in\{a,b\}$, to obtain the type of $\sh{S_a}$ and $\sh{S_b}$.  
   From the shadow region types, two pieces of
   information is easily obtained; firstly, we determine which of the Apollonius vertices 
   $v_{ijka},v_{ikja}$ and $v_{ijkb},v_{ikjb}$ actually exist and
   secondarily, if both $v_{ijkn}$ and $v_{ikjn}$ exist for 
   some $n\in\{a,b\}$, then their ordering on the oriented 
   trisector is also retrieved. Such deductions derive from the 
   study of the shadow region, as shown in Section~\ref{sub:The_classic_configuration} (see Lemma~\ref{lemma:phi_chi}).  
   For example, if $\sh{S_a}=(\chi,\phi)$, then both $v_{ijka}$ and 
   $v_{ikja}$ exist and appear on the oriented trisector in this order:
   $v_{ikja} \prec v_{ijka}$. 

   Now that the existence and partial ordering of the Apollonius vertices
   $v_{ijka}$ and $v_{ikja}$ ( resp., $v_{ijkb}$ and $v_{ikjb}$) is known, 
   we must provide a way of ``merging'' them into a complete 
   ordering. For this reason, we examine all 
   possible complete orderings of the Apollonius vertices  on the
   oriented trisector $\tri{ijk}$. The study of these orderings is seen in 
   the inverted plane \wspace. In Section~\ref{ssub:the_yspace_analysis}, 
   we present the strong geometric relationship that holds 
   between the spheres of the original \zspace and their images in the 
   inverted plane. The observations we make regarding the connection
   of the two spaces allow us to interpret geometric configurations on one 
   space to equivalent ones on the other. 
   A full analysis of how we tackle all possible configurations is 
   presented in  Section~\ref{sub:the_order_predicate_analysis}. 

   In our analysis, we prove that the \order predicate, in the worst case, 
   amounts to evaluate up to 4 \insphere predicates plus some auxiliary 
   tests of lesser algebraic cost. We have therefore proven the 
   following lemma. 

   \begin{lemma}
   The \order predicate can be evaluated by determining the sign 
   of quantities of algebraic degree at most 10 (in the input quantities). 
   \end{lemma}

 \section{The main algorithm} 
  \label{sub:the_main_algorithm}
  In this section, we describe in detail how the predicate    
  \conflict$(S_i,S_j,S_k,S_l,S_m,S_q)$ is resolved 
  with the use of the subpredicates \incone, \tritype, 
  \exist, \shadow and \order. 

 We begin by determining the type of the trisector 
 $\tri{ijk}$; this is done via the call of the \tritype$(S_i,S_j,S_k)$
 predicate. Recall that in the scope of this chapter, it is assumed that
 the  $\tri{ijk}$ is a hyperbola (or a line) and that none of the 
 subpredicates called return a degenerate answer.

 To answer the \conflict predicate, one must determine 
 which ``part'' of the edge $\edge{ijklm}$ remains in the Voronoi
 diagram after the insertion of the site $S_q$. This is plausible 
 by identifying the set of points of $\edge{ijklm}$ that still remain 
 in the updated Voronoi Diagram; each of these points must satisfy the 
 ``empty-sphere property'': a sphere, centered at that point and 
 tangent to the spheres $S_i,S_j,S_k$, must not intersect any other sites 
 of the Voronoi Diagram. As an immediate result, 
 a point $p$ of the edge $\edge{ijklm}$ in 
 $\mathcal{VD}(\mathcal{S})$ remains in 
 $\mathcal{VD}(\mathcal{S}\cup\{S_q\})$ if and only if $\tts{p}$ does 
 not intersect $S_q$. Since the shadow region of the sphere $S_q$ 
 with respect to the trisector $\tri{ijk}$ consists of all points 
 $p$ such that $\tts{p}$ intersects $S_q$, it must hold that the 
 part of the edge $\edge{ijklm}$ that no longer remains in  
 $\mathcal{VD}(\mathcal{S}\cup\{S_q\})$ is actually 
 $\edge{ijklm}\cap\sh{S_q}$ (see Figure~\ref{fig:02}) . In conclusion, the result of 
 the \conflict predicate is exactly the set $\edge{ijklm}\cap\sh{S_q}$
 seen as an interval or union of intervals of $\RR$.

 To determine the intersection type of $\edge{ijklm}\cap\sh{S_q}$, 
 we first take into account that the finite edge $\edge{ijklm}$ consists 
 of all points $p$ on the oriented trisector $\tri{ijk}$ bounded by the 
 points $v_{ijkl}$ and $v_{ikjm}$ from left and right respectively 
 (see Section~\ref{sub:voronoi_edges}). Next, we consider the 
 type of $\sh{S_q}$ which can be evaluated as shown in 
 Section~\ref{sub:the_shadowregion_predicate_analysis} and  
 is one of the following: 
 $(-\infty,\phi)$, $(\chi,+\infty)$, $(\chi,\phi)$, 
 $(-\infty,\phi)\cup(\chi,+\infty)$, $\emptyset$ or $\RR$. 

 If the edge 
 $\edge{ijklm}$ is seen as the interval $(\lambda_1,\mu_2)$, 
 evidently the intersection type of $E'=\edge{ijklm}\cap\sh{S_q}$ 
 must be one of the following 6 types, each corresponding to a different 
 answers of the \conflict predicate. 

 \begin{itemize}
  \item 
  If $E'$ is of type $\emptyset$, the predicate returns \noconflict.
  \item 
  If $E'$ is of type $\edge{ijklm}$, the predicate returns \fullconflict.
  \item 
  If $E'$ is of type $(\lambda_1,\phi)$, the predicate returns \leftvertex.
  \item 
  If $E'$ is of type $(\chi,\mu_2)$, the predicate returns \rightvertex.
  \item 
  If $E'$ is of type $(\lambda_1,\phi)\cup(\chi,\mu_2)$, the 
  predicate returns \verticesconflict. 
  \item
  If $E'$ is of type $(\chi,\phi)$, the predicate 
  returns \middleconflict.
 \end{itemize}

 This observation suggests that, if we provide a way to identify the 
 type of $E'$, we can answer the \conflict predicate. Taking into
 consideration that
 \begin{itemize}
  \item 
  $\lambda_1$ and $\mu_2$ correspond to $v_{ijkl}$ and $v_{ikjm}$, 
  respectively as shown in Section~\ref{sub:voronoi_edges}, and 
  \item
  $\chi,\phi$ correspond to $v_{ikjq}$ and $v_{ijkq}$, respectively 
  as stated in Lemma~\ref{lemma:phi_chi} that we prove in 
  Section~\ref{sub:The_classic_configuration},
 \end{itemize}
 it becomes apparent that if we order all Apollonius vertices 
 $v_{ijkl}, v_{ikjm}$ and any of the existing among
 $v_{ijkq}, v_{ikjq}$, bearing in mind the type of $\sh{S_q}$, 
 we can deduce the type of $E'$. 

 For example, let us assume that
 $\sh{S_q}$ type is $(-\infty,\chi)\cup(\phi,+\infty)$. If
 $v_{ijkl} \prec v_{ikjq} \prec v_{ikjm} \prec v_{ijkq}$ on 
 the oriented trisector $\tri{ijk}$, or equivalently
 $\lambda_1<\chi<\mu_2<\phi$, 
 we can conclude that $E'$ is of type $(\lambda_1,\phi)$ and the
 \conflict predicate would return \leftvertex. 
 
 Therefore, it is essential that we are able to provide an ordering 
 of the Apollonius vertices $v_{ijkl}, v_{ikjm}$ and any of 
 the existing among $v_{ijkq}, v_{ikjq}$. Such a task is 
 accomplished via the call of the 
 \order$(S_i,S_j,S_k,S_a,S_b)$ predicate 
 $(a,b)=(l,q)$ and $(m,q)$. The outcomes of these predicates 
 consist of the orderings of all possible Apollonius vertices of 
 the sites $S_i,S_j,S_k,S_a$ and $S_i$, $S_j$, $S_k$, $S_b$ on the 
 trisector $\tri{ijk}$. 
 These partial orderings can then be merged into a complete ordering,
 which contains the desired one. 
 The results' combination principle is identical to the one 
 used when we have to order a set of numbers but we can only 
 compare two at a time. 

 A detailed algorithm that summarizes the analysis of this Section 
 and can be followed to answer the 
 $\text{\conflict}(S_i,S_j,S_k,S_l,S_m;S_q)$ is described in the following steps. Note that the trisector $\tri{ijk}$ is assumed to be 
 a branch of a hyperbola or a line, \ie, the outcome 
 of \tritype$(S_i,S_j,S_k)$ is ``hyperbolic''.


  \begin{description}
  \item[Step 1] We evaluate $SRT(q)=\text{\shadow}(S_i,S_j,S_k,S_q)$. 
  If $SRT(q)=\emptyset$ or $\RR$, we return \noconflict or 
  \fullconflict respectively. Otherwise, if 
  $SRT(q)$ is
  $(-\infty,\phi)$, $(\chi,+\infty)$, $(\chi,\phi)$ or $
  (-\infty,\phi)\cup(\chi,+\infty)$, we go to Step 2a, 2b, 2c or 2d, 
  respectively. Note that $\phi$ and
  $\chi$ correspond to the Apollonius vertices $v_{ijkq}$ and 
  $v_{ikjq}$ respectively.
  \item[Step 2a]  
  We evaluate $I_1=\text{\insphere}(S_i,S_j,S_k,S_l,S_q)$. 
  If $I_1=+$ then $v_{ijkl}\not\in\sh{S_q}$ or equivalently 
  $v_{ijkq}\prec v_{ijkl}\prec v_{ikjm}$; in this case the predicate's 
  outcome is \noconflict.
  Otherwise, if $I_1=-$, we know that $v_{ijkl}\in\sh{S_q}$ and we 
  have to evaluate $I_2=\text{\insphere}(S_i,S_k,S_j,S_m,S_q)$.
  If $I_2=+$ then $v_{ikjm}\not\in\sh{S_q}$ or equivalently 
  $v_{ijkq}\prec v_{ikjm}$; since 
  $v_{ijkl} \prec v_{ijkq}\prec v_{ikjm}$ the predicate's outcome is 
  \leftvertex.  
  If $I_2=-$ then $v_{ikjm}\in\sh{S_q}$ and subsequently 
  $v_{ikjm}\prec v_{ijkq}$; since 
  $v_{ijkl} \prec v_{ikjm} \prec v_{ijkq}$ the predicate's outcome is 
  \fullconflict.
  \item[Step 2b]
  We evaluate $I_1=\text{\insphere}(S_i,S_j,S_k,S_l,S_q)$. 
  If $I_1=-$ then $v_{ijkl}\in\sh{S_q}$ or equivalently 
  $v_{ikjq}\prec v_{ijkl}\prec v_{ikjm}$; in this case the predicate's 
  outcome is \fullconflict.
  Otherwise, if $I_1=-$, we know that $v_{ijkl}\not\in\sh{S_q}$ and we 
  have to evaluate $I_2=\text{\insphere}(S_i,S_k,S_j,S_m,S_q)$.
  If $I_2=-$ then $v_{ikjm}\in\sh{S_q}$ or equivalently 
  $v_{ikjq}\prec v_{ikjm}$; since 
  $v_{ijkl} \prec v_{ikjq}\prec v_{ikjm}$ the predicate's outcome is 
  \rightvertex.  
  If $I_2=+$ then $v_{ikjm}\not\in\sh{S_q}$ and subsequently 
  $v_{ikjm}\prec v_{ikjq}$; since 
  $v_{ijkl} \prec v_{ikjm} \prec v_{ikjq}$ the predicate's outcome is 
  \noconflict.
  \item[Step 2c]
  We evaluate $I_1=\text{\insphere}(S_i,S_j,S_k,S_l,S_q)$ and 
  $I_2=\text{\insphere}(S_i,S_k,S_j,S_m,S_q)$. 
  \begin{itemize}
  \item If $(I_1,I_2)=(-,+)$ then $v_{ijkl}\in\sh{S_q}$ or equivalently 
  $v_{ikjq}\prec v_{ijkl}\prec v_{ijkq}$ and  
  $v_{ikjm}\not\in\sh{S_q}$. Since $v_{ijkl}\prec v_{ikjm}$ it must 
  hold that  $v_{ikjq}\prec v_{ijkl}\prec v_{ijkq}\prec v_{ikjm}$ and 
  therefore the predicate's outcome is \leftvertex. 
  \item If $(I_1,I_2)=(+,-)$ then $v_{ikjm}\in\sh{S_q}$ or equivalently 
  $v_{ikjq}\prec v_{ikjm}\prec v_{ijkq}$ and  
  $v_{ijkl}\not\in\sh{S_q}$. Since $v_{ijkl}\prec v_{ikjm}$ it must 
  hold that  $v_{ijkl} \prec v_{ikjq} \prec v_{ikjm} \prec v_{ijkq}$ and 
  therefore the predicate's outcome is \rightvertex. 
  \item If $(I_1,I_2)=(-,-)$ 
  then $v_{ijkl},v_{ikjm}\in\sh{S_q}$. Since $v_{ijkl}\prec v_{ikjm}$ 
  it must hold that $v_{ikjq}\prec v_{ijkl}\prec v_{ikjm} \prec v_{ijkq}$;
  in this case the predicate's outcome is \fullconflict. 
  \item If $(I_1,I_2)=(+,+)$ then there are three possible cases, due 
  to $v_{ijkl},v_{ikjm}\not\in\sh{S_q}$: either 
  $v_{ikjq} \prec v_{ijkq} \prec v_{ijkl}\prec v_{ikjm}$, 
  $v_{ijkl} \prec v_{ikjq} \prec v_{ijkq} \prec v_{ikjm}$ or
  $v_{ijkl} \prec v_{ikjm} \prec v_{ikjq} \prec v_{ijkq} $. 
  To distinguish among the cases, we first call the 
  $\text{\order}(S_i,S_j,S_k,S_l,S_q)$ predicate. 
  If it returns that $v_{ikjq} \prec v_{ijkq} \prec v_{ijkl}$ then 
  we are in the first case and the predicate returns \noconflict. 
  Otherwise, it will necessarily return that 
  $v_{ijkl} \prec v_{ikjq} \prec v_{ijkq}$ and we also have to call 
  the $\text{\order}(S_i,S_j,S_k,S_m,S_q)$ predicate. 
  If it returns that 
  $v_{ikjq} \prec v_{ijkq} \prec v_{ikjm}$ then we are in the second 
  case and the predicate returns \middleconflict. 
  Otherwise, it will return $v_{ikjm} \prec v_{ikjq} \prec v_{ijkq}$
  and therefore we are in the third case thus the predicate's outcome 
  is \noconflict.
  \end{itemize}
  \item[Step 2d]
  We evaluate $I_1=\text{\insphere}(S_i,S_j,S_k,S_l,S_q)$ and 
  $I_2=\text{\insphere}(S_i,S_k,S_j,S_m,S_q)$. 
  \begin{itemize}
  \item If $(I_1,I_2)=(-,+)$ then $v_{ikjm}\not\in\sh{S_q}$ or equivalently 
  $v_{ikjq}\prec v_{ikjm}\prec v_{ijkq}$ and  
  $v_{ijkl}\in\sh{S_q}$. Since $v_{ijkl}\prec v_{ikjm}$ it must 
  hold that  $ v_{ijkl}\prec v_{ijkq}\prec v_{ikjm} \prec v_{ikjq}$ and 
  therefore the predicate's outcome is \leftvertex. 
  \item If $(I_1,I_2)=(+,-)$ then $v_{ijkl}\not\in\sh{S_q}$ or equivalently 
  $v_{ikjq}\prec v_{ijkl}\prec v_{ijkq}$ and  
  $v_{ikjm}\in\sh{S_q}$. Since $v_{ijkl}\prec v_{ikjm}$ it must 
  hold that  $v_{ijkq} \prec  v_{ijkl} \prec v_{ikjq} \prec v_{ikjm} $ and 
  therefore the predicate's outcome is \rightvertex. 
  \item If $(I_1,I_2)=(+,+)$ 
  then $v_{ijkl},v_{ikjm}\not\in\sh{S_q}$. Since $v_{ijkl}\prec v_{ikjm}$ 
  it must hold that $v_{ijkq}\prec v_{ijkl}\prec v_{ikjm} \prec v_{ikjq}$;
  in this case the predicate's outcome is \noconflict. 
  \item If $(I_1,I_2)=(-,-)$ then there are three possible cases, due 
  to $v_{ijkl},v_{ikjm}\in\sh{S_q}$: either 
  $v_{ijkq} \prec v_{ikjq} \prec v_{ijkl}\prec v_{ikjm}$, 
  $v_{ijkl} \prec v_{ijkq} \prec v_{ikjq} \prec v_{ikjm}$ or
  $v_{ijkl} \prec v_{ikjm} \prec v_{ijkq} \prec v_{ikjq} $. 
  To distinguish among the cases, we first call the 
  $\text{\order}(S_i,S_j,S_k,S_l,S_q)$ predicate. 
  If it returns that $v_{ijkq} \prec v_{ikjq} \prec v_{ijkl}$ then 
  we are in the first case and the predicate returns \fullconflict. 
  Otherwise, it will necessarily return that 
  $v_{ijkl} \prec v_{ijkq} \prec v_{ikjq}$ and we also have to call 
  the $\text{\order}(S_i,S_j,S_k,S_m,S_q)$ predicate. 
  If it returns that 
  $v_{ijkq} \prec v_{ikjq} \prec v_{ikjm}$ then we are in the second 
  case and the predicate returns \verticesconflict. 
  Otherwise, it will return $v_{ikjm} \prec v_{ijkq} \prec v_{ikjq}$
  and therefore we are in the third case thus the predicate's outcome 
  is \fullconflict.
  \end{itemize}
  \end{description}

\begin{figure}[tbp]
   \centering
   \includegraphics[width=0.85\textwidth]{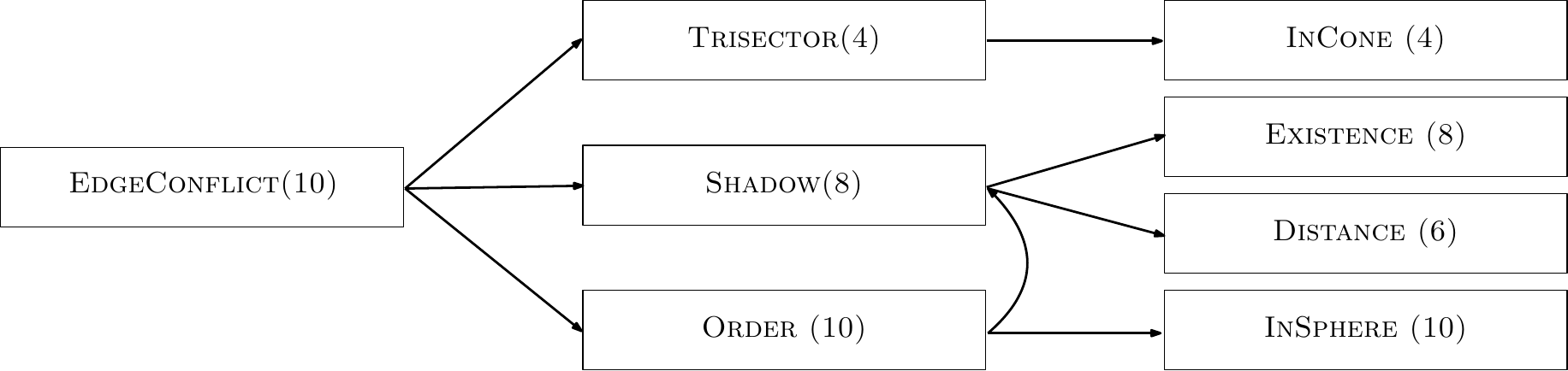}
   \caption[The layout of (sub)predicates used to 
     answer the \conflict predicate in the case of a hyperbolic trisector.]{The layout of predicates and their subpredicates used to 
     answer the \conflict predicate in the case of a hyperbolic trisector. The number next to each predicate corresponds to its algebraic degree. It is assumed that every
     subpredicate returns a non-degenerate answer.}
   \label{fig:predicates}
  \end{figure}

  A sketch of the subpredicates used when answering the 
  \conflict predicate is shown in Figure \ref{fig:predicates}. 
  Since the highest algebraic degree needed in the evaluation of the 
  subpredicates used is 10, we have proven the following theorem.

  \begin{theorem}
  The \conflict predicate for hyperbolic trisectors can be 
  evaluated by determining the sign of quantities of algebraic degree 
  at most 10 (in the input quantities).
  \end{theorem}

  \section{The EdgeConflict Predicate for infinite hyperbolic edges} 
  \label{sec:the_edgeconflict_predicate_for_infinite_hyperbolic_edges}
    

    A variation of the algorithm presented in the previous section
    can be used to answer the following predicates:
    \begin{itemize}
    \item \textsc{InfiniteRightEdgeConflict}$(S_i,S_j,S_k,S_l,S_q)$: 
    describes the intersection type of an infinite Voronoi edge 
    $e$ that lies on the hyperbolic trisector $\tri{ijk}$ and is 
    bounded on the left by $v_{ijkl}$. The existence of such a 
    Voronoi edge is equivalent to the fact that either 
    only $v_{ijkl}$ exist among $\{v_{ijkl}, v_{ikjl}\}$ or both exist 
    and $v_{ikjl} \prec v_{ijkl}$, based on the remarks of 
    previous sections. 
    \item \textsc{InfiniteLeftEdgeConflict}$(S_i,S_j,S_k,S_m,S_q)$: 
    describes the intersection type of an infinite Voronoi edge 
    $e$ that lies on the hyperbolic trisector $\tri{ijk}$ and is 
    bounded on the right by $v_{ikjm}$. The existence of such a 
    Voronoi edge is equivalent to the fact that either 
    only $v_{ikjm}$ exist among $\{v_{ijkm}, v_{ikjm}\}$ or both exist 
    and $v_{ikjm} \prec v_{ijkm}$, based on the remarks of 
    previous sections. 
    \end{itemize} 

    These two predicates are also called in the scope of a randomized 
    incremental algorithm that constructs the 3D Apollonius diagram, 
    similarly with the \conflict predicate. The algorithms that decides 
    them have minor differences than the algorithm presented in 
    Section~\ref{sub:the_main_algorithm} to reflect the fact that 
    the Voronoi edge $e$ has only one finite bound. The possible 
    non-degenerate outcomes of these predicates are the same with 
    the \conflict predicate with the following differences. 

    \begin{itemize}
    \item In $\textsc{InfiniteRightEdgeConflict}$, the edge $e$ is 
    only bounded on the left and therefore the outcome \rightvertex 
    denotes that only a part of $e$ adjacent to $v_{ijkl}$ remains on 
    the updated Voronoi diagram after the insertion of $S_q$. Moreover, 
    \verticesconflict denotes that only a finite part of $e$ that is not 
    adjacent to $v_{ijkl}$ remains on the updated Voronoi diagram.
    \item In $\textsc{InfiniteLeftEdgeConflict}$, the edge $e$ is 
    only bounded on the right and subsequently the outcome \leftvertex 
    denotes that only a part of $e$ adjacent to $v_{ikjm}$ remains on 
    the updated Voronoi diagram after the insertion of $S_q$. Furthermore, 
    \verticesconflict denotes that only a finite part of $e$ that is not 
    adjacent to $v_{ikjm}$ remains on the updated Voronoi diagram.
    \end{itemize}

    The main idea behind the following algorithms is the same with the 
    one presented in Section~\ref{sub:the_main_algorithm}. We 
    consider the topological form of the edge $e$ as an 
    interval $(\lambda_1,+\infty)$ or $(-\infty,\mu_2)$ and the 
    outcome of the predicate reflects the intersection of this interval 
    with the respective interval that represents the shadow region of
    $S_q$.

    \paragraph*{Algorithm for 
    $\textsc{InfiniteRightEdgeConflict}(S_i,S_j,S_k,S_l;S_q)$.}
    
    \begin{description}
    \item[Step 1] We evaluate $SRT(q)=\text{\shadow}(S_i,S_j,S_k,S_q)$. 
    If $SRT(q)=\emptyset$ or $\RR$, we return \noconflict or 
    \fullconflict respectively. Otherwise, if 
    $SRT(q)$ has the form
    $(-\infty,\phi)$, $(\chi,+\infty)$, $(\chi,\phi)$ or $
    (-\infty,\phi)\cup(\chi,+\infty)$, then we go to Step 2a, 2b, 2c or 
    2d respectively. Note that $\phi$ and
    $\chi$ correspond to the Apollonius vertices $v_{ijkq}$ and 
    $v_{ikjq}$ respectively. 
    \item[Step 2a]  
    We evaluate $I=\text{\insphere}(S_i,S_j,S_k,S_l,S_q)$. If $I=+$ 
    then $v_{ijkl}\not\in\sh{S_q}$ or equivalently $v_{ijkq}\prec v_{ijkl}$;
    in this the predicate's outcome is \noconflict. Otherwise, if 
    $I=-$ then $v_{ijkl}\in\sh{S_q}$ or equivalently 
    $v_{ijkl}\prec v_{ijkq}$ hence the predicate's outcome is 
    \leftvertex. 
    \item[Step 2b]
    We evaluate $I=\text{\insphere}(S_i,S_j,S_k,S_l,S_q)$. If $I=+$ 
    then $v_{ijkl}\not\in\sh{S_q}$ or equivalently $v_{ijkl}\prec v_{ikjq}$;
    in this the predicate's outcome is \rightvertex. Otherwise, if 
    $I=-$ then $v_{ijkl}\in\sh{S_q}$ or equivalently 
    $v_{ikjq}\prec v_{ijkl}$ hence the predicate's outcome is 
    \fullconflict. 
    \item[Step 2c]
    We evaluate $I=\text{\insphere}(S_i,S_j,S_k,S_l,S_q)$. If $I=-$ 
    then $v_{ijkl}\in\sh{S_q}$ or equivalently 
    $v_{ikjq}\prec v_{ijkl}\prec v_{ijkq}$; 
    in this the predicate's outcome is \leftvertex. 
    Otherwise, if $I=+$ we call the $\text{\order}(S_i,S_j,S_k,S_l,S_q)$
    and we can determine whether $v_{ijkl}\prec v_{ikjq}\prec  v_{ijkq}$
    or $v_{ikjq}\prec  v_{ijkq} \prec v_{ijkl}$. In the former case, 
    the predicate returns \middleconflict otherwise, in the latter case 
    it returns \noconflict.
    \item[Step 2d]
    We evaluate $I=\text{\insphere}(S_i,S_j,S_k,S_l,S_q)$. If $I=+$ 
    then $v_{ijkl}\not\in\sh{S_q}$ or equivalently 
    $v_{ikjq}\prec v_{ijkl}\prec v_{ijkq}$; 
    in this the predicate's outcome is \rightvertex. 
    Otherwise, if $I=-$ we call the $\text{\order}(S_i,S_j,S_k,S_l,S_q)$
    and we can determine whether $v_{ijkl}\prec v_{ikjq}\prec  v_{ijkq}$
    or $v_{ikjq}\prec  v_{ijkq} \prec v_{ijkl}$. In the former case, 
    the predicate returns \verticesconflict otherwise, in the latter case 
    it returns \fullconflict.
    \end{description}

    \paragraph*{Algorithm for 
    $\textsc{InfiniteLeftEdgeConflict}(S_i,S_j,S_k,S_m;S_q)$.}
    
    \begin{description}
    \item[Step 1] We evaluate $SRT(q)=\text{\shadow}(S_i,S_j,S_k,S_q)$. 
    If $SRT(q)=\emptyset$ or $\RR$, we return \noconflict or 
    \fullconflict respectively. Otherwise, if 
    $SRT(q)$ has the form
    $(-\infty,\phi)$, $(\chi,+\infty)$, $(\chi,\phi)$ or $
    (-\infty,\phi)\cup(\chi,+\infty)$, then we go to Step 2a, 2b, 2c or 
    2d respectively. Note that $\phi$ and
    $\chi$ correspond to the Apollonius vertices $v_{ijkq}$ and 
    $v_{ikjq}$ respectively. 
    \item[Step 2a]  
    We evaluate $I=\text{\insphere}(S_i,S_k,S_j,S_m,S_q)$. If $I=+$ 
    then $v_{ikjm}\not\in\sh{S_q}$ or equivalently $v_{ijkq}\prec v_{ikjm}$;
    in this the predicate's outcome is \leftvertex. Otherwise, if 
    $I=-$ then $v_{ikjm}\in\sh{S_q}$ or equivalently 
    $v_{ikjm}\prec v_{ijkq}$ hence the predicate's outcome is 
    \fullconflict. 
    \item[Step 2b]
    We evaluate $I=\text{\insphere}(S_i,S_k,S_j,S_m,S_q)$. If $I=+$ 
    then $v_{ikjm}\not\in\sh{S_q}$ or equivalently $v_{ikjm}\prec v_{ikjq}$;
    in this the predicate's outcome is \noconflict. Otherwise, if 
    $I=-$ then $v_{ikjm}\in\sh{S_q}$ or equivalently 
    $v_{ikjq}\prec v_{ikjm}$ hence the predicate's outcome is 
    \rightvertex. 
    \item[Step 2c]
    We evaluate $I=\text{\insphere}(S_i,S_k,S_j,S_m,S_q)$. If $I=-$ 
    then $v_{ikjm}\in\sh{S_q}$ or equivalently 
    $v_{ikjq}\prec v_{ikjm}\prec v_{ijkq}$; 
    in this the predicate's outcome is \rightvertex. 
    Otherwise, if $I=+$ we call the $\text{\order}(S_i,S_j,S_k,S_m,S_q)$
    and we can determine whether $v_{ikjm}\prec v_{ikjq}\prec  v_{ijkq}$
    or $v_{ikjq}\prec  v_{ijkq} \prec v_{ikjm}$. In the former case, 
    the predicate returns \noconflict otherwise, in the latter case 
    it returns \verticesconflict.
    \item[Step 2d]
    We evaluate $I=\text{\insphere}(S_i,S_k,S_j,S_m,S_q)$. If $I=+$ 
    then $v_{ikjm}\not\in\sh{S_q}$ or equivalently 
    $v_{ikjq}\prec v_{ikjm}\prec v_{ijkq}$; 
    in this the predicate's outcome is \leftvertex. 
    Otherwise, if $I=-$ we call the $\text{\order}(S_i,S_j,S_k,S_m,S_q)$
    and we can determine whether $v_{ikjm}\prec v_{ikjq}\prec  v_{ijkq}$
    or $v_{ikjq}\prec  v_{ijkq} \prec v_{ikjm}$. In the former case, 
    the predicate returns \fullconflict otherwise, in the latter case 
    it returns \verticesconflict.
    \end{description}

  Since the evaluation of \textsc{InfiniteRightEdgeConflict} and 
  \textsc{InfiniteLeftEdgeConflict} only require the call of 
  the \shadow, \insphere and \order predicates we have proven the 
  following theorem.  

  \begin{theorem}
  The \inflconflict and \infrconflict predicates for hyperbolic trisectors 
  can be evaluated by determining the sign of quantities of algebraic degree
  at most 10 (in the input quantities).
  \end{theorem}






\section{Design and Analysis of SubPredicates} 
 \label{sec:algebraic_analysis}

 In this Section, we provide a detailed description on how to answer every 
 subpredicate involved in the algorithm presented in 
 Section~\ref{sub:the_main_algorithm}. For each primitive, besides
 analyzing how we derive the outcome, we also compute its algebraic 
 degree, \ie, the maximum algebraic degree of all quantities that have 
 to be evaluated to obtain the subpredicate's result.


\subsection{The \incone predicate} 
  \label{sub:the_incone_predicate_analysis}

  To answer the \incone$(S_a,S_b,S_c)$ predicate 
  we first determine the number of possible tangent planes 
  to the sites $S_a, S_b,$ and $S_c$ that leave them all 
  on the same side; there can be either 0, 1, 2 or 
  $\infty$ such planes. Bear in mind that the \incone predicate can
  be called only if no one of the spheres $S_a$ and $S_b$ are contained 
  inside another.

  If $S_a$ and $S_b$ have different radii, then 
  $\cone$ will denote the cone that contains and is tangent to these spheres, whereas $\cone^-$ will symbolize the symmetric cone with the same 
  axis and apex. Let us no consider each of the four possible cases 
  regarding the number of cotagent planes, since it is indicative of the 
  relative position of the three spheres.

  \begin{enumerate}
  \item 
  If no such plane exists, there are four cases to consider; 
  \begin{itemize}
   \item 
   $S_c$ lies strictly inside the cone $\cone$;  
   the predicate returns $\inside$.
    \item 
   $S_c$ lies strictly inside the cone $\cone^-$;
   the predicate returns  $\outside$.
   \item 
   $S_c$ fully intersects the cone $\cone$ in the sense that 
   there is a circle on $S_c$ that is outside $\cone$. In this case, 
   the predicate returns  $\outside$.
   \item There is a circle on $S_c$ that is outside $\cone$; $S_c$ could lie strictly inside the cone $\cone^-$ or fully intersect the cone $\cone$, 
   $\cone^-$ or both. In all these cases, the predicate returns $\outside$.
  \end{itemize}

  \item 
  If there is only one such plane, then ${S_c}$ touches ${\cone}$ in 
  a single point. There are three cases to consider; 
  \begin{itemize}
   \item $S_c$ lies strictly inside the cone $\cone$; the predicate 
   returns \ptouch.
   \item $S_c$ fully intersects the cone $\cone$ (there is a circle in 
   $S_c$ that is outside $\cone$). In this case, the predicate returns  
   $\outside$.
   \item $S_c$ is tangent to the cone $\cone^-$ at a single point.
   In this case, the predicate returns $\outside$.
  \end{itemize}

  \item 
  If there are two such planes, the spheres must lie in convex position,
  hence the predicate returns  $\outside$.
  \item 
  If there are infinite such planes, the spheres $S_a,S_b$ and $S_c$
  have collinear centers and the points of tangency of each sphere 
  with the cone is a single circle. The predicate returns $\ctouch$ 
  in this scenario. 
  \end{enumerate}

  In the case no cotagent plane to all sites $S_a,S_b$ and $S_c$ exist, 
  we must be able to tell if $S_c$ lies inside the cone $\cone^-$. 
  However, this check is only needed in the case $r_a\neq r_b$; if 
  $r_a=r_b$, the cone $\cone$ degenerates into a cylinder and 
  $\cone^-$ does not exist. 

  If $r_a=r_b=r$ we can immediately answer the predicate if $r_c>r$; 
  $S_c$ has a point outside the cylinder and the outcome is \outside. 
  Otherwise, we initially consider the case where $C_a,C_b$ and 
  $C_c$ are collinear; this is equivalent to the cross product of the 
  vectors $\ov{C_aC_c}$ and $\ov{C_aC_b}$ being zero. Since 
  $\ov{C_aC_c}\times\ov{C_aC_b}=(-D^{yz}_{abc},D^{xz}_{abc},-D^{xy}_{abc})$,
  it holds that $\ov{C_aC_c}\times\ov{C_aC_b}=\ov{0}$ if and only if
  $D^{xz}_{abc}=D^{xz}_{abc}=D^{xz}_{abc}=0$, which is a 2-degree demanding 
  operation. In the collinear scenario, we can easily answer the 
  \incone predicate based on the sign of $r_c-r$; if it is  
  negative or zero, the answer is  \inside or \ctouch, respectively. 

  If $r_a=r_b=r$ but the centers of $C_a,C_b$ and 
  $C_c$ are not collinear, then if $r_c=r$ we immediately answer 
  \outside. Otherwise, if $r_c-r<0$, we consider the sign of the quantity
  $A=d(C_c,\ell)+r_c-r$, where $\ell$ denotes the line going through 
  $C_a$ and $C_b$ and $d(C_c,\ell)$ denotes the Euclidean distance of 
  $C_c$ from $\ell$. If $A$ is positive or negative, 
  then the outcome of \incone predicate is \outside or \inside, respectively.
  Otherwise, if $A$ equals zero, the answer is \ptouch.
  Notice that $d(C_c,\ell)=|\ov{C_aC_c}\times\ov{C_aC_b}|/|\ov{C_aC_b}|$
  and therefore

  \begin{align}
  A
  &= \sgn(d(C_c,\ell)+r_c-r)
   = \sgn(|\ov{C_aC_c}\times\ov{C_aC_b}|/|\ov{C_aC_b}|+r_c-r) \\
  &= \sgn(|\ov{C_aC_c}\times\ov{C_aC_b}|+(r_c-r)|\ov{C_aC_b}|)
   = \sgn(|\ov{C_aC_c}\times\ov{C_aC_b}|^2-(r_c-r)^2|\ov{C_aC_b}|^2)\\
  &= \sgn(((D^{xz}_{abc})^2+(D^{xz}_{abc})^2+(D^{xz}_{abc})^2)
   -(r_c-r)^2((x_b-x_a)^2+(y_b-y_a)^2+(z_b-z_a)^2))
  \end{align}
  \noindent
  since $r_c-r<0$.

  Let us now consider the case $r_a\neq r_b$ in detail. First, observe that 
  we can assume without loss of generality that $r_a<r_b$ as this follows 
  from the definition of the \incone predicate. Indeed, since 
  \incone$(S_a,S_b,S_c)$ and \incone$(S_b,S_a,S_c)$ represent the same 
  geometric inquiry, we can exchange the notation of the spheres $S_a$ 
  and $S_b$ in case $r_a>r_b$. 

  Taking this into consideration, we denote 
  $K$ to be the apex of the cone $\cone$ and $\Pi_K$ to be the plane that 
  goes through $K$ and perpendicular to axis $\ell$ of the cone. We also 
  denote by $\Pi_K^+$ the half-plane defined by the plane $\Pi_K$ and 
  the centers $C_a$ and $C_b$, whereas its compliment half-plane is 
  denoted by $\Pi_K^-$.

  It is obvious that if the center of the sphere $S_c$ does not lie 
  in $\Pi_K^+$ the predicate must return $\outside$ since $S_c$ has 
  at least one point outside the cone $\cone$. Moreover, if the center 
  $C_c$ lies in $\Pi_K^+$ then $S_c$ cannot lie inside $\cone^-$ and this 
  case is ruled out. To check if $C_c$ lies on $\Pi_K^+$, we first 
  observe that $C_a$ and $C_b$ define the line $\ell$ hence for 
  every point $P$ of $\ell$ stands that $\ov{OP}=\ov{OC_a}+t\ov{C_aC_b}$ 
  for some $t\in\RR$.
  It is clear that, for a sphere with center $P(t)$ to be tangent 
  to the cone it must have radius $r(t)$ that 
  is linearly dependent with $t$, \ie, $r(t) = k_1t+k_0$. To evaluate 
  $k_1$ and $k_0$, we observe that for $t=0$, $P(0)\equiv C_a$ hence 
  $r(0)=r_a$ and respectively for $t=1$, $P(1)\equiv C_b$ hence 
  $r(1)=r_b$. We conclude that 
  $r(t)=t\cdot (r_b-r_a)+r_a, \ t\in\RR$.
  The cone apex lies on $\ell$ so
  $\ov{OK} = \ov{OC_a}+t_c\ov{C_aC_b}$
  for $t_c\in\RR$ such that
  $r(t_c)=0 $ or equivalently $ t_c = r_a/(r_a-r_b)$. 
  In this way we have evaluated the cone apex coordinates which derive 
  from the relation
  $\ov{OK} = \ov{OC_a}+\ov{C_bC_a}\cdot{r_a}/(r_b-r_a)$. 
  Since $\Pi_K$ is perpendicular to $\ell$ and therefore to $\ov{C_aC_b}$, 
  and $\ov{C_aC_b}$ points towards the positive side of $\Pi_K$, 
  the point $C_c$ lies on the positive half plane $\Pi_K^+$ iff the quantity
  $M = \ov{C_aC_b}\cdot\ov{KC_c}$ is strictly positive.

  To evaluate the sign of $M$ we have 
  \begin{align} \sgn(M) &= 
  \sgn(\ov{C_aC_b}\cdot\ov{KC_c})=\sgn(\ov{C_aC_b}\cdot(\ov{OC_c}-\ov{OK}))\\
  &= \sgn(\ov{C_aC_b}\cdot\ov{OC_c}-\ov{C_aC_b}\cdot\ov{OK})\\
  &= 
  \sgn(\ov{C_aC_b}\cdot(\ov{OC_c}-\ov{OC_a}-\dfrac{r_a}{r_b-r_a}\ov{C_bC_a})) \\
  &= \sgn(\ov{C_aC_b}\cdot((r_b-r_a)\ov{C_aC_c}+r_a\ov{C_aC_b}))\sgn(r_b-r_a)\\
  &= 
  \sgn((r_b-r_a)\ov{C_aC_c}\cdot\ov{C_aC_b}+r_a\ov{C_aC_b}\cdot\ov{C_aC_b})
  \end{align}
  \noindent
  so determining $\sgn(M)$ requires operations of degree 3, since 
  $r_b-r_a$ is strictly positive.

  If $M<0$ or $M=0$ and $r_c>0$ the predicate immediately 
  returns $\outside$. In the special case where $M=0$ and $r_c=0$, 
  the sphere $S_k$ is essentially the apex of the cone $\cone$ and 
  the predicate returns $\ctouch$. 

  Lastly and for the rest of this section, we consider the case $M>0$ in detail, as further analysis is required to answer the \incone 
  predicate. We break down our analysis depending on the collinearity 
  of the centers $C_a,C_b$ and $C_c$.

  \subsubsection{The Centers \texorpdfstring{$C_a,C_b,C_c$}{Ci,Cj,Ck} 
    are Collinear} 
  \label{ssub:incone_collinear_centers}

  If $C_a,C_b$ and $C_c$ are collinear, they all lie on the line $\ell$, 
  and therefore  $\ov{OC_c} = \ov{OC_a}+t_o\ov{C_aC_b}$
  for some $t_o\in\RR$. Equivalently, we get that
  $t_o\ov{C_aC_b} = \ov{C_aC_c}$ and since $C_a$ and $C_b$ cannot be 
  identical we can evaluate $t_o=X/Y$ where 
  $(X,Y)=(x_c-x_a,x_b-x_a)$ or $(y_c-y_a,y_b-y_a)$ or $(z_c-z_a,z_b-z_a)$,
  if $x_b-x_a\neq 0$ or $y_b-y_a\neq 0$ or $z_b-z_a\neq 0$ respectively. 

  Denote $r(t)$ as before, we evaluate the sign $S$ of $r_c-r(t_o)$, 
  \begin{align}
  S &= \sgn(r_c-r(t_o)) = \sgn(r_c-r_a-(r_b-r_a)t_o) \\
  &= \sgn(r_c-r_a-(r_b-r_a)X/Y)\\
  &= \sgn(Y)\sgn((r_c-r_a)Y-(r_b-r_a)X)\\
  &= -\sgn(Y)\sgn( (r_a-r_c)Y +(r_b-r_a)X )
  \end{align}
  \noindent
  which requires operations of degree at most 2.

  We can now answer the predicate because if $r_c<r(t_o)$, \ie, $S$
  is negative, then $S_c$ lies 
  strictly inside the cone; otherwise, if $r_c>r(t_o)$,  \ie, $S$ is 
  positive, then $S_c$ intersects the cone. 
  If $r_c=r(t_c)$ or equivalently $S$ is zero, then $S_c$ touches $\cone$ in a circle.
  In conclusion, we get that 
  \begin{equation*}
  \text{\incone}(S_a,S_b,S_c) = 
  \begin{cases}
  \inside, &  \text{if } S<0,\\
  \ctouch, &  \text{if } S=0,\\
  \outside, & \text{if } S>0,
  \end{cases}
  \end{equation*}

  \subsubsection{Non-Collinear Centers} 
  \label{ssub:incone_non_collinear_centers}
  If $C_a,C_b$ and $C_c$ are not 
  collinear and must examine the number of possible tritangent 
  planes to the sites $S_n$ for $n\in\{a,b,c\}$.  
  Denote $\Pi: \kappa x+\lambda y+\mu z+\nu=0$ a plane tangent to 
  $S_a,S_b$ and $S_c$ that leaves the spheres on the same half-plane,
  and assume without loss of generality that $\kappa^2+\lambda^2+\mu^2=1$. Since the sphere $S_n$ for 
  $n\in\{i,j,k\}$ touches the plane $\Pi$, we get that
  $\delta(S_n,\Pi)=\kappa x_n+\lambda y_n+\mu z_n+\nu=r_n$. 

  We examine the resulting system of equations
   \begin{align}
  \kappa x_a+\lambda y_a+\mu z_a &= r_a-\nu\\
  \kappa x_b+\lambda y_b+\mu z_b &= r_b-\nu\\
  \kappa x_c+\lambda y_c+\mu z_c &= r_c-\nu\\
  \kappa^2+\lambda^2+\mu^2 &= 1
  \end{align}

  \noindent
  and distinguish the following cases
  \begin{itemize}
  \item 
  if $D^{xyz}_{abc}\neq 0$, we can express $\kappa,\lambda$ and $\mu$ linearly in terms of $\nu$.
  Substituting these expressions in the last equation, we 
  get a quadratic equation that vanishes at $d$; 
  the sign of the discriminant $\Delta'$
  of this quadratic reflects the number
  of possible $\nu$'s and therefore tangent planes to the spheres 
  $S_a,S_b$ and $S_c$. 
  \item 
  if $D^{xyz}_{abc}=0$ and since the centers of the spheres $S_a,S_b$ 
  and $S_c$ are not collinear, one of the quantities 
  $D^{xy}_{abc},D^{xz}_{abc}$ or $D^{yz}_{abc}$
  is non-zero, without loss of generality assume $D^{xy}_{abc}\neq 0$.
  In this case, we can express $\kappa$ and $\lambda$ linearly in terms 
  of $\mu$, 
  whereas
  $\nu = D^{xyr}_{abc}/D^{xy}_{abc}$. From the last equation, we get 
  a quadratic equation that vanishes at $\mu$, 
  and the sign of the discriminant $\Delta''$ again reflects 
  the number of possible $\mu$'s and therefore tangent planes to the 
  spheres $S_a,S_b$ and $S_c$. 
  \end{itemize}

  \noindent
  Writing down the expressions of the discriminants, we finally evaluate 
  that $\Delta' = 4(D^{xyz}_{abc})^2\Delta$ 
  and $ \Delta''=4(D^{xy}_{abc})^2\Delta$,  where 
  \begin{equation}
  \Delta = (D^{xy}_{abc})^2+(D^{xz}_{abc})^2+(D^{yz}_{abc})^2
   -(D^{xr}_{abc})^2-(D^{yr}_{abc})^2-(D^{zr}_{abc})^2.
  \end{equation}
  \noindent
  Since the signs of the discriminants $\Delta',\Delta''$ and $\Delta$
  are identical, we evaluate $\sgn(\Delta)$ and proceed as follows:
  \begin{enumerate}
  \item 
  If $\Delta>0$, there are two planes tangent to all three spheres 
  $S_a, S_b$ and $S_c$; the predicate returns $\outside$.
  \item 

  If $\Delta=0$, there is a single plane tangent to the spheres 
  $S_a,S_b$ and $S_c$ and we have to determine if $S_c^o$ lies strictly inside $\cone$ or not. In the former case, the predicate 
  returns $\ptouch$ whereas in the latter, $S_c^o$ intersects the 
  exterior of $\cone$ and the predicate returns $\outside$.

  Notice that, only if we are in the former case, there exists a proper
  $\epsilon>0$ such that the inflated sphere $\tilde{S_c}$, with 
  radius $\tilde{r_c}=r_c+\epsilon$, will point-touch the cone $\cone$.
  (Note that the tangency points of the cone $\cone$ and the
  spheres $S_c$ and $\tilde{S_c}$  are not the same!) Therefore, if we 
  consider the analysis of the predicate \incone$(S_a,S_b,\tilde{S_c})$, 
  we will conclude that the ``perturbed'' discriminant 
  $\tilde{\Delta}$ vanishes for this $\epsilon>0$.


  For the evaluation of $\tilde{\Delta}$, we simply substitute 
  $r_c$ in $\Delta$ with $\tilde{r_c}=r_c+\epsilon$
  and rewrite $\tilde{\Delta}=\tilde{\Delta}(\epsilon)$ as a polynomial 
  in terms of $\epsilon$:
  $\tilde{\Delta}(\epsilon)=\Delta_2\epsilon^2+\Delta_1\epsilon+\Delta_0$,
  where $\Delta_0 = \Delta=0$ and 
  \begin{align}
  \Delta_2 &= -[(x_b-x_a)^2+(y_b-y_a)^2+(z_b-z_a)^2] ( <0 ),\\
  \Delta_1 &= -2\big((x_b-x_a)D^{xr}_{abc}+(y_b-y_a)D^{yr}_{abc}
    +(z_b-z_a)D^{zr}_{abc}\big).
  \end{align}

  Since $\epsilon=0$  is a root of the quadratic (in terms of $\epsilon$) 
  $\tilde{\Delta}(\epsilon)$, a simple use of Vieta's formula shows that 
  $\tilde{\Delta}(\epsilon)$ has a positive root, if and only if 
  $\sgn(\Delta_1)$ is strictly positive. In conjunction with our previous 
  remarks, the predicate should return $\ptouch$ 
  if $\Delta_1>0$; otherwise it should return $\outside$. 
  \item 
  If $\Delta<0$, there is no plane tangent to the 
  spheres $S_a,S_b$ and $S_c$. Since $S_c$ lying inside the cone $\cone^-$ is 
  ruled out (due to the sign of $M$ being positive), 
  we have to distinguish between the two possible cases; 
  $S_c$ lies strictly inside $\cone$ or $S_c$ intersects $\vartheta\cone$.

  It follows that, either $S_c$ lies strictly within the cone and the
  predicate must return $\inside$, or $S_c$ fully intersects the cone 
  and the predicate must return $\outside$. 
  Using the same analysis as in case $\Delta=0$, we observe that
  if we inflate (or deflate) $S_c$, the perturbed sphere 
  $\tilde{S}_c$ with radius $\tilde{r_c}=r_c+\epsilon$ will
  touch the cone $\cone$ for two different values $\epsilon_1$ and 
  $\epsilon_2$. 
  The predicate must return $\inside$ if we must inflate $S_c$ 
  to touch $\cone$,  \ie, if $\epsilon_1,\epsilon_2>0$ whereas 
  the predicate must return $\outside$ if we must deflate $S_c$, 
  to point-touch $\cone$,  \ie, if $\epsilon_1,\epsilon_2<0$. 
  As shown in the case $\Delta=0$, the perturbed discriminant that will 
  appear during the evaluation of \incone$(S_a,S_b,\tilde{S_c})$ is
  $\tilde{\Delta}(\epsilon)=\Delta_2\epsilon^2+\Delta_1\epsilon+\Delta_0$
  where
  \begin{align}
  \Delta_2 &= -[(x_b-x_a)^2+(y_b-y_a)^2+(z_b-z_a)^2] ( <0 ),\\
  \Delta_1 &= -2\big((x_b-x_a)D^{xr}_{abc}+(y_b-y_a)D^{yr}_{abc}
    +(z_b-z_a)D^{zr}_{abc}\big),\\
  \Delta_0 &= \Delta < 0.
  \end{align}
  
  Since the sphere $\tilde{S_c})$ point-touches the cone $\cone$ for 
  $\epsilon=\epsilon_1,\epsilon_2$, the discriminant 
  $\tilde{\Delta}(\epsilon)$ must vanish for these epsilons, as mentioned 
  in the previous case. Therefore, $\epsilon_1$ and $\epsilon_2$ 
  are the roots of the quadratic (in terms of $\epsilon$)
  $\Delta_2\epsilon^2+\Delta_1\epsilon+\Delta_0$ and we know, using 
  Vieta's rule and the fact that $\Delta_0<0$, that 
  $\epsilon_1,\epsilon_2$ are both negative (resp., positive) if 
  and only if $\Delta_1$ is negative (resp., positive). In summary, 
  if $\Delta_1$ is positive or negative, the predicate returns 
  \inside or \outside, respectively.
  \end{enumerate}

  The analysis of this section summarizes to the following algorithm 
  that answers the $\text{\incone}(S_a,S_b,S_c)$ predicate.

  \begin{description}
  \item[Step 1] If $C_c=C_n$ for $n\in\{a,b\}$ return 
  \outside, \inside or \ctouch if the quantity $r_c-r_n$ is positive, negative or zero, respectively. Otherwise, go to step 2
  \item[Step 2] If $r_a=r_b$ go to Step 3, otherwise go to Step 5.
  \item[Step 3] If $r_c>r_a$ return \outside. Otherwise, evaluate 
  the quantities $D^{xz}_{abc}, D^{xz}_{abc}$ and $D^{xz}_{abc}$. 
  If all equal zero go to Step 4a, otherwise go to Step 4b.
  \item[Step 4a] If $r_c=r_a$ return \ctouch otherwise, if 
  $r_c<r_a$ return \inside.
  \item[Step 4b] If $r_c=r_a$ return \outside, otherwise 
  evaluate $A=\sgn(((D^{xz}_{abc})^2+(D^{xz}_{abc})^2+(D^{xz}_{abc})^2)-(r_c-r)^2((x_b-x_a)^2+(y_b-y_a)^2+(z_b-z_a)^2))$. If $A$ is positive, 
  negative or zero, return \outside, \inside or \ptouch, respectively.
  \item[Step 5] If $r_b<r_a$ exchange the notation
  of $S_a$ and $S_b$. Evaluate $M=\sgn(\ov{C_aC_c}\cdot\ov{C_aC_b}(r_b-r_a)+r_a\ov{C_aC_b}\cdot\ov{C_aC_b})$ and if $M<0$ or $M=0$ and $r_c>0$
  return \outside. Otherwise, if $M=r_c=0$ return \ctouch. In any other
  case, go to Step 6. 
  \item[Step 6] Evaluate the quantities $D^{xz}_{abc}, D^{xz}_{abc}$ 
  and $D^{xz}_{abc}$. If all equal zero, go to Step 7, otherwise go 
  to step 8.
  \item[Step 7] Evaluate $S= -\sgn(Y)\sgn( (r_a-r_c)Y +(r_b-r_a)X )$
  where $(X,Y)=(x_c-x_a,x_b-x_a)$ or $(y_c-y_a,y_b-y_a)$ or
  $(z_c-z_a,z_b-z_a)$, if $x_b-x_a\neq 0$ or $y_b-y_a\neq 0$ or 
  $z_b-z_a\neq 0$, respectively. If $S$ is positive, negative or zero, 
  we return \outside, \inside or \ctouch, respectively.
  \item[Step 8] Evaluate $\Delta = (D^{xy}_{abc})^2+(D^{xz}_{abc})^2+(D^{yz}_{abc})^2 -(D^{xr}_{abc})^2-(D^{yr}_{abc})^2-(D^{zr}_{abc})^2$. 
  If $\Delta$ is positive, we return \outside. Otherwise, if 
  $\Delta$ is zero or negative, we go to Step 9a or 9b, respectively.
  \item[Step 9a] Evaluate $\Delta_1 = -2\big((x_b-x_a)D^{xr}_{abc}+(y_b-y_a)D^{yr}_{abc} +(z_b-z_a)D^{zr}_{abc}\big)$. If $\Delta_1>0$ we return 
  \ptouch, otherwise we return \outside. 
  \item[Step 9b] Evaluate $\Delta_1 = -2\big((x_b-x_a)D^{xr}_{abc}+(y_b-y_a)D^{yr}_{abc} +(z_b-z_a)D^{zr}_{abc}\big)$. If $\Delta_1>0$ we return 
  \inside, otherwise we return \outside. 
  \end{description}

  Note that, since the algorithm that decides the \incone 
  predicate demands the evaluation of quantities of  
  degree at most 4 (in the input quantities), we have proven 
  the following lemma.

  \begin{lemma}
  The \incone predicate can be evaluated by determining 
  the sign of quantities of algebraic degree at most 4 
  (in the input quantities).
  \end{lemma}

 \subsection{The \tritype predicate} 
  \label{sub:the_tritype_predicate_analysis}
\begin{figure}[htbp]
   \centering
   \includegraphics[width=0.4\textwidth]{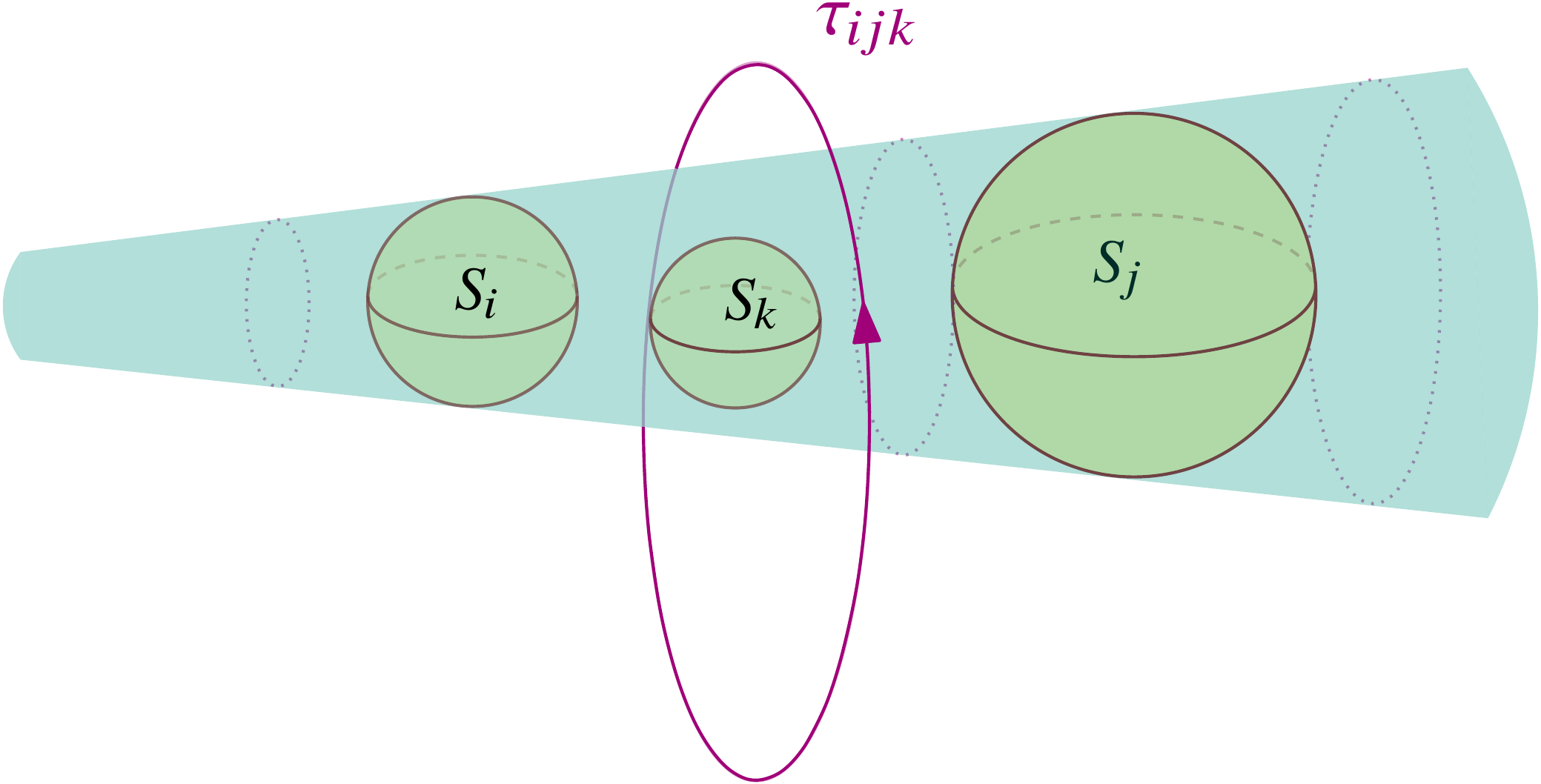} \hspace{1cm}
   \includegraphics[width=0.4\textwidth]{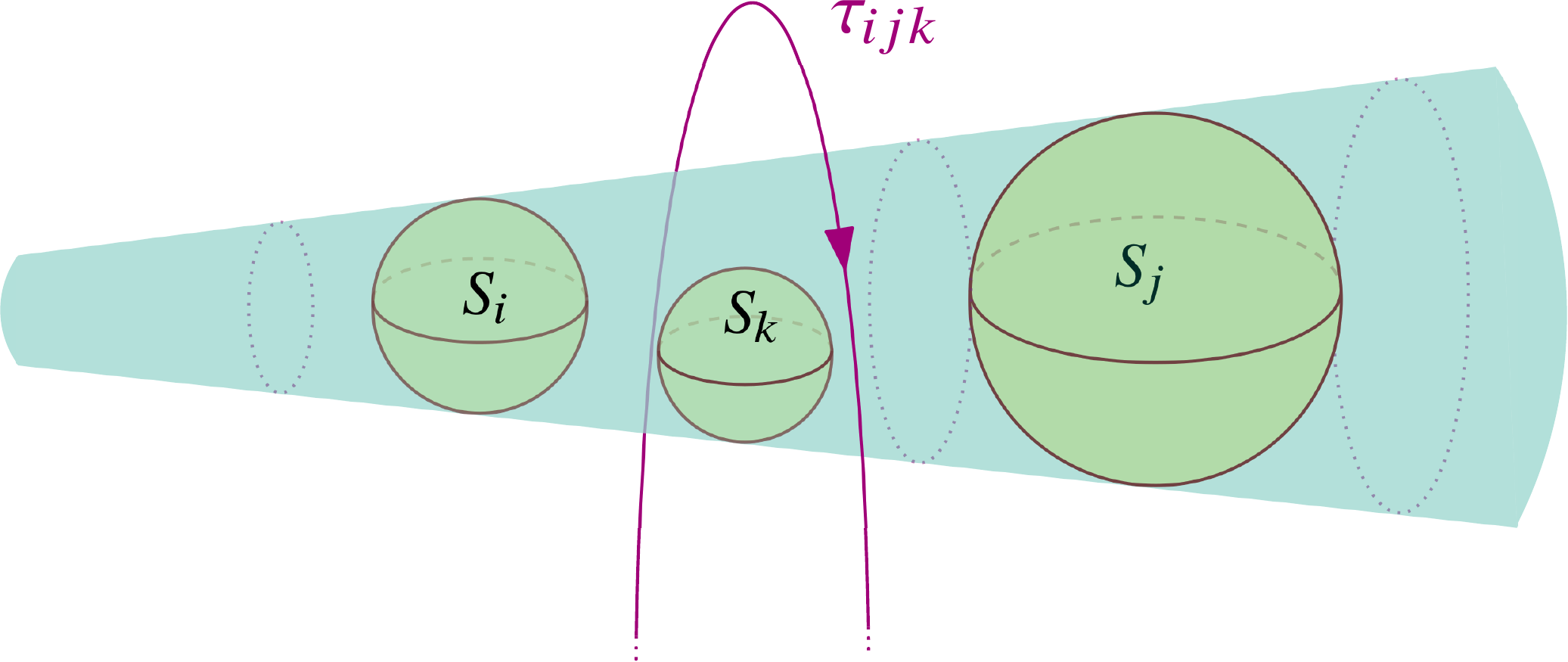}
   \caption[An elliptic (left) and a parabolic (right) trisector.]{The case of an \emph{elliptic} (Left) and a 
   \emph{parabolic} trisector (Right). }
   \label{fig:07}
  \end{figure}

  Assuming the trisector $\tri{ijk}$ exists and has Hausdorff dimension 1,
  observe that the trisector's type and the relative position of the 
  spheres are closely related.
  Specifically, 
  \begin{itemize}
  \item 
  If the spheres are in \emph{convex position}, \ie there exist two 
  distinct commonly tangent planes that leave them on the same side, 
  then their trisector can either be a hyperbola or a line, in the special 
  case $r_i=r_j=r_k$. The predicate returns ``hyperbolic'' in this scenario.
  \item 
  If the spheres are in in \emph{strictly non-convex position}, \ie 
  one of them lies strictly inside the cone defined by the other two, 
  then their trisector can either be an ellipse or a circle, 
  in the special case where $C_i,C_j$ and $C_k$ are collinear.
  The predicate returns ``elliptic'' in this scenario (see 
  Figure~\ref{fig:07}, Left).
  \item 
  If the spheres are in \emph{degenerate non-convex position}, \ie 
  they are in non-convex position and the closure of all three touch 
  their convex hull, then their trisector is a parabola and the 
  predicate returns ``parabolic''(see 
  Figure~\ref{fig:07}, Right).
  \end{itemize} 

  Therefore, one can answer the \tritype predicate if the relative position of the spheres three input spheres is identified. We accomplish such 
  task by combining the  outcomes of the three \incone predicates with inputs
  $(S_i,S_j,S_k)$, $(S_i,S_k,S_j)$ and $(S_j,S_k,S_i)$.
  \begin{itemize}
    \item If at least one outcome is $\inside$, 
  then the spheres are in \emph{strictly non-convex position} and the 
  trisector's type is ``elliptic''.
    \item If at least one outcome is 
  $\ptouch$, then the spheres are in \emph{degenerate non-convex 
  position} and the \tritype predicate returns ``parabolic''.
    \item Finally, if all three outcomes are $\outside$, then the sites are
  in \emph{convex position} and the trisector's type is ``hyperbolic''.
  \end{itemize}

  We have argued that the \tritype predicate can be resolved by calling 
  the \incone predicate at most three times (for example, if the first 
  \incone returns $\inside$ the trisector must be ``elliptic''). Since
  the \incone predicate is a 4-degree demanding operation in the input 
  quantities, we have proven the following lemma. 

  \begin{lemma}
  The \tritype predicate can be evaluated by determining 
  the sign of quantities of algebraic degree at most 4 
  (in the input quantities).
  \end{lemma}

 \subsection{The \exist predicate} 
  \label{sub:the_existence_predicate_analysis}

  To answer the \exist$(S_i,S_j,S_k,S_a)$ predicate, we 
  first break up our analysis depending on whether the radii of 
  all sites $S_n$, for $n\in\{i,j,k,a\}$, are equal; if $r$ is 
  indeed their common radius, we deflate them by $r$. 
  The existence of a sphere tangent to the original spheres amounts 
  to the existence of a sphere tangent to the centers of the 
  sites where the latter cotangent sphere is the former inflated by $r$.
  Regarding the existence of the Apollonius sphere of the centers
  $C_n$ for $n\in\{i,j,k,a\}$, we consider the tetrahedron
  $C_iC_jC_kC_a$ formed and check it for flatness; this is a 
  3-degree demanding operation. 
  If it is not flat, there are is a 
  single sphere $S$ passing through all the centers, hence there is only 
  one external Apollonius sphere and therefore the \exist predicate 
  returns 1. 
  If the tetrahedron is flat, \ie, the centers lie on the same 
  plane $\Pi_{ijka}$, we check them for co-circularity
  (this is a 5-degree demanding operation); 
  if co-circular, there are infinite number of spheres tangent to 
  the sites (either internal or external), hence the \exist
  predicate returns $\infty$ (degenerate answer). If they are not 
  co-circular, there can not be a sphere tangent to all sites and the \exist
  predicate returns 0. Note that if the \exist
  predicate would return $\infty$ we apply a QSP scheme to resolve the 
  degeneracy.

  Let us now consider the case where the radii of the sites 
  $S_n$ for $ n\in\{i,j,k,a\}$ are not all equal.
  Since a reordering of the sites does not affect the \exist
  predicate, we reorder them such that 
  $r_a=\min\{r_i,r_j,r_k,r_a\}$. 
  We now deflate all spheres by $r_a$ and then invert all sites 
  with respect to the point $C_a$; we call this ``inversion 
  through the sphere $S_a$''. 

  In the inverted \wspace, a plane tritangent to the inverted 
  spheres $\inv{S_i},\allowbreak\inv{S_j},\allowbreak\inv{S_k}$  
  amounts to a sphere 
  tangent to all sites $S_i, S_j, S_k $ and $S_a$ in the original 
  \zspace.
  For the corresponding sphere to be an external Apollonius sphere 
  of the sites,  the following conditions must stand:
  \begin{enumerate}
  \item 
  The plane must leave all inverted spheres on one side, called
  the positive side of the plane, and 
  \item 
  The origin $\OO=(0,0,0)$ of the \wspace, that corresponds to 
  the ``point at infinity'' in the \zspace, must also lie on the 
  positive side of the plane. Geometrically, this means that we 
  are looking for an external Apollonius sphere. 
  \end{enumerate}

  Considering Condition 1, we denote $\inv{\Pi}_{ijk}: au+bv+cw+d=0$ 
  a plane tangent to 
  $\inv{S_i},\inv{S_j}$ and $\inv{S_k}$ in the inverted space 
  and assume without loss of generality that $a^2+b^2+c^2=1$. 
  Since the signed Euclidean distance of a point $P(u_p,v_p,w_p)$ 
  from the plane $\inv{\Pi}_{ijk}$ is 
  $\delta(P,\Pi_{ijk})=au_p+bv_p+cw_p+d$ and the sphere 
  $\inv{S_n}$, for $ n\in\{i,j,k\}$, touches $\inv{\Pi}_{ijk}$ and lies on 
  its positive side, we get that 
  $\delta(\inv{C_n},\Pi_{ijk})=au_n+bv_n+cw_n+d=\rho_n$. 

  A tuple $(a,b,c,d)$ that satisfies the resulting 
  system of equations, 
  amounts to a tangent plane in \wspace and an Apollonius 
  sphere in \zspace. 
  In order for condition 2 to be valid, 
  the point $\OO=(0,0,0)$ of the \wspace must lie on the positive
  side of $\Pi_{ijk}$, \ie, the signed distance of $\OO$ from the plane 
  $\Pi_{ijk}$ must be positive. Equivalently, we want 
  $\delta(\mathcal{O},\Pi_{ijk})= d $ to be
  positive, hence we are only interested in 
  the solutions $(a,b,c,d)$ that satisfy $d>0$. 

  The rest of this section is devoted to the algebraic analysis of the 
  aforementioned system of equations to determine the number of such 
  solutions with the minimum algebraic cost. The conclusion of our analysis
  is that such a task is possible by evaluating expressions of algebraic
  degree at most 8 (in the input quantities) yielding the lemma at the 
  end of this section. 

  Our main tool is Crammer's rule and therefore two major cases rise 
  during the analysis of the system
  \begin{align}
  au_i+bv_i+cw_i &= \rho_i-d,\\
  au_j+bv_j+cw_j &= \rho_j-d,\\
  au_k+bv_k+cw_k &= \rho_k-d,\\
  a^2+b^2+c^2 &= 1.
  \end{align}

  If $D^{uvw}_{ijk}\neq 0$, we can express $a,b$ and $c$ in
  terms of $d$ as follows
  \begin{equation}
  a = \dfrac{D^{vw\rho}_{ijk}-d D^{vw}_{ijk}}{D^{uvw}_{ijk}}, \quad
  b = -\dfrac{D^{uw\rho}_{ijk}-d D^{uw}_{ijk}}{D^{uvw}_{ijk}}, \quad
  c = \dfrac{D^{uv\rho}_{ijk}-d D^{uv}_{ijk}}{D^{uvw}_{ijk}}.
  \end{equation} 

  We will then substitute the expressions of $a,b$ and $c$ to the 
  equation $a^2+b^2+c^2=1$ and conclude that $d$ is a root
  of $M(d)=M_2d^2+M_1d+M_0$, where 
 
  \begin{align}
  M_2 &= (D^{uv}_{ijk})^2+(D^{uw}_{ijk})^2+(D^{vw}_{ijk})^2, \\
  M_1 &= D^{vw\rho}_{ijk}D^{vw}_{ijk}+D^{uw\rho}_{ijk}D^{uw}_{ijk}
      +D^{uv\rho}_{ijk}D^{uv}_{ijk},\\
  M_0 &= (D^{vw\rho}_{ijk})^2+(D^{uw\rho}_{ijk})^2+(D^{uv\rho}_{ijk})^2
      -(D^{uvw}_{ijk})^2.
  \end{align}

  The signs of $M_1$ and $M_0$ are determined using the  
  the following equalities
  \begin{align}
  \sgn(M_1) 
  &= -\sgn(D^{vw\rho}_{ijk}D^{vw}_{ijk}+D^{uw\rho}_{ijk}D^{uw}_{ijk}
    +D^{uv\rho}_{ijk}D^{uv}_{ijk})\\
  &= -\sgn(E^{yzr}_{ijk}E^{yzp}_{ijk}+E^{xzr}_{ijk}E^{xzp}_{ijk}
    +E^{xyr}_{ijk}E^{xyp}_{ijk}),\\
  \sgn(M_0) 
  &= \sgn((D^{vw\rho}_{ijk})^2+(D^{uw\rho}_{ijk})^2+(D^{uv\rho}_{ijk})^2
    -(D^{uvw}_{ijk})^2)\\
  &= \sgn((E^{yzr}_{ijk})^2+(E^{xzr}_{ijk})^2+(E^{xyr}_{ijk})^2
    -(E^{xyz}_{ijk})^2),
  \end{align}
  \noindent
  where $M_2$ is always positive unless $E^{xyp}_{ijk}, E^{xzp}_{ijk}$ 
  and $E^{yzp}_{ijk}$ are all zero; in this case $M_2=0$.
  The expressions that appear in the evaluation of $M_2, M_1$ and 
  $M_0$ have maximum algebraic degree 7 in the input quantities.


  First, we shall consider the case $M_2\neq 0$; 
  this is geometrically equivalent to the non-collinearity of the 
  inverted centers $\inv{C_n}$ for $ n\in\{i,j,k\}$. Since $M_2$ 
  is assumed to be strictly positive, the quadratic polynomial $M(d)$ has 
  0, 1 or 2 real roots depending on whether the sign of the discriminant
  $\Delta_M = M_1^2-4M_2M_0$ is negative, zero or positive 
  respectively. We evaluate the discriminant $\Delta_M$ of $M(d)$ to be
  \begin{equation}
  \Delta_M = 4 (D^{uvw}_{ijk})^2 \big((D^{uv}_{ijk})^2+(D^{uw}_{ijk})^2
   +(D^{vw}_{ijk})^2-(D^{u\rho}_{ijk})^2-(D^{v\rho}_{ijk})^2-(D^{w\rho}_{ijk})^2
   \big),
  \end{equation}
  hence determining $\sgn(\Delta_M)$ requires 8-fold algebraic operations
  since
  \begin{equation}
  \sgn(\Delta_M) = 
  \sgn \left( (E^{xyp}_{ijk})^2+(E^{xzp}_{ijk})^2+(E^{yzp}_{ijk})^2
   -(E^{xrp}_{ijk})^2-(E^{yrp}_{ijk})^2-(E^{zrp}_{ijk})^2 \right).
  \end{equation}

  If $\Delta_M<0$, the predicate returns ``0'', whereas if $\Delta_M = 0$
  we get that $M(d)$ has a double root $d = -M_1/(2M_2)$; 
  the predicate returns ``1 double'' if $\sgn(d) = -\sgn(M_1)$ is positive, 
  otherwise it returns ``0''. Finally, if $\Delta_M>0$, $M(d)$ 
  has two distinct roots $d_1<d_2$, whose sign we
  check for positiveness. Using Vieta's rules and 
  since $\sgn(M_1) = -\sgn(d_1+d_2)$ and
  $\sgn(M_0) = \sgn(d_1)\sgn(d_2)$ we conclude that, 
  if $M_0$ is negative the predicate returns ``1''. 
  Otherwise, the predicate's outcome is ``0'' if $\sgn(M_1)$
  is positive or ``2'' if $\sgn(M_1)$ is negative. 

  We now analyse the case where $M_2=0$. Since the quantities 
  $D^{uv}_{ijk}, D^{uw}_{ijk}$ and $D^{vw}_{ijk}$ are all zero, we 
  get that $a = D^{vw\rho}_{ijk}/D^{uvw}_{ijk}$
  $b = D^{uw\rho}_{ijk}/D^{uvw}_{ijk}$ and 
  $c = D^{uv\rho}_{ijk}/D^{uvw}_{ijk}$. Therefore, since 
  $d = \rho_i-(au_i+bv_i+cw_i)$ we can evaluate the sign of $d$ as
  \begin{align}
   \sgn(d) &=\sgn(D^{uvw}_{ijk})
   \sgn(D^{uvw}_{ijk}\rho_i-D^{vw\rho}_{ijk}u_i-D^{uw\rho}_{ijk}v_i-D^{uv\rho}_{ijk}w_i)\\
   &= \sgn(E^{xyz}_{ijk})\sgn(E^{xyz}_{ijk}r^{\star}_i-E^{yzr}_{ijk}x^{\star}_i-E^{xzr}_{ijk}y^{\star}_i-E^{xyr}_{ijk}z^{\star}_i)
  \end{align}
 
  The evaluation of the sign of $d$ therefore demands operations of algebraic degree 4 (in the input quantities). If $d>0$ the predicate 
  returns ``1'' otherwise it returns ``0''.

  Last, we consider the case $D^{uvw}_{ijk}=0$. We can safely assume that 
  at least one of the quantities $D^{uv}_{ijk},D^{uw}_{ijk}$ 
  and $D^{vw}_{ijk}$ is non-zero since otherwise the centers 
  $\inv{C_n}$ for $ n\in\{i,j,k\}$ would be collinear
  \footnote{If $D^{uv}_{ijk}=D^{uw}_{ijk}=D^{vw}_{ijk}=0$, 
  the projections of the points $\inv{C_n}$, $n\in\{i,j,k\}$, on all
  three planes $w=1$, $v=1$ and $u=1$ would form a flat triangle. 
  For each projection, this is equivalent to either 
  some of the projection points coinciding or all three being collinear. 
  Since the original centers $\inv{C_n}$ are distinct points for 
  $n\in\{i,j,k\}$, they must be collinear for such 
  a geometric property to hold.}
  ,yielding a contradiction.
  Assume without loss of generality that $D^{uv}\neq 0$, we can solve the 
  system of equations in terms of $c$ and we get that
  \begin{equation}
  a = \dfrac{-D^{v\rho}_{ijk}+cD^{vw}_{ijk}}{D^{uv}_{ijk}},\quad
  b = \dfrac{D^{u\rho}_{ijk}-cD^{uw}_{ijk}}{D^{uv}_{ijk}},\quad
  d = \dfrac{D^{uv\rho}_{ijk}}{D^{uv}_{ijk}}.
  \end{equation}

  If we substitute $a$ and $b$ in the equation $a^2+b^2+c^2=1$, 
  we get that $c$ is a root of a quadratic polynomial
  $L(c)=L_2c^2+L_1c+L_0$, where

  \begin{align}
  L_2 &= (D^{uv}_{ijk})^2+(D^{uw}_{ijk})^2+(D^{vw}_{ijk})^2,\\
  L_1 &= -2(D^{v\rho}_{ijk}D^{vw}_{ijk}+D^{u\rho}_{ijk}D^{uw}_{ijk}),\\
  L_0 &= (D^{u\rho}_{ijk})^2+(D^{v\rho}_{ijk})^2-(D^{uv}_{ijk})^2.
  \end{align}
  We evaluate the discriminant of $L(c)$ to be
  \begin{equation}
  \Delta_L = 4 (D^{xy}_{ijk})^2 \left[(D^{xy}_{ijk})^2+(D^{xz}_{ijk})^2+(D^{yz}_{ijk})^2
   -(D^{xr}_{ijk})^2-(D^{yr}_{ijk})^2-(D^{zy}_{ijk})^2 \right],
  \end{equation}
  and therefore $\sgn(\Delta_L)=\sgn(\Delta_M)$. The evaluation of  
  $\sgn(\Delta_M)$ is known to require 8-fold algebraic operations 
  as shown in a previous case.

  If $\Delta_M<0$, the predicate returns ``0''; there is no tangent plane
  in the inverted space. Otherwise, we determine the sign of $d$ to be
  \begin{equation}
  \sgn(d)=\sgn\big(D^{uv\rho}_{ijk}/D^{uv}_{ijk}\big)
         =\sgn\big(E^{xyr}_{ijk}/E^{xyp}_{ijk}\big)
         = \sgn(E^{xyr}_{ijk})\sgn (E^{xyp}_{ijk}) 
  \end{equation}
  and the following cases arise
  \begin{enumerate}
  \item 
  If $\Delta_M=0$, the predicate returns ``0'' if $d<0$ or ``1'' if $d>0$.
  \item 
  If $\Delta_M>0$, the predicate returns ``0'' if $d<0$ or ``2'' if $d>0$.
  \end{enumerate}

  Note that in the last case, the algebraic degrees of $\Delta_M$ and 
  $d$ are 8 and 4 respectively and that, in every possible scenario, 8 
  was the maximum algebraic degree of any quantity we had to evaluate. 
  
  \begin{lemma}
  The \exist predicate can be evaluated by determining 
  the sign of quantities of algebraic degree at most 8 
  (in the input quantities).
  \end{lemma}


 \subsection{The \distance predicate} 
  \label{sub:the_distance_predicate_analysis}

 In this section, we provide a detailed analysis regarding the evaluation of 
 the $\text{\distance}(S_i,\allowbreak S_j,\allowbreak S_k,\allowbreak S_a)$ 
 predicate, as this was defined 
 in Section~\ref{ssub:the_distance_predicate}. As stated there, the outcome 
 of this primitive is the tuple 
 $(\sgn{(\delta(S_a,\Pi_{ijk}^{-}))},\sgn{(\delta(S_a,\Pi_{ijk}^{+}))})$, 
 where the planes $\Pi_{ijk}^{-}$ and $\Pi_{ijk}^{+}$ are commonly tangent 
 to the sites $S_i,S_j$ and $S_k$. The existence of these planes is 
 guaranteed since the trisector $\tri{ijk}$ is assumed to be ``hyperbolic'' 
 (see Section~\ref{ssub:the_incone_and_tritype_predicates}). Also take into
 consideration that in the scope of this chapter, this subpredicate 
 never returns a degenerate answer, \ie, neither 
 $(\sgn{(\delta(S_a,\Pi_{ijk}^{-}))}$ nor 
 $\sgn{(\delta(S_a,\Pi_{ijk}^{+}))})$ can equal zero.

 We shall now consider such a plane $\Pi: ax+by+cz+d=0$ tangent to all
 sites $S_i,S_j$ and $S_k$, that leaves them all on the same side 
 (this site is denoted as the \emph{positive} side). 
 If we assume without loss of generality
 that $a^2+b^2+c^2=1$, it must stand that
 $\delta(C_n,\Pi)=ax_n+by_n+cz_n+d=r_n$ for $ n\in\{i,j,k\}$, where 
 $\delta(C_n,\Pi)$ denotes the signed Euclidean of the center 
 $C_n$ from the plane $\Pi$. If we consider the distance  
 $\epsilon=\delta(S_a,\Pi)=\delta(C_a,\Pi)-r_a$ of the sphere $S_a$ from 
 this plane, then the following system of equations must hold.

 \begin{align}
 ax_i+by_i+cz_i +d &= r_i, \\
 ax_j+by_j+cz_j +d &= r_j, \\
 ax_k+by_k+cz_k +d &= r_k, \\
 ax_a+by_a+cz_a +d &= r_a+\epsilon, \\
 a^2+b^2+c^2=1
 \end{align}

 Due to the initial assumption of a hyperbolic trisector, 
 only two such planes, $\Pi^{-}_{ijk}$ and $\Pi^{+}_{ijk}$, are cotangent 
 to the spheres $S_i,S_j$ and $S_k$ and therefore
 algebraicly satisfy the system of equations above.
 In other words, there exist only two distinct algebraic solutions
 $(a_\nu,b_\nu,c_\nu,d_\nu,\epsilon_\nu)$ for $\nu\in\{1,2\}$, 
 and apparently $\{\epsilon_1,\epsilon_2\}= \{\delta_a^+,\delta_a^-\}$, where $\delta_a^+=\delta(S_a,\Pi^{+}_{ijk})$ and $\delta_a^-=\delta(S_a,\Pi^{-}_{ijk})$.

 Bearing in mind that the answer to the \distance predicate is 
 actually the tuple $(\delta_a^+,\delta_a^-)$, 
 we want to determine the signs of $\epsilon_1$ and $\epsilon_2$ and 
 correspond them to $\delta_a^+$ and $\delta_a^-$. 
 Regarding the signs of $\{\epsilon_1,\epsilon_2\}$ 
 we algebraicly study the system of equations above. 

 First, we consider the case $D^{xyz}_{ijka}\neq 0$ in detail. Under this assumption and with the use of Crammer's rule, we may express 
 $a, b$ and $c$ with respect to $\epsilon$ as
 \begin{align}
 a = \dfrac{D^{yzr}_{ijka}-\epsilon D^{yz}_{ijk}}{D^{xyz}_{ijka}}, \quad
 b = \dfrac{-D^{xzr}_{ijka}+\epsilon D^{xz}_{ijk}}{D^{xyz}_{ijka}},\quad
 c = \dfrac{D^{xyr}_{ijka}+\epsilon D^{xy}_{ijk}}{D^{xyz}_{ijka}}. \quad
 \end{align}

 After substituting these expressions in the equation $a^2+b^2+c^2=1$,
 we obtain that 
 $\Lambda(\epsilon)=\Lambda_2\epsilon^2+\Lambda_1\epsilon+\Lambda_0=0$, 
 \begin{align}
 \Lambda_2 &= (D^{xy}_{ijk})^2+(D^{yz}_{ijk})^2+(D^{xz}_{ijk})^2, \\
 \Lambda_1 &= -2(D^{yzr}_{ijka}D^{yz}_{ijk}+D^{xzr}_{ijka}D^{xz}_{ijk}
              -D^{xyr}_{ijka}D^{xy}_{ijk} ), \\
 \Lambda_0 &= (D^{xyr}_{ijka})^2+(D^{xzr}_{ijka})^2+(D^{yzr}_{ijka})^2
              -(D^{xyz}_{ijka})^2.
 \end{align}

 Take into consideration that $\Lambda_2$ cannot be zero since otherwise
 the centers $C_i,C_j$ and $C_k$ would be collinear, yielding 
 a contradiction; we have assumed that $\tri{ijk}$ is 
 hyperbolic. Since $\Lambda(\epsilon)$, which is definitely a 
 quadratic in terms of $\epsilon$, has the aforementioned 
 $\epsilon_1$ and $\epsilon_1$ as roots, we may use Vieta's formula 
 to determine the signs of $\epsilon_1,\epsilon_2$. All we need is
 the signs of $\Lambda_1$ and $\Lambda_0$, quantities of algebraic degree 
 5 and 6 respectively (we already proved $\Lambda_2$ is 
  positive).

 If both roots are positive, negative or zero, the predicate returns 
 $(+,+)$, $(-,-)$ or $(0,0)$ respectively. If the sign
 of the roots differ, we must consider a way of distinguishing which 
 of them corresponds to $\delta_a^+$ and $\delta_a^-$. Since the 
 $\epsilon_1$ and $\epsilon_2$ have different signs, the set 
 $\{\epsilon_1,\epsilon_2\}=\{\delta_a^-,\delta_a^+\}$ is one of the 
 following: $\{+,-\},\{0,-\}$ or $\{0,+\}$. 

 The cases we now consider are products of geometric observations 
 based on the three possible configurations of the centers $C_i, C_j, C_k$
 and $C_a$:
 \begin{itemize}
  \item  If $D^{xyz}_{ijka}$ is positive, then $C_a$ lies 
 on the positive side of the plane $\Pi_{ijk}$. 
 In this case, only the geometric configurations
 $(\delta_a^-,\delta_a^+)= (+,-),(0,-)$ 
 or $(+,0)$ are possible. 
 \item  If $D^{xyz}_{ijka}$ is negative, then $C_a$ lies
 on the negative side of the plane $\Pi_{ijk}$. 
 In this case, only the geometric configurations
 $(\delta_a^-,\delta_a^+)= (-,+),(-,0)$ 
 or $(0,+)$ are possible. 
 \end{itemize}

 For example, if the roots of $\Lambda(\epsilon)$ turn out to be 
 one positive and one negative, the predicate will return $(+,-)$ if 
 $D^{xyz}_{ijka}$ is positive or $(-,+)$ if 
 $D^{xyz}_{ijka}$ is negative. 
 
 Lastly, we consider the case $D^{xyz}_{ijka}=0$, where the 
 centers $C_n$, for 
 $n\in\{\allowbreak i,\allowbreak j,\allowbreak k,\allowbreak a\}$, 
 are coplanar. In this context, 
 it is both algebraicly and geometricaly apparent that 
 $\epsilon_1=\epsilon_2=\epsilon$; the sphere $S_a$ either intersects, is tangent or does not intersect both planes $\Pi_{ijk}^-$ and $\Pi_{ijk}^+$.
 Specifically, if $D^{xyz}_{ijk}\neq 0$, then 
 $\epsilon= -{D^{xyzr}_{ijka}}/{D^{xyz}_{ijk}}$ and 
 we immediately evaluate 
 $\sgn(\epsilon)=-\sgn(D^{xyzr}_{ijka})\cdot\allowbreak\sgn(D^{xyz}_{ijk})$.
 If $D^{xyz}_{ijk}=0$ then at least one of the quantities 
 $D^{xy}_{ijk},D^{xz}_{ijk},D^{yz}_{ijk}$ does not equal zero, 
 since the centers $C_n$ for $n\in\{i,j,k\}$ are not collinear. 
 Assume without loss of generality that $D^{xy}_{ijk}\neq 0$, then
 $\epsilon={D^{xyr}_{ijka}}/{D^{xy}_{ijk}}$ and 
 we evaluate $\sgn(\epsilon)=\sgn(D^{xyr}_{ijka})\sgn(D^{xy}_{ijk})$. 
 In every case the predicate returns $(\sgn(\epsilon),\sgn(\epsilon))$. 
 
 Taking into consideration that the evaluation of $\Lambda_1$ is the most 
 degree-demanding operation to resolve the \distance predicate, we 
 have proven the following lemma.
 
 \begin{lemma}
 The \distance predicate can be evaluated by determining 
 the sign of quantities of algebraic degree at most 6 
 (in the input quantities).
 \end{lemma}


 \subsection{The \shadow Predicate} 
 \label{sub:the_shadowregion_predicate_analysis}
 In this section, we provide a way of resolving the 
 \shadow$(S_i,S_j,S_k,S_\alpha)$ predicate as this was described in 
 Section~\ref{ssub:the_shadow_predicate}. Assuming that  the trisector
 $\tri{ijk}$ is ``hyperbolic'' and no degeneracies occur, we have shown 
 that the outcome, which is the topological structure of $\sh{S_a}$ on 
 $\tri{ijk}$, is one of the following: $\emptyset$, $(-\infty,\infty)=\RR$, 
 $(-\infty,\phi)$, $(\chi,+\infty)$, $(\chi,\phi)$, or 
 $(-\infty,\phi)\cup(\chi,+\infty)$, where $\phi,\chi\neq\pm\infty$.

 Initially, the predicates $\text{\exist}(S_i,S_j,S_k,S_\alpha)$ and
 $\text{\distance}(S_i,S_j,S_k,S_\alpha)$ are called; let us 
 denote by $E$ and $(\sigma_1,\sigma_2)$ their respective outcomes. 
 The geometric interpretation of these two quantities leads to the 
 resolution of the shadow region $\sh{S_\alpha}$ with respect to 
 $\tri{ijk}$. 

 Regarding the meaning of the signs $\sigma_1$ and $\sigma_2$, 
 assume that $\sigma_1=+$. In this case, the sphere $S_\alpha$ does not 
 intersect the plane $\Pi_{ijk}^-$ which is in fact the Apollonius sphere
 $\tts{\invmap{-\infty}}$. In other words, $\invmap{-\infty}$ 
 does not belong to $\sh{S_\alpha}$, \ie, $-\infty$ does not ``show up'' in the predicate's outcome. 
 Using similar arguments, if $\sigma_2=+$ then $+\infty$ 
 does not show up in the outcome whereas, and if $\sigma_1$ or
 $\sigma_2$ are negative then $-\infty$ or 
 $+\infty$ shows up, respectively. Note that 
 $\sigma_1,\sigma_2\in\{+,-\}$ under the assumption of no degeneracies.

 Regarding the geometric interpretation of $E$, we have mentioned that 
 the boundary points of the closure of $\sh{S_\alpha}$ correspond to 
 centers of finite
 Apollonius spheres of the sites $S_n$, for $n\in\{i,j,k,\alpha\}$
 that are not centered at infinity. 
 We have shown in Section~\ref{sub:the_existence_predicate_analysis} that the
 cardinality of these Apollonius spheres is in fact $E$ and assuming no 
 degeneracies there are either 0, 1 or 2. 

 The combined information of $\sigma_1, \sigma_2$ and $E$ is used to 
 determine the type of $\sh{S_\alpha}$, as follows:
 \begin{itemize}
 \item 
 If $E=0$, the shadow region $\sh{S_\alpha}$ has no boundary 
 points, hence it's type is either $(-\infty,+\infty)$ 
 or $\emptyset$. If $\sigma_1=-$ (or $\sigma_2=-$), the predicate 
 returns $(-\infty,+\infty)$ otherwise, if $\sigma_1=+$ 
 (or $\sigma_2=+$), it returns $\emptyset$.
 \item 
 If $E=1$, then the closure of $\sh{S_\alpha}$ has one finite boundary 
 point, hence it's type is either $(-\infty,\phi)$ 
 or $(\chi,+\infty)$. If $\sigma_1=-$ (or $\sigma_2=+$), the
 predicate returns $(-\infty,\phi)$ otherwise, 
 if $\sigma_1=+$ (or $\sigma_2=-$) it returns $(\chi,+\infty)$).
 \item 
 Finally, if $E=2$ then the closure of $\sh{S_\alpha}$ has two 
 finite boundary points, hence it's type is either 
 $(\chi,\phi)$ or $(-\infty,\phi)\cup(\chi,+\infty)$. If
 $\sigma_1=+ $ (or $\sigma_2=+$) the predicate returns 
 $(\chi,\phi)$ otherwise, if $\sigma_1=-$ (or $\sigma_2=-$), it 
 returns $(-\infty,\phi)\cup(\chi,+\infty)$.
 \end{itemize}

  Not that all other combinations of the outcomes of the \exist and 
  \distance predicates correspond to degenerate shadow regions; these 
  are handled in detail in 
  Section~\ref{sub:the_perturbed_shadowregion_predicate}.

 Since the evaluation of the \shadow predicate only requires the call 
 of the \distance and the \exist predicates, which demand operations 
 of maximum algebraic degree 6 and 8 respectively, we proved the 
 following lemma.
 
 \begin{lemma}
 The \shadow predicate can be evaluated by determining 
 the sign of quantities of algebraic degree at most 8 
 (in the input quantities).
 \end{lemma}



 \subsection{The \order predicate} 
 \label{sub:the_order_predicate_analysis}

  The major subpredicate called during the evaluation of the 
  \conflict predicate via the algorithm described in 
  Section~\ref{sub:the_main_algorithm} is the so called \order predicate.
  As already described in Section~\ref{ssub:the_order_predicate}, 
  the \order$(S_i,S_j,S_k,S_a,S_b)$ returns the order of appearance 
  on the oriented trisector $\tri{ijk}$ of any Apollonius vertices 
  defined by the sites $S_i,S_j,S_k$ and $S_a$ or $S_b$. This primitive
  needs to be evaluated only if both $\sh{S_a}$ and $\sh{S_b}$ are 
  not of the form $\emptyset$ or $(-\infty,+\infty)$, with $\{a,b\}\in\{l,m,q\}$. 
  Indeed, for $a\in\{l,m\}$ we know that $v_{ijkl}$ and $v_{ikjm}$ exist and therefore $\sh{S_l}$ and $\sh{S_m}$ can not be of that type. Moreover, 
  $\sh{S_q}$ cannot be of the form $\emptyset$ or $(-\infty,+\infty)$; 
  the \order predicate is not necessary if this is the case.
  To conclude, the predicates \shadow$(S_i,S_j,S_k,S_n)$ for $n\in\{a,b\}$ are called in advance and their outcomes are considered to be known for the rest of the analysis. As already mentioned, $\tri{ijk}$ is also assumed to be a ``hyperbolic'' trisector.

  The analysis of the \order predicate is the major contribution of 
  this thesis. We demonstrate the usefulness 
  of the inversion technique by proving the strong connection between the original and the inverted space. By exploiting this relation, we are able 
  to create useful tools which can also be used in both 2D and 3D Apollonius
  diagrams to improve existing results. 

  In the rest of this Section, we introduce the inverted 
  space and make some initial observations on how it is connected with the
  original space. Afterwards, 
  we define  a 2-dimensional sub-space of the inverted space, 
  to make useful geometric observations effortless. Lastly, we break up 
  our analysis of the \order predicate 
  according to the shadow region types $\sh{S_a}$ and $\sh{S_b}$. Ultimately, we prove the following lemma.

  \begin{lemma}
  The \order predicate can be 
  evaluated by determining the sign of quantities of algebraic 
  degree at most 10 (in the input quantities).
  \end{lemma}

  \subsubsection{The \texorpdfstring{\wspace}{W-space}} 
  \label{ssub:the_w_space}

  The original space where the sites $S_i,S_j,S_k,S_a$ and 
  $S_b$ lie is called the \zspace. However, most of our analysis
  is carried in the inverted \wspace, as defined in 
  Section~\ref{sub:inversion} with a slight modification.

  Specifically, observe that the definition of the \wspace 
  depends on the choice of the sphere $S_I$. Since a cyclic 
  permutation of the sites $S_i,S_j$ and $S_k$ does not alter the 
  outcome of the \order predicate, we assume that 
  $k<i,j$; otherwise we reorder the sites so it does, 
  hence $r_k$ is not larger than $r_i$ and $r_j$. We now select to 
  invert \zspace
  ``through the sphere $S_k$'', \ie, we reduce the radii of 
  all initial sites by $r_k$ (and obtain $\mathcal{Z}^\star$-space)
  and then invert all points with $C_k$ as the pole. 

  Notice that 
  when we reduce the sites $S_a$ or $S_b$ by 
  $r_k$ we may end up with a sphere of negative radius if 
  $r_a<r_k$ or $r_b<r_k$. Although the existence (or not) of 
  spheres with negative radius in \wspace makes the geometric 
  configurations quite different to handle, the algebraic methods 
  we present here can handle both cases without modifications. 
  For example, if a sphere with a negative radius is tangent 
  to the positive side of a plane, then its center must 
  lie on the negative side of the plane; geometrically, this is  
  confusing but algebraicly there is no difference with the respective
  positive-radius scenario.
  For this reason, we shall assume for the rest of the Section that 
  all sites lying in \wspace have positive radii.

  The analysis that follows is based on the strong relation that 
  holds between the geometric configuration of the sites 
  $S_i,S_j,S_k,S_a$ and $S_b$ in \zspace and the
  corresponding configuration of the inverted sites 
  $\inv{S_i},\inv{S_j},\inv{S_a}$ and $\inv{S_b}$ in \wspace. 

  In \wspace, $\OO$ denotes the point $(0,0,0)$ which is the  
  image of the ``point at infinity'' of \zspace.  
  Given a point $p$ that lies on $\tri{ijk}$, $\tts{p}$ 
  denotes the external Apollonius sphere tangent to the
  sites  $S_i,S_j$ and $S_k$, centered at $p$.  
  Such a sphere $\tts{p}$ in \zspace corresponds  
  to a plane in \wspace, denoted $\itp{p}$, that is tangent to the 
  inverted sites $\inv{S_i}$ and $\inv{S_j}$ and therefore 
  tangent to the cone defined by them. 
  Notice that, since $\inv{S_i}$ and $\inv{S_j}$ are distinct 
  spheres of \wspace due to their pre-images $S_i$ and $S_j$ also 
  being distinct in \zspace, the cone $\cone(\inv{S_i},\inv{S_j})$ 
  is well defined. For the rest of this Section, we define $\wcone$
  to be the semi cone (or cylinder if $\rho_i=\rho_j$) 
  that contains $\inv{S_i}$ and $\inv{S_j}$.

  Let us observe what happens in \wspace when we consider this point 
  $p$ moving on $\tri{ijk}$ such that $\map{p}$ goes 
  from $-\infty$ towards $+\infty$. The corresponding plane
  $\itp{p}$ rotates while remaining tangent to $\wcone$, with starting 
  and ending positions the planes denoted by $\itp{-\infty}$ 
  and $\itp{+\infty}$ respectively.

  It is obvious that these two planes correspond to the two 
  Apollonius spheres of $S_i,S_j$ and $S_k$ ``at infinity'' 
 , \ie, the planes $\{\Pi_{ijk}^-,\Pi_{ijk}^+\}$.
  These planes of \zspace must be distinct in the case of a 
  ``hyperbolic'' trisector $\tri{ijk}$ as 
  shown in Sections~\ref{sub:the_incone_predicate_analysis} and 
  \ref{sub:the_distance_predicate_analysis}, and, as a result, 
  their images in \wspace must also be distinct. 

  Moreover, each of $\itp{-\infty}$ 
  and $\itp{+\infty}$ must go through $\OO$ because the their 
  pre-images are planes that go through the ``point at infinity'' 
  in \zspace. Combining these last two remarks, we conclude that 
  the points $\inv{C_i},\inv{C_j}$ and $\OO$ are not collinear
  and  $\OO$ lies strictly outside the semi cone $\wcone$. 
  It is of great importance to understand that the last fact holds only 
  because $\tri{ijk}$ is ``hyperbolic''; if we were studying the 
  ``elliptic'' trisector type, $\OO$ 
  would lie strictly inside the semi cone $\wcone$ and 
  in the degenerate case of a ``parabolic'' trisector, 
  $\OO$ would lie on the boundary of $\wcone$.

  For every point $p\in\tri{ijk}\backslash\{\pm\infty\}$, the sphere 
  $\tts{p}$ is an external Apollonius sphere and therefore 
  does not contain the ``point at infinity'' in \zspace. Correspondingly,
  its image in \wspace, \ie, the plane $\itp{p}$,  must leave 
  the point $\OO$ and the centers of the spheres $\inv{S_i}$ and 
  $\inv{S_j}$ on the same side. The side of the plane $\itp{p}$ that 
  contains $\OO$ is  
  called \emph{positive} whereas the other is referred 
  to as \emph{negative}.

  Let us now consider the plane $\inv{\Pi}$ that goes through 
  the points $\inv{C_i},\inv{C_j}$ and the point $\OO$ of \wspace.
  The well-definition of $\inv{\Pi}$ follows from the  
  non-collinearity of the three points we proved earlier.
  This plane turns out to be the image of the plane $\Pi_{ijk}$
  that goes through the centers $C_i, C_j$ and $C_k$ 
  (and apparently the point at infinity) in \zspace. 

  The latter plane separates \zspace into two half-spaces, 
  $\HH_+$ and $\HH_-$, where $\HH_+$ (resp., $\HH_-$) denotes 
  the set of points $N$ such that \orient$(N,C_i,C_j,C_k)$ 
  is positive (resp., negative). The plane $\inv{\Pi}$ also separates
  \wspace into two half-spaces, $\inv{\HH}_+$ and $\inv{\HH}_-$, 
  where $\inv{\HH}_+$ (resp., $\inv{\HH}_-$) denotes 
  the set of points $M$ such that \orient$(M,\inv{C_i},\inv{C_j},\OO)$
  is positive (resp., negative).

  If we now consider a point $C_n=(x_n,y_n,z_n)$ of \zspace and 
  its inversion image $\inv{C_n}$ in \zspace,we can easily prove that
  \begin{align}
  \text{\orient}(\inv{C_n},\inv{C_i},\inv{C_j},\OO)&=
  \sgn(\inv{p_i}\inv{p_j}\inv{p_n}D^{uvw}_{nij})=
  \sgn(D^{xyz}_{nijk})\\
  &= \text{\orient}(C_n,C_i,C_j,C_k).
  \end{align}

  \noindent
  since $\inv{C_n}=(u_n,v_n,w_n)$, where 
  $(u_n,v_n,w_n)=(\inv{x_n},\inv{y_n},\inv{z_n})(\inv{p_n})^{-1}$ and 
  \begin{equation}
  \inv{x_n}=x_n-x_k,\ \ \inv{y_n}=y_n-y_k,\ \ \inv{z_n}=z_n-z_k,\ \
  \inv{p_n}=(\inv{x_n})^2+(\inv{y_n})^2+(\inv{z_n})^2.   
  \end{equation}

  This result indicates that $\inv{\HH_+}$ and $\inv{\HH_-}$ are 
  in fact the inversion images of the open semi-spaces 
  $\HH_+$ and $\HH_-$. 

  This simple correspondence of semi-spaces yields a remarkable
  result. If we consider a point $p\in\tri{ijk}^+$ (resp., $\tri{ijk}^-$),
  then the tangency points of the 
  Apollonius sphere $\tts{p}$ with the spheres $S_i$ and $S_j$ 
  must lie on  $\HH_+$ (resp., $\HH_-$).
  If this fact is considered through inversion, the tangency points 
  of the plane $\itp{p}$ with
  the semi cone $\wcone$ must lie on  $\inv{\HH}_+$ 
  (resp., $\inv{\HH}_-$). Furthermore, if we let $p$ move on 
  $\tri{ijk}^+$ (resp., $\tri{ijk}^-$) such that $\map{p}$ goes 
  to $+\infty$ (resp., $-\infty$), we can deduce that the plane
  $\itp{+\infty}$ (resp., $\itp{-\infty}$) 
  goes through the point $\OO$ and its tangency 
  points with the spheres $\inv{S_i}$ and $\inv{S_j}$ lie on 
  $\inv{\HH}_+$ (resp., $\inv{\HH}_-$).

  \subsubsection{The \texorpdfstring{\yspace}{Y-space}} 
  \label{ssub:the_yspace_analysis}

\begin{figure}[tbp]
   \centering
   \includegraphics[width=0.95\textwidth]{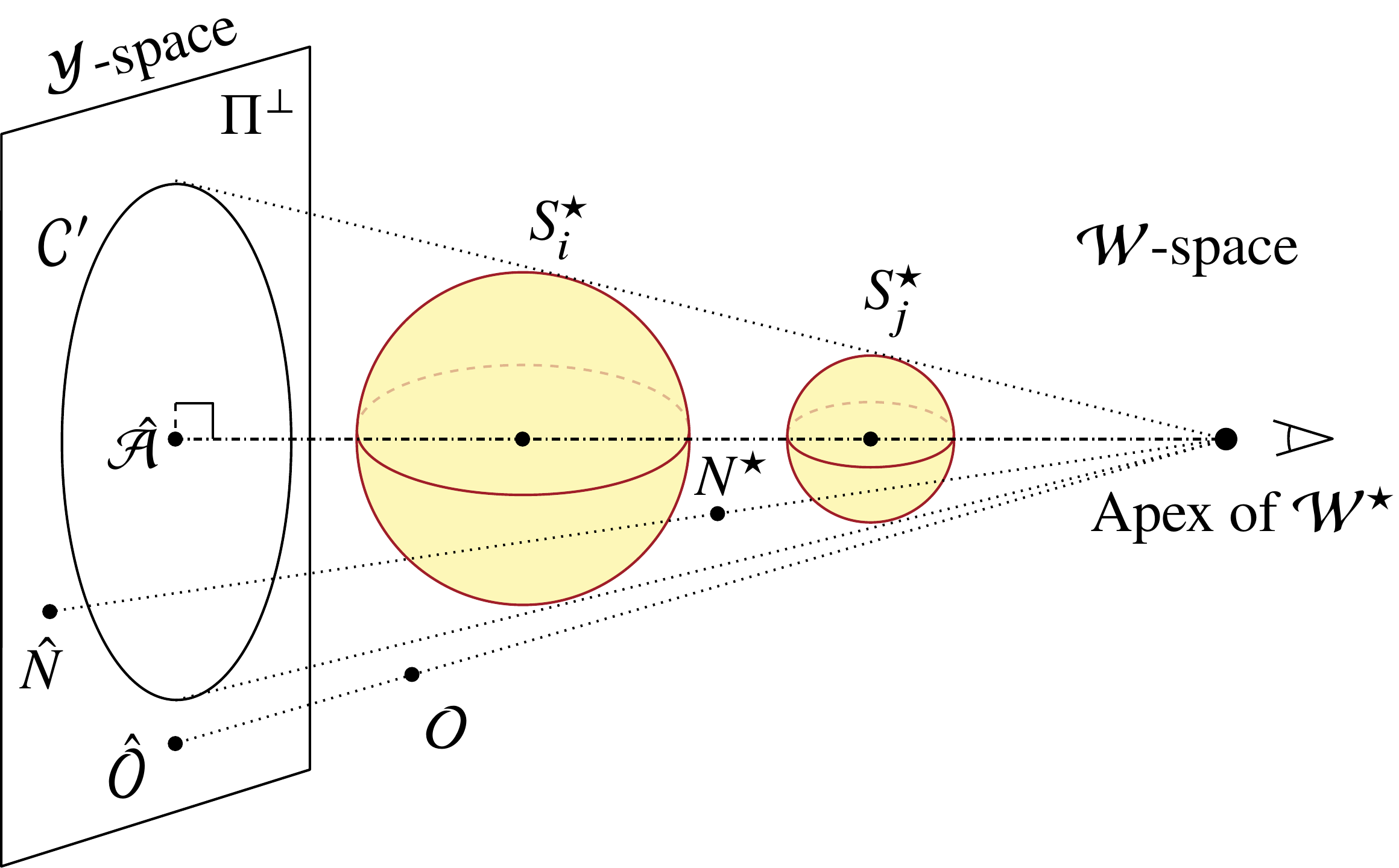}
   \caption[The connection between the \wspace and the \yspace.]{The \yspace, where most of the analysis of the 
   \order predicate is carried on, is in essence a projection of 
   \wspace to a plane $\Pi^\bot$ via the apex of the cone $\wcone$.  
   Since \yspace is a 2-dimensional space, an observation 
   regarding a \wspace geometric configuration is made easier 
   if we consider the corresponding configuration in \yspace. }
   \label{fig:08}
  \end{figure}

  All these observations are indicative of the strong connection 
  of the original and and the inverted space. However, since a 
  three-dimensional space such as the \wspace makes observations and 
  case breakdowns too complex, we will now consider a sub-space to 
  carry out our analysis. For this reason, we consider a (random) plane 
  $\Pi^\bot$ in \wspace that is perpendicular to the axis of the 
  semi-cone $\wcone$ at point $\yinv{\mathcal{A}}$ and 
  intersects it at a full circle $\ycone$;
  the intersection of \wspace and $\Pi^\bot$ is called the \yspace. 
  Such a plane $\Pi^\bot$ exist both when $\inv{\rho_i}\leq 
  \inv{\rho_j}$ and $\inv{\rho_i}> \inv{\rho_j}$ and therefore 
  \yspace is well-defined in both cases. 
  Notice that in every figure representing the \yspace, we always 
  depict the $\Pi^\bot$ plane such that the vector 
  $\ov{\inv{C_i}\inv{C_j}}$ points ``towards'' the reader 
  (see Figure~\ref{fig:08}). 

  In \yspace, we will use the following notation:
  \begin{itemize}
  \item  
  $\ill{p},\ill{\pm\infty}$ 
  and $\ill{\oo}$  denote the intersection of the plane $\Pi^\bot$ with the planes 
  $\itp{p},\inv{\Pi}(\pm\infty)$ and $\inv{\Pi}(\oo)$ 
  respectively.
  \item 
  $\yinv{\eta},\yinv{\theta},\yinv{o}$ and $\yinv{p}$ denote 
  the points of tangency of $\ycone$ and the lines 
  $\ill{-\infty}$, $\ill{+\infty}$, $\ill{\oo}$ and $\ill{p}$
  respectively (for $p\in\tri{ijk}$).
  \item 
  The intersection of $\inv{\HH}_+$ 
  (resp., $\inv{\HH}_-$) with the $\Pi^\bot$ plane is called 
  the positive (resp., negative) half-plane $\yinv{\HH}_+$
  (resp., $\yinv{\HH}_-$).
  \item
  The positive (resp., negative) side of the line
  $\ill{p}$ for a point $p\in\tri{ijk}$ to be the side that 
  contains (resp., does not contain) the point $\yinv{\mathcal{A}}$.  
  \item
  $\yinv{\OO}$ denotes the point of intersection of 
  the lines $\ill{-\infty}$ and $\ill{+\infty}$. 
  \end{itemize}

  We shall now define an equivalency relation between the trisector 
  $\tri{ijk}$ and an arc of $\ycone$, which is the biggest idea 
  upon which the rest of our analysis is based.
  If a point $p$ moves on $\tri{ijk}$ such 
  that $\map{p}$ goes from $-\infty$ to $+\infty$ then, in \yspace, 
  the corresponding point $\yinv{p}$ moves on $\ycone$ from the 
  point $\yinv{\eta}$ to the point $\yinv{\theta}$, going
  through the point $\yinv{o}$. Observe that there is a 
  1-1 correspondence between the oriented trisector $\tri{ijk}$ and 
  the oriented arc $(\yinv{\eta} \yinv{o} \yinv{\theta})$. We denote this 
  1-1 and onto mapping from $\tri{ijk}$ to
  the arc $(\yinv{\eta} \yinv{o} \yinv{\theta})$ by $\arc{\cdot}$, such that $\arc{p}=\yinv{p}$. 
  
\begin{figure}[tbp]
  \centering
  \includegraphics[width=0.9\textwidth]{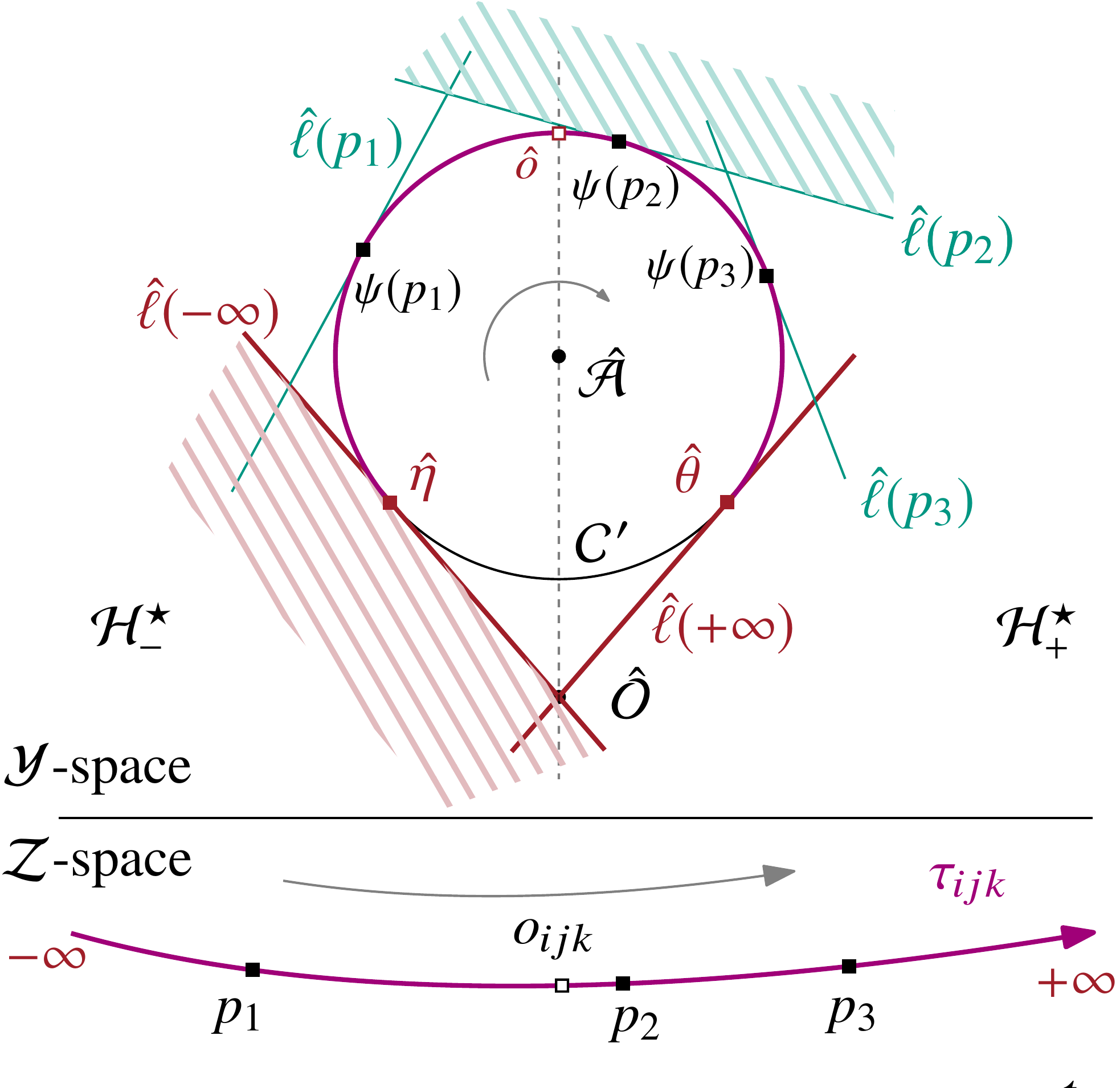}
  \caption[The strong relation between the \zspace 
  and the \yspace.]{There is a strong relation between the \zspace 
  and the \yspace. When the point $p_i$ of \zspace traverses 
  the trisector $\tri{ijk}$ towards its positive direction, the corresponding point $\arc{p_i}$ of 
  \zspace traverses the arc $(\arc{\eta}, \arc{o}, \arc{\theta})$. 
  At the same time, $\ill{p_i}$ rotates in \yspace, remaining tangent 
  to $\ycone$, with starting and ending positions the lines 
  $\ill{-\infty}$ and $\ill{+\infty}$ respectively. The red and green 
  area symbolize the negative side of the lines $\ill{-\infty}$ 
  and $\ill{p_2}$, respectively.}
  \label{fig:zspace_yspace_equivalency}
  \end{figure}

  What naturally follows is that the order of appearance of the 
  vertices $v_{ijka}$, $v_{ikja}$, $v_{ijkb}$ and $v_{ikjb}$ on the
  oriented trisector amounts to the order of appearance of the points 
  $\arc{v_{ijka}}$, $\arc{v_{ikja}}$, $\arc{v_{ijkb}}$ and $\arc{v_{ikjb}}$ 
  on the oriented arc $(\arc{\eta},\arc{o},\arc{\theta})$
  (see Figure~\ref{fig:zspace_yspace_equivalency}). 
  Consider, however, that we only need to order the Apollonius vertices that 
  actually exist among $v_{ijka},v_{ikja},v_{ijkb}$ and $v_{ikjb}$. 

  \begin{lemma}
  \label{lemma:ordering_equivalence}
  There is a 1-1 correspondence between the 
  order of appearance of the existing vertices among
  $v_{ijka}, v_{ikja}, v_{ijkb}$ and $v_{ikjb}$ on the oriented 
  hyperbolic trisector $\tri{ijk}$ and the order of appearance of the existing 
  points $\arc{v_{ijka}}$, $\arc{v_{ikja}}$, $\arc{v_{ijkb}}$ 
  and $\arc{v_{ikjb}}$ on the oriented arc 
  $(\arc{\eta},\arc{o},\arc{\theta})$.
  \end{lemma}

  The lemma suggests that the outcome of the \order predicate could 
  return the order of appearance of the images of the aforementioned Apollonius vertices on the arc $(\arc{\eta},\arc{o},\arc{\theta})$ 
  instead of the order of the original vertices on the trisector 
  $\tri{ijk}$.

  Towards our goal of obtaining the ordering of the inverted 
  Apollonius vertices, we denote 
  the circle $\yinv{S_n}\in\text{\yspace}$ 
  for $n\in\{a,b\}$ which will be considered as the image 
  of $\inv{S_n}$ of \wspace. We need to define the image of these 
  spheres in a proper way such that they carry their geometric 
  properties from \wspace to \yspace. For this reason, we consider 
  the center $\yinv{C_n}$ of $\yinv{S_n}$ to be the 
  intersection of $\Pi^{\bot}$ with the line that goes through 
  the apex of the cone $\wcone$ and the center $\yinv{C_n}$. 
  Observe that such a line is well defined since the latter
  two points cannot coincide; if they did then $\sh{S_n}$ would be 
  $\emptyset$ or $\RR$ yielding a contradiction
  \footnote{In such a geometric configuration, there would not 
  exist a plane in \wspace co-tangent to all spheres 
  $\inv{S_i}$,$\inv{S_j}$ and $\inv{S_n}$. Equivalently, 
  in \zspace there would not exist an Apollonius sphere of 
  the sites $S_i$,$S_j$,$S_k$ and $S_n$ hence $\sh{S_n}$ would be 
  either $\emptyset$ or $\RR$ based on the analysis of 
  Section~\ref{sub:the_shadowregion_predicate_analysis}. An equivalent arguements is that, if the apex of $\wcone$ and  $\yinv{C_n}$ were 
  coinciding then this would mean that $S_n$ either contains 
  or is contained in $S_k$, which is ruled out.}. Finally, 
  the radius of $\yinv{C_n}$ is such that $\yinv{C_n}$ is tangent 
  to each of the existing lines $\ill{v_{ijkn}}$ and $\ill{v_{ikjn}}$ 
  (at least one of them exists due to $\sh{S_n}$ not being $\emptyset$ 
  or $\RR$).

  Another crucial property we want to point out derives from the 
  inversion mapping we used to go from $\mathcal{Z}^\star$-space to 
  \wspace. The mapping $W(z)$ we used is known to be 
  \emph{inclusion preserving}, \ie, the relative position of 
  two spheres in the original space is preserved in the inverted one. 
  For example, consider a sphere $S_\mu$ that 
  intersects the (existing) Apollonius sphere
  $\tts{v_{ikjn}}$ (resp., $\tts{v_{ijkn}}$), for $n\in\{a,b\}$ in 
  \zspace. After reducing both spheres by $r_k$, their images 
  in $\mathcal{Z}^\star$ retain the same relative position, \ie, they 
  intersect. 
  Applying the inversion mapping, it must stand that 
  the sphere $\inv{S_\mu}$ must intersect the negative side of
  $\inv{\Pi}(v_{ikjn})$ (resp., $\inv{\Pi}(v_{ijkn})$) since this 
  half space is precisely the inversion image of the interior of 
  $\tts{v_{ikjn}}$ (resp., $\tts{v_{ijkn}}$). Finally, 
  if we consider this configuration in \yspace, we deduce that
  $\yinv{S_\mu}$ must intersect the negative side of $\ill{v_{ikjn}}$
  (resp., $\ill{v_{ijkn}}$). 

\begin{figure}[htbp]
   \centering
   \includegraphics[width=0.7\textwidth]{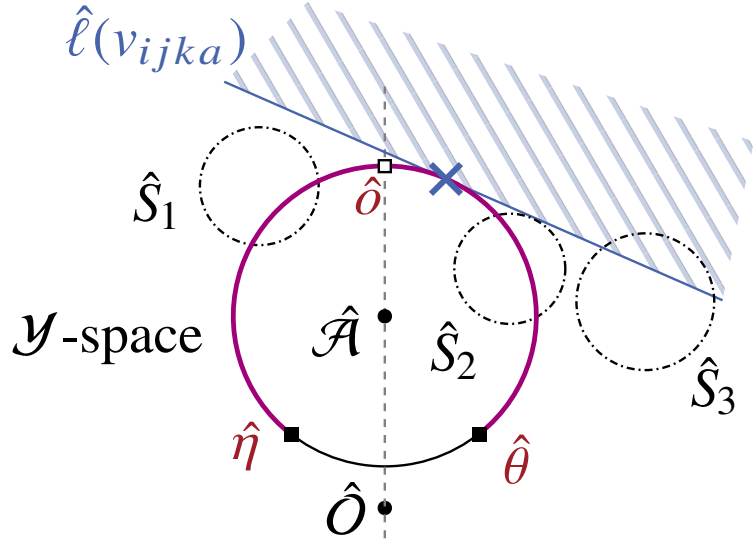}
   \caption[The \insphere predicate in \yspace.]{As Lemma~\ref{lemma:yinv_insphere_equiv} suggests, 
   the \insphere predicate with input $(S_i,S_j,S_k,S_a,S_n)$ 
   is positive, zero or negative for $n=0,1$ or $2$ respectively.}
   \label{fig:09b}
  \end{figure}

  In a similar way we can show that if $S_\mu$ 
  is tangent or does not intersect the Apollonius sphere
  $\tts{v_{ikjn}}$ (resp., $\tts{v_{ijkn}}$) then, in \yspace, 
  $\yinv{S_\mu}$ is tangent to $\ill{v_{ikjn}}$
  (resp., $\ill{v_{ijkn}}$) or does not intersect its negative side (see Figure~\ref{fig:09b}).
  A fact tightly connected with these observations is that 
  the relative position of 
  $S_\mu$ and $\tts{v_{ikjn}}$ (resp., $\tts{v_{ijkn}}$) is provided
  by the \insphere predicate. Specifically,
  \begin{itemize}
   \item if \insphere $(S_i,S_k,S_j,S_n,S_m)$ is $-$, $0$ or $+$ then
   $S_m$ intersects, is tangent to or does not intersect the Apollonius
   sphere $\tts{v_{ikjn}}$ and,
   \item if \insphere $(S_i,S_j,S_k,S_n,S_m)$ is $-$, $0$ or $+$ then
   $S_m$ intersects, is tangent to or does not intersect the Apollonius
   sphere $\tts{v_{ijkn}}$.
  \end{itemize}
  Since these \insphere predicates can be evaluated as shown in 
  Section~\ref{sub:the_insphere_predicate}, 
  the relative position of $\yinv{S_\mu}$ with respect to
  any of the existing lines $\ill{v_{ikjn}}$ and $\ill{v_{ijkn}}$ can 
  be determined in \yspace, for $n\in\{a,b\}$.

  \begin{lemma}
  \label{lemma:yinv_insphere_equiv}
  The circle $\yinv{S_m}$ intersects, is tangent to or does 
  not intersect the negative side of $\ill{v_{ikjn}}$ 
  (resp., $\ill{v_{ijkn}}$) if and only if 
  the \insphere predicate with input $(S_i$,$S_k$,$S_j$,$S_n$,$S_m)$
   (resp., $(S_i$,$S_j$,$S_k$,$S_n$, $S_m)$ ) is negative, zero or 
  positive respectively. 
  \end{lemma}

  \subsubsection{The classic configuration} 
  \label{sub:The_classic_configuration}

  When the $\text{\order}(S_i, S_j, S_k, S_a, S_b)$ predicate is 
  called, we initially determine the shadow region types of 
  $\sh{S_a}$ and $\sh{S_b}$ via the appropriate \shadow predicates. 
  If the type of each shadow region is $(\chi,\phi)$ or $(\chi,+\infty)$ 
  (not necessary the same), we say that  we are in 
  \emph{classic configuration}. In such a setup, we can distinguish 
  simpler cases regarding the ordering the images of the Apollonius vertices 
  on the oriented arc $(\yinv{\eta} \yinv{o} \yinv{\theta})$.

  We therefore break up the analysis the \order predicate depending on 
  whe\-ther we are in a classic 
  (Section~\ref{sub:The_classic_configuration}) or 
  non-classic configuration (Section~\ref{sub:ordering_in_a_non_classic_configuration}), the latter 
  being reduced to the former using various observations.
  Let us now study in more detail what kind of information derives 
  from the fact that the sites $S_a$ and $S_b$ satisfy the conditions of a
  classic configuration. 

  Suppose $S_n$, for $n=a$ or $b$, is $(\chi,\phi)$ and therefore, 
  the endpoints $\{\chi,\phi\}$ must correspond to the two 
  Apollonius vertices $\{\map{v_{ijkn}},\map{v_{ikjn}})\}$ on
  the trisector $\tri{ijk}$ based on the remarks of 
  Section~\ref{ssub:the_shadow_predicate}. Let us consider
  $\chi$ and $\phi$ as $\map{v_n}$ and $\map{v_n^\prime}$ respectively, 
  with $\{v_n,v_n^\prime\}=\{v_{ijkn},v_{ikjn}\}$. Then, for every 
  $p\in\tri{ijk}$ such that $v_n\prec p\prec v_n^\prime$,
  the sphere $\tts{p}$ must intersect $S_n$ as this follows from the 
  definition of $\sh{S_n}$. 

  If we consider this point $p$ moving on $\tri{ijk}$, initialy starting 
  from the left-endpoint position $v_n$, then we observe that the 
  Apollonius sphere $\tts{v_n}$ intersects $S_n$ if we move its center
  infinitesimally towards the positive direction of $\tri{ijk}$. 
  Taking a closer look at the tangency points $T_i,T_j,T_k$ and $T_n$ 
  of $\tts{v_n}$ with the spheres $S_i,S_j,S_k$ and $S_n$ respectively, 
  and since the orientation of $\tri{ijk}$ is based in such a way 
  on the orientation of $C_i,C_j$ and $C_k$, it must hold that 
  $T_n$ must lie with respect to the plane formed by $T_i$, $T_j$ and 
  $T_k$ such that $T_iT_jT_kT_n$ be negative oriented. For that reason, 
  $v_n$ is in fact $v_{ikjn}$ and subsequently $v_n^\prime$ is $v_{ijkn}$.

  The same argument can be used to prove that if $\sh{S_n}$ is 
  $(\chi,+\infty)$ then $\chi$ corresponds to $\map{v_{ikjn}}$. 
  If we apply a similar analysis in all shadow region types that 
  contain a finite endpoint, it will lead to the following lemma.

  \begin{lemma}
  \label{lemma:phi_chi}
  If the type of the shadow region $\sh{S_n}$ of a sphere $S_n$ 
  on a hyperbolic trisector is one of the following: 
  $(-\infty,\phi)$, $(\chi,+\infty)$, $(\chi,\phi)$, or 
  $(-\infty,\phi)\cup(\chi,+\infty)$ where 
  $\phi,\chi\neq\pm\infty$, then $\chi\equiv\map{v_{ikjn}}$ 
  and $\phi\equiv\map{v_{ijkn}}$.
  \end{lemma}

  As the lemma suggest, in a classic configuration, 
  for $n\in\{a,b\}$,
  \begin{itemize}
   \item if $\sh{S_n}=(\chi,\phi)$, then 
   both $v_{ijkn}$ and $v_{ikjn}$ exist and $v_{ikjn}\prec v_{ijkn}$, 
   whereas
   \item if $\sh{S_n}=(\chi,+\infty)$, then 
   $v_{ikjn}$ exists while $v_{ijkn}$ does not. 
  \end{itemize}

\begin{figure}[htbp]
   \centering
   \includegraphics[width=0.7\textwidth]{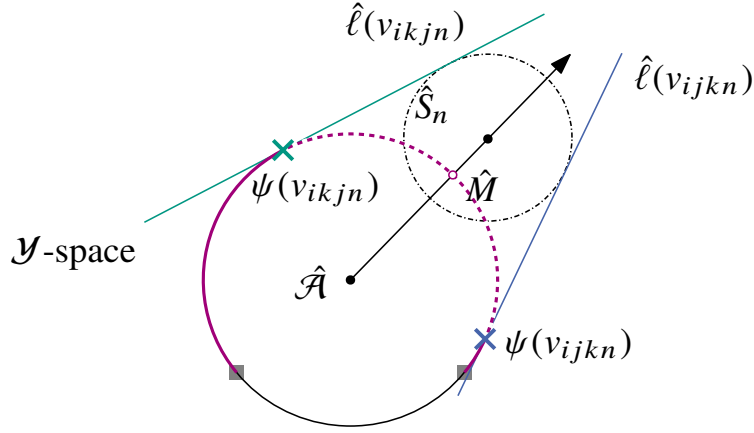}
   \caption[The location of $\yinv{S_n}$ if both $v_{ikjn}$ and $v_{ijkn}$ exist.]{If both $v_{ikjn},v_{ijkn}$ exist and $\sh{S_a}$ 
   is $(\chi,\phi)$ with respect to the trisector $\tri{ijk}$, 
   $\yinv{C_n}$ must lie on the ray $(\yinv{\mathcal{A}},\yinv{M})$. 
   The dotted arc represent the image of $\sh{S_n}$ in 
   \yspace; it is obvious that, for any $\arc{p}$ in this arc, 
   $\yinv{S_n}$ intersects the negative side of the line 
   $\ill{p}$. As of Lemma~\ref{lemma:yinv_insphere_equiv},
   $S_n$ must intersect the sphere $\tts{p}$ which is equivalent 
   to $p\in\sh{S_n}$.}
   \label{fig:10a}
  \end{figure}

  An equally important result arises when pondering of the possible positions 
  of the circle $\yinv{S_n}$, for $n\in\{a,b\}$ with any of the 
  existing lines $\ill{v_{ikjn}}$ and $\ill{v_{ijkn}}$. Firstly, let us consider the scenario where $\sh{S_n}=(\chi,\phi)$
  and in consequence, both lines exist. In this case, both points 
  $\arc{v_{ikjn}}$ and $\arc{v_{ijkn}}$ exist on the oriented arc 
  such that $\arc{v_{ikjn}}\prec\arc{v_{ijkn}}$, as this follows from 
  all previous remarks. From the definition of  
  $\sh{S_n}=(\chi,\phi)$, it derives as a result that for a point 
  $p$ on the trisector $\tri{ijk}$ such that 
  $v_{ikjn}\prec p v_{ijkn}$, the sphere $\tts{p}$ intersects 
  with $S_n$. Using the ``inclusion preserving'' argument, it must 
  stand that, in \yspace, $\yinv{S_n}$ intersects with the negative side 
  of $\ill{p}$. Therefore, if $\yinv{M}$ is the midpoint $\arc{v_{ikjn}}$ 
  and $\arc{v_{ijkn}}$ on the arc $(\eta o \theta)$
  and $\mathbb{V}$ denotes the open ray from $\yinv{\mathcal{A}}$ towards
  $\yinv{M}$, then the circle $\yinv{S_n}$ must be centered at a point
  on $\mathbb{V}$, \ie, $\yinv{C_n}\in\mathbb{V}$ 
  (see Figure~\ref{fig:10a}).

\begin{figure}[htbp]
   \centering
   \includegraphics[width=0.7\textwidth]{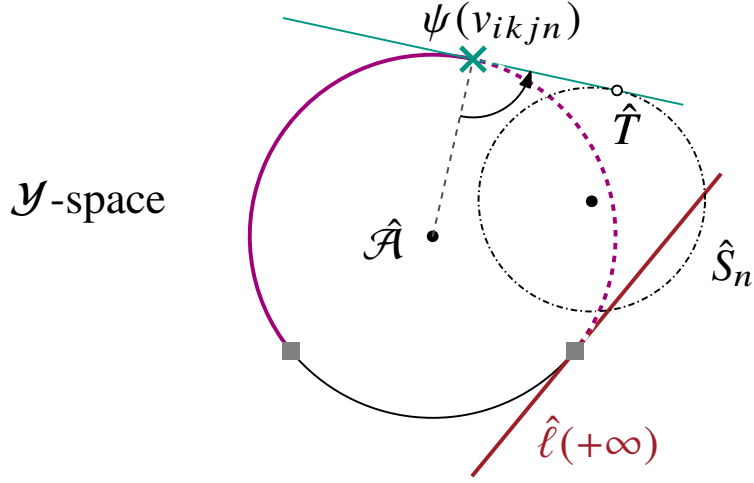}
   \caption[The location of $\yinv{S_n}$ if $v_{ikjn}$ exist and 
   $v_{ijkn}$ does not. ]{If $v_{ikjn}$ exists and $v_{ijkn}$ does not, 
   the shadow region of $S_n$ is known to be $(\chi,+\infty)$.
   This means that the tangency point $\yinv{T}$ of $\yinv{S_n}$
   with the line $\ill{v_{ikjn}}$ is on the side of the line 
   that forms a $90^\circ$ angle with the vector 
   $(\arc{v_{ikjn}},\yinv{\mathcal{A}})$. The image of 
   $\sh{S_n}$ in \yspace in this case is the dotted arc which is 
   indeed of the form $(\chi,+\infty).$ }
   \label{fig:10b}
  \end{figure}

  Lastly, let us examine the case where $\sh{S_n}=(\chi,+\infty)$. We begin 
  by observing that $S_n$ must intersect with $\Pi_{ijk}^+$ due to 
  the definition of the shadow region and this amounts, 
  in \yspace, to $\yinv{S_n}$ intersecting the negative 
  side of the line $\ill{+infty}$. Moreover, following a similar 
  analysis as in the case of $\sh{S_n}=(\chi,\phi)$, we come to 
  the conclusion that $\yinv{S_n}$ must be tangent to 
  $\ill{v_{ikjn}}$ at a point $\yinv{T}$ such that the (counterclowise)
  angle $(\yinv{\mathcal{A}},\arc{v_{ikjn}},\yinv{T})$ is $90^\circ$ 
  (not $270^\circ$, see Figure~\ref{fig:10b}). 
  This fact must hold for the line 
  $\ill{p}$ to intersect $\yinv{S_n}$, for every point 
  $p\in\tri{ijk}$ with 
  $\arc{v_{ikjn}}\prec \arc{p} \prec \yinv{\theta}$.

 \subsubsection{Ordering the Apollonius vertices in a classic configuration} 
  \label{sub:ordering_in_a_classic_configuration}
  
  In a classic configuration, we take for granted that, for 
  $n\in\{a,b\}$, either $\sh{S_n}=(\chi,\phi)$ and therefore 
  $v_{ikjn}\prec v_{ijkn}$ on the trisector $\tri{ijk}$ 
  or $\sh{S_n}=(\chi,+\infty)$ and 
  only $v_{ikjn}$ exists on $\tri{ijk}$. To order all of these 
  existing Apollonius vertices, we break down our analysis into 
  four sub-configurations.
  \begin{description}
   \item[Case A.] All vertices $v_{ikja},v_{ijka},v_{ikjb}$ and 
   $v_{ijkb}$ exist, \ie, both $\sh{S_a}$ and $\sh{S_b}$ are 
   of type $(\chi,\phi)$.
   \item[Case B.] Only the vertices $v_{ikja}$ and 
   $v_{ikjb}$ exist, \ie, both $\sh{S_a}$ and $\sh{S_b}$ are 
   of type $(\chi,+\infty)$.
   \item[Case C.] Only the vertices $v_{ikja},v_{ijka}$ and 
   $v_{ikjb}$ exist, \ie, the type of $\sh{S_a}$ and $\sh{S_b}$ are 
   $(\chi,\phi)$ and $(\chi,+\infty)$ respectively.
   \item[Case D.] Only the vertices $v_{ikjb},v_{ijkb}$ and 
   $v_{ikja}$ exist, \ie, the type of $\sh{S_a}$ and $\sh{S_b}$ 
   are $(\chi,+\infty)$ and $(\chi,\phi)$ respectively.
  \end{description}

  The last Case D is identical with the Case C if we name exchange the 
  spheres $S_a$ and $S_b$. Therefore, if Case D arises, we 
  evaluate \order$(S_i$, $S_j$, $S_k$, $S_b$, $S_a)$ instead, 
  which falls in Case C,
  and return the resulting ordering of $v_{ikjb},v_{ijkb}$ and $v_{ikja}$.
  Consequently, we only need to consider the Cases 
  A, B and C; the analysis of each configuration is deployed 
  separately in the following sections. 

  \paragraph*{\textbf{Analysis of Case A}}
   Given that all Apollonius vertices $v_{ijka},v_{ikja},v_{ijkb},v_{ikjb}$
   exist on $\tri{ijk}$ and  $v_{ikja}\prec v_{ijka}$ as well as 
   $v_{ikjb}\prec v_{ijkb}$ , the list of all possible orderings (and thus ouctomes of the \order predicate) is the following
   \begin{description}
   \item[\answer 1.] $v_{ikja}\prec v_{ijka}\prec v_{ikjb}\prec v_{ijkb}$,
   \item[\answer 2.] $v_{ikja}\prec v_{ikjb}\prec v_{ijka}\prec v_{ijkb}$,
   \item[\answer 3.] $v_{ikjb}\prec v_{ikja}\prec v_{ijka}\prec v_{ijkb}$,
   \item[\answer 4.] $v_{ikjb}\prec v_{ikja}\prec v_{ijka}\prec v_{ijka}$,
   \item[\answer 5.] $v_{ikjb}\prec v_{ijkb}\prec v_{ikja}\prec v_{ijka}$,
   \item[\answer 6.] $v_{ikja}\prec v_{ikjb}\prec v_{ijkb}\prec v_{ijka}$.
   \end{description}

  \begin{figure}[tbp]
   \centering
   \includegraphics[width=0.28\textwidth]{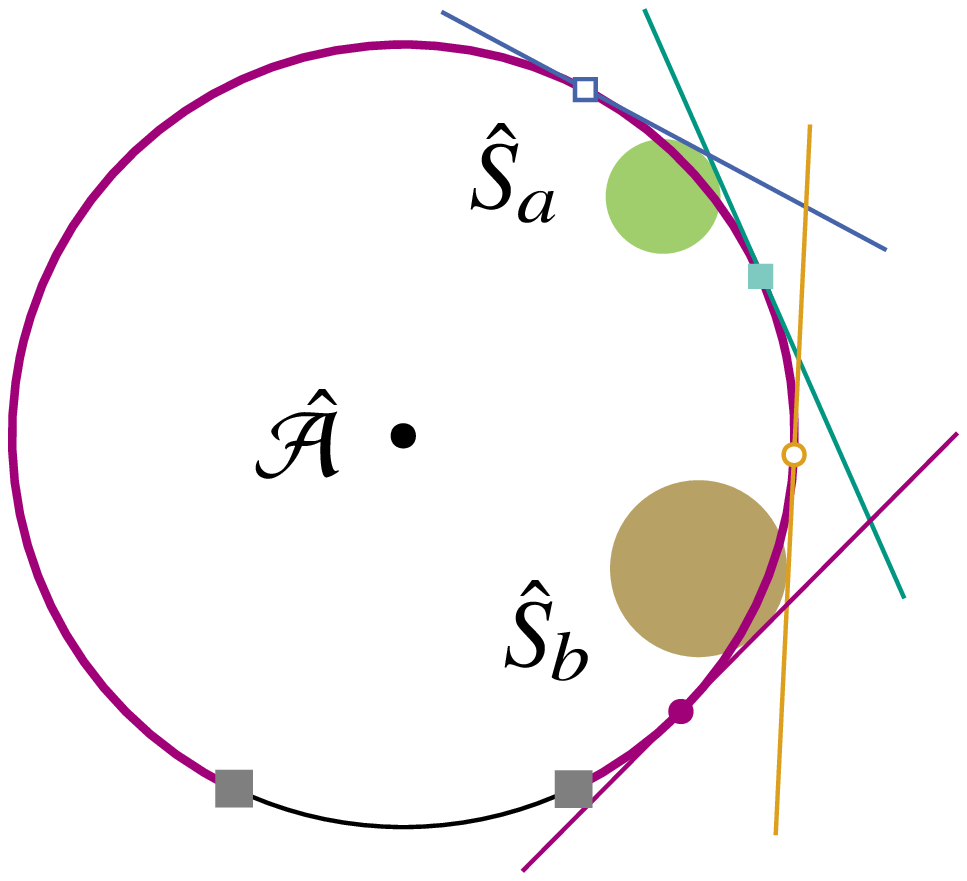}\hspace{1 cm}
   \includegraphics[width=0.28\textwidth]{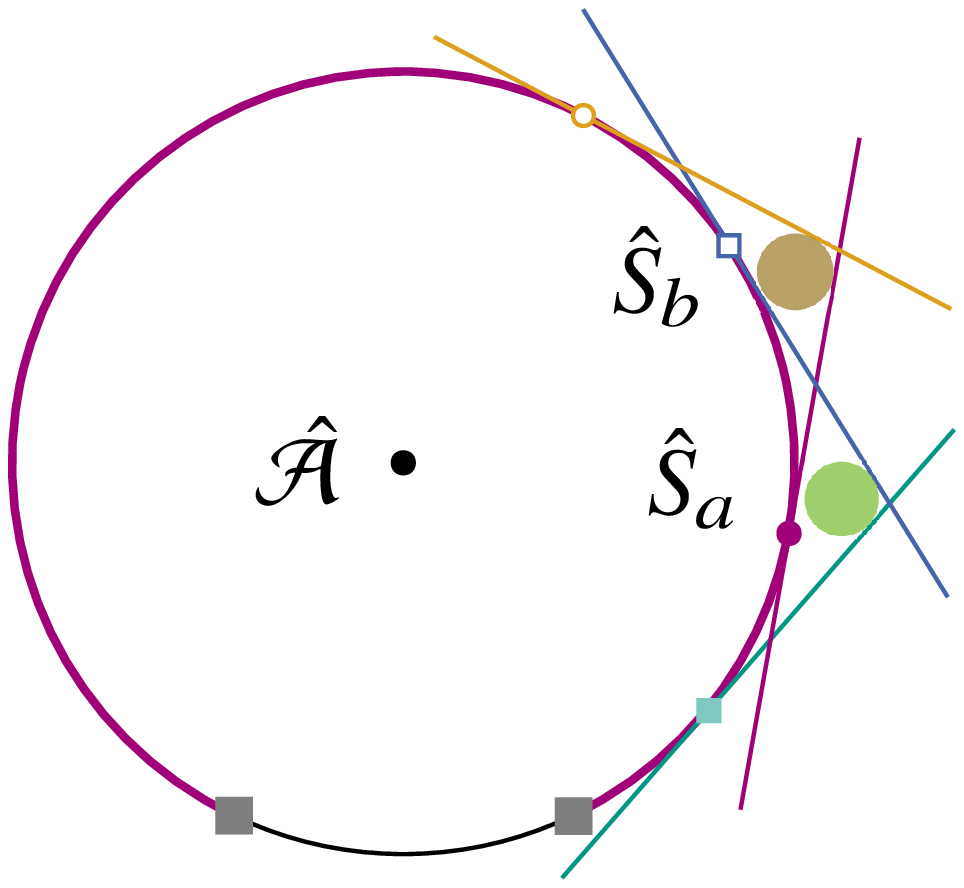}\\
   \includegraphics[width=0.28\textwidth]{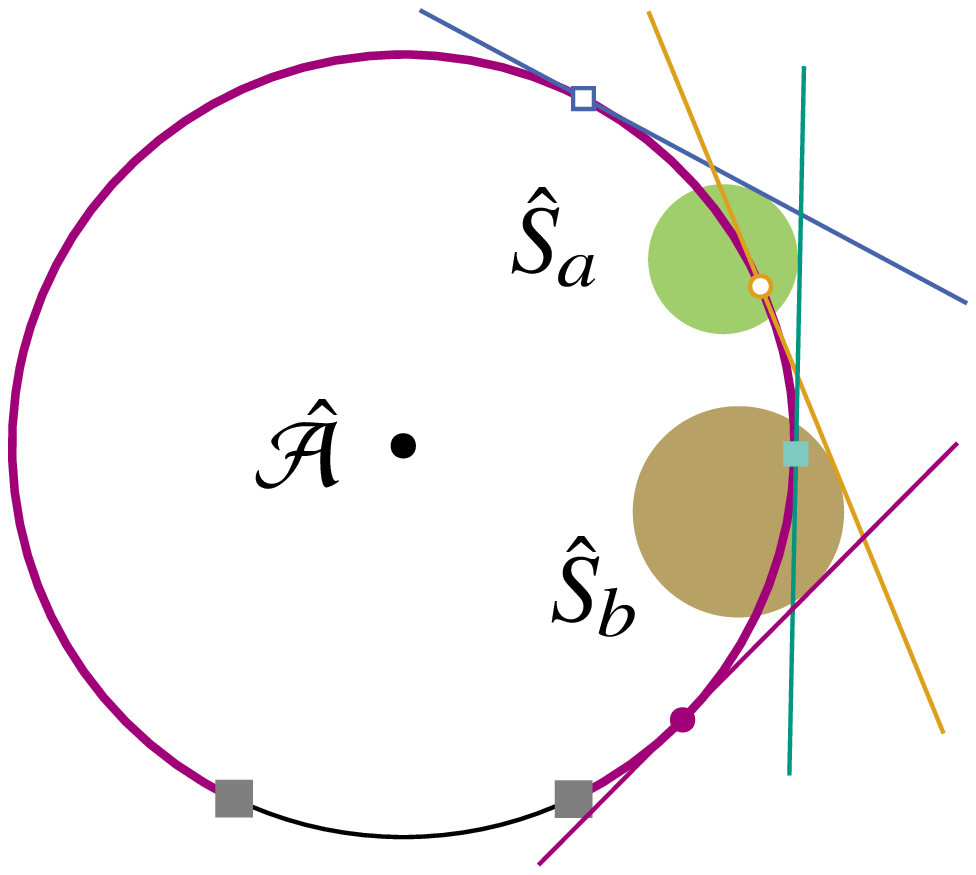}\hspace{1 cm}
   \includegraphics[width=0.28\textwidth]{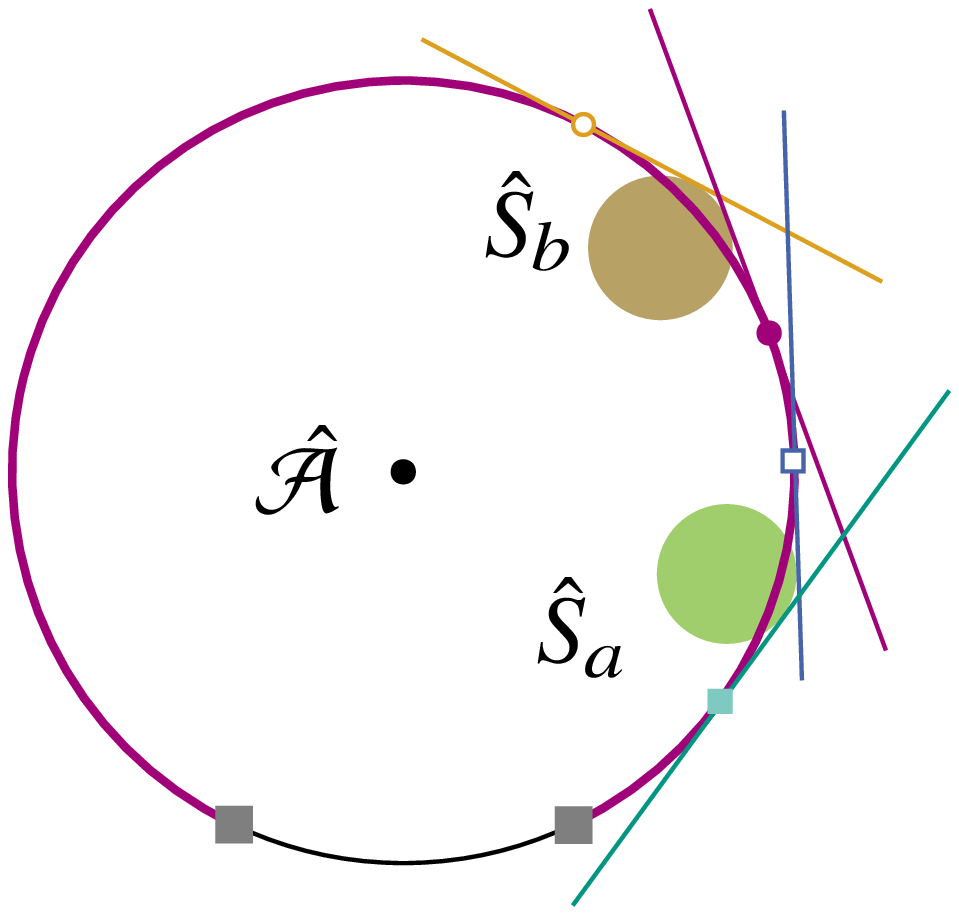}\\
   \hspace{5 mm}\includegraphics[width=0.31\textwidth]{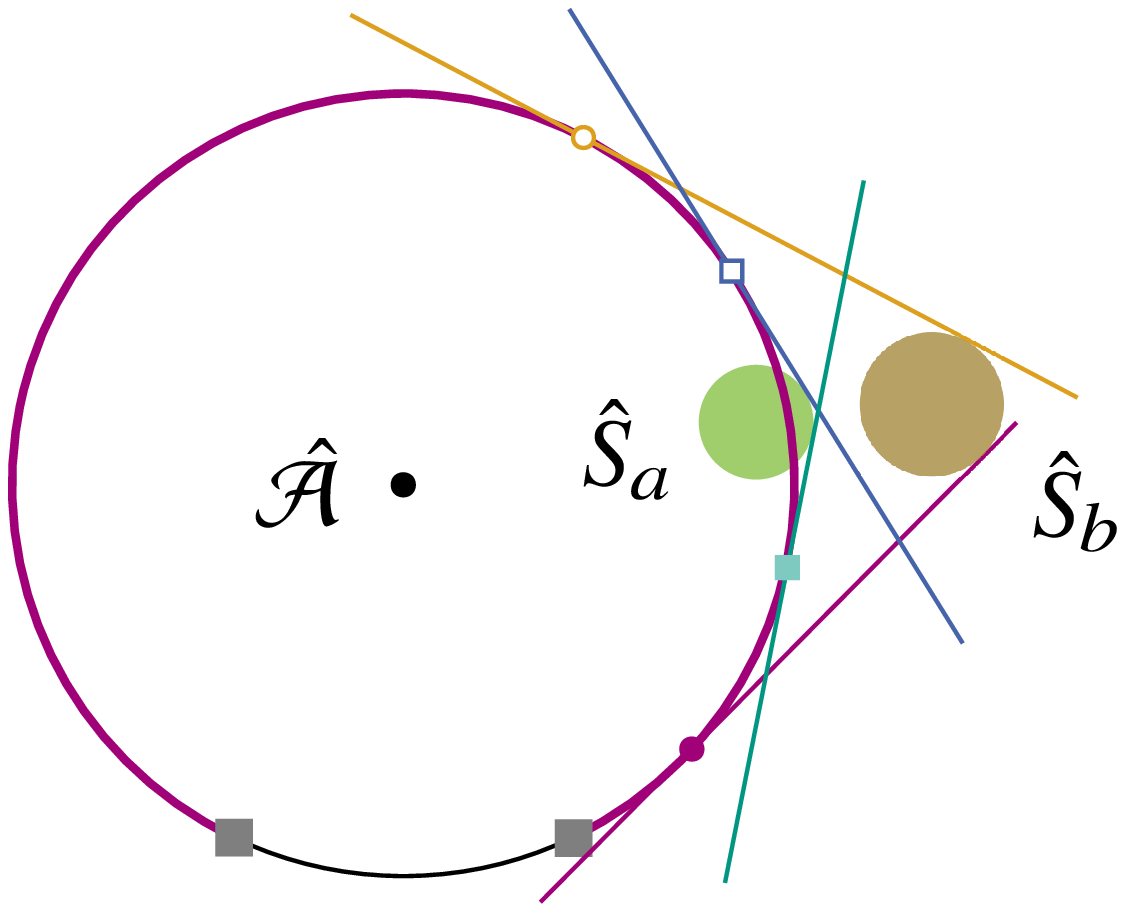}\hspace{5 mm}
   \includegraphics[width=0.31\textwidth]{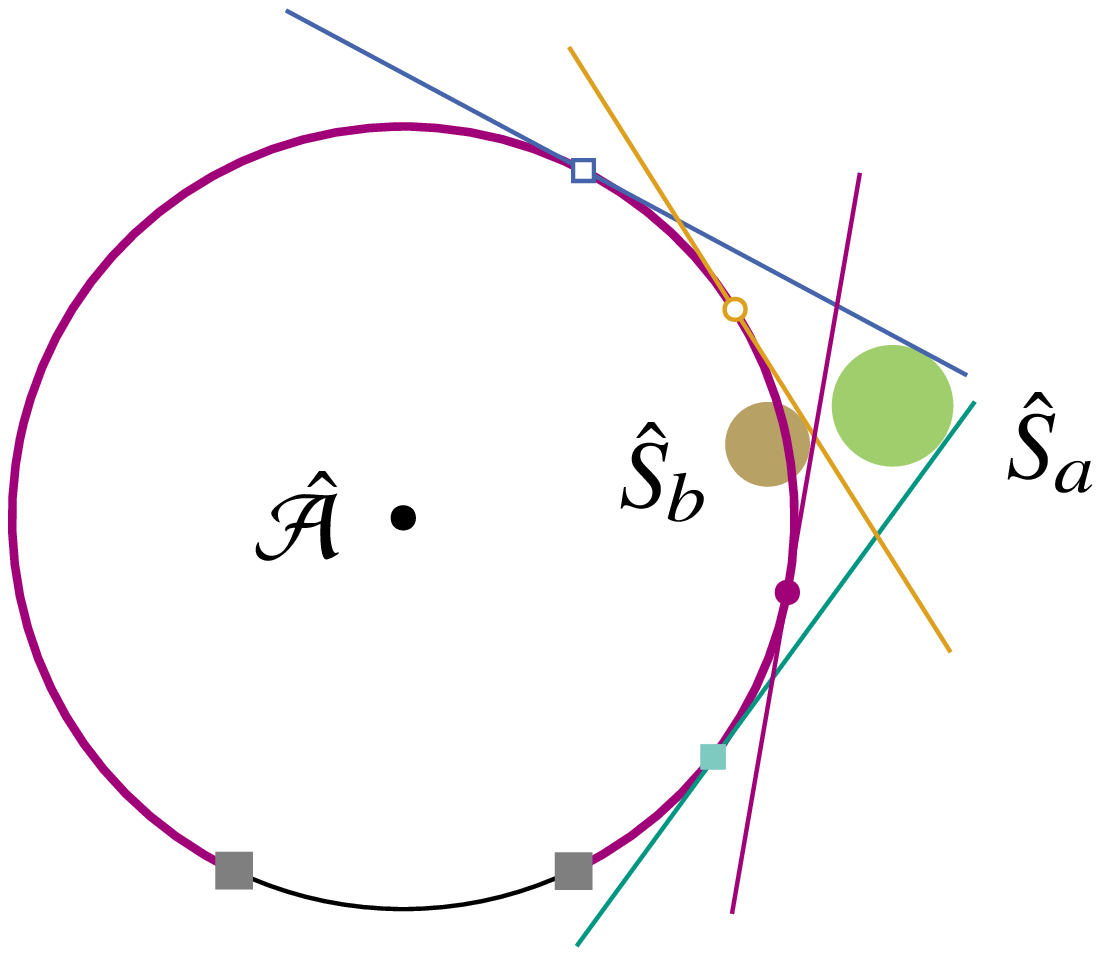}\\ \vspace{10pt}
   \includegraphics[width=0.8\textwidth]{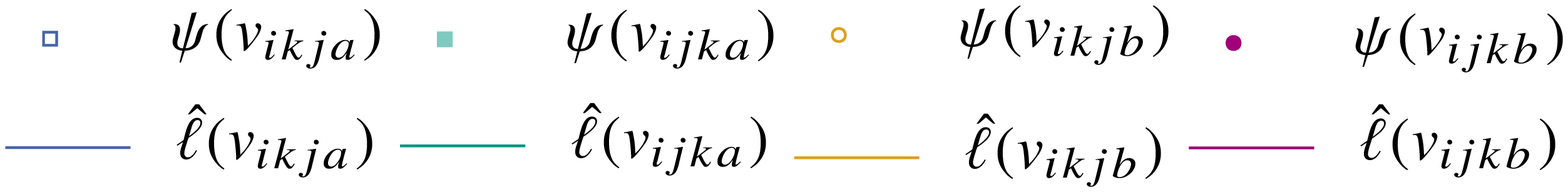}
   \caption[Possible orderings of the images of $v_{ikja}, v_{ijka},v_{ikjb}$ and $v_{ijkb}$ in \yspace.]{Under the assumption that all Apollonius vertices 
   $v_{ikja}, v_{ijka},v_{ikjb}$ and $v_{ijkb}$ exist on the trisector
   $\tri{ijk}$, we consider all possible orderings of these vertices. 
   As of Lemma~\ref{lemma:ordering_equivalence}, each of these 
   orderings is equivalent to a respective ordering of the 
   points $\arc{v_{ikja}}, \arc{v_{ijka}},\arc{v_{ikjb}}$ and 
   $\arc{v_{ijkb}}$ on the oriented arc 
   $(\arc{\eta},\arc{o},\arc{\theta})$ of \yspace. For every 
   possible ordering a possible location of $\yinv{S_a}$ and 
   $\yinv{S_b}$ is considered, such that the shadow regions 
   $\sh{S_n}$ for $n\in\{a,b\}$ is of type $(\chi,\phi)$ since
   we examine a classic configuration. 
   From top to bottom, Left:  \answer 1, 2, 3.
   From top to bottom, Right: \answer 4, 5, 6. }
   \label{fig:11a_11f}
  \end{figure}

   Any of these ordering on the trisector is 
   equivalent to the corresponding ordering of 
   $\arc{v_{ikja}}$, $\arc{v_{ikjb}}$, $\arc{v_{ijkb}}$ and $\arc{v_{ijka}}$
   on the arc $(\yinv{\eta}, \yinv{o}, \yinv{\theta})$ as stated in Lemma~\ref{lemma:ordering_equivalence}.
   
   We now study separately all these possible cases in \yspace 
   following the same approach. Firstly, we place the images of the 
   Apollonius vertices on the arc according to the \answer we are 
   examining. Then, we consider a possible location for each of the circles 
   $\yinv{S_a}$ and $\yinv{S_b}$ taking into consideration the remarks 
   made in the Section~\ref{sub:The_classic_configuration}. Lastly, we draw some conclusions 
   regarding the relative position of $\yinv{S_a}$ and $\yinv{S_b}$ 
   with the lines $\arc{v_{ikjb}}, \arc{v_{ijkb}}$ and 
   $\arc{v_{ijka}},\arc{v_{ikja}}$ respectively. The later observations
   are then translated as \insphere test's results based on 
   Lemma~\ref{lemma:yinv_insphere_equiv}. 
   
   Let us consider one case in detail, for example the \answer 2 configuration; a similar approach will be applied to each \answer. 
   In Figure~\ref{fig:11a_11f} (Left Column, 2nd Row), 
   we consider a 
   random\footnote{In Figure~\ref{fig:11a_11f}, the circles 
   $\yinv{S_a}$ and $\yinv{S_b}$ always appear to 
   be centered on the same side of the line going through 
   $\yinv{\mathcal{A}}$  and $\yinv{\OO}$. This was done for 
   reasons of consistency and does not always 
   correspond to reality, 
   since it would be equivalent to $\inv{C_a}$ and $\inv{C_b}$ 
   always lying on the same side of the plane going through the 
   points $\inv{C_i}, \inv{C_j}$ and $\inv{C_k}$. }  
   layout of the points  $\arc{v_{ikja}}$, $\arc{v_{ikjb}}$, 
   $\arc{v_{ijkb}}$ and $\arc{v_{ijka}}$ (and the respective 
   tangent planes at these points) that appear in the order 
   \answer 2 dictates. 
   In the same figure, we provide
   a possible location of $\yinv{S_n}$, for $n\in\{a,b\}$ with respect
   to the selected layout;  $\yinv{S_n}$ 
   must be tangent to both $\ill{v_{ikjn}}$ and $\ill{v_{ijkn}}$, and
   centered according to the analysis of Section~\ref{sub:The_classic_configuration}. 

   Finally, we inspect the relative position of $\yinv{S_a}$ 
   (resp., $\yinv{S_b}$) with the lines $\arc{v_{ikjb}}$ and 
   $\arc{v_{ikjb}}$ (resp., $\arc{v_{ikja}}$ and 
   $\arc{v_{ikja}}$). In any such random layout, it must hold that
   \begin{itemize}
    \item 
    $\yinv{S_a}$ intersects the negative side of $\arc{v_{ikjb}}$
    but does not intersect the negative side of $\arc{v_{ijkb}}$ and, 
    \item 
    $\yinv{S_b}$ intersects the negative side of $\arc{v_{ijka}}$
    but does not intersect the negative side of $\arc{v_{ijka}}$. 
   \end{itemize}
   
   Another way of proving this, is by looking at the shadow regions 
   of $S_a$ and $S_b$ on the arc. For example, in a
   \answer 2 configuration, $v_{ikja}\prec v_{ikjb}\prec v_{ijka}$ 
   and subsequently  $v_{ikjb}\in\sh{S_a}$, since $\sh{S_a}$ 
   consists of all points $p\in\tri{ijk}$ with 
   $v_{ikja}\prec p\prec v_{ijka}$. As a result the sphere 
   $\tts{v_{ikjb}}$ must intersect the sphere $S_a$, \ie, 
   $\yinv{S_a}$ intersects the negative side of $\ill{v_{ikjb}}$. 

   Lastly, we translate the obtained relative positions of circles 
   and lines of \yspace to \insphere tests outcomes. For example, if 
   $\yinv{S_a}$ intersects the negative side of $\ill{v_{ikjb}}$, 
   we conclude that \insphere$(S_i,S_k,S_j,S_b,S_a)$ is negative, as
   an immediate result of Lemma~\ref{fig:zspace_yspace_equivalency}. 
   In conclusion, we get that if the Apollonius vertices we seek to 
   order appear as in \answer 2, then 
   \begin{itemize}
    \item  
    \insphere$(S_i,S_k,S_j,S_b,S_a)=-$ and 
    \insphere$(S_i,S_j,S_k,S_b,S_a)=+$,
    \item  
    \insphere$(S_i,S_k,S_j,S_a,S_b)=+$ and 
    \insphere$(S_i,S_j,S_k,S_a,S_b)=-$.
   \end{itemize}
   
   \begin{table}[tbp]
   \begin{center}
   \begin{tabular}{|c||c|c|c|c|c|c|}
   \hline
   \answer: &  \: 1 \: & \: 2 \: & \: 3 \: & \: 4 \: & \: 5 \: & \: 6 \: \\
   \hline\hline
   $\text{\insphere}(S_i,S_k,S_j,S_b;S_a)$&$+$&$-$&$+$&$+$&$+$&$-$\\ \hline
   $\text{\insphere}(S_i,S_j,S_k,S_b;S_a)$&$+$&$+$&$+$&$-$&$+$&$-$\\ \hline
   $\text{\insphere}(S_i,S_k,S_j,S_a;S_b)$&$+$&$+$&$-$&$-$&$+$&$+$\\ \hline
   $\text{\insphere}(S_i,S_j,S_k,S_a;S_b)$&$+$&$-$&$-$&$+$&$+$&$+$\\ \hline
   \end{tabular}
   \end{center}
   \caption[Signs of \insphere predicates in the 2VS2 scenario.]{Case A: Signs of all possible \insphere tests 
   that follow from the analysis of each \answer. 
   Notice that only 
   the rows that correspond to \answer 1 and \answer 5 are identical 
   and therefore we only need the signs of these \insphere predicates 
   to determine the \answer most of the time. If all predicates return 
   positive, we require some auxiliary tests to distinguish between the 
   two cases.}
   \label{tab:2VS2}
   \end{table}

   Ultimately, we create a table of the four possible \insphere 
   outcomes that hold in each of the \answer's 1 to 6 
   (see Table~\ref{tab:2VS2}). A simple way of distinguishing 
   the ordering of the Apollonius vertices becomes clear now, due 
   to tuple of outcomes being so different in most \answer's. 
   Indeed, if $\mathcal{Q}=(Q_1,Q_2,Q_2,Q_4)$ 
   denotes the ordered tuple of the \insphere predicate outcomes, where
   \begin{align*}
   Q_1 &= \text{\insphere}(S_i,S_k,S_j,S_b,S_a), \ \ 
   Q_2 = \text{\insphere}(S_i,S_j,S_k,S_b,S_a),\\
   Q_3 &= \text{\insphere}(S_i,S_k,S_j,S_a,S_b), \ \ 
   Q_4 = \text{\insphere}(S_i,S_j,S_k,S_a,S_b),
   \end{align*} 

   \noindent
   then the \order predicate returns :
   \begin{itemize}
    \item 
     the ordering of \answer 2, if $\mathcal{Q}=(-,+,+,-)$ or,
    \item 
     the ordering of \answer 3, if $\mathcal{Q}=(+,+,-,-)$ or,
    \item 
     the ordering of \answer 4, if $\mathcal{Q}=(+,-,-,+)$ or,
    \item 
     the ordering of \answer 6, if $\mathcal{Q}=(-,-,+,+)$.
   \end{itemize}

\begin{figure}[htbp]
    \centering
    \includegraphics[width=0.95\textwidth]{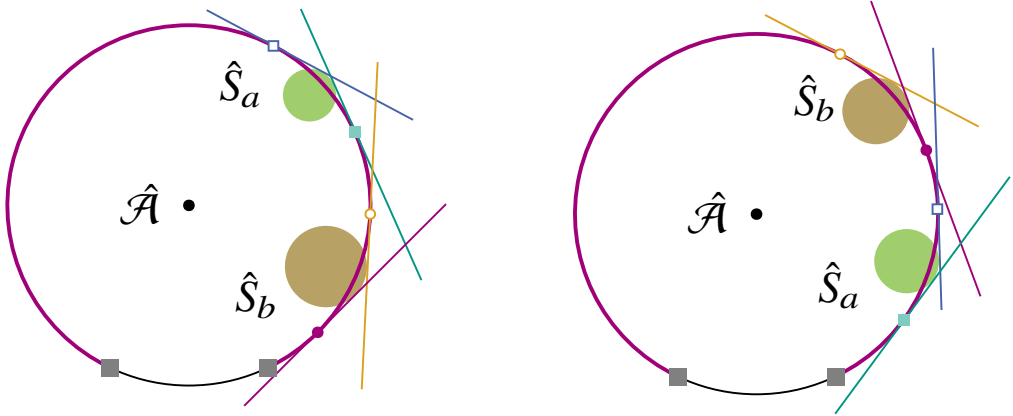}
    \caption[The dilemma between \answer 1 and \answer 5.]{If $\mathcal{Q}=(+,+,+,+)$ then we must determine 
     if the ordering of the Apollonius vertices correspond to \answer 1 
     (Left) or 5 (Right). It is apparent that we are in the first 
     case if and only if the ray $(\mathcal{A},t)$ ``meets'' 
     $\yinv{C_a}$ first as $t$ traverses the arc 
     $(\arc{\eta},\arc{o},\arc{\theta})$.}
    \label{fig:2VS2_N1_vs_N5}
   \end{figure}

   Finally, if $\mathcal{Q}=(+,+,+,+)$ then either \answer 1 or 
   \answer 5 is the correct ordering of the vertices
   (see Figure~\ref{fig:2VS2_N1_vs_N5}). To resolve this 
   dilemma, we distinguish cases depending on the ordering 
   of the midpoints $M_a$ and $M_b$ of the arcs 
   $(\arc{v_{ikja}},\arc{v_{ijka}})$ 
   and $(\arc{v_{ikjb}},\arc{v_{ijkb}})$ respectively. Since it must 
   either hold that 
   $\{v_{ikja}\prec v_{ijka}\}\prec \{v_{ikjb}\prec v_{ijkb}\}$ 
   (\answer 1) or 
   $\{v_{ikjb}\prec v_{ijkb}\}\prec \{v_{ikja}\prec v_{ijka}\}$ 
   (\answer 5), then we are obviously in the former case if 
   $M_a\prec M_b$ or in the latter if $M_b\prec M_a$. 

   To determine the ordering of $M_a$ and $M_b$ on the arc
   $(\yinv{\eta},\yinv{o},\yinv{\theta})$, we shall use the auxiliary
   point $\yinv{o}$. Initially, we reflect on the fact that, 
   for $n\in\{a,b\}$, $\yinv{C_n}$ is known to lie on the open ray
   from $\mathcal{A}$ towards $M_n$. 
   It is also apparent that the points $\OO$, $\mathcal{A}$ 
   and $\yinv{o}$ are collinear and appear in this order on the 
   line $\yinv{\ell}$ they define. 

   Based on the definition of \yspace and the remarks of 
   Section~\ref{ssub:the_yspace_analysis}, the midpoint $M_n$, 
   for $n\in\{a,b\}$, satisfies

   \begin{itemize}
   \item 
   $M_n\prec\yinv{o}$ if and only if
   $\text{\orient}(\inv{C_n},\inv{C_i},\inv{C_j},\OO)<0$,
   \item
   $\yinv{o}\prec M_n$ if and only if
   $\text{\orient}(\inv{C_n},\inv{C_i},\inv{C_j},\OO)>0$,
   \item
   $M_n\equiv\yinv{o}$ if and only if
   $\text{\orient}(\inv{C_n},\inv{C_i},\inv{C_j},\OO)=0$.
   \end{itemize}
   
   Lastly, we notice that 
   $\text{\orient}(\inv{C_b},\inv{C_i},\inv{C_j},\inv{C_a})<0$ is 
   equivalent to $\yinv{C_b}$ lying on the ``right side'' of the 
   oriented line going from $\mathcal{A}$ to $\yinv{C_a}$.
   
   Ultimately, we determine the relative position of $M_a$ and $M_b$ 
   by combining all the information extracted of the \orient 
   predicates mentioned, using the following algorithm. 

\begin{figure}[htbp]
    \centering
    \includegraphics[width=0.95\textwidth]{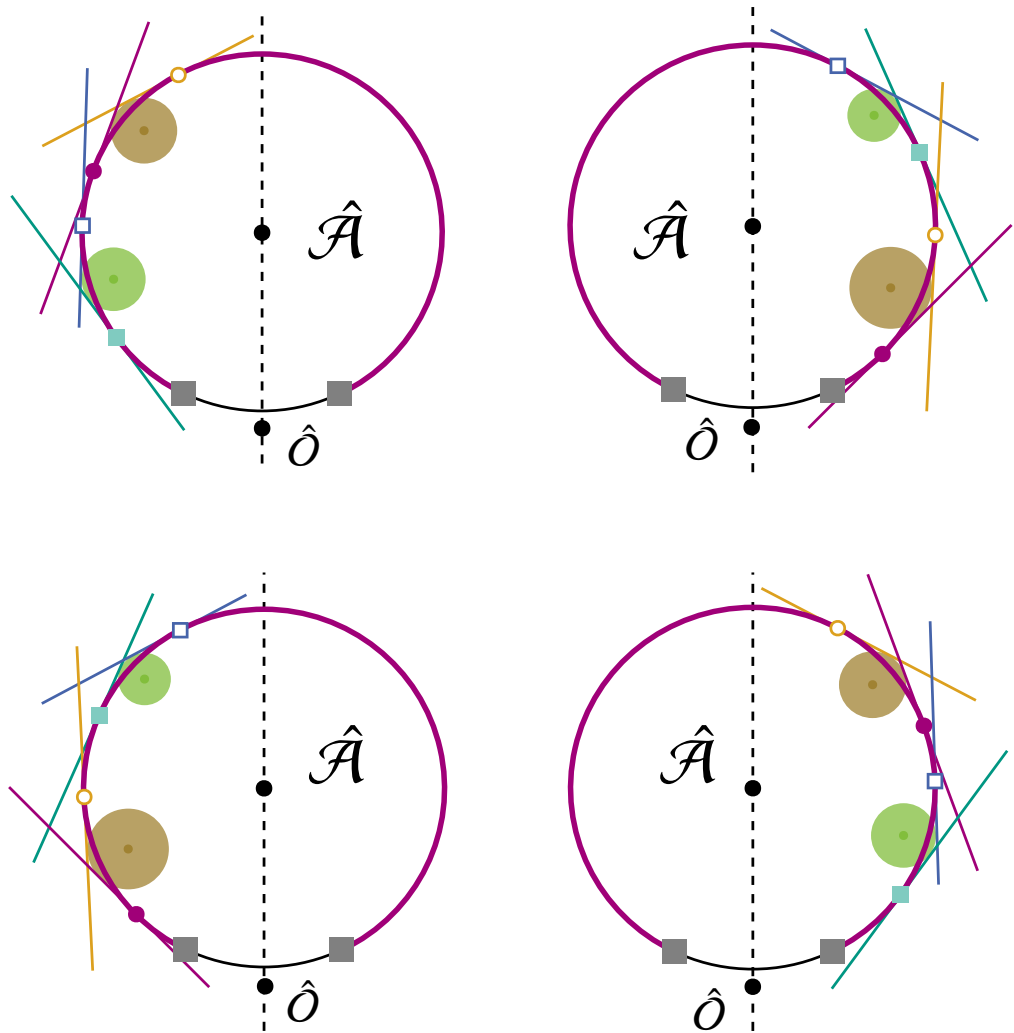}
    \caption[Deciding between \answer 1 and \answer 5 when $o_1\cdot o_2>0$.]{If $o_1\cdot o_2>0$, the 
    centers $\yinv{C_a}$ and $\yinv{C_b}$ must lie on the same side of 
    the line that goes through $\yinv{\mathcal{A}}$ and $\yinv{\OO}$.
    No matter which side the centers lie on, if 
    $o_3<0$ 
    or equivalently $\yinv{C_a}$ lies on the left side of the 
    oriented line that goes from $\yinv{\mathcal{A}}$ to  $\yinv{C_b}$, 
    (Top 2 Figures) then we obtain the ordering described 
    in \answer 1. Otherwise, we obtain the ordering described 
    in \answer 5 (Bottom 2 Figures).}
    \label{fig:2vs2_step2a}
   \end{figure}

\begin{figure}[htbp]
    \centering
    \includegraphics[width=0.95\textwidth]{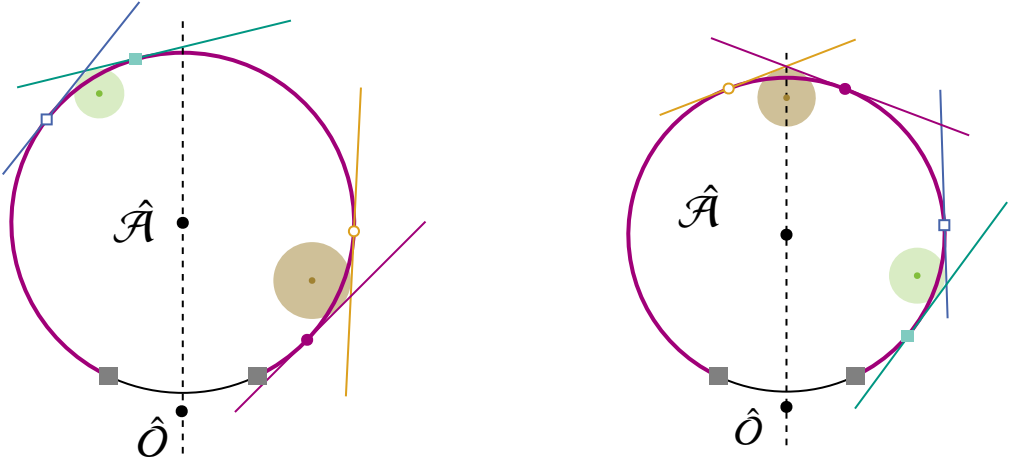}
    \caption[Deciding between \answer 1 and \answer 5 when $o_1\cdot o_2\leq 0$.]{If $o_1\cdot o_2\leq 0$, the 
    centers $\yinv{C_a}$ and $\yinv{C_b}$ lie on different sides of 
    the line that goes through $\yinv{\mathcal{A}}$ and $\yinv{\OO}$ 
    (Left) or only one of them lies on the line (since we are in 
    either \answer 1 or 5).
    No matter which side the centers lie on, if 
    $o_1< o_2$
    we obtain the ordering described in \answer 1. Otherwise, 
    we obtain the ordering described in \answer 5 (Bottom 2 Figures).
    }
    \label{fig:2vs2_step2b}
   \end{figure}

   \begin{description}
    \item[Step 1.] 
    We evaluate 
    $o_1=\text{\orient}(\inv{C_a},\inv{C_i},\inv{C_j},\OO)$, 
    $o_2\allowbreak=\allowbreak\text{\orient}\allowbreak(\allowbreak\inv{C_b},\allowbreak\inv{C_i},\inv{C_j},\OO)$ and $\Pi=\sgn(o_1)\sgn(o_2)$. 
    If $\Pi>0$ go to Step 2a, otherwise go to Step 2b.
    \item[Step 2a.] 
    Either $M_a,M_b\prec \yinv{o}$ or $\yinv{o}\prec M_a,M_b$. 
    In either case, we evaluate
    $o_3=\text{\orient}\allowbreak(\allowbreak\inv{C_b},\allowbreak\inv{C_i}, \inv{C_j}, \inv{C_a})$. 
    If $o_3<0$ then $M_a\prec M_b$, and the \order predicate returns the 
    ordering of \answer 1. Otherwise, $M_b\prec M_a$ and the ordering of
    \answer 5 is returned.
    (see Figure~\ref{fig:2vs2_step2a}).
    \item[Step 2b.] 
    Either $\yinv{o}$ lies in-between $M_a$ and $M_b$ or 
    is identical with one of them.
    In both cases, if 
    $o_1<o_2$ then $M_a\prec M_b$ and the \order predicate return the ordering of \answer 1, otherwise, $M_b\prec M_a$ and the ordering 
    of \answer 5 is returned.
    (see Figure~\ref{fig:2vs2_step2b}).
   \end{description} 

   \paragraph*{\textbf{Analysis of Case B}}
   Given that $\sh{S_a}$ and $\sh{S_b}$ are both of the form 
   $(\chi,+\infty)$ and therefore only the Apollonius vertices
   $v_{ikja}$ and $v_{ikjb}$ exist on $\tri{ijk}$, the ordering
   of these vertices on $(\yinv{\eta}, \yinv{o}, \yinv{\theta})$ 
   is either
   \begin{description}
   \item[\answer 1.] $v_{ikja}\prec v_{ikjb}$ or,
   \item[\answer 2.] $v_{ikjb}\prec v_{ikja}$.
   \end{description}

   A similar analysis with the Case A is used to resolve the predicate 
   in Case B; we create a table regarding the possible outcomes 
   of the \insphere tests with inputs 
   $(S_i$,$S_k$,$S_j$,$S_a$,$S_b)$ and 
   $(S_i$,$S_k$,$S_j$,$S_b$,$S_a)$. Recall that the outcome of 
   $Q_1=\text{\insphere}(S_i,S_k,S_j,S_a,S_b)$ 
   (resp., $Q_2=\text{\insphere}(S_i,S_k,S_j,S_a,S_b)$) 
   is $-$, $0$ or $+$ if the circle $\yinv{S_b}$ (resp., $\yinv{S_a}$)
   intersects, is tangent to or does not intersect the negative side 
   of $\ill{v_{ikja}}$ (resp., $\ill{v_{ikjb}}$).

   Using a simpler approach, we observe that 
   \begin{itemize}
    \item 
    in \answer 1, $v_{ikja}$ does not belong to the shadow region 
    of the sphere $S_b$ on $\tri{ijk}$ and therefore $\tts{v_{ikja}}$
    does not intersect $S_b$ or equivalently $Q_1=+$. Moreover,
    in this case,  $v_{ikjb}$ belongs to the shadow region 
    of the sphere $S_a$ on $\tri{ijk}$ and therefore $\tts{v_{ikjb}}$  
    intersects  $S_a$ or equivalently $Q_2=-$. 
    \item 
    In \answer 2, $v_{ikja}$ belongs to the shadow region 
    of the sphere $S_b$ on $\tri{ijk}$ and therefore $\tts{v_{ikja}}$  
    intersects  $S_b$ or equivalently $Q_1=-$. Furthermore, 
    $v_{ikjb}$ does not belong to the shadow region 
    of the sphere $S_a$ on $\tri{ijk}$ and therefore $\tts{v_{ikjb}}$   
    does not intersects $S_a$ or equivalently $Q_2=+$. 
   \end{itemize}

   In conclusion we can answer the 
   \text{\order}$(S_i,S_j,S_k,S_a,S_b)$ predicate in case B by 
   evaluating $Q_1$; if $Q_1=+$ then return \answer 1 otherwise,
   if $Q_1=-$ return \answer 2. 
   Equivalently, we could evaluate $Q_2$ instead of $Q_1$; 
   if $Q_2=-$ then return \answer 1 otherwise
   if $Q_2=-$ return \answer 2. The following equivalencies are 
   depicted in Table~\ref{tab:1VS1} and this concludes the analysis of 
   Case B. 

   \begin{table}[bt]
   \begin{center}
   \begin{tabular}{|c||c|c|}
   \hline
   & OrderCase 1 & OrderCase 2 \\
   \hline\hline
   $\text{\insphere}(S_i,S_k,S_j,S_b;S_a)$ & $+$ & $-$ \\ \hline
   $\text{\insphere}(S_i,S_k,S_j,S_a;S_b)$ & $-$ & $+$ \\ 
   \hline
   \end{tabular}
   \end{center}
   \caption[Signs of \insphere predicates in the 1VS1 scenario.]{Case B: Signs of all possible \insphere tests 
   that follow from the analysis of each \answer. Notice that each 
   column is distinct and therefore we can determine the  \answer
   after the outcomes of the  \insphere predicates.}
   \label{tab:1VS1}
   \end{table}

   \begin{figure}[tbp]
   \centering
   \includegraphics[width=0.4\textwidth]{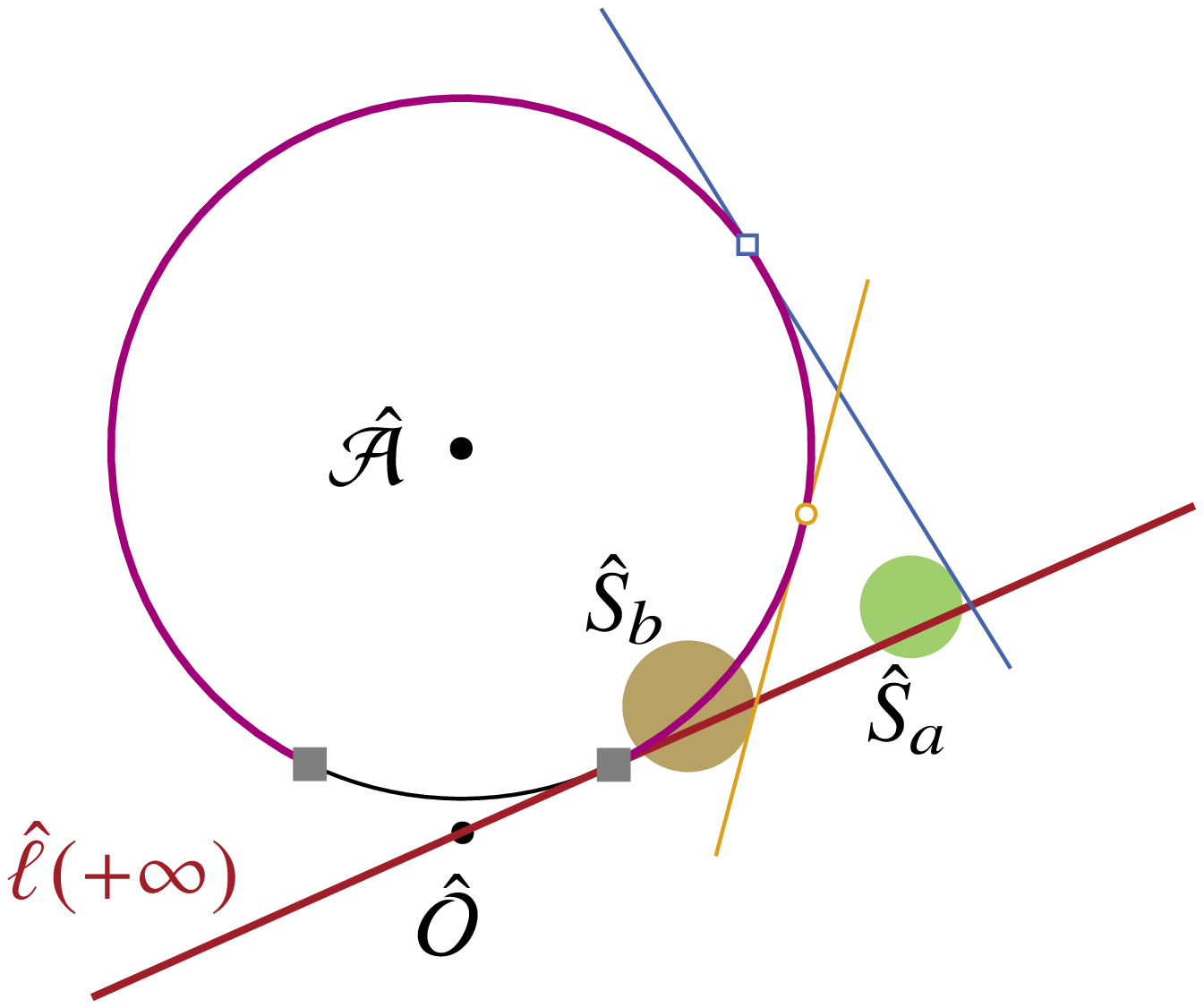}\hspace{1 cm}
   \includegraphics[width=0.4\textwidth]{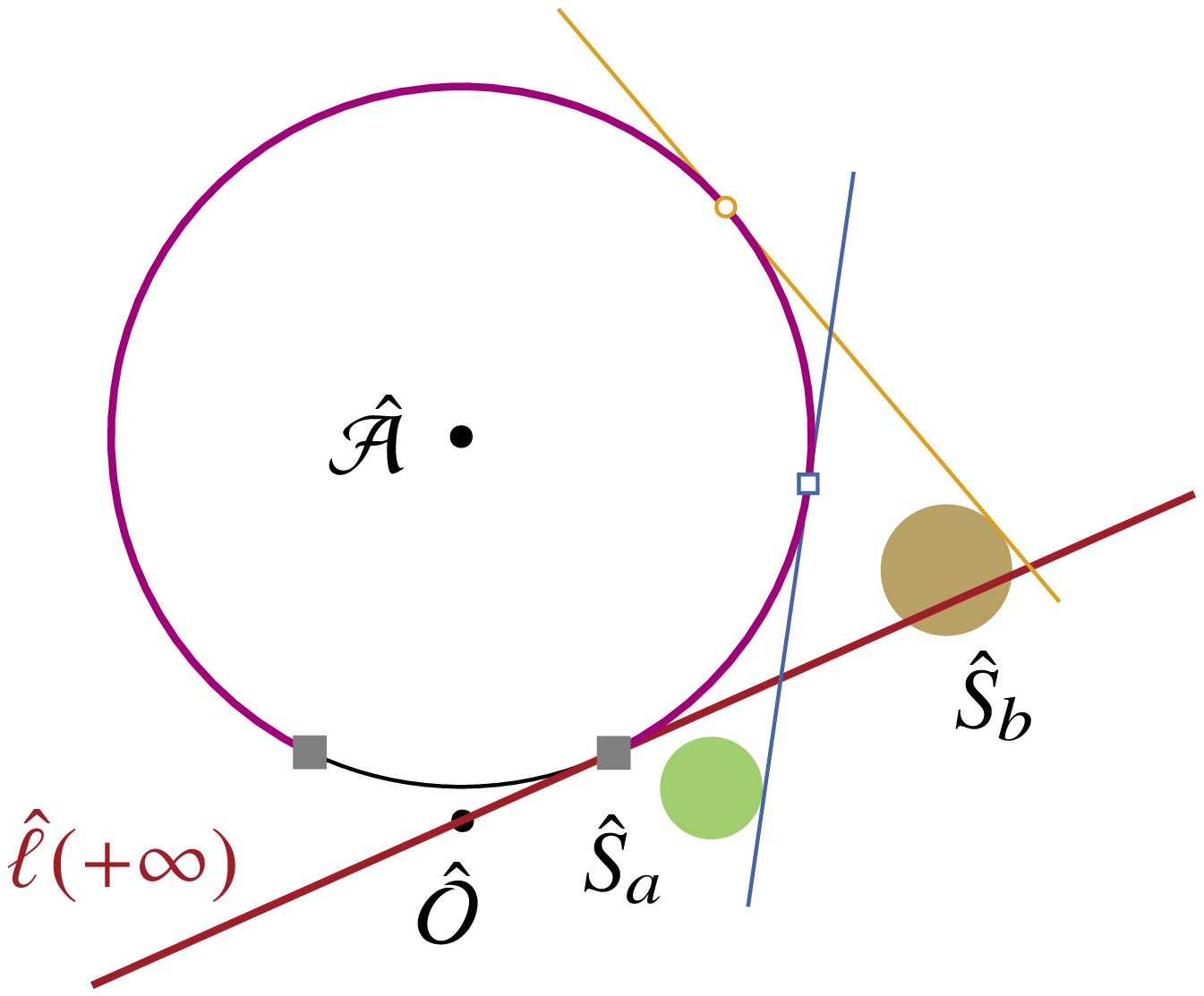}
   \caption[Possible orderings of the images of $v_{ikja}$ and $v_{ijkb}$ in \yspace.]{In Case B, it is assumed that only the Apollonius vertices 
   $v_{ikja}$ and $v_{ijkb}$ exist on the trisector
   $\tri{ijk}$. We consider the two possible orderings of these vertices:
   \answer 1 (Left) and \answer 2 (Right). Similar with Case A, 
   we consider the corresponding ordering of the points  
   $\arc{v_{ikja}}$ and $\arc{v_{ijkb}}$ on the arc 
   $(\arc{\eta},\arc{o},\arc{\theta})$. A possible location for 
   the circles $\yinv{S_a}$ and $\yinv{S_b}$ is drawn based on 
   the analysis of Section~\ref{sub:the_main_algorithm}.}
   \label{fig:12a_12b}
   \end{figure}

   \paragraph*{\textbf{Analysis of Case C}}

   In Case C, it is assumed that $\sh{S_a}=(\chi,\phi)$ hence
   $v_{ikja}\prec v_{ijka}$ while $\sh{S_b}=(\chi,+\infty)$ and 
   consequently only $v_{ikjb}$ exists on the arc 
   $(\yinv{\eta}, \yinv{o}, \yinv{\theta})$. All three possible 
   orderings of these three Apollonius vertices on the 
   arc are 
   \begin{description}
   \item[\answer 1.] $v_{ikja}\prec v_{ijka}\prec v_{ikjb}$,
   \item[\answer 2.] $v_{ikja}\prec v_{ikjb}\prec v_{ijka}$ and 
   \item[\answer 3.] $v_{ikjb}\prec v_{ikja}\prec v_{ijka}$.
   \end{description}

   The analysis of this Case uses the same tools and analysis 
   presented in the previous two cases with small adjustments, 
   since $\sh{S_a}=(\chi,\phi)$ and $\sh{S_b}=(\chi,+\infty)$ in 
   the case studied. 
   Let us denote by $Q_1,Q_2$ and $Q_3$ the results of the 
   \insphere predicates with inputs $(S_i,S_k,S_j,S_b,S_a)$, 
   $(S_i,S_k,S_j,S_a,\allowbreak S_b)$ and $(S_i,S_j,S_k,S_b,S_a)$ respectively.

   Notice now that
   \begin{itemize}
   \item 
   in \answer 1, $v_{ikjb}$,$ v_{ikja}$ and $v_{ijka}$ 
   do not belong to the shadow region of $S_a$, $S_b$ and $S_b$ 
   respectively and therefore it must stand that $Q_1=Q_2=Q_3=+$.
   \item 
   in \answer 2, $v_{ikjb}$ and $v_{ijka}$ 
   belong to the shadow region of $S_a$ and $S_b$ 
   respectively and for this reason $Q_1=-$ and $Q_3=-$.
   On the other hand, $v_{ikja}$ does not belong to 
   the shadow region of $S_b$ and therefore $Q_2=+$. Finally,
   \item 
   in \answer 3, both $v_{ikja}$ and $v_{ijka}$ 
   belong to the shadow region of $S_b$ and 
   consequently $Q_2=-$ and $Q_3=-$ whereas,
   $v_{ikjb}$ does not belong to the shadow region of $S_a$
   and therefore $Q_1=+$.
   \end{itemize}

   Since the tuple $\mathcal{Q}=(Q_1,Q_2,Q_3)$ is different 
   in each \answer 1 to 3, we can answer the predicate by 
   evaluating the three \insphere predicates hence $\mathcal{Q}$ 
   and correspond it the respective ordering (also see 
   Table~\ref{tab:1VS2}):
   \begin{itemize}
    \item if $\mathcal{Q}=(+,+,+)$, return the ordering of \answer 1 or,
    \item if $\mathcal{Q}=(-,+,-)$, return the ordering of \answer 2 otherwise,
    \item if $\mathcal{Q}=(+,-,-)$, return the ordering of \answer 3.
   \end{itemize}

   \begin{table}[tbp]
   \begin{center}
   \begin{tabular}{|c||c|c|c|}
   \hline
   & OrderCase 1 & OrderCase 2 & OrderCase 3 \\
   \hline\hline
   $\text{\insphere}(S_i,S_k,S_j,S_b;S_a)$ & $+$ & $-$ & $+$ \\ \hline
   $\text{\insphere}(S_i,S_k,S_j,S_a;S_b)$ & $+$ & $+$ & $-$ \\ \hline
   $\text{\insphere}(S_i,S_j,S_k,S_a;S_b)$ & $+$ & $-$ & $-$ \\ \hline
   \end{tabular}
   \end{center}
   \caption[Signs of \insphere predicates in the 1VS2 scenario.]{Case C: Signs of all possible \insphere tests 
   that follow from the analysis of each \answer. Notice that each 
   column is distinct and therefore we can determine the  \answer
   after the outcomes of the \insphere predicates, as in Case B.}
   \label{tab:1VS2}
   \end{table}

   \begin{figure}[tbp]
   \centering
   \includegraphics[width=0.4\textwidth]{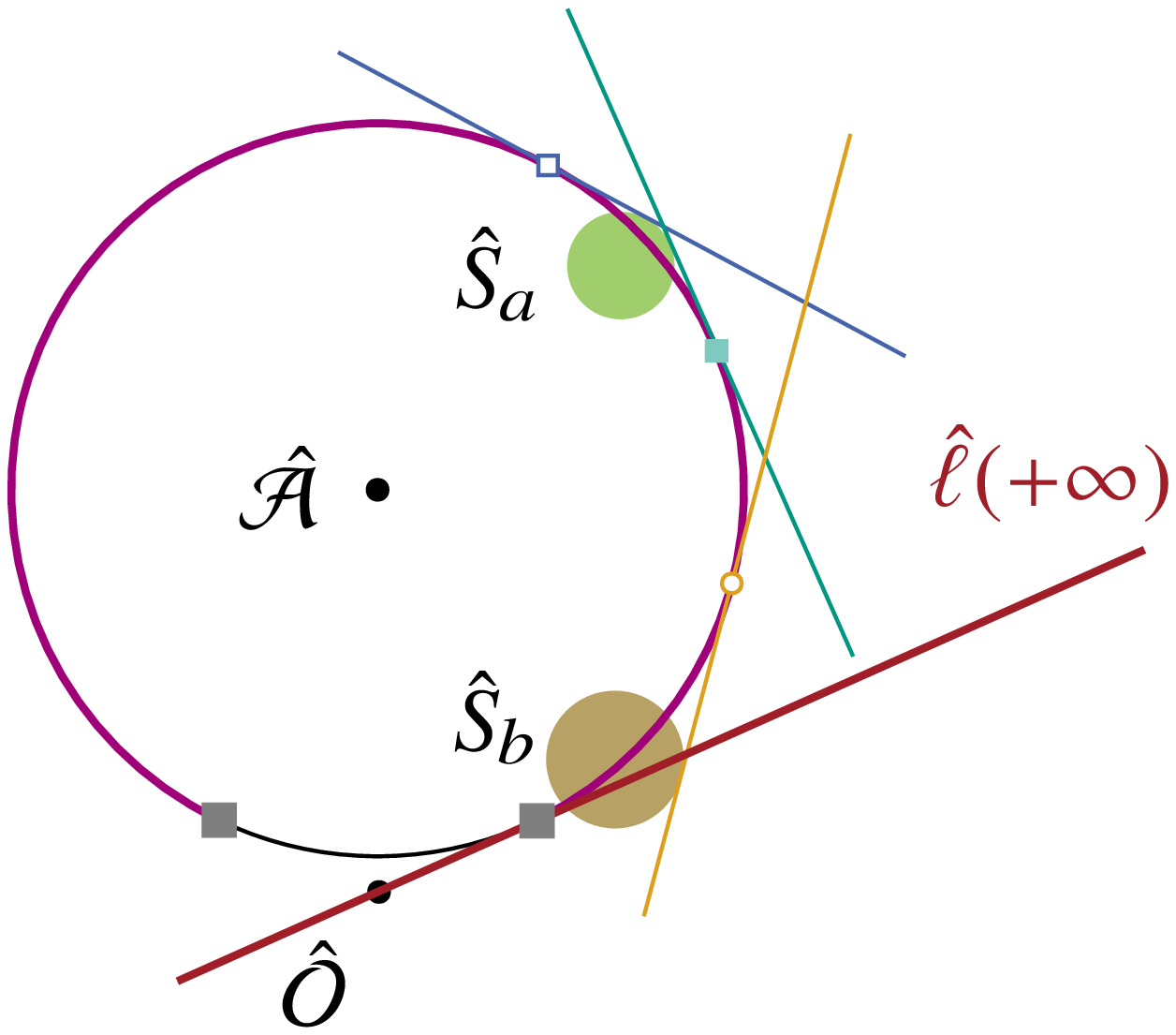}\hspace{1 cm}
   \includegraphics[width=0.4\textwidth]{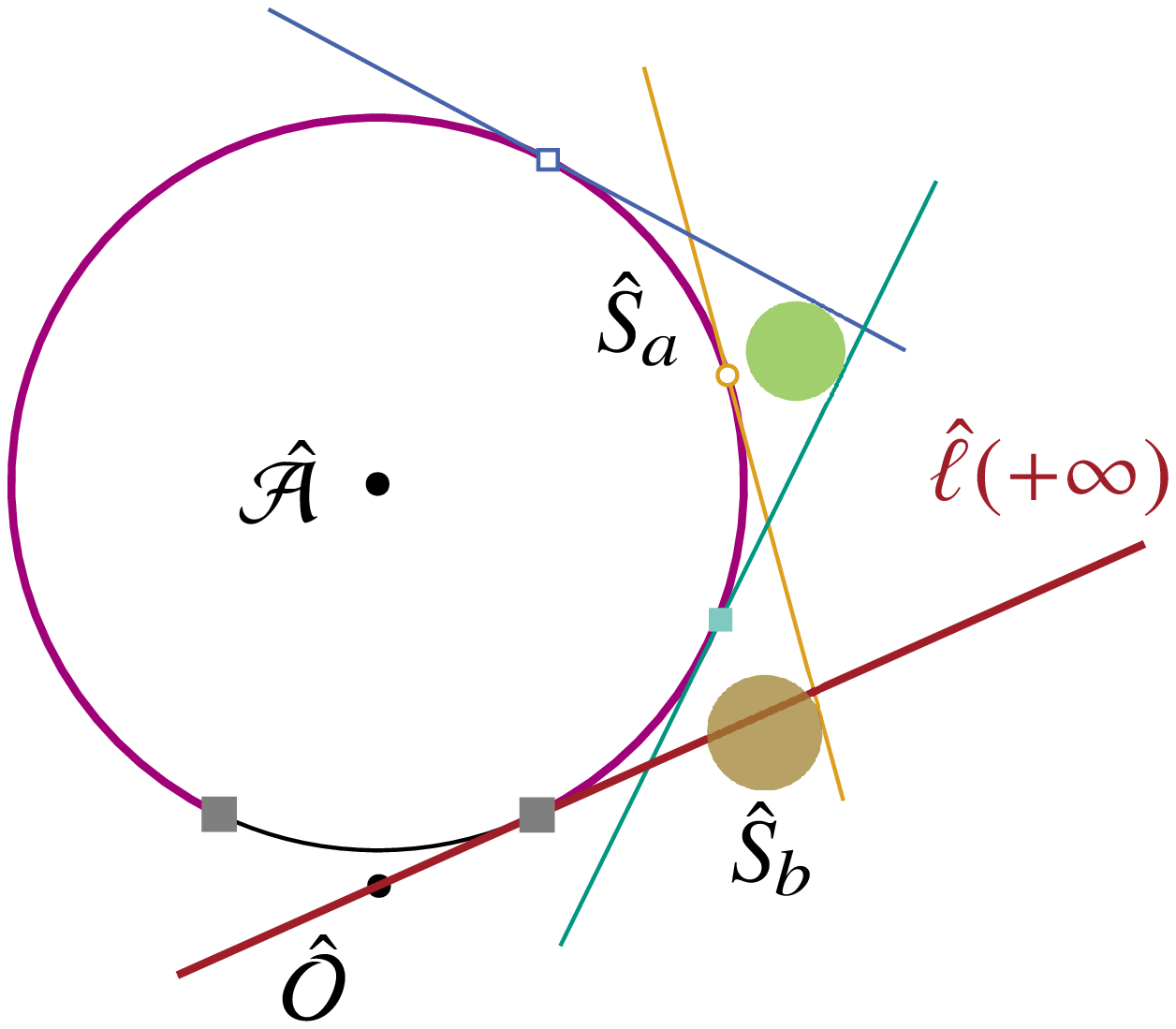}\\
   \includegraphics[width=0.4\textwidth]{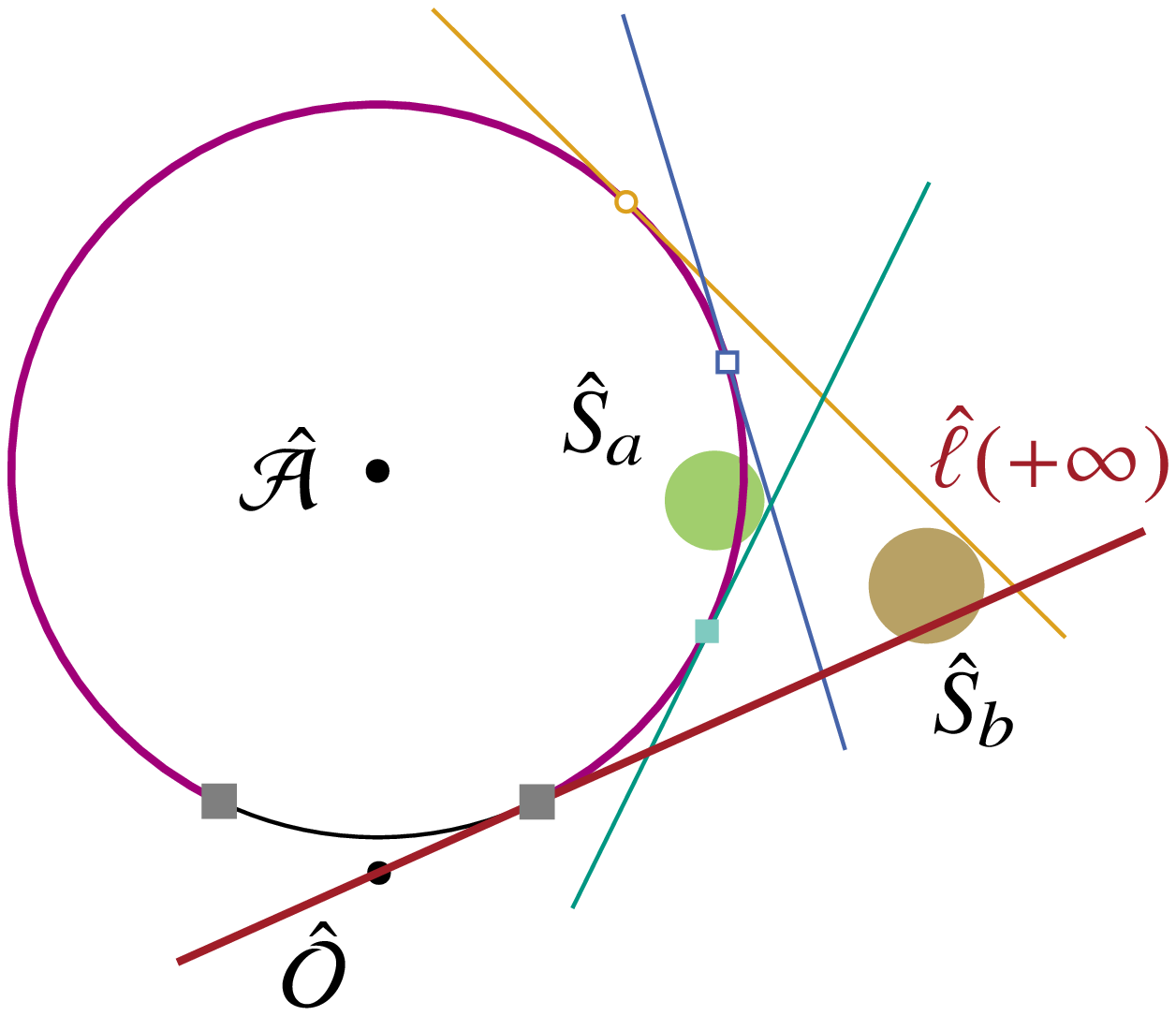}
   \caption[Possible orderings of the images of $v_{ikja}, v_{ijka}$ 
   and $v_{ijkb}$ in \yspace.]{In Case C, it is assumed that only the Apollonius vertices 
   $v_{ikja}, v_{ijka}$ and $v_{ijkb}$ exist on the trisector
   $\tri{ijk}$. We consider the three possible orderings of these vertices:
   \answer 1 (Top Left), \answer 2 (Top Right) and \answer 3 (Bottom). 
   Similar with Case A and B, 
   we consider the corresponding ordering of the points  
   $\arc{v_{ikja}}$ and $\arc{v_{ijkb}}$ on the arc 
   $(\arc{\eta},\arc{o},\arc{\theta})$. A possible location for 
   the circles $\yinv{S_a}$ and $\yinv{S_b}$ is drawn based on 
   the analysis of Section~\ref{sub:the_main_algorithm}.}
   \label{fig:13a_13c}
   \end{figure}

   \paragraph*{\textbf{Algebraic Cost to resolve the Cases A, B or C}}
   The analysis of the Cases A, B and C showed that the answer 
   of the $\text{\order}(S_i,S_j,S_k,S_a,S_b)$ predicate
   in a classic configuration ultimately amounts to determining 
   the outcomes of up to four \insphere predicates and, if needed, 
   some auxiliary \orient tests.

   To answer any of the \insphere predicates that may require 
   evaluation, we must 
   perform operations of maximum algebraic degree 10 (in the input 
   quantities), as mentioned in Section~\ref{sub:the_insphere_predicate}. 

   Regarding the auxiliary \orient primitives, we observe that
   \begin{align}
    \text{\orient}(\inv{C_b},\inv{C_i},\inv{C_j},\inv{C_a})
      &= \sgn(D^{uvw}_{bija})
      = \sgn(\inv{p_i}\inv{p_j}\inv{p_a}\inv{p_b})\sgn(E^{xyzp}_{bija})\\
      &= \sgn(E^{xyzp}_{bija}),  
   \end{align}
   where the quantity $E^{xyzp}_{bija}$ and  is an expression of 
   algebraic degree 5 on the input quantities. The expression
   $\text{\orient}(\inv{C_n}, \inv{C_i}, \inv{C_j}, \OO)$, for $n\in\{a,b\}$ 
   can be evaluated as shown in Section~\ref{ssub:the_yspace_analysis},
   \begin{align}
   \text{\orient}(\inv{C_n},\inv{C_i},\inv{C_j},\OO)&=
   \sgn(\inv{p_i}\inv{p_i}\inv{p_n}D^{uvw}_{nij})=
   \sgn(D^{xyz}_{nijk})\\
   &= \text{\orient}(C_n,C_i,C_j,C_k)
   \end{align}

   \noindent and therefore its evaluation requires operations of
   algebraic degree 4 (in the input quantities). 

   In conclusion, since the evaluation of the \insphere predicates 
   is the most degree-demanding operation throughout the 
   evaluation of the \order predicate in a classic configuration, we 
   have proven the following lemma. 

   \begin{lemma}
   The \order predicate in a classic configuration can be evaluated 
   by determining the sign of quantities of algebraic 
   degree at most 10 (in the input quantities).
   \end{lemma}

   \subsubsection{Ordering the Apollonius vertices in 
   a non-classic configuration} 
   \label{sub:ordering_in_a_non_classic_configuration}
   
   In the previous section we presented a way to resolve the 
   \text{\order}$(S_i,S_j,S_k,S_a,S_b)$ predicate under the 
   assumption that $\sh{S_a}$ and $\sh{S_b}$ were 
   either $(\chi,+\infty)$ or $(\chi,\phi)$ (not necessary the same); 
   we called this \emph{a classic configuration}. 
   In this section, we will assume we are in a non-classic 
   configuration, \ie, at least one of $\sh{S_a}$ or $ \sh{S_b}$  
   is $(-\infty,\phi)$ or $(-\infty,\phi)\cup(\chi,\phi)$. 
   For convenience, these last two forms of a shadow region are 
   labelled as non-classic whereas the classic forms are 
   $(\chi,+\infty)$ and $(\chi,\phi)$.

   If $\sh{S_n}$ has a non-classic type, for $n=a$ or $b$, then we claim 
   that there exist a sphere $S_N$, for $N=A$ or $B$ respectively,
   such that:
   \begin{itemize}
    \item 
    if $\sh{S_n}=(-\infty,\phi)$ then $\sh{S_N}=(\chi,+\infty)$ 
    and $v_{ijkn}\equiv v_{ikjN}$ or,
    \item 
    if $(-\infty,\phi)\cup(\chi,\phi)$ then $\sh{S_N}=(\chi,\phi)$
    and $v_{ijkn}\equiv v_{ikjN}$ as well as $v_{ikjn}\equiv v_{ijkN}$.
   \end{itemize}

   \noindent If these conditions hold, we will say that 
   $S_n$ and $S_N$ are \emph{equivalent spheres}. Notice 
   that if $\sh{S_n}$ has a non-classic type then the shadow region
   of its equivalent sphere has a classic type and vice versa. The utility 
   of this equivalency is that it enable us to make a connection 
   between a classic and a non-classic configuration in the following 
   way. 

   When the predicate $\text{\order}(S_i,S_j,S_k,S_a,S_b)$ is 
   called then 
   \begin{enumerate}
    \item 
    if $\sh{S_a}$ and $\sh{S_b}$ have a classic type,
    we are in a classic and therefore, we resolve the predicate
    based on the analysis of 
    Section \ref{sub:ordering_in_a_classic_configuration}.
    \item 
    If $\sh{S_a}$ has a classic type and $\sh{S_b}$ does not, 
    then we call \order$(S_i,S_j,S_k,S_a,\allowbreak S_B)$. Since both
    $\sh{S_a}$ and $\sh{S_B}$ have a classic type, this predicate 
    can be evaluated using analysis of 
    Section~\ref{sub:ordering_in_a_classic_configuration} with 
    some adjustments. The predicate's outcome would be the 
    ordering of $v_{ikja}$, $v_{ikjB}$ and any of the existing
    $v_{ijka}$ or $v_{ijkB}$. Using the property of equivalent 
    spheres, we could answer the initial predicate by 
    substituting $v_{ikjB}$ with $v_{ijkb}$ and, if it exists, 
    $v_{ijkB}$ with $v_{ikjb}$.
    \item 
     If $\sh{S_b}$ has a classic type and $\sh{S_a}$ does not,
     then we follow a similar analysis with the previous case. 
     We evaluate \order$(S_i$, $S_j,S_k,S_A,S_b)$ and 
     in the resulting ordering of the Apollonius vertices 
     $v_{ikjA}$, $v_{ikjb}$ and any of the existing
     $v_{ijkA}$ or $v_{ijkb}$, we will substitute  
     $v_{ikjA}\equiv v_{ijka}$ and if necessary, 
     $v_{ijkA}\equiv v_{ikja}$, to obtain the answer to the 
     initial \order predicate. 
     \item 
     Finally, if both $\sh{S_a}$ and $\sh{S_b}$ do not have 
     a classic type we evaluate $\text{\order}(S_i,\allowbreak S_j,\allowbreak S_k,\allowbreak S_A,\allowbreak S_B)$.
     As before, we substitute $v_{ikjA}\equiv v_{ijka}$, 
     $v_{ikjB}\equiv v_{ijkb}$ and if necessary, 
     $v_{ijkA}\equiv v_{ikja}$ and/or $v_{ijkB}\equiv v_{ikjb}$, 
     and the acquired ordering is the answer of the initial 
     \order predicate.
   \end{enumerate}

   The evaluation of the \order predicate called in any of these 
   4 cases will eventually require determining \insphere or 
   \orient predicates with inputs that involve the 
   sites $S_i,S_j,S_k,S_A$ (or $S_a$) and $S_B$(or $S_b$). 
   The list of all possible predicates that must be 
   evaluated, in the worst case scenario and assuming a classic 
   configuration, would be:
   \begin{itemize}
   \item \insphere$(S_i,S_k,S_k,S_a,S_b)$, 
   \item \insphere$(S_i,S_j,S_k,S_a,S_b)$, 
   \item \insphere$(S_i,S_k,S_j,S_b,S_a)$, 
   \item \insphere$(S_i,S_j,S_k,S_b,S_a)$, 
   \item \orient$(C_a,C_i,C_j,C_k)$,  
   \item \orient$(C_b,C_i,C_j,C_k)$ and 
   \item \orient$(\inv{C_a},\inv{C_i},\inv{C_j},\inv{C_b})$.
   \end{itemize}

   \noindent 
   It is apparent that we must be able to answer these predicates 
   when either one or both of $S_a$ and $S_b$ are substituted by 
   $S_A$ and $S$ respectively. 

   Firstly, we present a way of defining an equivalent sphere 
   $S_N$ when $\sh{S_n}$ has a non-classic type, 
   for $N=A$ or $B$ and $n=a$ or $b$ respectively. 
   Since $\yinv{C_n}$ cannot coincide with
   $\yinv{\mathcal{A}}$ (because there are either 1 or 2  
   cotangent lines to $\wcone$ and $\yinv{S_n}$), these points 
   define a line $\yinv{\ell_n}$. If a random point $\yinv{C_N}$ 
   is selected on $\yinv{\ell_n}$ such that $\yinv{\mathcal{A}}$ 
   lies in-between $\yinv{C_N}$ and $\yinv{C_n}$, then we 
   may choose an appropriate radius such that a circle $\yinv{S_n}$,
   centered at $\yinv{C_N}$, is tangent to any of the existing 
   lines $\ill{v_{ikjn}}$ and $\ill{v_ijkn}$. 

\begin{figure}[htbp]
    \centering
    \includegraphics[width=0.7\textwidth]{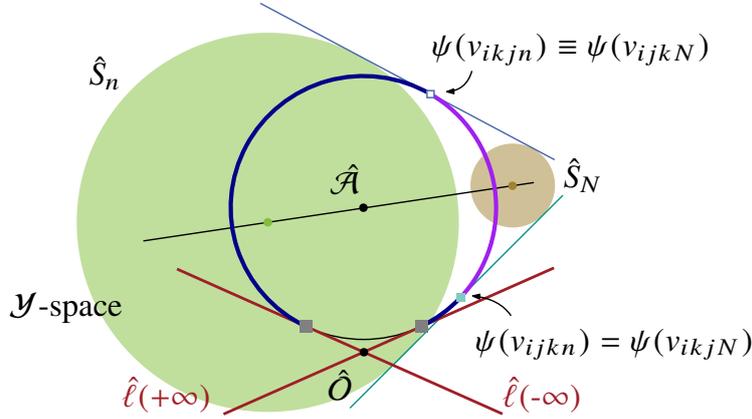}
    \caption[Equivalent spheres in \yspace when their shadow regions have two finite boundary points.]{The shadow region of $S_n$ is 
    $(-\infty,\phi)\cup(\chi,+\infty)$ as its image in \yspace is 
    the blue area of the arc. Notice that the 
    respective image of $\sh{S_N}$ is the purple area and therefore 
    $\sh{S_n}$ must equal $(\chi,\phi)$. Since the endpoints of the 
    two shadow regions coincide and based on Lemma~\ref{lemma:phi_chi}, 
    it must hold that $v_{ijkN}\equiv v_{ikjn}$ and 
    $v_{ikjN}\equiv v_{ijkn}$. Therefore, $\yinv{S_N}$ and $\yinv{S_n}$ 
    are equivalent.}
    \label{fig:14a}
   \end{figure}

   Notice that any sphere $S_N$ of \wspace whose corresponding image 
   in \yspace is the circle $\yinv{S_N}$ has the desired properties 
   of an equivalent sphere of $S_n$. Indeed, if $\sh{S_n}$ is 
   $(\chi,\phi)$ then it must stand that 
   $\sh{S_N}=(-\infty,\phi)\cup(\chi,+\infty)$ and 
   specifically the actual endpoints of these shadow regions on 
   the trisector $\tri{ijk}$ coincide. To prove this argument, 
   we only need observe in \yspace that the circle $\yinv{C_n}$ 
   intersects the negative side of a line $\ill{p}$ only for 
   $v_{ikjn}\prec v_{ijkn}$ whereas, these are the only family of 
   lines $\ill{p}$ for $p\in\tri{ijk}$ that do not intersect 
   $\yinv{S_N}$. As a conclusion the shadow region of $S_n$ and 
   $S_N$ must be complementary, \ie, 
   $\sh{S_N}=(-\infty,\phi)\cup(\chi,+\infty)$. From 
   Lemma~\ref{lemma:phi_chi}, we deduce that $v_{ijkN}\prec v_{ikjN}$ and 
   since these endpoints coincide with the endpoints of 
   $\sh{S_n}$ it must hold that $v_{ijkN}\equiv v_{ikjn}$ and 
   $v_{ikjN}\equiv v_{ijkn}$, since $v_{ikjn}\prec v_{ijkn}$ (see Figure~\ref{fig:14a}). 

\begin{figure}[htbp]
    \centering
    \includegraphics[width=0.7\textwidth]{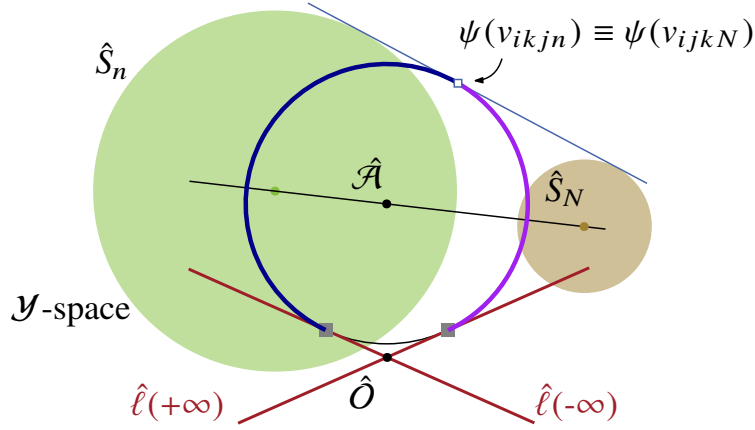}
    \caption[Equivalent spheres in \yspace when their shadow regions have one finite boundary point.]{The shadow region of $S_n$ is 
    $(-\infty,\phi)$ as its image in \yspace is 
    the blue area of the arc. Notice that the 
    respective image of $\sh{S_N}$ is the purple area and therefore 
    $\sh{S_n}$ must equal $(\chi,+\infty)$. Since the endpoints of the 
    two shadow regions coincide and based on Lemma~\ref{lemma:phi_chi}, 
    it must hold that $v_{ijkN}\equiv v_{ikjn}$. Therefore, $\yinv{S_N}$ 
    and $\yinv{S_n}$ are equivalent.}
    \label{fig:14b}
   \end{figure}

   Using a similar analysis, one can consider an equivalent 
   sphere $S_N$ of $S_n$, when $\sh{S_n}$ is assumed to be 
   $(-\infty,\phi)$. The center of the respective circle 
   $\yinv{C_N}$ is selected in the same way as above, and the radius 
   of $\yinv{S_N}$ is chosen such that the circle is tangent 
   to $\ill{v_{ijkn}}$. Again, we can conclude that 
   $\sh{S_N}$ and $\sh{S_n}$ are complementary since the 
   family of lines $\ill{p}$ for $\yinv{\eta}\prec p$ 
   are the locus of lines $\ill{p}$, with 
   $\yinv{p}\in (\yinv{\eta},\yinv{o},\yinv{\theta})$, whose negative
   side is intersected by $\yinv{S_n}$ and simultaneously, 
   whose negative side is not intersected by $\yinv{S_N}$ (see Figure~\ref{fig:14b}).

   An interesting observation is that $S_N$ is not uniquely defined 
   in the sense that we do not provide its exact coordinates expressed 
   as a function of the input quantities. This is a consequence of the  fact
   that there are infinite spheres $S_n$ that all share the same Apollonius vertices $v_{ikjn}$ and $v_{ijkn}$.

\begin{figure}[htbp]
    \centering
    \includegraphics[width=0.7\textwidth]{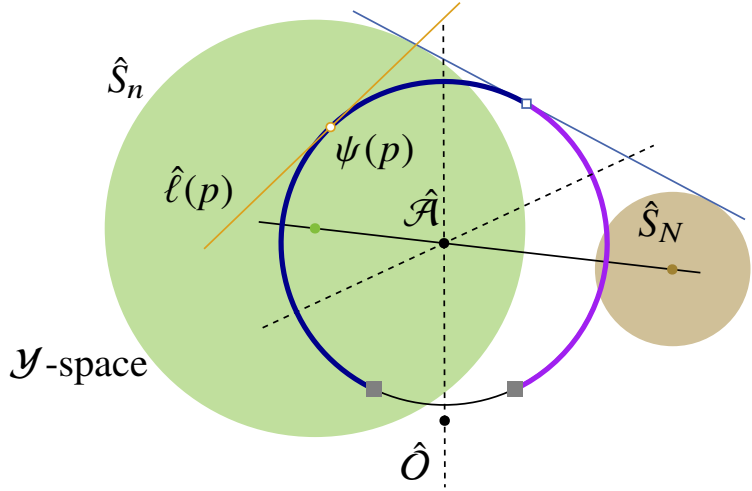}
    \caption[Remarks on the centers of equivalent spheres in \yspace.]{If $S_N$ is an equivalent sphere of $S_n$, then 
    it must hold that the centers $\yinv{C_n}$, $\yinv{C_N}$ and 
    $\yinv{\mathcal{A}}$ are collinear and the former two points 
    lie on opposite sides with respect to the latter. Observe that 
    they also lie on opposite sides with respect to any line that 
    goes through $\yinv{\mathcal{A}}$. Lastly, it is apparent that 
    a point $\arc{p}$ on the arc $(\arc{\eta},\arc{o},\arc{\theta})$
    must lie on the image of the shadow region of either $S_n$ or 
    $S_N$.}
    \label{fig:15}
   \end{figure}

   Resuming the analysis of the properties of the equivalent sphere, 
   we notice that if a point $p\in\tri{ijk}$ lies on the shadow region 
   of $S_n$ then it must not lie on the shadow region 
   of $S_N$ and vice versa. An equivalent statement would be that 
   a sphere $\tts{p}$, for $p\in\tri{ijk}$, intersects $S_n$ 
   if and only if it does not intersect $S_N$ (see Figure~\ref{fig:15}). 
   If $p$ is chosen to 
   be either $v_{ikjm}$ or $v_{ijkm}$, where 
   $m\in\{a,b\}\backslash\{n\}$, we get the following relations
   \begin{align}
   \text{\insphere}(S_i,S_j,S_k,S_m,S_N) &=
     -\text{\insphere}(S_i,S_j,S_k,S_m,S_n), \\
   \text{\insphere}(S_i,S_k,S_j,S_m,S_N) &=
     -\text{\insphere}(S_i,S_k,S_j,S_m,S_n).
   \end{align}
   
   Moreover, if $S_m$ has a non-classic type and 
   $S_M$ is an equivalent sphere, where
   $M=B$ if $m=b$ or $M=A$ if $m=a$, it is known that 
   $v_{ikjM}\equiv v_{ijkm}$ and, if $v_{ikjm}$ also exists, 
   then $v_{ijkM}\equiv v_{ikjm}$. Therefore, using the previous 
   observation for $p=v_{ijkM}$ or $v_{ikjM}\}$, 
   we obtain the following expressions,
   \begin{align}
   \text{\insphere}(S_i,S_j,S_k,S_M,S_N) 
    &= -\text{\insphere}(S_i,S_j,S_k,S_M,S_n) \\
     &=-\text{\insphere}(S_i,S_k,S_j,S_m,S_n), \\
   \text{\insphere}(S_i,S_k,S_j,S_M,S_N) 
     &=-\text{\insphere}(S_i,S_k,S_j,S_M,S_n) \\
     &=-\text{\insphere}(S_i,S_j,S_k,S_m,S_n).
   \end{align}

   These last four equalities can be used to evaluate any \insphere 
   predicate that arises during the evaluation of the \order 
   predicate in the case of a non-classic configuration. 

   Regarding 
   the respective \orient predicates that may have to be evaluated, 
   we consider the fact that $\yinv{\mathcal{A}}$, $\yinv{C_n}$
   and $\yinv{C_N}$ are collinear and the latter two lie on 
   opposite sides of $\yinv{\mathcal{A}}$. Subsequently, 
   it is also true that $\yinv{C_n}$
   and $\yinv{C_N}$ must lie on opposite sides with respect to any line
   $\yinv{\lambda}$ that goes through $\yinv{\mathcal{A}}$ (see 
   Figure~\ref{fig:15}).

   If we choose $\yinv{\lambda}$  to be the line $\yinv{\ell}$ 
   that goes through $\yinv{\OO}$ and bear in mind that the
   position of a point of \yspace with respect to this line 
   corresponds to the position of its pre-image in \zspace against 
   the plane $\Pi_{ijk}$, we infer that $C_n$ and $C_N$ lie on 
   different sides of $\Pi_{ijk}$ and therefore
   \[\text{\orient}(C_N,C_i,C_j,C_k)=-\text{\orient}(C_n,C_i,C_j,C_k).\]

   If  $\yinv{\lambda}$ is chosen to be the line that goes 
   through $\yinv{C_m}$ for $m\in\{a,b\}\backslash\{n\}$ then 
   it must hold in \yspace that $\inv{C_n}$ and  $\inv{C_N}$ 
   lie on different sides with respect to the plane that goes through
   $\inv{C_i}$, $\inv{C_j}$ and $\inv{C_m}$, which is equivalent to 
   \begin{align}
   \text{\orient}(\inv{C_N},\inv{C_i},\inv{C_j},\inv{C_m}) &=
     -\text{\orient}(\inv{C_n},\inv{C_i},\inv{C_j},\inv{C_m}),\\
   \text{\orient}(\inv{C_m},\inv{C_i},\inv{C_j},\inv{C_N}) &=
   -\text{\orient}(\inv{C_m},\inv{C_i},\inv{C_j},\inv{C_n}) .
   \end{align}
   \noindent Finally, combining the last two equations, we obtain that 
   \begin{align}
   \text{\orient}(\inv{C_N},\inv{C_i},\inv{C_j},\inv{C_M}) &=
   - \text{\orient}(\inv{C_n},\inv{C_i},\inv{C_j},\inv{C_M}) \\
   &=\text{\orient}(\inv{C_n},\inv{C_i},\inv{C_j},\inv{C_m}).
   \end{align}

   In conclusion, we have shown that 
   the evaluation of all 7 \insphere or \orient 
   predicates, that may involve one or two equivalent spheres, 
   can be amounted to the evaluation of respective predicates 
   that contain only the original spheres $S_a$ and $S_b$ instead.
   Ultimately, we proved that the algebraic cost of the 
   \order predicate in a non-classic configuration is the same 
   as in a classic configuration, yielding the following lemmas.

   \begin{lemma}
   The \order predicate in a non-classic configuration can be 
   evaluated by determining the sign of quantities of algebraic 
   degree at most 10 (in the input quantities).
   \end{lemma}

   \begin{lemma}
   The \order predicate can be 
   evaluated by determining the sign of quantities of algebraic 
   degree at most 10 (in the input quantities).
   \end{lemma}


%


\chapter{Non-Degenerate Case Analysis for Elliptic Trisectors} 
  \label{sec:non_degenerate_elliptic}

    In this chapter, our goal is to answer the 
    $\text{\conflict}(S_i,S_j,S_k,S_l,S_m,S_q)$ predicate under the 
    assumption that the trisector of the first three input sites is 
    either an ellipse or a circle and no degeneracies occur. 
    If the trisector $\tri{ijk}$ has one of 
    the aforementioned types, we will say that we are in an 
    \emph{elliptic trisector} case. Note that for the respective 
    non-degenerate \emph{hyperbolic case}, where $\tri{ijk}$ is a branch of a
    hyperbola or a line, the \conflict predicate 
    is analysed thoroughly in Chapter~\ref{sec:non_degenerate_hyperbolic}.

    These two cases share a lot of similarities in the sense that,
    most of the analysis and the subpredicates presented for the 
    hyperbolic trisector type can be used with little or 
    no modifications for the elliptic trisector type. 
    In Section~\ref{sub:differences_elliptic_hyperbolic}, we present 
    the major differences that are found among the two cases, with the 
    most important being spotted when trying to order Voronoi vertices 
    on the oriented trisector. The main algorithm that answers
    the \conflict predicate in an elliptic-trisector scenario
    is presented in Section~\ref{sub:conflict_elliptic}.


    \section{Differences between the elliptic and the hyperbolic case} 
    \label{sub:differences_elliptic_hyperbolic}
    
      \subsubsection{Orienting an elliptic trisector} 
      \label{ssub:elliptic_trisector_orientation}
      
        If the trisector of the sites $S_i, S_j$ and $S_k$ is an ellipse 
        or a circle, it necessarily holds that one of the three 
        sphere lies strictly inside the convex hull defined by the other 
        two \cite{WillThesis}. Since a cyclic permutation of the sites 
        $S_i,S_j$ and $S_k$ does not alter the outcome of the 
        \conflict predicate, we may assume that $S_k$ has the minimum 
        radius and therefore lies in the convex hull of $S_i$ and $S_j$.

        It is apparent that a naive way of ordering the trisector 
        $\tri{ijk}$ would be via the use of the ``right-hand rule'': 
        if our thumb points from $C_j$ towards $C_i$, the direction 
        pointed by our hand, if it ``wraps'' $\tri{ijk}$, is the 
        positive direction. Variations of this rule would include 
        pointing our thumb from $C_i$ towards $C_j$ or even considering 
        the radii of the spheres $S_i$ and $S_j$ in order to choose 
        among the two possible orientations.

        However, there is a more clever way of defining the 
        orientation of this elliptic trisector within the context of 
        what we are trying to achieve, \ie, answer the \conflict 
        predicate. Firstly, we must understand that  
        orienting the trisector $\tri{ijk}$ was initially needed
        to properly define a Voronoi edge $e_{ijklm}$. In the case of 
        a hyperbolic trisector type, we were able to use a simple 
        ordering $\prec$ that derived by the orientation 
        of $\tri{ivk}$ via the ``right-hand rule'' to 
        observe that 
        $e_{ijklm}:=\{\tau\in\tri{ijk}: v_{ijkl}\prec\tau\prec v_{ikjm} \}$. 
        Therefore, in order to be consistent with the analysis and 
        notation of Chapter~\ref{sec:non_degenerate_hyperbolic}, 
        we must define an orientation that will preserve this fact.

        To accomplish this, we begin by considering the shadow region 
        of $S_l$ and $S_m$ on the trisector $\tri{ijk}$. It is clear 
        that both these shadow regions must have at least one 
        boundary point, corresponding to the center of the respective 
        Apollonius sphere of $S_i,S_j,S_k$ and either $S_l$ or 
        $S_m$ respectively. Assuming no degeneracies and according 
        to the analysis that we will present in the subsection 
        ``Shadow Regions'' below, if at least one of the Apollonius 
        vertices $\{v_{ikjn}, v_{ijkn}\}$ exist, for some 
        $n\in\{l,m,q\}$, then both must exist. As a consequence, 
        it must hold that all vertices $v_{ijkl},v_{ikjl},v_{ijkm}$ and 
        $v_{ikjm}$ exist on the trisector $\tri{ijk}$ since 
        $e_{ijklm}$ is assumed to exist. 

        We now orient $\tri{ijk}$ such that, when traversing the 
        shadow region of $S_l$ on $\tri{ijk}$ following the 
        positive orientation, we ``start'' from $v_{ikjl}$
        and ``end'' at $v_{ijkl}$. Since $e_{ijklm}$ is a valid 
        Voronoi edge, it is now certain that, while 
        traversing the rest of the trisector from $v_{ijkl}$ 
        to $v_{ikjl}$, we will encounter $v_{ikjm}$ first among 
        $\{v_{ijkm},v_{ikjm}\}$. Furthermore, for all points 
        $\tau\in e_{ijklm}$, it holds that 
        $v_{ijkl}\prec \tau \prec v_{ikjm}$, where $a\prec b \prec c$ 
        denotes that we will encounter the points $a,b$ and $c$ on the
        trisector in this order while traversing it in its positive 
        direction. 

\begin{figure}[htbp]
        \centering
        \includegraphics[width=0.7\textwidth]{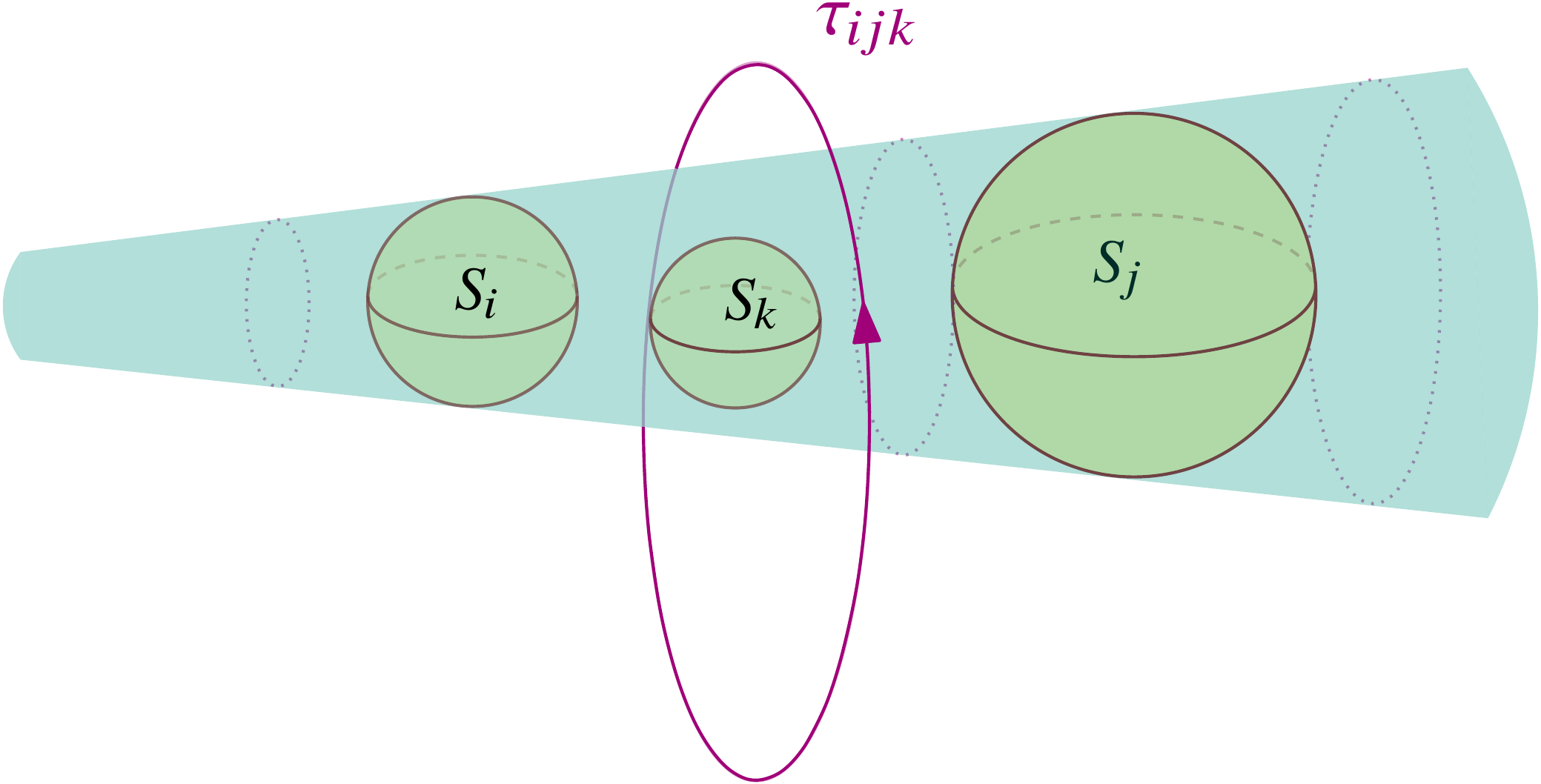}
        \caption[The orientation of an elliptic trisector.]{Orienting  an elliptic trisector, 
        using the ``right-hand rule''; if our thumb points
        from $C_j$ towards $C_i$, the direction pointed by our hand, if  
        it ``wraps'' $\tri{ijk}$, is the positive direction. }
        \label{fig:07_again}
        \end{figure}

        Although, this way of orienting $\tri{ijk}$ helps us 
        preserve most of the analysis of 
        Section~\ref{sub:the_shadowregion_predicate_analysis} as well 
        as Lemma~\ref{lemma:phi_chi}, it is 
        quite complicated. We may simplify this task as follows.
        If $\mathcal{A}=\ap{t}$ is the Apollonius sphere of the sites 
        $S_i, S_j$ and $S_k$, centered at $t\in\tri{ijk}$, let $T_n$ 
        be the tangency point $\mathcal{A}$ and $S_n$, 
        for $n\in\{i,j,k\}$. If we fold our right hand to follow the 
        points $T_i,T_j$ and $T_k$ (in that order),
        our thumb will be showing the positive direction of $\tri{ijk}$.
        However, since we can safely assume 
        that $S_k$ lies in the convex hull of $S_i$ and $S_j$ after a 
        proper name exchange, the above right-hand rule is equivalent
        to an even simpler rule. The orientation of $\tri{ijk}$ is 
        the one provided by the original ``naive'' right-hand rule: 
        if our thumb points from $C_j$ towards $C_i$, the direction 
        pointed by our hand, if it ``wraps'' $\tri{ijk}$, is the 
        positive direction (see Figure~\ref{fig:07_again}).

      \subsubsection{Differences in \wspace and \yspace} 
      \label{ssub:differences_in_wspace_and_yspace}
      
        Following the same analysis presented in 
        Section~\ref{ssub:the_w_space}, \wspace is the plane we will 
        end up after reducing all original sites $S_n$, 
        for $n\in\{i,j,k,l,m,q\}$, by radius $r_k=\min\{r_i,r_j,r_k\}$ 
        and inverting (see Section~\ref{sub:inversion}) with
        $C_k$ as the pole. As before, we can also define the 2-dimensional
        \yspace, where geometric remarks are easier to take place. 

        However, there is a crucial difference between the observations of 
        Section~\ref{ssub:the_w_space} that hold for hyperbolic trisector 
        types and the respective elliptic case we want to study. In 
        the former case, there exist two Apollonius spheres-planes of 
        the sites $S_i,S_j$ and $S_k$, centered at infinity, whereas, in 
        the latter case, no such sphere exist. The equivalency of this 
        fact in \wspace is that, the image of all these Apollonius 
        spheres are planes, commonly tagent to the spheres 
        $\inv{S_i}$ and $\inv{S_j}$, that leave the point $\OO$ on the 
        same side as the latter two spheres. Moreover, there can not 
        exist a plane cotagent to the spheres $\inv{S_i}$ and $\inv{S_j}$
        such that $\inv{S_i}$ and $\inv{S_j}$ lie on the one half-plane 
        and $\OO$ on the other. Subsequently, the point $\OO$ must lie 
        strictly inside the semi-cone $\wcone$ defined by $\inv{S_i}$ and 
        $\inv{S_j}$ (or the respective cylinder if these two spheres 
        have equal radii). 

        Respectively, the point $\yinv{\OO}$ in \yspace must lie 
        inside the circle $\ycone$. Using the same arguements as in 
        Section~\ref{ssub:the_yspace_analysis}, one can show that  
        the image of $\ap{t}$, \ie, the Apollonius sphere of $S_i, S_j$ 
        and $S_k$ centered at $t$, is a plane in \wspace. As the 
        point $t$ moves on the oriented trisector $\tri{ijk}$, the 
        plane rotates remaining tangent to $\yinv{S_i}$ and $\yinv{S_j}$. 
        In \yspace, the corresponding image is that of a line $\ill{t}$ 
        that rotates remaining tangent to the circle $\ycone$. The main
        difference with the hyperbolic case is that there are no 
        maximal positions in these rotations; the rotating plane in 
        \wspace (resp., line in \yspace) never goes through the point 
        $\OO$ (resp., $\yinv{\OO}$). 

\begin{figure}[tbp]
        \centering
        \includegraphics[width=0.8\textwidth]{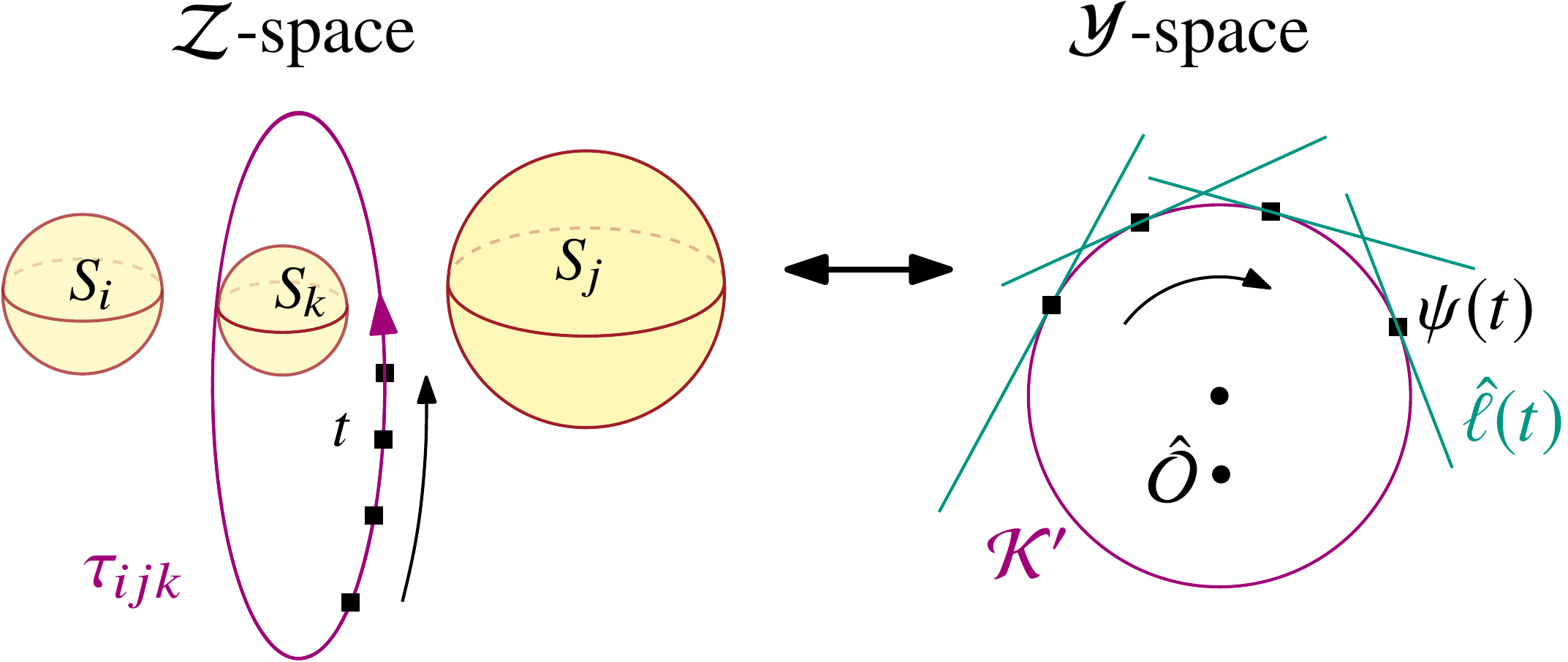}
        \caption[The elliptic trisector in \yspace]{As 
        $t$ moves on the elliptic trisector $\tri{ijk}$ following its 
        positive orientation, its image $\arc{t}$ moves on the oriented 
        circle $\ycone$. The image of the Apollonius sphere $\ap{t}$ 
        in \yspace is the line $\ill{t}$ that remains tangent to 
        $\ycone$. Since all $\ap{t}$ are finite, $\ill{t}$ must 
        leave $\yinv{\OO}$ on its positive side, therefore $\yinv{\OO}$
        must lie inside $\ycone$. Apparently, there is a 1-1 and onto
        correspondence between the trisector $\tri{ijk}$ in \zspace and 
        $\ycone$ in \wspace.}
        \label{fig:29}
        \end{figure}

        Note that we can appropriately orient the circle $\ycone$ using 
        the ``right-hand rule'' again: if our thumb points from 
        $\yinv{C_j}$ towards $\yinv{C_i}$, the direction pointed by 
        our hand, if it “wraps” $\ycone$, is the positive direction. 
        (see Figure~\ref{fig:29})
        Observe that there is a 
        1-1 correspondence between the oriented trisector $\tri{ijk}$ and 
        the oriented circle $\ycone$. We denote this 
        1-1 and onto mapping from $\tri{ijk}$ to
        the circle $\ycone$ by $\arc{\cdot}$, such that $\ap{t}$ 
        of \zspace maps to the line $\ill{t}$ of \yspace that is tangent to 
        $\ycone$ at the point $\arc{t}$. As a result of the 
        orientations defined on $\ycone$ and 
        $\tri{ijk}$ along with the remarks of the 
        Section~\ref{ssub:the_yspace_analysis}, we can easily deduct 
        the following lemma, equivalent to 
        lemma~\ref{lemma:ordering_equivalence} for hyperbolic trisectors.

        \begin{lemma}
        \label{lemma:ordering_equivalence_elliptic}
        There is a 1-1 correspondence between the 
        order of appearance of the existing vertices among
        $v_{ijka}, v_{ikja}, v_{ijkb}$ and $v_{ikjb}$ on the oriented 
        elliptic trisector $\tri{ijk}$ and the order of appearance of 
        the existing points $\arc{v_{ijka}}$, $\arc{v_{ikja}}$, 
        $\arc{v_{ijkb}}$ and $\arc{v_{ikjb}}$ on the oriented circle 
        $\ycone$.
        \end{lemma}

        This lemma enables us to follow similar strategy as the one 
        presented in Section~\ref{ssub:the_order_predicate} to 
        order the Apollonius vertices $v_{ijkl}, v_{ikjm}, v_{ijkq}$ 
        and $v_{ikjq}$ on the oriented $\tri{ijk}$. Then, using the 
        remarks of Section~\ref{sub:the_main_algorithm}, we 
        deduce the intersection type of $e_{ijklm}$ and $\sh{S_q}$ and 
        therefore answer the \conflict predicate.


      \subsubsection{Shadow Regions on elliptic trisectors} 
      \label{ssub:shadow_regions_on_elliptic_trisectors}
      
        Let us assume that the trisector of the spheres $S_i, S_j$ 
        and $S_k$ is elliptic and that $k<i,j$. 
        After reducing the radii of all 
        sites $S_i,S_j,S_k$ and $S_n$, for $n\in\{l,m,q\}$, and then 
        inverting through the point $C_k$, we consider the position 
        of $\yinv{S_n}$ against the semicone $\wcone$ defined by 
        $\yinv{S_i}$ and $\yinv{S_j}$. The cases where $\yinv{S_n}$ 
        point-touches or is full tangent to 
        the cone from the inside are considered degenerate and are 
        beyond the scope of this chapter. Therefore, we can assume that
        either 0 or 2 distinct planes commonly tangent to the spheres 
        $\yinv{S_i}, \yinv{S_j}$ and $\yinv{S_n}$ may exist. Due to the 
        location of $\OO$, any such plane will correspond to an 
        external Apollonius sphere of $S_i, S_j, S_k$ and $S_n$ in 
        \zspace. As an immediate result, either 0 or 2 such 
        Apollonius spheres must exist. 

        Since the existence of $v_{ijkl}$ (resp., $v_{ikjm}$) derives from 
        the assumption that $e_{ijklm}$ is a valid Voronoi edge, 
        at least one Apollonius sphere of the sites 
        $S_i, S_j, S_k$ and $S_l$ (resp., $S_m$) exist. Applying the 
        previous remark, there must also exist another one;
        this sphere must be centered at the Apollonius vertex 
        $v_{ikjl}$ (resp., $v_{ijkm}$). We have therefore concluded 
        that all vertices $\{v_{ijkl}, v_{ikjl}, v_{ijkm}, v_{ikjm}\}$ 
        must exist and moreover, using the same argument, that either both 
        or neither of $\{v_{ijkq}, v_{ikjq}\}$ exist. 

        According to the analysis of 
        Section~\ref{ssub:the_shadow_predicate}, the number of 
        boundary points of $\sh{S_n}$ represent how many of the vertices
        $\{v_{ijkn}, v_{ikjn}\}$ exist, \ie, 0 or 2 based on our remarks. 
        It is straightforward to deduce that, when seen as an interval, 
        $\sh{S_n}$ must either be $\emptyset$ or $\RR$, in case of 0 
        boundary points, or $(\chi,\phi)$, in case of 2 boundary points. 
        The notation we use for elliptic trisectors differs a little from 
        the one used for hyperbolic ones:
        \begin{itemize}
          \item $\emptyset$ indicates that $\sh{S_n}$ does not
          contain any point of $\tri{ijk}$,
          \item $\RR$ symbolizes that $\sh{S_n}$ consists of all points
          of $\tri{ijk}$ and 
          \item $(\chi,\phi)$ denotes that $\sh{S_n}$ is an arc on 
          $\tri{ijk}$. This arc contains all points $t$ of the 
          trisector such that $v_{ikjn}\prec t\prec v_{ijkn}$; this 
          fact derives from the proof of lemma~\ref{lemma:phi_chi} and 
          the proper orientation of $\tri{ijk}$. 
        \end{itemize}

        All these remarks sum up to the equivalent of 
        lemma~\ref{lemma:phi_chi} for elliptic trisectors.

        \begin{lemma}
        \label{lemma:phi_chi_elliptic}
        The type of the shadow region $\sh{S_n}$ of a sphere $S_n$ 
        on an elliptic trisector is one of the following: 
        $\emptyset, \RR$ or $(\chi,\phi)$. In the latter case, 
        $\chi$ and $\phi$ correspond to the points $v_{ikjn}$  
        and $v_{ijkn}$ of the oriented trisector $\tri{ijk}$ 
        respectively. 
        \end{lemma}



    \section{The \conflict predicate for Elliptic Trisectors} 
    \label{sub:conflict_elliptic}
      
      In this section, we provide a detailed algorithm that is used 
      to answer the \conflict$(S_i,S_j,S_k,S_l,S_m,S_q)$ predicate 
      assuming that $\tri{ijk}$ is an elliptic trisector. The main 
      idea behind this algorithm is already presented in 
      Section~\ref{sub:the_main_algorithm}; our primary goal is to 
      determine the intersection type of the edge $e_{ijklm}$ 
      (or simply $e$) and the shadow region $\sh{S_q}$.

      If the topological form of $\sh{S_q}$, denoted by $SRT(q)$, 
      is either $\emptyset$ or 
      $\RR$ then we can immediately return \noconflict or \fullconflict
      respectively. However if $SRT(q)$ is $(\chi,\phi)$, 
      then both $v_{ijkq}$ and $v_{ikjq}$ must exist and we may have 
      to determine the order of appearance of the vertices 
      $\{v_{ijkq}, v_{ikjq}, v_{ijkl}, v_{ikjm}\}$ on the oriented 
      trisector. The ordering of these Apollonius vertices on 
      $\tri{ijk}$ is acquired from the respective ordering of their
      images on $\ycone$, in \yspace. Using this information, 
      we can ultimately deduce the intersection type of 
      $e_{ijklm}\cap\sh{S_q}$ and return it as the 
      answer of the \conflict predicate (see 
      Section~\ref{sub:the_main_algorithm}). 

      At this point, it is wise to recall that the possible answers to 
      the \conflict predicate, as presented in 
      Section~\ref{sub:problem_outline} and 
      \ref{sub:the_main_algorithm}, are the following:

      \begin{itemize}
      \item 
      \noconflict : no portion of $e$ is destroyed by the insertion 
      of $S_q$ in the Apollonius diagram of the five sites.
      \item 
      \fullconflict : the entire edge $e$ is destroyed by the addition 
      of $S_q$ in the Apollonius diagram of the five sites.
      \item 
      \leftvertex : a subsegment of $e$ adjacent to its origin 
      vertex ($v_{ijkl}$) disappears in the Apollonius diagram 
      of the six sites.
      \item 
      \rightvertex : is the symmetric case of the \leftvertex case; 
      a subsegment of $e$ adjacent to the vertex $v_{ikjm}$ disappears 
      in the Apollonius diagram of the six sites.
      \item 
      \verticesconflict : subsegments of $e$ adjacent to its two 
      vertices disappear in the Apollonius diagram of the five sites.
      \item
      \middleconflict : a subsegment in the interior of $e$ disappears 
      in the Apollonius diagram of the five sites.
      \end{itemize}

      Lastly, it is quite convinient that the \exist$(S_i,S_j,S_k,S_q)$ 
      predicate can be evaluated as described in 
      Section~\ref{sub:the_existence_predicate_analysis} without any 
      modifications for elliptic trisectors.
      This subpredicate will return whether 0 or 2 
      of the vertices $\{v_{ikjq},v_{ijkq}\}$ exist; 
      assuming no degeneracies, these are the only possible answers.

      \begin{description}
      \item[Step 1] We evaluate $E=\text{\exist}(S_i,S_j,S_k,S_q)$, as
      described in Section~\ref{sub:the_existence_predicate_analysis}. 
      The predicate's outcome is either 0 or 2, assuming no degeneracies. 
      If $E=0$ go to Step 2 otherwise, if $E=2$ go to step 3.
      \item[Step 2] We 
      evaluate $I=\text{\insphere}(S_i,S_j,S_k,S_l,S_q)$ 
      (or \insphere$(S_i,S_k,S_j,S_m,S_q)$); if $I=+$, the 
      \conflict predicate's outcome is \noconflict otherwise, if  
      $I=-$, the outcome is \fullconflict. 

      {\textbf{Explanation:}} 
      Since none of $\{v_{ijkq},v_{ikjq}\}$ exist, the 
      topological form $SRT(q)$ of $\sh{S_q}$ is either $\emptyset$ or 
      $\RR$. In the first case, the \conflict predicate must 
      return \noconflict
      and in the latter \fullconflict. Moreover, in the case where 
      $SRT(q)=\emptyset$, none of the points of $\tri{ijk}$, including 
      $v_{ijkl}$ and $v_{ikjm}$, belong to $\sh{S_q}$, \ie, the Apollonius
      sphere centered at these points do not intersect $S_q$. Therefore, 
      in this case both \insphere outcomes must be $+$. In the opposite 
      case, both Apollonius vertices $v_{ijkl}$ and $v_{ikjm}$ would belong
      to $\sh{S_q}$ and equivalently, the corresponding Apollonius spheres
      would intersect $S_q$.
      \item[Step 3] We evaluate $I_1=\text{\insphere}(S_i,S_j,S_k,S_l,S_q)$ 
      and $I_2=\text{\insphere}(S_i,S_k,S_j,S_m,S_q)$. 
      If $(I_1,I_2)=(-,+)$ or $(+,-)$ return \leftvertex or \rightvertex 
      respectively. Otherwise, if  $(I_1,I_2)=(+,+)$ or $(-,-)$ go to 
      Step 4. 

      {\textbf{Explanation:}}
      Since both $v_{ijkq}$ and $v_{ikjq}$ exist, the 
      topological form $SRT(q)$ of $\sh{S_q}$ is $(\chi,\phi)$. If 
      $(I_1,I_2)=(-,+)$ then the ``left'' vertex $v_{ijkl}$ of the edge
      $e_{ijklm}$ lies inside $\sh{S_q}$ and the ``right'' vertex $v_{ikjm}$ 
      does not. Due to the simple topological forms of $\sh{S_q}$ and the 
      Voronoi edge, we deduce that $\sh{S_q}$ intersects 
      only the left part of the edge hence the \conflict predicate's outcome
      is \leftvertex. In the symmetric case where $(I_1,I_2)=(+,-)$, 
      the shadow region $\sh{S_q}$ intersects only the right part of the 
      edge and the predicate's answer is \rightvertex.
      \item[Step 4] We evaluate $I_3=\text{\insphere}(S_i,S_j,S_k,S_m,S_l)$. 
      If $I_3=-$ go to Step 5 otherwise, if $I_3=+$ go to Step 6.

      {\textbf{Explanation:}}
      We break down our analysis into two cases A and B, which we 
      study in Steps 5 and 6 respectively. In case A (resp., B), 
      $v_{ijkm}$ belongs (resp., does not belong) to the shadow 
      region $\sh{S_l}$ on the trisector since the Apollonius sphere 
      centered at $v_{ijkm}$ intersects $S_l$ as the outcome of the 
      \insphere predicate suggests. Since 
      $\sh{S_l}:=\{t\in\tri{ijk}: v_{ikjl}\prec t\prec v_{ijkl} \}$,
      a quick conclusion to be drawn is that,
      \begin{itemize}
      \item in case A, $v_{ijkl}\prec v_{ikjm}\prec v_{ikjl}\prec v_{ijkm}$
      (see Figure~\ref{fig:30}(left)), 
      whereas 
      \item in case B, $v_{ijkl}\prec v_{ikjm}\prec v_{ijkm}\prec v_{ikjl}$
      (see Figure~\ref{fig:30}(right)). 
      \end{itemize}
      As stated before, $a\prec b \prec c$ denotes that you will 
      meet $a,b$ and $c$ in this order while \emph{positively} 
      traversing the trisector, starting from point $a$. 

      \textbf{Important notice:} For the sake of clarity, the notation 
      of the figures of this Section was simplified. Although all figures
      represent configurations in \yspace, the Apollonius vertices are 
      denoted by $v_{\alpha\beta\gamma\delta}$ instead of 
      $\arc{v_{\alpha\beta\gamma\delta}}$, 
      where $\{\alpha,\beta,\gamma,\delta\}\in\{i,j,k,l,m,q\}$. 
      Moreover, the image of the shadow region of $S_\alpha$ 
      is denoted by $\sh{S_a}$ instead of $\arc{\sh{S_a}}$ and the 
      image of the Voronoi edge is denoted by $e_{ijklm}$. 
      Lastly, although the images of the shadow regions 
      of $S_l,S_m$ and $S_q$ lie on the circle $\ycone$, sometimes 
      they are drawn as if they lied on a concentric inner circle.

      \begin{figure}[tbp]
      \centering
      \includegraphics[width=0.8\textwidth]{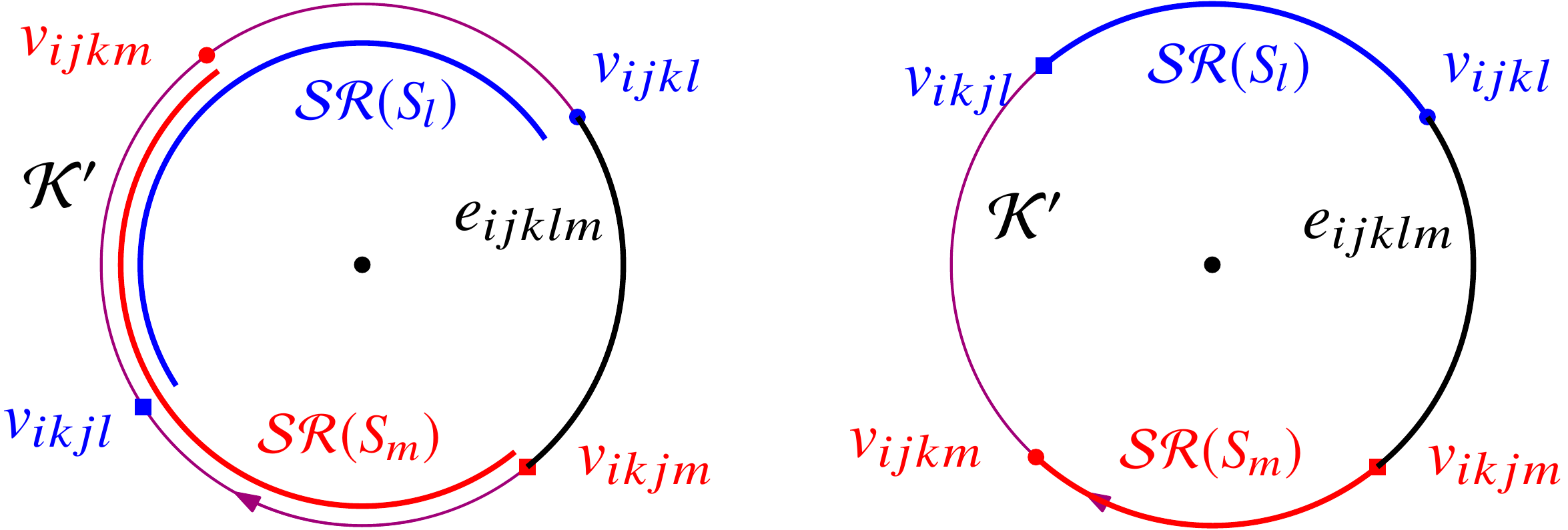}
      \caption[Possible orderings of $v_{ijkl},v_{ikjm},v_{ikjl}$ and 
      $v_{ijkm}$ on $\ycone$]{ Since the Voronoi edge $e_{ijklm}$ is 
      bounded by $v_{ijkl}$ and $v_{ikjm}$ on the oriented circle 
      trisector $\tri{ijk}$, the points of $e_{ijklm}$ can not 
      lie in the shadow regions of $S_l$ and $S_m$. Since the latter 
      two are bounded by $v_{ikjl}$ and $v_{ikjm}$ on the ``left'' and
      $v_{ijkl}$ and $v_{ijkm}$ on the ``right'' respectively, the only 
      possible orderings of these vertices are the ones described in 
      Case A(left figure) or Case B(right figure).  }
      \label{fig:30}
      \end{figure}
      \item[Step 5] We determine if all of the following predicates 
      return $+$:
      \begin{itemize}
      \item $I_4=\text{\insphere}(S_i,S_k,S_j,S_q,S_l)$,
      \item $I_5=\text{\insphere}(S_i,S_k,S_j,S_q,S_m)$,
      \item $I_6=\text{\insphere}(S_i,S_j,S_k,S_q,S_l)$ and,
      \item $I_7=\text{\insphere}(S_i,S_j,S_k,S_q,S_m)$.
      \end{itemize}
      If all $I_4$ to $I_7$ are positive then 
      \begin{itemize}
      \item return \verticesconflict, if $I_1=I_2=-$ or
      \item return \middleconflict, if $I_1=I_2=+$.
      \end{itemize}
      Otherwise, if at least one of $I_4$ to $I_7$ is negative, then
      \begin{itemize}
      \item return \fullconflict, if $I_1=I_2=-$ or
      \item return \noconflict, if $I_1=I_2=+$.
      \end{itemize}

      {\textbf{Explanation:}}
      If $I_1=I_2=-$, then both parts adjecent to the endpoints 
      of the Voronoi edge will no longer exist in the updated Voronoi
      diagram. In this scenario, either a part in the middle will still 
      remain (\verticesconflict) or all of the edge will cease to 
      exist (\fullconflict). In the former case, both 
      $v_{ikjq}$ and $v_{ijkq}$ ought to lie between $v_{ijkl}$ and
      $v_{ikjm}$ on the trisector (see Figure~\ref{fig:31}(left)). However, since we are in Step  
      5, it holds that 
      $v_{ijkl}\prec v_{ikjm}\prec v_{ikjl}\prec v_{ijkm}$. As 
      a result, an equivalent expression of 
      $v_{ijkl}\prec v_{ikjq},v_{ijkq}\prec v_{ikjm}$ would be 
      that $v_{ikjq},v_{ijkq}\not\in\sh{S_l}$ and 
      $v_{ikjq},v_{ijkq}\not\in\sh{S_m}$. Lastly, these four expressions 
      amount to all \insphere outcomes $I_4$ to $I_7$ being positive.

      In a similar way, if $I_1=I_2=+$ then both parts adjecent 
      to the endpoints of the Voronoi edge will remain in the 
      updated Voronoi diagram. In such scenario, either a part in 
      the middle will cease to exist (\middleconflict) or 
      all of the edge will remain intact (\noconflict). The former 
      case is equivalent to both $v_{ikjq}$ and 
      $v_{ijkq}$ lying between $v_{ijkl}$ and $v_{ikjm}$ on the trisector
      (see Figure~\ref{fig:31}(right)).
      As stated above, such a geometric configuration amounts to all 
      \insphere outcomes $I_4$ to $I_7$ being positive.

      \begin{figure}[tbp]
      \centering
      \includegraphics[width=0.8\textwidth]{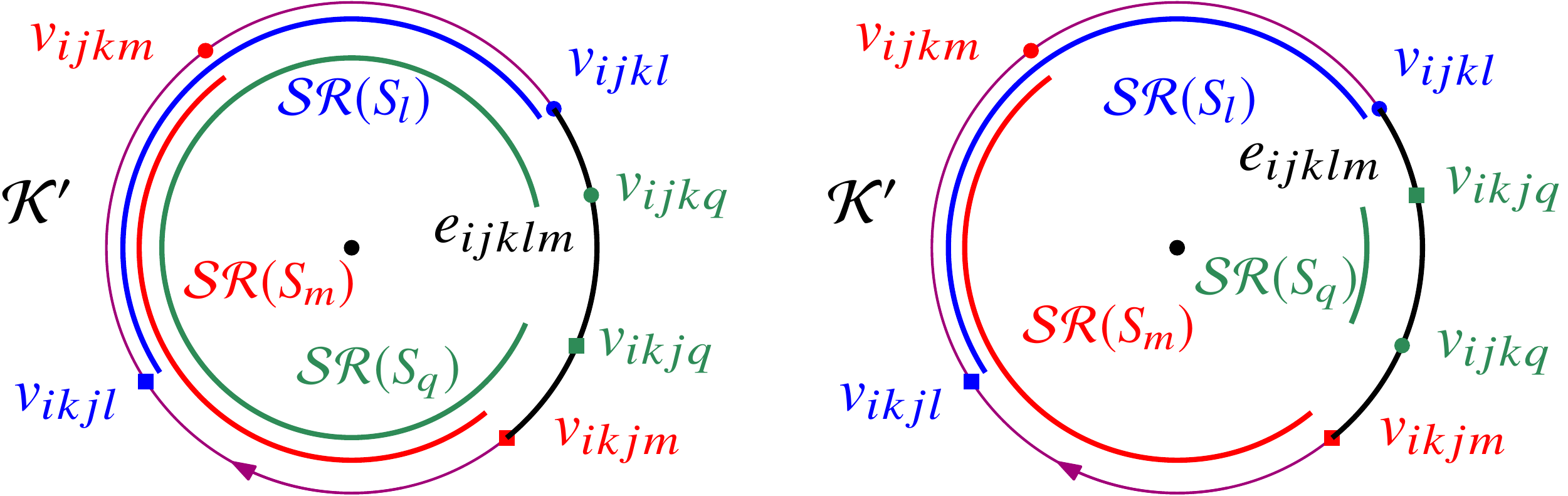}
      \caption[Resolving Step 5]{If all \insphere predicate $I_4$ to 
      $I_7$ are positive then the vertices $v_{ijkq}$ and $v_{ikjq}$ do 
      not lie on neither $\sh{S_l}$ nor $\sh{S_m}$ and therefore must 
      lie on the edge $e_{ijklm}$. In this configuratoin, 
      if $I_1=I_2=-$, both endpoints of 
      the edge $e_{ijklm}$ lie on $\sh{S_q}$ and the \conflict predicate
      must return \verticesconflict (left figure). If $I_1=I_2=+$, 
      both endpoints of the edge $e_{ijklm}$ lie outside $\sh{S_q}$ 
      and the \conflict predicate must return \middleconflict 
      (right figure). }
      \label{fig:31}
      \end{figure}

      \item[Step 6] 
      We determine if all of the following predicates 
      return $+$:
      \begin{itemize}
      \item $I_4=\text{\insphere}(S_i,S_k,S_j,S_q,S_l)$,
      \item $I_5=\text{\insphere}(S_i,S_k,S_j,S_q,S_m)$,
      \item $I_6=\text{\insphere}(S_i,S_j,S_k,S_q,S_l)$ and,
      \item $I_7=\text{\insphere}(S_i,S_j,S_k,S_q,S_m)$.
      \end{itemize}
      If at least one of them is negative, go to Step 6a otherwise 
      go to Step 6b.

      {\textbf{Explanation:}}
      For Steps 6,6a and 6b, it holds that 
      $v_{ijkl}\prec v_{ikjm}\prec v_{ijkm}\prec v_{ikjl}$ 
      hence all \insphere predicates $I_4$ to $I_7$ being positive 
      implies that the vertices $v_{ikjq}$ and $v_{ijkq}$ lie either 
      between $v_{ijkl}$ and $v_{ikjm}$ or between $v_{ijkm}$ 
      and $v_{ikjl}$ (see Figure~\ref{fig:33}).

      \begin{figure}[btp]
      \centering
      \includegraphics[width=0.4\textwidth]{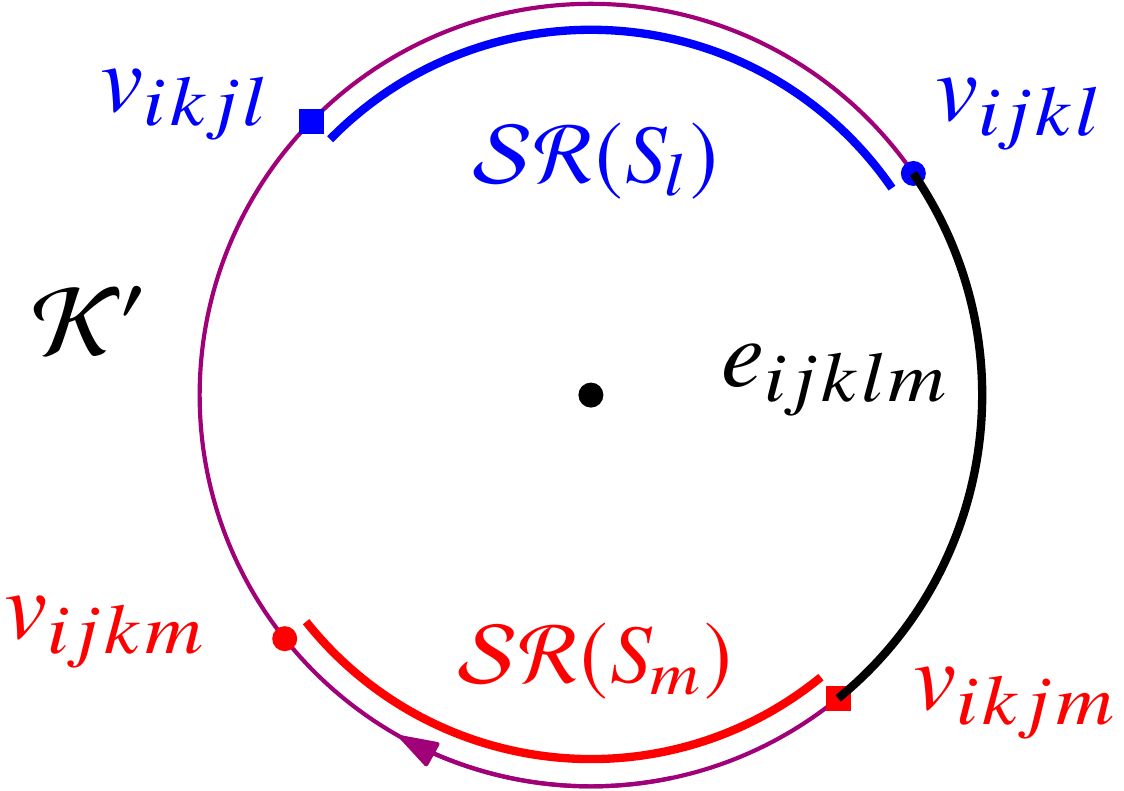}
      \caption[The dilemma of Step 6]{If all \insphere predicate $I_4$ to 
      $I_7$ are positive then the vertices $v_{ijkq}$ and $v_{ikjq}$ do 
      not lie on neither $\sh{S_l}$ nor $\sh{S_m}$. In Case B, this 
      is equivalent to $v_{ijkq}$ and $v_{ikjq}$ lying either on the 
      edge $e_{ijklm}$ or in-between $v_{ijkm}$ and $v_{ikjl}$. Further 
      analysis is required to determine in which part of $\wcone$ they 
      actually lie. }
      \label{fig:33}
      \end{figure}
      \item[Step 6a]
      If $I_1=I_2=-$ return \fullconflict otherwise, 
      if $I_1=I_2=+$ return \noconflict.

      {\textbf{Explanation:}}
      In this Step, $v_{ijkq}$ and 
      $v_{ikjq}$ can not both lie in-between the endpoints
      $v_{ijkl}$ and $v_{ikjm}$
      of the Voronoi edge $e_{ijklm}$ and therefore the \conflict 
      predicate's answers \verticesconflict and \middleconflict are 
      not feasible.
      As a result, if $I_1=I_2=-$ 
      the \conflict predicate returns \fullconflict whereas,
      if $I_1=I_2=+$ the predicate returns \noconflict.
      \item[Step 6b] Go to Step 6bA.

      \textbf{Analysis behind Step 6:}
      In this Step, we have to determine if both or 
      none of $v_{ijkq}$ and $v_{ikjq}$ lie in-between $v_{ijkl}$ and 
      $v_{ikjm}$ on the oriented trisector. 
      In order to distinguish among the two possible scenarios of 
      Step 6b, we follow a similar analysis with the one presented in 
      Section~\ref{sub:ordering_in_a_classic_configuration} when we 
      wanted to decide between OrderCase 1 and 6. 

      Since we are in Step 6b, all \insphere outcomes $I_4$ to $I_7$ are
      positive and $I_1=I_2$. A geometric consequence of these results 
      is that while traversing the trisector the encountered
      Apollonius vertices will appear consecutively in pairs 
      $\{v_{ijkl},v_{ikjl}\}$, $\{v_{ijkm},v_{ikjm}\}$ 
      and $\{v_{ijkq},v_{ikjq}\}$. Equivalently, both vertices 
      $v_{ijkq}$ and $v_{ikjq}$ appear either in-between 
      $v_{ijkl}$ and $v_{ikjm}$ (\ie, on the Voronoi edge) or 
      in-between $v_{ijkm}$ and $v_{ikjl}$. This remark proves 
      that the order of appearance of these ``pairs'' is crucial 
      for our analysis. 

      To obtain this ordering, we will have to study the corresponding 
      configuration in \yspace. Let us denote by $M_n$, 
      for $n\in\{l,m,q\}$, the middle point of the image of $\sh{S_n}$
      in \yspace. The order of appearance of the aforementioned pairs 
      is equivalent to the order of appearance of $M_l,M_m$ and $M_q$ 
      on $\ycone$. 

      Given the ordering of these points and the \insphere outcomes 
      $I_1$ and $I_2$ we can ultimately determine if $v_{ijkq}$ and 
      $v_{ikjq}$ lie or not in-between $v_{ijkl}$ and $v_{ikjm}$.
      For example, assume we evaluate that $M_l\prec M_q \prec M_m$ and 
      that $I_1=I_2=+$. Due to $I_1$ and $I_2$ being positive, it is 
      known that the shadow region $\sh{S_q}$ either lies in the middle 
      part of the Voronoi edge (and the outcome is \middleconflict) or 
      it does not intersect the edge at all (and the outcome is 
      \noconflict). In the former case both $v_{ijkq}$ and 
      $v_{ikjq}$ would lie in-between $v_{ijkl}$ and $v_{ikjm}$ and 
      therefore $M_l\prec M_q \prec M_m$. In the latter case, 
      both $v_{ijkq}$ and $v_{ikjq}$ would have to lie 
      in-between $v_{ijkm}$ and $v_{ikjl}$ hence 
      it would hold that $M_l\prec M_m \prec M_q$.  

      Using the same arguements, we can deduce that 
      \begin{itemize}
      \item $v_{ijkq}$ and $v_{ikjq}$ lie in-between $v_{ijkl}$ 
      and $v_{ikjm}$ if either $I_1=I_2=+$ and 
      $M_l\prec M_q \prec M_m$ (see Figure~\ref{fig:34}(left)) 
      or $I_1=I_2=-$ and 
      $M_l\prec M_m \prec M_q$ (see Figure~\ref{fig:34}(right)) whereas,
      \item $v_{ijkq}$ and $v_{ikjq}$ do not lie in-between $v_{ijkl}$ 
      and $v_{ikjm}$ if either $I_1=I_2=+$ and 
      $M_l\prec M_m \prec M_q$ (see Figure~\ref{fig:36}(left)) 
      or $I_1=I_2=-$ and $M_l\prec M_q \prec M_m$ 
      (see Figure~\ref{fig:36}(right)) .
      \end{itemize}

      \begin{figure}[tbp]
      \centering
      \includegraphics[width=0.8\textwidth]{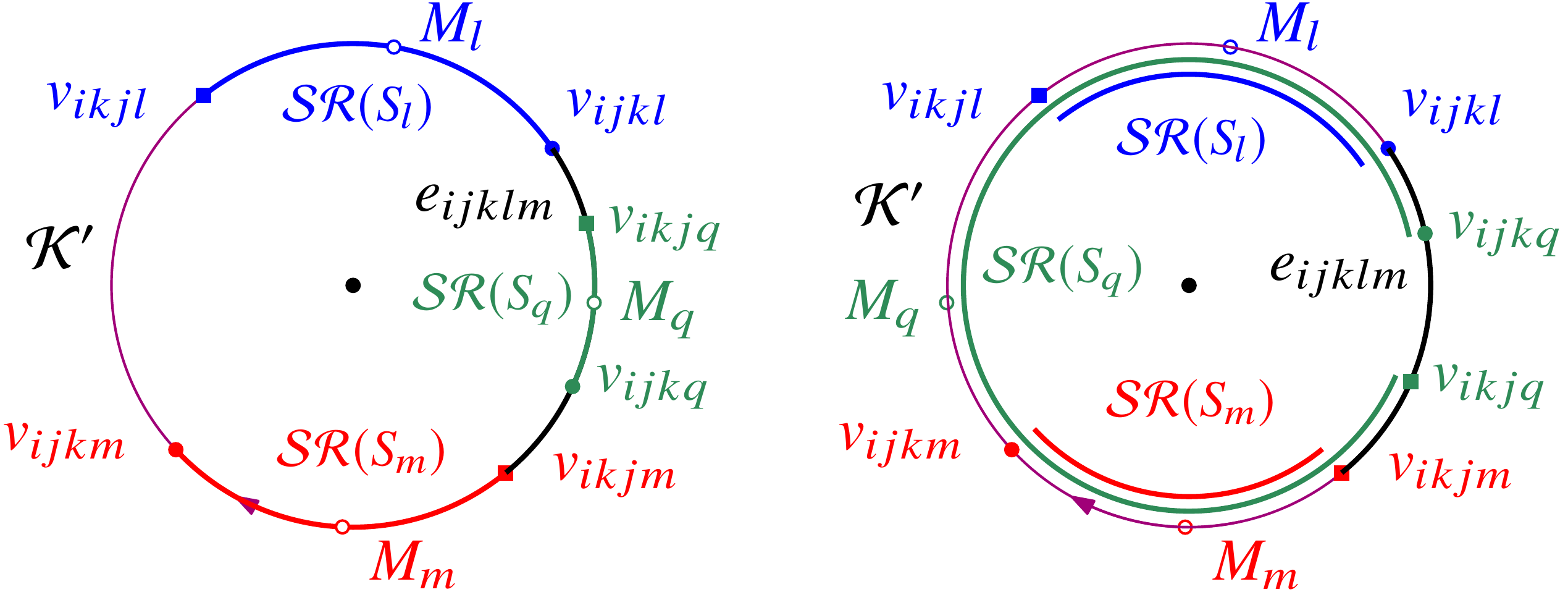}
      \caption[Step 6b if both $v_{ijkq},v_{ikjq}$ lie on the 
      edge $e_{ijklm}$]{ If both $v_{ijkq},v_{ikjq}$ lie on the 
      edge $e_{ijklm}$, \ie, in-between the points $v_{ijkl}$ and 
      $v_{ikjm}$, then either a \middleconflict (left figure) or 
      a \verticesconflict (right figure) configuration arises. We 
      can distinguish which is the actually case if can determine the 
      order of appearance of $M_l$, $M_m$ and $M_q$, which are 
      the centers of the corresponding shadow regions $\sh{S_l}$, 
      $\sh{S_m}$ and $\sh{S_q}$. In the \middleconflict scenario, 
      $M_l\prec M_q \prec M_m$ and $I_1=I_2=+$ whereas in the 
      \verticesconflict case, $M_l\prec M_m \prec M_q$ 
      and $I_1=I_2=-$.}
      \label{fig:34}
      \end{figure}

      \begin{figure}[tbp]
      \centering
      \includegraphics[width=0.8\textwidth]{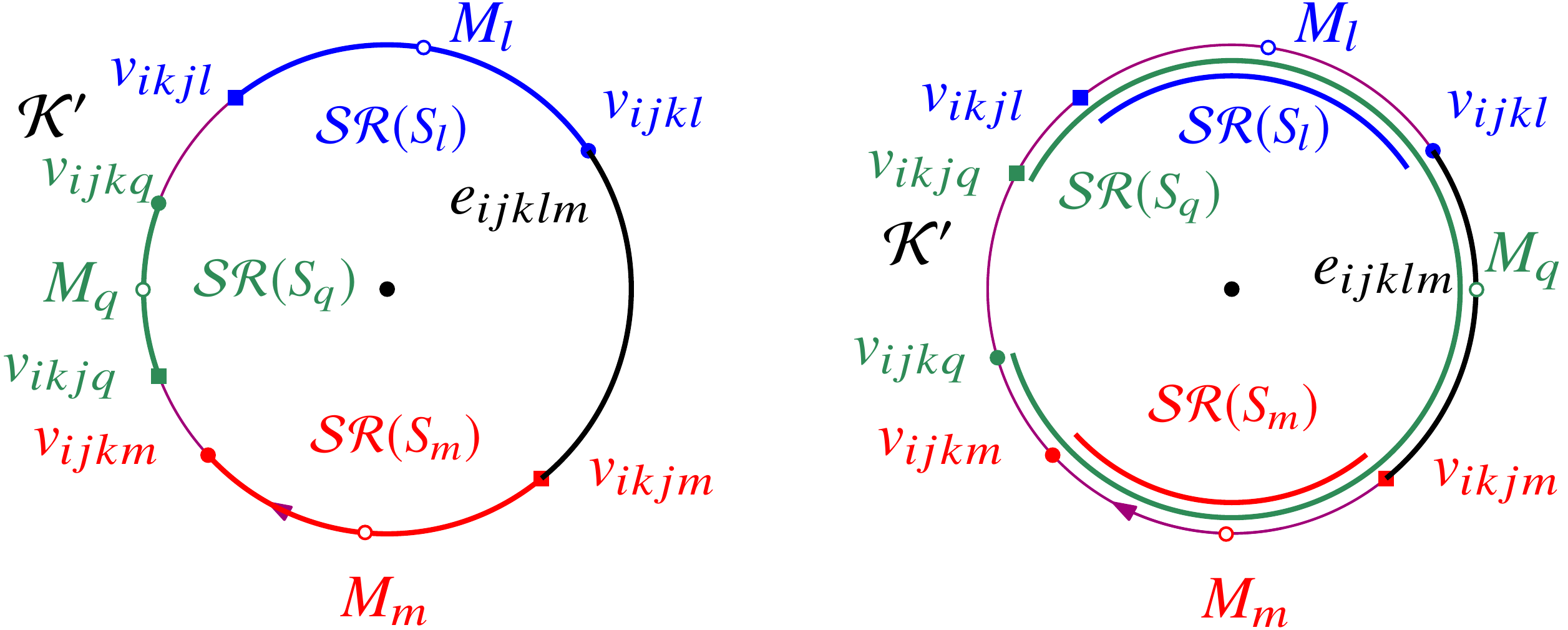}
      \caption[Step 6b if both $v_{ijkq},v_{ikjq}$ do not lie on the 
      edge $e_{ijklm}$]{ If both $v_{ijkq},v_{ikjq}$ lie 
      in-between the points $v_{ijkm}$ and 
      $v_{ikjl}$, then either a \noconflict (left figure) or 
      a \fullconflict (right figure) configuration arises. We 
      can distinguish which is the actually case if can determine the 
      order of appearance of $M_l$, $M_m$ and $M_q$, which are 
      the centers of the corresponding shadow regions $\sh{S_l}$, 
      $\sh{S_m}$ and $\sh{S_q}$. In the \noconflict scenario, 
      $M_l\prec M_m \prec M_q$ and $I_1=I_2=+$ whereas in the 
      \verticesconflict case, $M_l\prec M_q \prec M_m$ 
      and $I_1=I_2=-$.}
      \label{fig:36}
      \end{figure}

      Let us now focus on how one may determine the order of $M_l,M_m$
      and $M_q$ on $\ycone$. Firstly, we reflect on the fact that the 
      the open ray from the center $\yinv{\mathcal{A}}$ of $\ycone$ 
      towards $M_n$, for $n\in\{l,m,q\}$,
      goes through the center $\yinv{C_n}$ of the circle $\yinv{S_n}$. 
      Consequently, the order of appearance of $M_l,M_m$
      and $M_q$ on $\ycone$ is the order of appearance of the corresponding
      open rays, if we apply the same orientation with $\tri{ijk}$ 
      (see Figure~\ref{fig:38}).
      The latter ordering in \yspace can be deduced using orientation 
      predicates of the points $\inv{C_\mu}$ of \wspace 
      for $\mu\in\{i,j,l,m,q\}$. The required predicates to 
      obtain the ordering are presented in Steps 6bA-C; the same 
      analysis is followed in 
      Section~\ref{sub:ordering_in_a_classic_configuration}.

      \begin{figure}[tbp]
      \centering
      \includegraphics[width=0.7\textwidth]{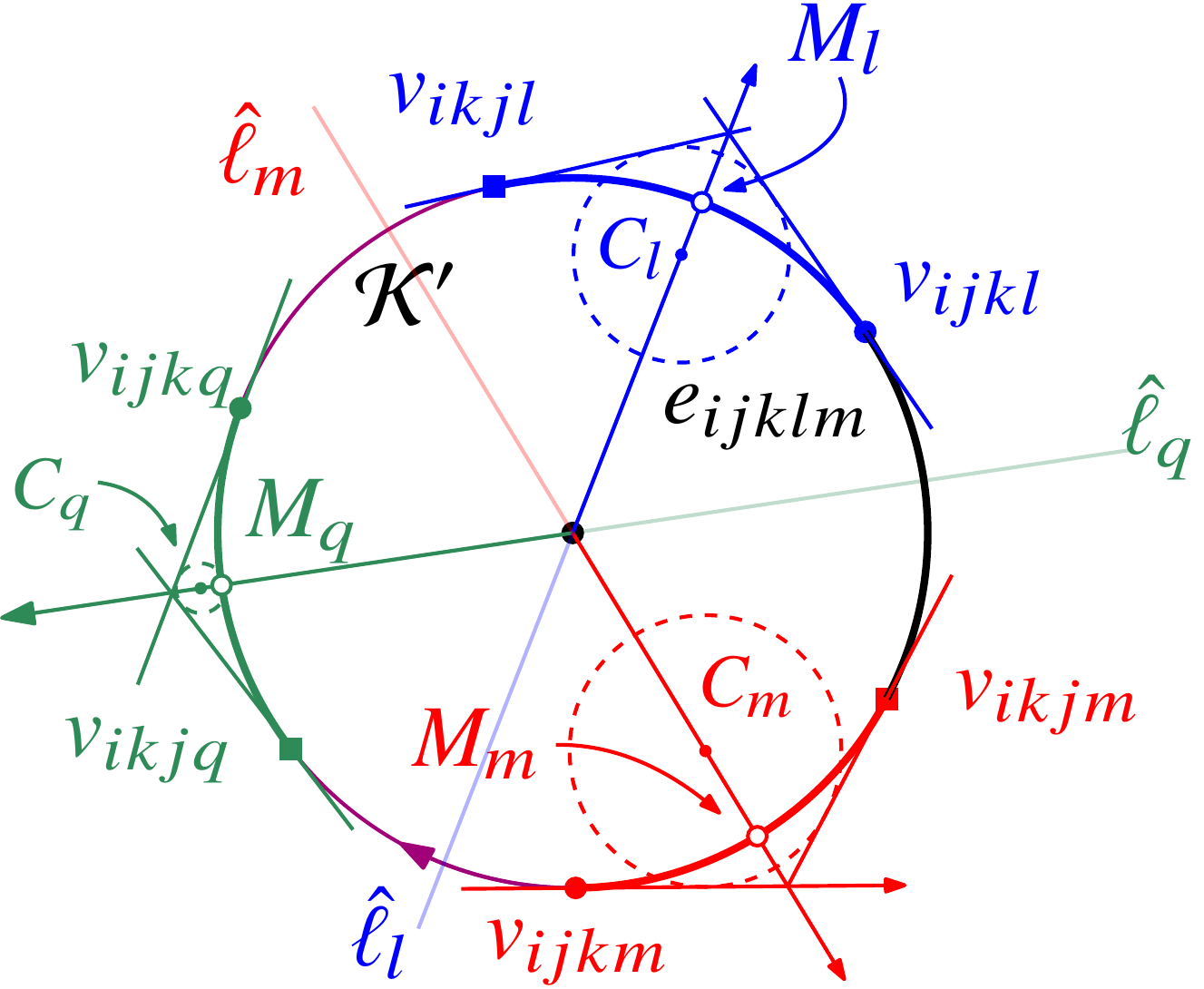}
      \caption[Ordering $M_l, M_m$ and $M_q$ on $\wcone$]{To order the 
      points $M_l, M_m$ and $M_q$ on $\wcone$, we consider some
      major facts. Firstly, as proven in 
      Section~\ref{sub:ordering_in_a_classic_configuration}, 
      the image of $C_n$, for $n\in\{l,m,q\}$,
      in \yspace must lie on the ray starting from the center of $\wcone$ 
      and going through $M_n$.
      As a consequence, the line $\yinv{\ell}_n$ of \yspace 
      defined by this ray 
      corresponds in \wspace to a plane containing the centers
      $\inv{C_i},\inv{C_i}$ and $\inv{C_n}$. We shall denote the 
      the halfspace of \yspace defined by $\yinv{\ell}_n$ and the point 
      $M_n$ if it moves infinitesimally 
      following (resp., opposite) the positive 
      orientation of $\tri{ijk}$ as the \emph{right} (resp., \emph{left}) 
      \emph{side of $\yinv{\ell}_n$}.
      Due to the orientation of $\wcone$, 
      for all points $\inv{N}$ of \wspace that satisfy
      $\text{\orient}(\inv{N},\inv{C_i},\inv{C_j},\inv{C_n})<0$, the 
      respective image $\yinv{N}$ in \yspace must lie on the right side of
      $\yinv{\ell}_n$. The \orient predicate is known to be equivalent 
      to \text{\orient}$(N,C_i,C_j,C_n)$ from previous sections. 
      Ultimately, we can safely conclude that $M_l\prec M_m \prec M_q$ 
      if $C_m$ and $C_q$ lie on the right and left side of 
      $\yinv{\ell}_l$ respectively. Otherwise, $C_m$ and $C_q$ must 
      lie on $\yinv{\ell}_l$ or on the same side; in either case, 
      checking if $C_q$ lies on the left or right side of $\yinv{\ell}_m$
      determines the ordering of $M_l,M_m$ and $ M_q$.}
      \label{fig:38}
      \end{figure}

      \begin{description}
      \item[Step 6bA] 
      We evaluate 
      $\Pi=O_1\cdot O_2 $, where 
      $O_1=\text{\orient}(\inv{C_q},\inv{C_i},\inv{C_j},\inv{C_l})$ and
      $O_2=\text{\orient}(\inv{C_m},\inv{C_i},\inv{C_j},\inv{C_l})$. 
      If $\Pi>0$ go to Step 6bB, otherwise go to Step 6bC.
      
      {\textbf{Explanation:}} 
      If $\Pi>0$ then both $M_q$ and $M_m$ lie on the same side 
      of $\ycone$ with respect to the line that goes through 
      $\yinv{\mathcal{A}}$ and $M_l$. 
      \item[Step 6bB.] 
      We evaluate
      $O_3=\text{\orient}(\inv{C_q}, \inv{C_i}, \inv{C_j}, \inv{C_m})$. 
      If $O_3<0$ then go to Step 6d if $I_1=I_2=+$ or go to Step 6c if 
      $I_1=I_2=+$. 
      Otherwise, if $O_3>0$ then go to Step 6c if $I_1=I_2=+$ or go 
      to Step 6d if $I_1=I_2=+$. 

      {\textbf{Explanation:}} 
      Both $M_q$ and $M_m$ lie on the same side 
      of $\ycone$ with respect to the line that goes through 
      $\yinv{\mathcal{A}}$ and $M_l$. If $O_3<0$, then in \yspace it 
      must hold that $M_q$ lies on the right half-plane defined by the 
      array that goes from $\yinv{\mathcal{A}}$ to $M_m$. In this 
      configuration, it must necessarily hold that $M_l\prec M_m\prec M_q$. 
      On the other hand, if $O_3>0$, we deduce that 
      $M_l\prec M_q\prec M_m$.

      \item[Step 6bC.] 
      If $\sgn(O_1)>\sgn(O_2)$ then go to Step 6d if $I_1=I_2=+$ 
      or go to Step 6c if $I_1=I_2=+$. 
      Otherwise, go to Step 6c if $I_1=I_2=+$ or go to Step 6d 
      if $I_1=I_2=+$. 

      {\textbf{Explanation:}} 
      It is known that $M_q$ and $M_m$ do not lie on the same side 
      of $\ycone$ with respect to the line that goes through 
      $\yinv{\mathcal{A}}$ and $M_l$. If $\sgn(O_1)>\sgn(O_2)$ then 
      $M_q$ (resp., $M_m$) must lie on the right (resp., left) 
      half-plane defined by the array that goes from $\yinv{\mathcal{A}}$ 
      to $M_l$. Therefore, we easily conclude that $M_l\prec M_m\prec M_q$.
      Using the same arguements, if $\sgn(O_1)>\sgn(O_2)$ does not hold, 
      we deduce that $M_l\prec M_q\prec M_m$.
      \end{description} 
      \item[Step 6c]
      If $I_1=I_2=+$ return \middleconflict otherwise, if $I_1=I_2=-$
      return \verticesconflict.
      {\textbf{Explanation:}}
      In this configuration, both $v_{ikjq}$ and $v_{ijkq}$ lie on 
      the Voronoi edge $e_{ijklm}$. 
      This implies that if $I_1=I_2=+$ then the shadow
      region of $S_q$ is an interval in the middle of $e_{ijklm}$ and 
      therefore the predicate returns \middleconflict. If $I_1=I_2=-$ 
      then $\sh{S_q}$ would be all the trisector except the aforementioned
      interval in the middle of of $e_{ijklm}$ and the predicate must 
      return \verticesconflict in this case.
      \item[Step 6d]
      If $I_1=I_2=+$ return \noconflict otherwise, if $I_1=I_2=-$
      return \fullconflict.
      {\textbf{Explanation:}}
      In this configuration, $v_{ikjq}$ and $v_{ijkq}$ can not both 
      lie on the Voronoi edge $e_{ijklm}$. As a result, if 
      $I_1=I_2=+$ then the predicate returns \noconflict since 
      \middleconflict is infeasible. Additionally, if $I_1=I_2=-$ 
      the predicate returns is \fullconflict as the alternative 
      answer, \verticesconflict, is not possible.
      \end{description}


    \section{Algebraic degrees and Conclusions} 
    \label{sub:algebraic_degrees_and_conclusions}

\begin{figure}[tbp]
      \centering
      \includegraphics[width=0.85\textwidth]{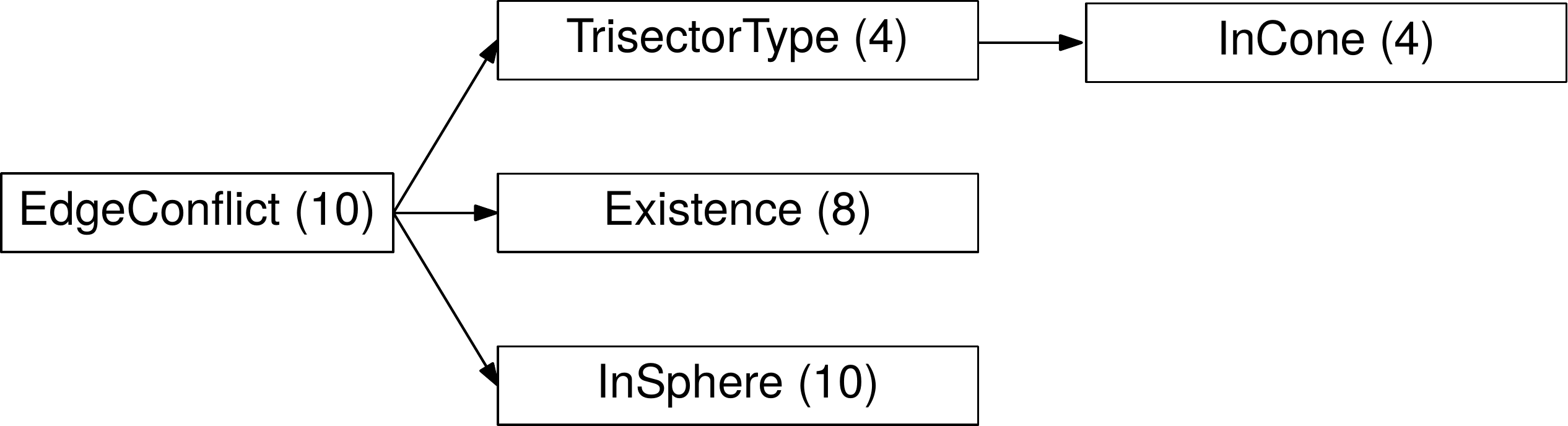}
      \caption[The (sub)predicates  used to 
      answer the \conflict in case of an elliptic trisector.]{The layout of predicates and their subpredicates used to 
      answer the \conflict predicate for elliptic trisectors. 
      The number next to each predicate corresponds to its algebraic 
      cost. It is assumed that every subpredicate returns a 
      non-degenerate answer.}
      \label{fig:predicates_elliptic}
      \end{figure}

      Using the algorithm presented in the previous section, 
      we can answer the \conflict predicate in case of elliptic trisectors,
      assuming all sub-predicates return a non degenerate answer. 
      A layout of all possible subpredicates used, 
      aside from primitives such as orientation tests, is 
      is shown in Figure \ref{fig:predicates_elliptic}.
      
      Note that the highest algebraic degree needed in the 
      evaluation of the subpredicates used is 10. Moreover, during 
      Steps 6bA-C we may have to evaluate orientation tests of the 
      form $\text{\orient}(\yinv{S_a},\yinv{S_i},\yinv{S_j},\yinv{S_b})$,
      for $a,b\in\{l,m,q\}$. In Section~\ref{sub:The_classic_configuration} 
      (paragraph \textit{Algebraic Cost to resolve the Cases A, B or C}), 
      we proved that 
      \begin{align}
      \text{\orient}(\inv{C_b},\inv{C_i},\inv{C_j},\inv{C_a})
      &= \sgn(D^{uvw}_{bija})
      = \sgn(\inv{p_i}\inv{p_j}\inv{p_a}\inv{p_b})\sgn(E^{xyzp}_{bija})\\
      &= \sgn(E^{xyzp}_{bija}),  
      \end{align}
      where $E^{xyzp}_{bija}$ is an expression of 
      algebraic degree 5 on the input quantities. Since the evaluation
      of the \insphere predicate is the most degree demanding operation 
      of the algorithm presented in Section~\ref{sub:conflict_elliptic},
      we have proven the following theorem.

      \begin{theorem}
      The \conflict predicate for hyperbolic trisectors can be 
      evaluated by determining the sign of quantities of algebraic degree 
      at most 10 (in the input quantities).
      \end{theorem}

\chapter{Degenerate Case Analysis} 
 \label{sec:degenerate_hyperbolic}

    In this chapter, the qualitative symbolic perturbation
    scheme that was introduced in Section~\ref{sub:qsp_introduction} 
    is applied to resolve the degeneracies for the 
    \incone, \distance, \shadow and \exist predicates in the 
    Sections~\ref{sub:the_perturbed_incone_predicate}, 
    \ref{ssub:the_perturbed_distance_predicate}, 
    \ref{sub:the_perturbed_shadowregion_predicate} and 
    \ref{sub:the_perturbed_existence_predicate} respectively. The  
    analysis of the respective predicates under the no-degeneracy 
    assumption can be found in 
    previous Sections~\ref{sub:the_incone_predicate_analysis}, 
    \ref{sub:the_distance_predicate_analysis}, 
    \ref{sub:the_shadowregion_predicate_analysis} and
    \ref{sub:the_existence_predicate_analysis} respectively.


  \section{The \texorpdfstring{\inconep}{perturbed Incone} Predicate} 
  \label{sub:the_perturbed_incone_predicate}

    As described in Section~\ref{ssub:the_incone_and_tritype_predicates}, 
    there are four possible outcomes of the predicate 
    $\text{\incone}(\allowbreak S_i,\allowbreak S_j,\allowbreak S_k)$. In case the predicate returns 
    one of the non-degenerate answers \outside or \inside, then 
    the outcome of $\text{\incone}(S_i,S_j,S_k)$ is trivialy the same.
    However, further analysis is required for the evaluation of 
    the \inconep predicate if \incone  returns
    one of the degenerate answers \ptouch or \ctouch. 
    
    If \incone returned \ptouch then 
    $S_k$ lies inside the semi-cone defined by $S_i$ and $S_j$ and 
    is also internally tangent to it at a single point. 
    To resolve the degeneracy, we perturb the spheres in the 
    order we defined in Section~\ref{sub:qsp_introduction}.
    If $k>i,j$ then the site $S_k$, after its infinitesimal 
    inflation, will properly intersect 
    the cone and therefore the perturbed predicate must return 
    \outside. 

    If $k$ has not the maximum index then either $i$ or $j$ has. 
    Since a name exchange of $S_i$ and $S_j$ does not affect 
    the \incone predicate, let us assume that $i>j,k$. 
    Under this assumption, we first need to examine if
    $t_k$ is identical with either $t_i$ or $t_j$, where $t_n$
    denotes the tangency point of the cone and the sphere 
    $S_n$, for $n\in\{i,j,k\}$. Indeed, if $t_k\equiv t_i$ or $t_j$, 
    then it must hold that $S_k$ lies within and is 
    internally tangent to $S_i$ or $S_j$ respectively. In this 
    scenario, if $S_i$ is inflated, $S_k$ would 
    lie strictly inside the cone and the perturbed predicate 
    would return \inside. To determine if this is the corresponding 
    geometric configuration, we only need to check if 
    $t_k\equiv t_n$, for $n\in\{i,j\}$. The last expression 
    is equivalent to $d(C_k,C_n)=r_n-r_k$ and
    $(x_k-x_n)^2+(y_k-y_n)^2+(z_k-z_n)^2=(r_k-r_n)^2$ and therefore 
    is a 2-degree demanding operation.

\begin{figure}[tbp]
    \centering
    \includegraphics[width=0.95\textwidth]{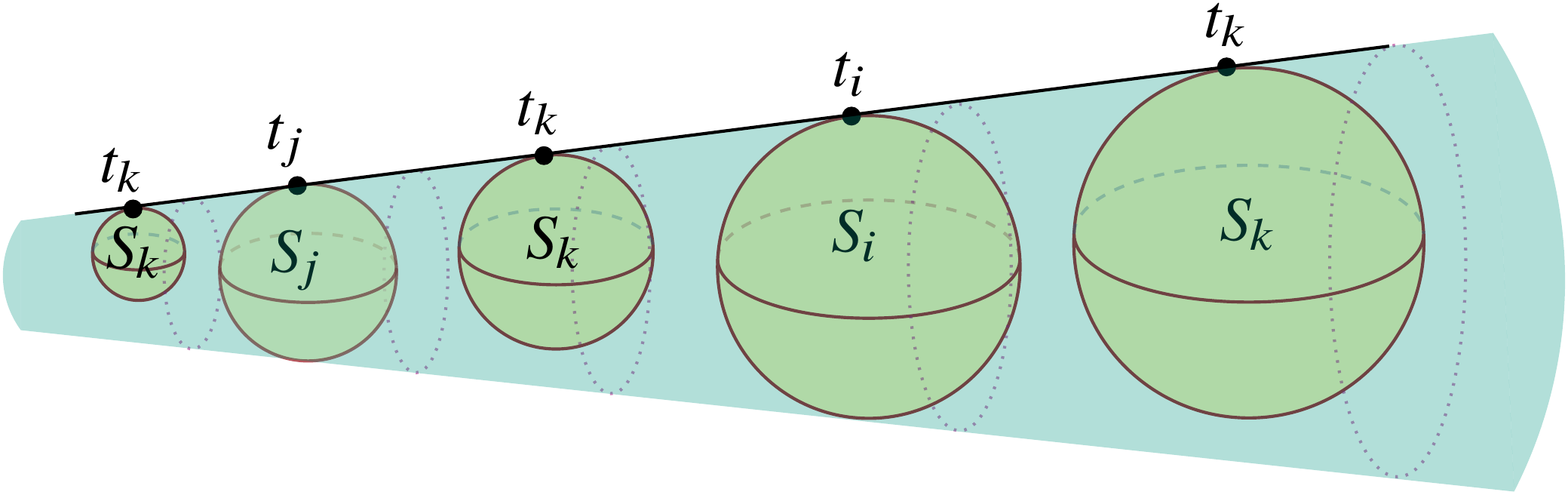}
    \caption[\incone degeneracy: \ptouch]{ The spheres $S_k$ in
    the center and on the right lie in such position that 
    $t_i$ and $t_k$ are on
    the same side with respect to $t_j$. After $S_i$ is perturbed, the 
    cone defined by $S_i$ and $S_j$ will no longer intersect such 
    spheres. The outcome of $\text{\incone}(S_k,S_j,S_i)$ is \ptouch 
    only for the sphere $S_k$ on the left; the respective
    outcome for the other two spheres is \outside.}
    \label{fig:16}
    \end{figure}

    Finally, if $t_i,t_j$ and $t_k$ are distinct points 
    they must be collinear and it must also hold that, after the 
    inflation of $S_i$, the sphere $S_k$ will lie strictly 
    inside the cone, thus the corresponding \inconep predicate 
    would return \inside,  if and only if $t_i$ and $t_k$ lie on the same 
    side with respect to $t_j$ (see Figure~\ref{fig:16}). 
    In this geometric configuration
    and since $S_k$ in known to be internally tangent to the cone 
    of $S_i$ and $S_j$, it is apparent that the outcome of 
    $\text{\incone}(S_k,S_j,S_i)$ is \outside. On the other hand, 
    if $t_j$ is in-between $t_i$ and $t_k$ then the respective 
    outcome would be \ptouch, since $i>j$. 
    Summarizing, the outcome of the examined \inconep predicate 
    is \inside or \outside if $\text{\incone}(S_k,S_j,S_i)$ 
    answers \outside or \ptouch respectively.

    Lastly, we resolve the \inconep predicate in the scenario that the 
    corresponding \incone predicate returned \ctouch. Notice that 
    the centers of the spheres $S_i, S_j$ and $S_k$ have to 
    be collinear in this case.
    Apparently, if $k>i,j$, the sphere $S_k$ would intersect the cone 
    after being inflated  and the outcome of the \inconep predicate 
    would be \outside. If this is not the case, we can 
    assume that $i>j,k$ since a name exchange of $S_i$ and $S_j$ 
    does not affect the predicates' answer. 

\begin{figure}[htbp]
    \centering
    \includegraphics[width=0.95\textwidth]{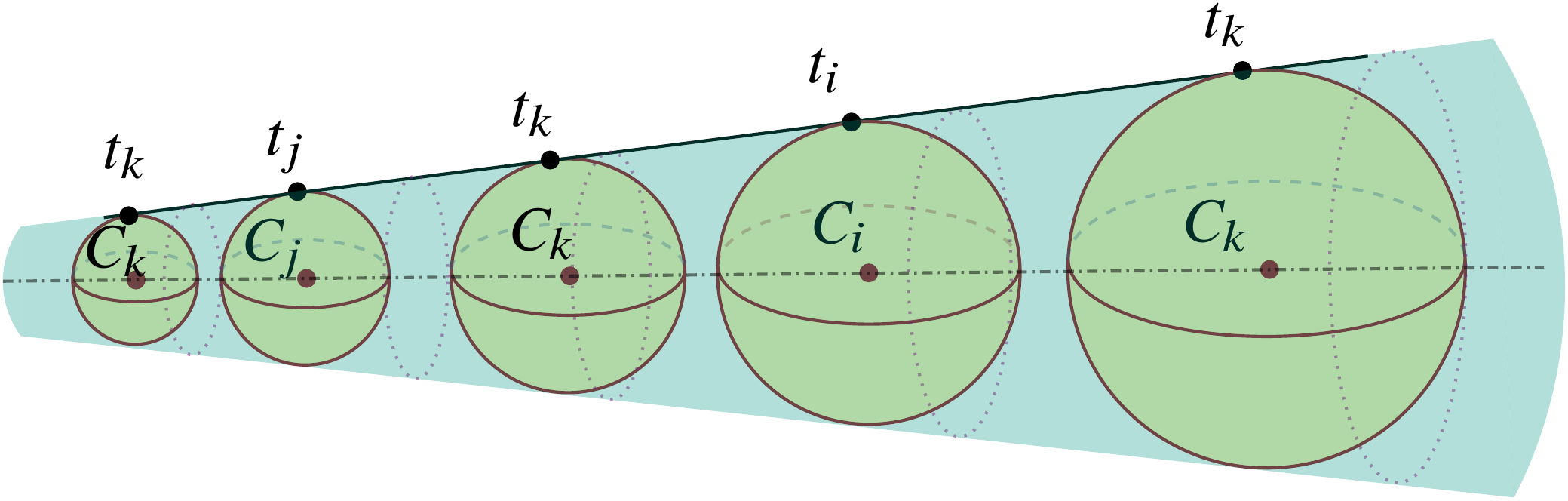}
    \caption[\incone degeneracy: \ctouch]{The spheres $S_k$ in
    the center and on the right lie in such position that 
    $t_i$ and $t_k$ are on
    the same side with respect to $t_j$. After $S_i$ is perturbed, the 
    cone defined by $S_i$ and $S_j$ will no longer intersect such 
    spheres. $t_i$ and $t_k$ are on the same side with respect to $t_j$ $C_j$if and only if $C_i$ and $C_k$ are on
    the same side with respect to $C_j$.}
    \label{fig:17}
    \end{figure}

    Due to the collinearity of the centers $C_i,C_j$ and $C_k$,
    the tangency points $t_i,t_j$ and $t_k$ must be distinct. Otherwise, 
    if any two of these tangency points coincide, the respective 
    spheres would also coincide, yielding a contradiction ; $S_i,S_j$ 
    and $S_k$ are assumed to be distinct. Lastly, we observe that, 
    after $S_i$ is perturbed, the sphere $S_k$ will intersect the cone,
    and therefore the corresponding \inconep predicate would return 
    \outside, if and only if $C_j$ lies in-between the points $C_i$ and 
    $C_k$ (see Figure~\ref{fig:17}). Otherwise, $S_k$ would lie completely inside the cone and 
    the \inconep predicate would return \inside .  
    Subsequently, \inconep predicate should return \outside 
    if the parallel vectors $\ov{C_jC_i}$ and $\ov{C_jC_k}$ have opposite 
    directions or equivalently
    $(x_i-x_j,y_i-y_j,z_i-z_j)=\lambda (x_k-x_j,y_k-y_j,z_k-z_j)$ 
    for some $\lambda<0$, which is a 1-degree demanding operation. 
    If the previous equality holds for some 
    $\lambda>0$ then the \inconep predicate should return \inside.

    The analysis of this section is summarized in the following 
    algorithm that answers the $\text{\inconep}(S_i,S_j,S_k)$ predicate.

    \begin{description}
    \item[Step 1] If $I=\text{\incone}(S_i,S_j,S_k)$ is \outside
    or \inside, return $I$. Otherwise go to Step 2.
    \item[Step 2] If $k>i,j$ return \outside. Otherwise, assume $i>j,k$. 
    If $I$ is \ptouch or \ctouch then go to Step 3 or 5 respectively.
    \item[Step 3] If $(x_k-x_n)^2+(y_k-y_n)^2+(z_k-z_n)^2=(r_k-r_n)^2$ 
    for some $n\in\{i,j\}$, return \inside. Otherwise go to Step 3.
    \item[Step 4] Evaluate $I_2=\text{\incone}(S_k,S_j,S_i)$; 
    if $I_2$ is \outside return \inside otherwise, return \outside.
    \item[Step 5] If all three equations $\sgn(x_i-x_j)=\sgn(x_k-x_j)$, 
    $\sgn(y_i-y_j)=\sgn(y_k-y_j)$ and $\sgn(z_i-z_j)=\sgn(z_k-z_j)$ hold, 
    return \inside. Otherwise, return \outside.
    \end{description}

    To summarize the analysis of this section, we have shown that
    in order to resolve the degeneracy of the 
    $\text{\incone}(S_i,S_j,S_k)$ predicate and determine
    $\text{\inconep}(S_i,S_j,S_k)$ we may have to evaluate  
    an additional \incone predicate or perform various operations 
    of lower algebraic cost. 
    Consequently, we have proven the following lemma.

    \begin{lemma}
    The \inconep predicate can be evaluated by determining 
    the sign of quantities of algebraic degree at most 4
    (in the input quantities).
    \end{lemma}


  \section{%
  The \texorpdfstring{\distancep}{perturbed Distance} 
  Predicate} 
  \label{ssub:the_perturbed_distance_predicate}
  
  When the predicate $\text{\distance}(S_i,S_j,S_k,S_a)$ is called, 
  it returns the tuple of signs $(d_1,d_2)$, where  $d_1$ and $d_2$
  correspond to the distance of the sphere $S_a$ from the planes 
  $\Pi_{ijk}^{-}$ and $\Pi_{ijk}^{+}$. As described in 
  Sections~\ref{ssub:the_distance_predicate} and \ref{sub:the_distance_predicate_analysis}, these planes are cotagent 
  to all sites $S_i,S_j$ and $S_k$, and both exist only in the case
  $\tri{ijk}$ is a hyperbolic trisector. 

  The outcome of the \distance predicate is consider 
  degenerate if either one or both of  
  $\{d_1,d_2\}$ equal 0. In these degenerate configurations, either 
  one (Case A) or both (Case B) of the planes 
  $\Pi_{ijk}^{-}$ and $\Pi_{ijk}^{+}$ are
  also tangent to $S_a$. We break down our analysis and 
  study these two cases separately.

  \textbf{Case A.} Let $\delta$ denote the $d_\nu$, for $\nu\in\{1,2\}$,
  that equals zero and $\delta'$ denotes the other one. 
  In the case studied, there exists only one plane 
  $\Pi:ax+by+cz+d=0$ that is commonly tangent to $S_n$ for all 
  $n\in\{i,j,k,a\}$.
  The answer of the respective \distancep predicate will 
  be the same as \distance if $\delta$ is replaced with 
  $\Delta\in\{-,+\}$. To decide if $\Delta=-$ or $+$, 
  we will have to determine whether $S_a$ intersects
  or not respectively the plane $\Pi$ after the perturbation 
  scheme is applied.

  Let $t_n$ denote the tangency point of the plane $\Pi$ with the 
  sphere $S_n$, for $n\in\{i,j,k,a\}$. If $a>i,j,k$ then, 
  after the perturbation, the inflated sphere $S_a$
  will intersect the plane $\Pi$ and therefore we shall return 
  $\Delta=-$. Otherwise, we cyclically
  permute the sites $S_i,S_j$ and $S_k$ such that $i>j,k,a$; 
  notice that such a name exchange does not alter the outcome of 
  the \distance predicate.

  Since $i>j,k,a$, $S_i$ is the sphere that will be initially 
  perturbed. Notice 
  that if $t_a\equiv t_i$, the perturbation will result in $S_a$ not
  intersecting $\Pi$, hence $\Delta=+$. In case 
  $t_a\equiv t_j$ then either $r_a>r_j$ or $r_a<r_j$; the radii $r_a$
  and $r_j$ can not be the same since the spheres $S_a$ and $S_j$ are
  considered to be distinct. If it holds that $r_a>r_j$ then $S_j$ must 
  lie inside and be tangent to $S_a$. In this configuration, $\Pi$ 
  will intersect $S_a$ after the perturbation hence $\Delta=-$.  
  Respectively, if $r_a<r_j$ then $S_a$  
  must lie inside and be tangent to $S_j$; in this configuration we 
  return $\Delta=+$ as $S_a$ does not intersect $\Pi$ after 
  the perturbation. The same analysis can be applied in the case 
  $t_a\equiv t_k$; it either holds that $r_a>r_k$ or $r_a<r_k$ 
  resulting in $\Delta=-$ or $\Delta=+$ respectively.

  Let us now consider the case where $t_j, t_k$ and $t_a$ are all 
  distinct and additionally collinear points. 
  Let us consider the relative position of $S_a$ and the 
  semi-cone defined by $S_j$ and $S_k$; the outcome of 
  $\text{\incone}(S_j,S_k,S_a)$ must be either \ptouch or \outside. 
  In the former case, $S_a$ will not
  intersect $\Pi$ after the perturbation and moreover, we know that 
  $\delta'=+$ as $S_a$ does not intersect any other plane cotagent 
  to $S_j$ and $S_k$. In the latter case, $S_a$ will 
  intersect $\Pi$ after the perturbation and we also know that 
  $\delta'=-$ since, in this configuration, $S_a$ 
  intersected all other planes cotagent to $S_j$ and $S_k$. Therefore, 
  in both scenarios, we should return $\Delta=\delta'$.
  
  Lastly, we consider the case where $t_i,t_j,t_k$ and $t_a$ are all
  distinct and $t_a$ does not lie on the line $\ell(t_j,t_k)$ defined 
  by $t_j$ and $t_k$. Since two distinct planes commonly tangent to 
  $S_i,S_j$ and $S_k$ exist, it also holds that $t_i$ does not lie on 
  $\ell(t_j,t_k)$. Therefore, on the plane $\Pi$, the points $t_i$ 
  and $t_a$ lie either on the same or on different sides with respect 
  to the line $\ell(t_j,t_k)$. In the former case, $S_a$ will not 
  intersect $\Pi$ after the perturbation, thus $\Delta=+$.  
  In the latter case respectively, we get $\Delta=-$ since $S_a$ 
  intersects $\Pi$ after applying the perturbation scheme.

  For the sake of clarity, we summarize the analysis of Case A as follows.

  \begin{description}
  \item[Step 1] 
  If $a>i,j,k$ then return $\Delta=-$. Otherwise, 
  assume that $i>j,k,a$ and go to Step 2.
  \item[Step 2] 
  If $t_a\equiv t_i$, return $\Delta=+$. Otherwise go to Step 2.
  \item[Step 3] 
  If $t_a\equiv t_j$,  return $\Delta=-$ if $r_a>r_j$, or 
  $\Delta=-$ if $r_a<r_j$. Otherwise go to Step 4.
  \item[Step 4] 
  If $t_a\equiv t_k$, then return $\Delta=-$ if $r_a>r_k$ or 
  $\Delta=+$ if $r_a<r_k$. Otherwise go to Step 5.
  \item[Step 5] 
  If $t_j,t_k$ and $t_a$ are collinear, return $\Delta=\Delta'$. 
  Otherwise go to Step 6.
  \item[Step 6] 
  If $t_i$ and $t_a$ are on the same side of 
  $\ell(t_j,t_k)$, return $\Delta=+$. Otherwise, if 
  $t_i$ and $t_a$ are on different sides of 
  $\ell(t_j,t_k)$, return $\Delta=-$.
  \end{description}

  For Steps 2 to 4, we have to determine whether $t_a\equiv t_n$, 
  for $n\in\{i,j,k\}$. This is equivalent to the spheres $S_a$ and $S_n$
  being internally tangent and therefore 
  $(x_a-x_n)^2+(y_a-y_n)^2+(z_a-z_n)^2=(r_a-r_n)^2$, which is a 2-degree
  demanding operation. 

  For Steps 5 and 6, the relative position of the points $t_a$ and $t_i$
  and the line $\ell(t_j,t_k)$ on $\Pi$ is required to resolve the 
  degeneracy. Initially, we consider the algebraic expression 
  $ax+by+cz+d=0$ of the plane $\Pi$ and we assume, without 
  loss of generality, that $a^2+b^2+c^2=1$. Since all sites $S_n$, for 
  $n\in\{i,j,k,a\}$ are tangent to $\Pi$, it must hold that 
  $ax_n+by_n+cz_n+d=r_n$. Using Crammer's rule, we evaluate the 
  vector $V=(a,b,c)=V'/D^{xyz1}_{ijka}$, where 
  $V'=(D^{ryz}_{ijka},D^{xrz}_{ijka},D^{xyr}_{ijka})$; it is well known 
  that $V'$ is perpendicular with the plane $\Pi$. 

  We now consider the plane $\Pi'$ that contains the line 
  $\ell(t_j,t_k)$ and is perpendicular the plane $\Pi$. 
  This plane is necessarily parallel to the vector $V'$ and must 
  contain both points $C_j$ and $C_k$. As an immediate result, we 
  deduce that $t_a$ is collinear with $t_j$ and $t_k$ if and only if 
  $C_a$ lies on the plane $\Pi'$, \ie, if 
  $C_j C_a \cdot (V'\times C_j C_k)=0$. The last expression can be 
  rewritten as $D^{ryz}_{ijka}D^{yz}_{akj}-D^{xrz}_{ijka}D^{xz}_{ajk}+
  D^{xyr}_{ijka}D^{xy}_{ajk}=0$ which is a 6-degree demanding operation.

  Lastly, for Step 6, we can determine whether $t_i$ and $t_a$ are 
  on the same side of $\ell(t_j,t_k)$, as this is equivalent to 
  $C_i$ and $C_a$ lying on the same side of the plane $\Pi'$. 
  In this case, the expressions $E_i=C_j C_i \cdot (V'\times C_j C_k)$ and 
  $E_a=C_j C_a \cdot (V'\times C_j C_k)$ must have the same sign. 
  Since $E_\nu=D^{ryz}_{ijk\nu}D^{yz}_{\nu kj}
  -D^{xrz}_{ijk\nu}D^{xz}_{\nu jk}+ D^{xyr}_{ijk\nu}D^{xy}_{\nu jk}$, for
  $\nu\in\{i,a\}$, we can determine if $\sgn(E_i)=\sgn(E_a)$ using 
  6-fold operations. 

  \textbf{Case B.} 
  In this configuration, $S_a$ it tangent to both planes $\Pi^{-}_{ijk}$ and 
  $\Pi^{+}_{ijk}$ and therefore it must hold that the centers of the
  spheres $S_n$, for $n\in\{i,j,k,a\}$, lie on the same plane,
  denoted by $\Pi'$. 
  The symmetry in this geometric configuration indicates 
  that, after the perturbation, the site $S_a$ will 
  intersect either both or none of the planes $\Pi^{-}_{ijk}$ and 
  $\Pi^{+}_{ijk}$ and as a subsequence, the result $R$ of \distancep 
  will be $(-,-)$ or $(+,+)$ respectively.

  We first consider the case where $a>i,j,k$. In this case, $S_a$ will 
  be perturbed first and as a result it will intersect both 
  $\Pi^{-}_{ijk}$ and $\Pi^{+}_{ijk}$. 
  It is apparent that the result of the perturbed
  predicate will be $(-,-)$ in this scenario.

  If it does not hold that $a>i,j,k$, we cyclically permute $S_i,S_j$ 
  and $S_k$ such that $i>j,k,a$. As stated before, this name exchange 
  does not alter the outcome of the \distance predicate. We now break 
  down our analysis depending on whether the centers $C_j,C_k$ and $C_a$
  are collinear or not. Note that these points are collinear if and only 
  if $C_j C_k \times C_j C_a = \vec{0}$ or equivalently
  $D_{akj}^{xy}=D_{akj}^{xz}=D_{akj}^{yz}=0$, which is a 2-degree 
  demanding operation. 

  If $C_j,C_k$ and $C_a$ are not collinear, then the sphere $S_a$ will 
  intersect both $\Pi^{-}_{ijk}$ and $\Pi^{+}_{ijk}$ after perturbing 
  $S_i$ if and only if both $C_a$ and $C_i$ lie, on the plane $\Pi'$, on 
  the same side with respect to the line $\ell(C_j,C_k)$. A simple 
  way to examine if the last property holds 
  is via the use of an auxiliary point $P$. 
  This point $P$ is selected among $\{(0,0,0),(0,0,1),(0,1,0),(1,0,0)\}$
  such that $O_i=\text{\orient}(P,C_j,C_k,C_i)$ does not equal zero 
 , \ie, $P$ does not lie on $\Pi'$. This is plausible since the possible 
  four candidates for $P$ are not coplanar and therefore at least one 
  does not lie on $\Pi'$. Observe now that $C_a$ and $C_i$ lie 
  on the same side with respect to the line $\ell(C_j,C_k)$ if and only if 
  the signs of $O_i$ and $O_a=\text{\orient}(P,C_j,C_k,C_a)$ are the same, 
  which is a 3-degree demanding operation.

  Lastly, we consider the case where $C_j,C_k$ and $C_a$ are 
  collinear. In this configuration, even after perturbing $S_i$, $S_a$
  still remains commonly tangent to both planes $\Pi^{-}_{ijk}$ and 
  $\Pi^{+}_{ijk}$. Since the degeneracy is not yet resolved, the
  perturbation of a second site among $S_j,S_k$ and $S_a$ is required. 
  If $a>j,k$, we will trivially have to return $(-,-)$ as $S_a$ will 
  intersect both cotagent planes. If $j>a,k$, $S_a$ will 
  intersect both $\Pi^{-}_{ijk}$ and $\Pi^{+}_{ijk}$ if and only if 
  $C_k$ lies in-between $C_j$ and $C_a$ or equivalently 
  $\vec{C_j C_k} = \lambda \cdot \vec{C_j C_a}$ for some $\lambda>0$. 
  Respectively, if $k>a,j$,
  $S_a$ will intersect both cotagent planes if and only if $C_j$ 
  lies in-between $C_k$ and $C_q$ or equivalently 
  $\vec{C_j C_k} = \lambda \cdot \vec{C_j C_a}$ for some $\lambda<0$. 
  Notice that since $C_j,C_k$ and $C_a$ are known to be collinear, 
  the sign of $\lambda$ in the last two cases is positive if and only if
  $\sgn(x_a-x_j)=\sgn(x_k-x_j)$, $\sgn(y_a-y_j)=\sgn(y_k-y_j)$ and 
  $\sgn(z_a-z_j)=\sgn(z_k-z_j)$, which is a 1-degree demanding operation.

  The analysis of Case B is summarized for clarity in the following steps.

  \begin{description}
  \item[Step 1] 
  If $a>i,j,k$ then return $(-,-)$. Otherwise, 
  assume that $i>j,k,a$ and go to Step 2. 
  \item[Step 2] Evaluate  
  $D_{akj}^{xy},D_{akj}^{xz}$ and $D_{akj}^{yz}$; if not all equal 0,
  go to Step 3 otherwise go to Step 4.
  \item[Step 3]Pick a point $P\in\{(0,0,0),(0,0,1),(0,1,0),(1,0,0)\}$
  such that $P\not\in\Pi(C_i,C_j,C_k)$. Evaluate
  $O_i=\text{\orient}(P,C_j,C_k,C_i)$ and 
  $O_a=\text{\orient}(P,C_j,C_k,C_a)$. Return $(+,+)$ if 
  $\sgn(O_i)=\sgn(O_a)$ otherwise, return $(-,-)$.
  \item[Step 4] If $a>j,k$ return $(-,-)$. Otherwise, if 
  $j>a,k$ go to Step 5 or else, if $k>a,j$ go to Step 6.
  \item[Step 5] 
  If all three equalities $\sgn(x_a-x_j)=\sgn(x_k-x_j)$, 
  $\sgn(y_a-y_j)=\sgn(y_k-y_j)$ and $\sgn(z_a-z_j)=\sgn(z_k-z_j)$ 
  hold, then return $(-,-)$, otherwise return $(+,+)$.
  \item[Step 6] 
  If all three equalities $\sgn(x_a-x_j)=\sgn(x_k-x_j)$, 
  $\sgn(y_a-y_j)=\sgn(y_k-y_j)$ and $\sgn(z_a-z_j)=\sgn(z_k-z_j)$ 
  hold, then return $(+,+)$, otherwise return $(-,-)$.
  \end{description}

  To summarize the analysis of this section, we have shown that
  in order to evaluate the 
  $\text{\distancep}(S_i,S_j,S_k,S_a)$ predicate we have 
  call the respective \distance predicate and perform  
  additional operations of algebraic cost at most 6.
  Consequently, we have proven the following lemma.

  \begin{lemma}
  The \distancep predicate can be evaluated by determining 
  the sign of quantities of algebraic degree at most 6
  (in the input quantities).
  \end{lemma}


  \section{%
  The \texorpdfstring{\shadowp}{perturbed ShadowRegion} Predicate for 
  Hyperbolic Trisectors} 
  \label{sub:the_perturbed_shadowregion_predicate}
  

  Based on the analysis of Sections~\ref{ssub:the_shadow_predicate} 
  and \ref{sub:the_shadowregion_predicate_analysis}, the outcome 
  of the $\text{\shadow}\allowbreak(S_i,\allowbreak S_j,\allowbreak S_k,\allowbreak S_a)$ predicate is the 
  topological form $SRT(S_a)$ of $\sh{S_a}$ on the hyperbolic 
  trisector $\tri{ijk}$ if seen as an interval or union of intervals.

  We have proven in previous sections that the boundary points 
  of $\sh{S_a}$ correspond to the Apollonius vertices $v_{ijka}$ and/or
  $v_{ikja}$ and that in non-degenerate configurations only 
  the following outcomes of the \shadow predicate are plausible:
  $\emptyset, \RR,\allowbreak (-\infty,\phi), \allowbreak (\chi,+\infty), 
  \allowbreak(\chi,\phi)$ and 
  $(-\infty,\phi)\cup(\chi,+\infty)$. A useful remark is stated in 
  Lemma~\ref{lemma:phi_chi}: \emph{the endpoints $\chi$ and $\phi$ 
    correspond to $v_{ikja}$ and $v_{ijka}$ respectively}.  As
  in previous sections, the notation ``$\sh{S_a}=\emptyset$'' will be 
  equivalent to ``$\sh{S_a}$ is of type $\emptyset$'' and 
  ``$SRT(S_a)=\emptyset$''.

  Regarding degenerate shadow regions, they can be classified into 
  two categories. The first category involves shadow regions where 
  $\chi$ and/or $\phi$ are allowed to coincide with $\pm\infty$ but 
  not with each other; we shall call these cases 
  \emph{degeneracies of type A}. The shadow regions where $\chi$ 
  and $\phi$, and therefore
  $v_{ikja}$ and $v_{ijka}$, coincide are called 
  \emph{degeneracies of type B}.

  To resolve degenerate shadow regions, we must first be able 
  to detect them and their type. 
  As shown in Section~\ref{sub:the_shadowregion_predicate_analysis}, 
  there is a close relation between the outcome of 
  $\text{\shadow}(S_i,S_j,S_k,S_a)$ and the outcomes of the 
  \exist and \distance predicates, given the same input. 
  Assuming no degeneracies, type of $\sh{S_a}$ is deduced
  by combining the latter two outcomes, as shown in 
  Table~\ref{tab:exist_distance_non_degenerate}.   
  If the \exist and \distance predicates return a combination that 
  does not appear in Table~\ref{tab:exist_distance_non_degenerate}, 
  then we have a degenerate shadow region $\sh{S_a}$ and our second 
  step is to determine its type.

  \begin{table}[ht]
  \begin{center}
  \begin{tabular}{|c|c||c|}
  \hline
  \exist & \distance & \multirow{2}{*}{$\sh{S_a}$} \\
  $(S_i,S_j,S_k,S_a)$ & $(S_i,S_j,S_k,S_a)$ & \\
  \hline\hline
  \multirow{2}{*}{0} 
  & $(+,+)$ & $\emptyset$ \\ \cline{2-3}
  & $(-,-)$ & $\RR$ \\ \cline{2-3}
  \hline
  \multirow{2}{*}{1} 
  & $(+,-)$ & $(\chi,+\infty)$ \\ \cline{2-3}
  & $(-,+)$ & $(-\infty,\phi)$ \\ \cline{2-3}
  \hline
  \multirow{2}{*}{2} 
  & $(+,+)$ & $(\chi,\phi)$ \\ \cline{2-3}
  & $(-,-)$ & $(-\infty,\phi)\cup(\chi,+\infty)$ \\ \cline{2-3}
  \hline
  \end{tabular}
  \end{center}
  \caption[The non-degenerate combinations of outcomes of the \exist and 
  the \distance predicates. ]{If the outcome of $\text{\exist}(S_i,S_j,S_k,S_a)$ 
  and $\text{\distance}(S_i,S_j,S_k,S_a)$ is provided, we 
  can safely deduce the form of the non-degenerate shadow 
  region $\sh{S_a}$ on the hyperbolic trisector $\tri{ijk}$ 
  as shown in this table.
  In all other possible combinations of these two predicates, 
  we have a degenerate $\sh{S_a}$.}
  \label{tab:exist_distance_non_degenerate}
  \end{table}

  Crucially, type B degeneracies occur only if $v_{ikja}\equiv v_{ijka}$. 
  In such cases, during the evaluation of the 
  $\text{\exist}(S_i,S_j,S_k,S_a)$ predicate as presented in 
  Section~\ref{sub:the_existence_predicate_analysis}, a double root of
  either $M(d)$ or $L(c)$ must appear. This double root indicates that 
  there is a ``double plane'' $\inv{\Pi}_{ijk}:au+bv+cw+d=0$ cotagent 
  to $\inv{S_i},\inv{S_j}$ and $\inv{S_a}$; this is the image  
  of the coinciding spheres $\ap{v_{ikja}}$ and $\ap{v_{ijka}}$ in \yspace. 
  
  Based on the analysis of 
  Section~\ref{sub:the_existence_predicate_analysis}, such double 
  roots may exist only if $\Delta_M=0$ in all possible cases. 
  Therefore, degenerate shadow regions are of type A if and only if 
  $\Delta_M$ does not equal zero otherwise, they are of type B. 
  Depending on its type, we follow the analysis of the corresponding 
  section to resolve the degeneracy.

  \paragraph*{Degeneracies of type A.}
  Firstly, we examine the degenerate shadow regions of type A. In 
  these cases, we reflect on all possible non-degenerate forms of 
  $\sh{S_a}$ that contain $\phi$ and/or $\chi$; these are 
  $(-\infty,\phi), (\chi,+\infty), (\chi,\phi)$ and 
  $(-\infty,\phi)\cup(\chi,+\infty)$. For each of these possible forms,
  we consider the outcome of the corresponding \distance and \exist 
  predicates if $\phi$ and/or $\chi$ coincided with $\pm\infty$ and 
  what will happen after applying the perturbation scheme. 
  Beforehand, we make three crucial remarks.

  {\bfseries Remark 1.} For all cases that we have to study, 
  one can observe that if $\phi$ and/or $\chi$ tend to 
  $-\infty$ (resp., $+\infty$)
  then $\Pi_{ijk}^{-}$ (resp., $\Pi_{ijk}^{+}$) becomes the Apollonius 
  sphere of $S_i,S_j,S_k$ and $S_a$. In this configuration,
  $S_a$ must necessarily be tangent to $\Pi_{ijk}^{-}$
  (resp., $\Pi_{ijk}^{+}$) yielding $d_1=0$ (resp., $d_2=0$), 
  where $(d_1,d_2)$ denotes the answer of 
  the $\text{\distance}(S_i,S_j,S_k,S_a)$ predicate.

  {\bfseries Remark 2.} After applying the perturbation scheme, 
  the endpoint(s) among $\{\phi,\chi\}$ that coincided with 
  $\pm\infty$ will move infinitesimally on the trisector $\tri{ijk}$. 
  Thus, if an endpoint initially coincided with $\pm\infty$ then, after
  the perturbation it will either move 
  infinitesimally towards $o_{ijk}$ and become ``finite'' or it  
  will move further away from $o_{ijk}$ and truly become ``infinite''. 
  In both cases, the degeneracy is resolved. To distinguish between the two
  possible scenarios, we have to consider the possible outcomes of the 
  perturbed \distance predicate and make a connection with the possible 
  shadow regions types that arise after the substition of the 
  endpoint(s) with its finite or infinite form(s).

  {\bfseries Remark 3.} The $\text{\exist}(S_i,S_j,S_k,S_a)$ predicate
  returns how mane of the Apollonius vertices $\{v_{ijka}, v_{ikja}\}$ 
  exist and correspond to \emph{finite} Apollonius spheres. Therefore, 
  although $\phi$ and/or $\chi$ correspond to the existence of $v_{ijka}$
  and $v_{ikja}$ respectively, the \exist predicate does not count 
  them if they coincide with $\pm\infty$.

  Using these three remarks, we may analyse all possible degenerate 
  shadow regions of type A and draw our conclusions. The following 
  two examples are indicative of the rest of the analysis that is 
  summarized in Table~\ref{tab:degeneracy A Type}.

  {\bfseries Example 1.} Let us consider the case where 
  $\sh{S_a}=(-\infty,\phi)$ and $\phi$ coincides with $+\infty$. 
  As a first observation, $\sh{S_a}$ has no ``finite'' boundary  and
  therefore $\text{\exist}(S_i,S_j,S_k,\allowbreak S_a)$ must return 0, based on 
  Remark 3. As of Remark 1, the $\text{\distance}(S_i,S_j,S_k,\allowbreak S_a)$ 
  predicate should return $(d_1,d_2)$, where $d_2=0$. Finally, using 
  Remark 2 we know that the perturbed shadow region \shp{$S_a$} will be
  either $(-\infty,\phi)$ or $(-\infty,+\infty)=\RR$. In the former 
  case, $\text{\distancep}(S_i,S_j,S_k,S_a)$, will be $(-,+)$ whereas
  in the latter, it will be $(-,-)$. We can now safely deduce that 
  $d_1=-$ in both cases and that \shp{$S_a$} can be identified by the 
  outcome of \distancep.

  {\bfseries Example 2.} We consider the case where 
  $\sh{S_a}=(\chi,\phi)$ and $\chi,\phi$ coincide with $-\infty,+\infty$
  respectively. As before, $\sh{S_a}$ has no ``finite'' boundary  and
  therefore $\text{\exist}\allowbreak(S_i,\allowbreak S_j,\allowbreak S_k,\allowbreak S_a)$ must return 0 
  based on Remark 3. The $\text{\distance}(S_i,S_j,S_k,S_a)$
  predicate should return $(0,0)$ as Remark 1 suggests.
  Using Remark 2, we know that $\phi$ (resp., $\chi$) will become either
  finite or infinite after the perturbation and therefore 
  the possible forms of \shp{$S_a$} are: $(-\infty,+\infty)=\RR, 
  (-\infty,\phi), (\chi,\infty)$ and $(\chi,\phi)$. However, since 
  the respective \distancep predicate is known to return 
  either $(+,+)$ or $(-,-)$ in this scenario (see 
  Section~\ref{ssub:the_perturbed_distance_predicate}), only the
  cases $\RR$ and $(\chi,\phi)$ are plausible.

  Using the Remarks 1-3 and similar analysis with the Examples 1 and 2, 
  we construct the Table~\ref{tab:degeneracy A Type} for all possible 
  degenerate shadow regions of type A. This table indicates that we
  can resolve a degeneracy of type A and determine the form 
  of \shp{$S_a$} using only the \exist, \distance and \distancep predicates.

  \begin{table}[ht]
  \begin{center}
  \begin{tabular}{|c|c|c||c|c|}
  \hline
  $E$ & $D$ & $D^{\epsilon}$ & \shp{$S_a$} & Original $\sh{S_a}$ \\
  \hline\hline
  \multirow{10}{*}{0}
  & \multirow{2}{*}{$(0,+)$ } 
  & $(-,+)$ & $(-\infty,\phi)$ & \multirow{2}{*}{$(-\infty,\phi)$ when $\phi\equiv -\infty$} \\ \cline{3-4}
  & & $(+,+)$ & $\emptyset$ & \\ \hhline{|~|=|=#==|} 
  & \multirow{2}{*}{$(0,-)$ } 
  & $(-,-)$ & $\RR$ & \multirow{2}{*}{$(\chi,+\infty)$ when $\chi\equiv -\infty$}\\ \cline{3-4}
  & & $(+,-)$ & $(\chi,+\infty)$ & \\ 
  \hhline{|~|=|=#==|} 
  & \multirow{2}{*}{$(+,0)$ } 
  & $(+,-)$ & $(\chi,+\infty)$ & \multirow{2}{*}{$(\chi,+\infty)$ when $\chi\equiv +\infty$} \\ \cline{3-4}
  & & $(+,+)$ & $\emptyset$ & \\ \hhline{|~|=|=#==|} 
  & \multirow{2}{*}{$(-,0)$ } 
  & $(-,-)$ & $\RR$ &  \multirow{2}{*}{$(-\infty,\phi)$ when $\phi\equiv +\infty$}\\ \cline{3-4}
  & & $(-,+)$ & $(-\infty,\phi)$ & \\ 
  \hhline{|~|=|=#==|} 
  & \multirow{2}{*}{$(0,0)$ } 
  & $(-,-)$ & $\emptyset$ & $(-\infty,\phi)\cup(\chi,+\infty)$\\ \cline{3-4}
  & & $(+,+)$ & $(-\infty,\phi)\cup(\chi,+\infty)$ &  when $\phi\equiv -\infty$ and $\chi\equiv +\infty$\\ 
  \hline \hline
  \multirow{8}{*}{1}
  & \multirow{2}{*}{$(0,+)$ } 
  & $(-,+)$ & $(-\infty,\phi)$ & \multirow{2}{*}{$(\chi,\phi)$ when $\chi\equiv -\infty$}\\ \cline{3-4}
  & & $(+,+)$ & $(\chi,\phi)$ & \\ 
  \hhline{|~|=|=#==|} 
  & \multirow{2}{*}{$(0,-)$ } 
  & $(-,-)$ & $(-\infty,\phi)\cup(\chi,+\infty)$ 
   & $(-\infty,\phi)\cup(\chi,+\infty)$  \\ \cline{3-4}
  & & $(+,-)$ & $(\chi,+\infty)$ & when $\phi\equiv -\infty$\\ \hhline{|~|=|=#==|} 
  & \multirow{2}{*}{$(+,0)$ } 
  & $(+,-)$ & $(\chi,+\infty)$ & \multirow{2}{*}{$(\chi,\phi)$ when $\phi\equiv +\infty$}\\ \cline{3-4}
  & & $(+,+)$ & $(\chi,\phi)$ & \\ 
  \hhline{|~|=|=#==|} 
  & \multirow{2}{*}{$(-,0)$ } 
  & $(-,-)$ & $(-\infty,\phi)\cup(\chi,+\infty)$ 
   & $(-\infty,\phi)\cup(\chi,+\infty)$ \\ \cline{3-4}
  & & $(-,+)$ & $(-\infty,\phi)$ & when $\chi\equiv +\infty$\\ 
  \hline
  \end{tabular}
  \end{center}
  \caption[Resolve of Degeneracies of type A using the \exist, \distance and \distancep predicates.]{$E,D$ and $D^{\epsilon}$ denote the outcomes
  of the \exist, \distance and \distancep predicates for input 
  $(S_i,S_j,S_k,S_a)$. The last column describes the original 
  degenerate shadow region $\sh{S_a}$ that was studied.}
  \label{tab:degeneracy A Type}
  \end{table}

  \paragraph*{Degeneracies of Type B.}
  Lastly, we consider degeneracies of type B. In these 
  degeneracies, the Apollonius vertices 
  $v_{ikja}$ and $v_{ijka}$ coincide; this ``double'' vertex is denoted
  by $V$. Evidently, the degenerate shadow region $\sh{S_a}$ of $S_a$ 
  must be either $(\chi,\phi)$ or 
  $(-\infty,\phi)\cup (\chi,+\infty)$, where $\phi$ and $\chi$ 
  essentially coincide. After the perturbation, these endpoints 
  infinitesimally move ``away'' from each other and their initial 
  position. Depending on the form of $\sh{S_a}$ and whether 
  $\phi$ moves following the positive orientation of the trisector or 
  not, we can deduce the perturbed shadow region \shp{$S_a$} of $S_a$.

  The analysis that allows us to determine \shp{$S_a$} is subdivided
  into two main cases. In Case 1, $V$ coincides with either $-\infty$ 
  or $+\infty$ on the trisector whereas in Case 2 it does not. 
  The outcome $(d_1,d_2)$ of $\text{\distance}(S_i,S_j,S_k,S_a)$ 
  indicates if we are in Case 1 or 2. When an Apollonius vertex
  coincides with $-\infty$ (resp., $+\infty$) on the trisector 
  then $d_1$ (resp., $d_2$) must equal 0, as $S_a$ is tangent to 
  $\Pi_{ijk}^{-}$ (resp., $\Pi_{ijk}^{+}$). Therefore, if $d_1$ or 
  $d_2$ equal 0, we follow the analysis of Case 1 otherwise, that 
  of Case 2.

  \textbf{Case 1.} If $d_1$ or $d_2$ equals 0, we immediately deduce 
  that $V$, and therefore both $\phi$ and $\chi$, coincide with 
  $-\infty$ or $+\infty$ respectively. The same strategy with 
  degeneracies of type A is followed here. Initially, we consider 
  all possible scenarios regarding the endpoints 
  ($\phi\equiv\chi\equiv -\infty$ or
  $\phi\equiv\chi\equiv +\infty$) and  
  the shadow regions types ($(\chi,\phi)$ 
  and $(-\infty,\phi)\cup (\chi,+\infty)$). For each of the four 
  possible subcases, we consider the perturbed shadow region
  \shp{$S_a$} that derives after $\phi$ moves infinitesimally, following the 
  positive or the negative orientation of the trisector. 
  Due to the symmetry of this configuration in \wspace and becoming
  apparent from the analysis of Case 2 below, at the same time that 
  $\phi$ moves infinitesimally on the trisector, the endpoint 
  $\chi$ also moves infinitesimally on the opposite direction. 

  For each subcase studied, we consider the resulting perturbed 
  shadow region \shp{$S_a$} along with the  outcomes of 
  the the respective \distance and \distancep predicates. Let us 
  consider two of these subcases; the analysis of the other two is 
  are essentially the same with the one provided with minor 
  modifications.

  \textbf{Subcase 1.} Assume that the original shadow region type 
  was $(\chi,\phi)$ and $\phi\equiv\chi\equiv -\infty$. If $\phi$ 
  moves following the positive orientation of $\tri{ijk}$, $\chi$ will 
  move on the opposite direction. In this scenario, $\chi$ will cease 
  to exist and $\phi$ will become ``finite''; the resulting 
  \shp{$S_a$} will be $(-\infty,\chi)$. In the scenario where 
  $\phi$ moves on the opposite direction of the one described above, 
  \shp{$S_a$} will essentially become $\emptyset$ as the right endpoint
  infinitesimally ``becomes'' $-\infty$. 

  Notice that the \distance predicate in this Subcase is $(0,d_2)$ 
  and therefore the \distancep predicate must be either $(-,d_2)$ or 
  $(+,d_2)$. If $\phi$ moves towards $+\infty$ or $-\infty$, then
  \shp{$S_a$} becomes $(-\infty,\chi)$ or $\emptyset$ respectively. 
  In these scenarios, it must hold that \distancep equals 
  $(-,+)$ or $(+,+)$ respectively. Since in both cases, 
  $+\infty$ is not part of the shadow region 
  we can conclude that $d_2=+$ in this Subcase 
  (see Section~\ref{sub:the_shadowregion_predicate_analysis}). 

  \textbf{Subcase 2.} Assume that the original shadow region type 
  was $(-\infty,\phi)\cup (\chi,+\infty)$ 
  and $\phi\equiv\chi\equiv +\infty$. If $\phi$ 
  moves following the positive orientation of $\tri{ijk}$, the perturbed 
  shadow region becomes $\RR$. Otherwise, if $\phi$ 
  moves on the opposite direction, it will become ``finite'' whereas 
  $\chi$ will move further towards $\+\infty$; the resulting 
  \shp{$S_a$} will be $(-\infty,\phi)$.

  Notice that the \distance predicate in this Subcase is $(d_1,0)$ 
  and therefore the \distancep predicate must be either $(d_1,-)$ or 
  $(d_1,+)$. If $\phi$ moves towards $+\infty$ or $-\infty$, then
  \shp{$S_a$} becomes $\RR$ or $(-\infty,\phi)$ respectively. 
  In these scenarios, it must hold that \distancep equals 
  $(-,-)$ or $(-,+)$ respectively. Since in both cases, 
  $-\infty$ is part of the shadow region 
  we can conclude that $d_1=-$ in this Subcase 
  (see Section~\ref{sub:the_shadowregion_predicate_analysis}).

  The results from the analysis of all four subcases is found in 
  Table~\ref{tab:degeneracy B Type Case 1}. Notice that since 
  no finite Apollonius vertices exist, the outcome of the 
  respective \exist predicate must equal 0. The results in this 
  table imply that, for the case studied, the type of 
  \shp{$S_a$} can be deduced from the combination of the outcomes of the 
  \distance and \distancep predicates alone. 

  \begin{table}[ht]
  \begin{center}
  \begin{tabular}{|c|c|c||c|c|}
  \hline
  $E$ & $D$ & $D^{\epsilon}$ & \shp{$S_a$} & Original $\sh{S_a}$ \\
  \hline\hline
  \multirow{8}{*}{0}
  & \multirow{2}{*}{$(0,+)$ } 
  & $(-,+)$ & $(-\infty,\phi)$ & \multirow{2}{*}{$(\chi,\phi)$ when 
  $\phi\equiv\chi\equiv -\infty$}\\ \cline{3-4}
  & & $(+,+)$ & $\emptyset$ & \\ 
  \hhline{|~|=|=#==|} 
  & \multirow{2}{*}{$(0,-)$ } 
  & $(-,-)$ & $\RR$ 
   & $(-\infty,\phi)\cup(\chi,+\infty)$  \\ \cline{3-4}
  & & $(+,-)$ & $(\chi,+\infty)$ & when $\phi\equiv\chi\equiv -\infty$\\ \hhline{|~|=|=#==|} 
  & \multirow{2}{*}{$(+,0)$ } 
  & $(+,-)$ & $(\chi,+\infty)$ & \multirow{2}{*}{$(\chi,\phi)$ when 
  $\phi\equiv\chi\equiv +\infty$}\\ \cline{3-4}
  & & $(+,+)$ & $\emptyset$ & \\ 
  \hhline{|~|=|=#==|} 
  & \multirow{2}{*}{$(-,0)$ } 
  & $(-,-)$ & $\RR$ 
   & $(-\infty,\phi)\cup(\chi,+\infty)$ \\ \cline{3-4}
  & & $(-,+)$ & $(-\infty,\phi)$ & when $\phi\equiv\chi\equiv +\infty$\\ 
  \hline
  \end{tabular}
  \end{center}
  \caption[Resolve of Degeneracies of type B, case 1, using the \exist, \distance and \distancep predicates.]{Degeneracies of type B, Case 1. 
  $E,D$ and $D^{\epsilon}$ denote the outcomes
  of the \exist, \distance and \distancep predicates for input 
  $(S_i,S_j,S_k,S_a)$. The last column describes the original 
  degenerate shadow region $\sh{S_a}$ that was studied.}
  \label{tab:degeneracy B Type Case 1}
  \end{table}

  \textbf{Case 2.} In this case, the ``double'' Apollonius vertex
  $V$ is necessarily finite, \ie, $V\in\tri{ijk}\backslash\{\pm\infty\}$.
  The possible shadow region forms of $\sh{S_a}$ are the same as before, 
  either $(\chi,\phi)$ (Subcase 1) or 
  $(-\infty,\phi)\cup(\chi,+\infty)$ 
  (Subcase 2), with $\phi$ and $\chi$ actually coinciding. 
  The same strategy that was used in Case 1 is also used for Case 2; 
  we consider the perturbed shadow
  region \shp{$S_a$} when $\phi$ and $\chi$ move infinitesimally 
  on the trisector, on opposite directions. Let us now consider in 
  detail these two subcases.

  \textbf{Subcase 1.} If $\sh{S_a}=(\chi,\phi)$ and $\phi$ moves 
  on the positive (resp., negative) direction of $\tri{ijk}$, 
  the non-degenerate shadow region
  \shp{$S_a$} of $S_a$ will become $(\chi,\phi)$ (resp., $\emptyset$). 
  Since $\phi\equiv\chi\neq\pm\infty$,  
  the \exist predicate returns 1 and both \distance and 
  \distancep predicates return $(+,+)$. 

  \textbf{Subcase 2.} If $\sh{S_a}=(-\infty,\phi)\cup(\chi,+\infty)$ and 
  $\phi$ moves on the positive (resp., negative) direction of 
  $\tri{ijk}$, the non-degenerate shadow region \shp{$S_a$} of $S_a$ 
  will become $\RR$ (resp., $(-\infty,\phi)\cup(\chi,+\infty)$). 
  Since $\phi\equiv\chi\neq\pm\infty$, 
  the \exist predicate returns 1 and  both \distance and 
  \distancep predicates return $(-,-)$. 

  The main difference with Case 1 that can easily be spotted is that
  we can no longer draw conclusions regarding \shp{$S_a$} from 
  the outcomes of the \distance and \distancep predicates. 
  Nevertheless, we are still able to distinguish the form of 
  $\sh{S_a}$; we are in Subcase 1 if \distance returned $(+,+)$ 
  otherwise, we are in Subcase 2.

  In either subcase, we have to apply the perturbation scheme to 
  resolve the degeneracy. Among the sites $S_i,S_j,S_k$ and $S_a$ 
  that may have to be perturbed, $S_k$ has the minimum radius among 
  the first three after a proper name exchange. Even if a tie 
  break arise, we select $S_k$ to be the sphere whose center is 
  lexicographically smaller than the rest; in all cases, it will 
  hold that $k<i,j$. Bearing that in mind and since 
  we will prove that perturbing only one site suffices to 
  resolve the degeneracy, $S_k$ can be considered as fixed. 
  Since the topology and and geometric objects of \wspace are 
  greatly affected by the sphere $S_k$, as it is the 
  sphere through which the inversion is made, we can safely consider
  that the perturbation of a sphere $S_n$ of \zspace, 
  with $n\neq k$, is equivalent to
  the symbolic perturbation of the respective site $\inv{S_n}$ 
  and $\yinv{S_n}$ in \wspace and \yspace respectively.

  For convinience, we carry the rest of the analysis of Case 2 in 
  the inverted \wspace. The \emph{double} Apollonius vertex 
  in \yspace will correspond to a \emph{double} plane $\inv{\Pi}$ 
  commonly tangent to all $\inv{S_i}, \inv{S_j}$ and $\inv{S_a}$ in \wspace.
  Evidently, the sphere $\inv{S_a}$ must be tangent to the semi-cone $\wcone$
  defined by the former two spheres. As a consequence of 
  Section~\ref{sub:the_order_predicate_analysis}, we are in Subcase 2 
  (resp., Subcase 1) if and only if all other planes tangent to 
  this cone intersect (resp., do not intersect) $\inv{S_a}$. 
  In \wspace, let $\omega_n$, for $n\in\{i,j,a\}$,  denote the tangency 
  point of $\inv{S_n}$ and the plane $\inv{\Pi}$. Since 
  $S_i$ and $S_j$ are considered to be distinct, $\omega_i$ and 
  $\omega_j$ must also be distinct.

  If $a>i,j,k$ then $S_a$ and therefore $\inv{S_a}$ will be initially
  perturbed. In Subcase 1, some planes tangent to the cone will now
  properly intersect the perturbed sphere $\invp{S_a}$ and therefore
  \shp{$S_a$} will become $(\chi,\phi)$. In Subcase 2, all planes tangent 
  to the cone will now properly intersect $\invp{S_a}$ hence 
  \shp{$S_a$} will become $\RR$. 

\begin{figure}[tbp]
  \centering
  \includegraphics[width=0.6\textwidth]{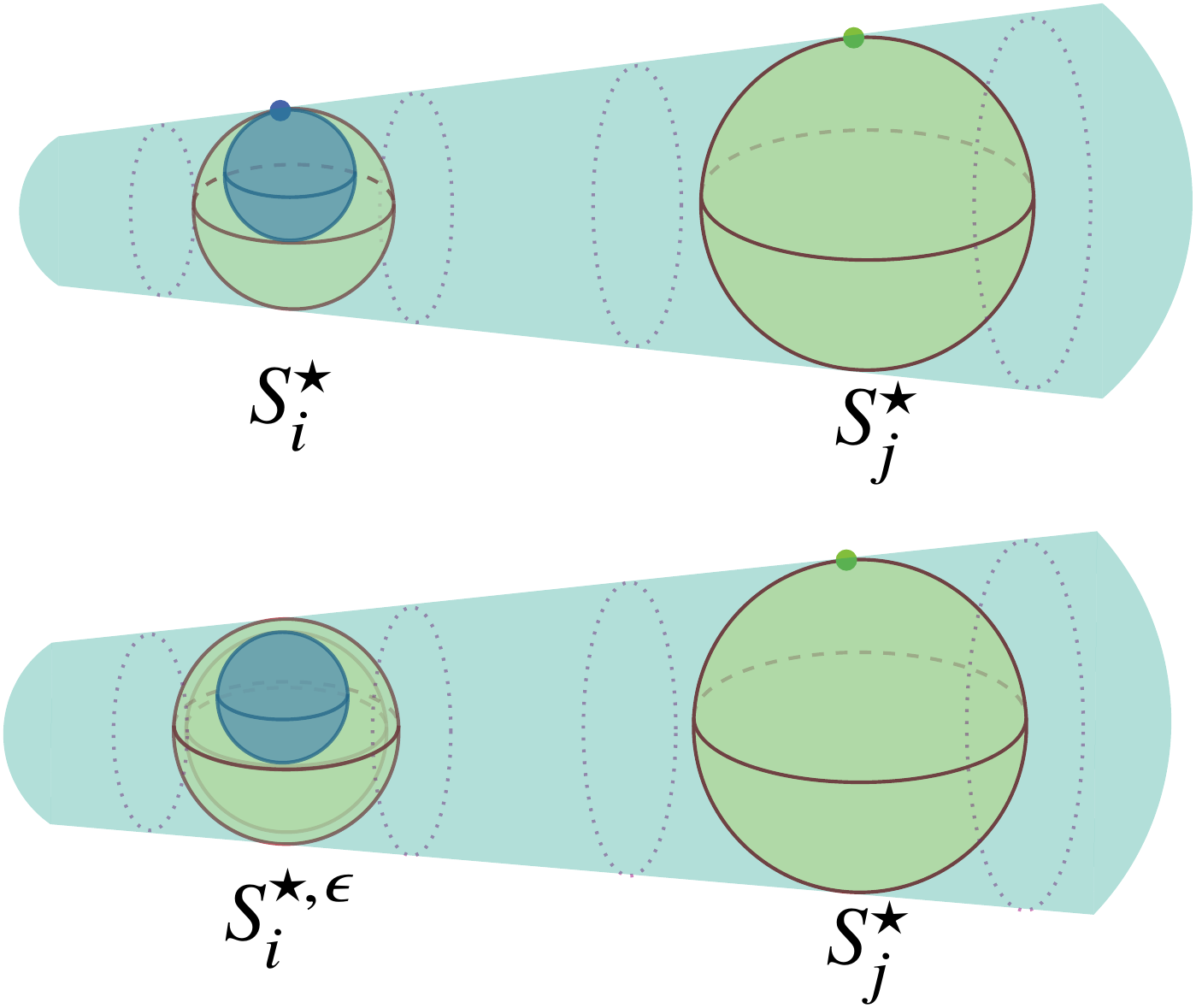}
  \caption[\shadow degeneracy when $t_i=t_a$ and $\sh{S_a}=(\chi,\phi)$]
  {In Subcase 1, if $\omega_a=\omega_i$ then $\inv{S_a}$ lies inside and is
  tangent to $\inv{S_i}$. After $\inv{S_i}$ is perturbed, the cone defined 
  by the perturbed sphere $\invp{S_i}$ and $\inv{S_j}$ will contain
  $\inv{S_a}$ and therefore \shp{$S_a$} will be $\emptyset$.}
  \label{fig:18}
  \end{figure}

\begin{figure}[tbp]
  \centering
  \includegraphics[width=0.6\textwidth]{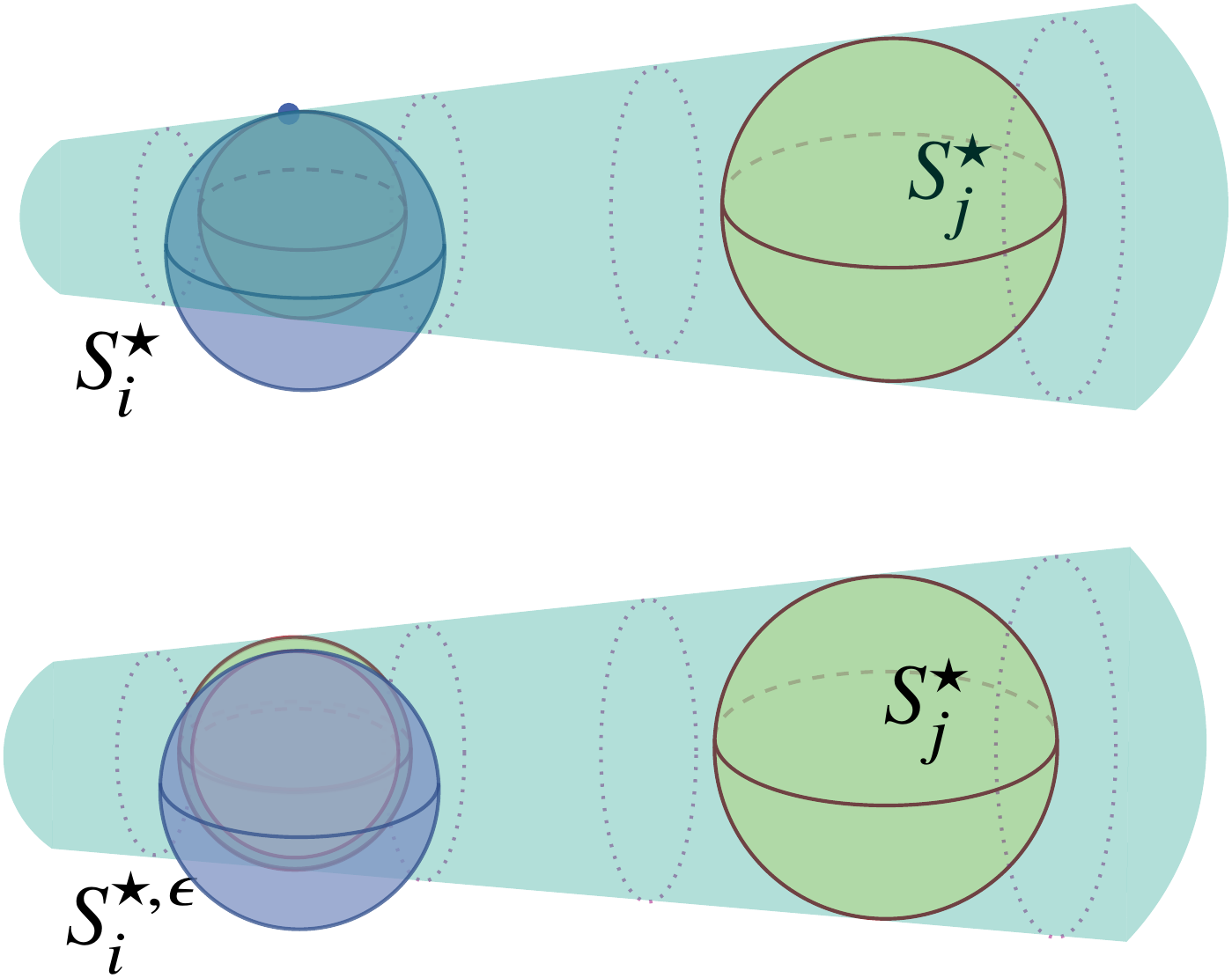}
  \caption[\shadow degeneracy when $t_i=t_a$ and 
  $\sh{S_a}=(-\infty,\phi)\cup(\chi,+\infty)$] {In Subcase 2, if $\omega_a=\omega_i$ then $\inv{S_i}$ lies inside and is 
  tangent to $\inv{S_a}$. After $\inv{S_i}$ is perturbed, some planes
  tangent to the perturbed sphere $\invp{S_i}$ and $\inv{S_j}$ 
  will not intersect $\inv{S_a}$ and therefore \shp{$S_a$} will be 
  $(-\infty,\phi)\cup(\chi,+\infty)$.}
  \label{fig:19}
  \end{figure}

\begin{figure}[tbp]
  \centering
  \includegraphics[width=0.6\textwidth]{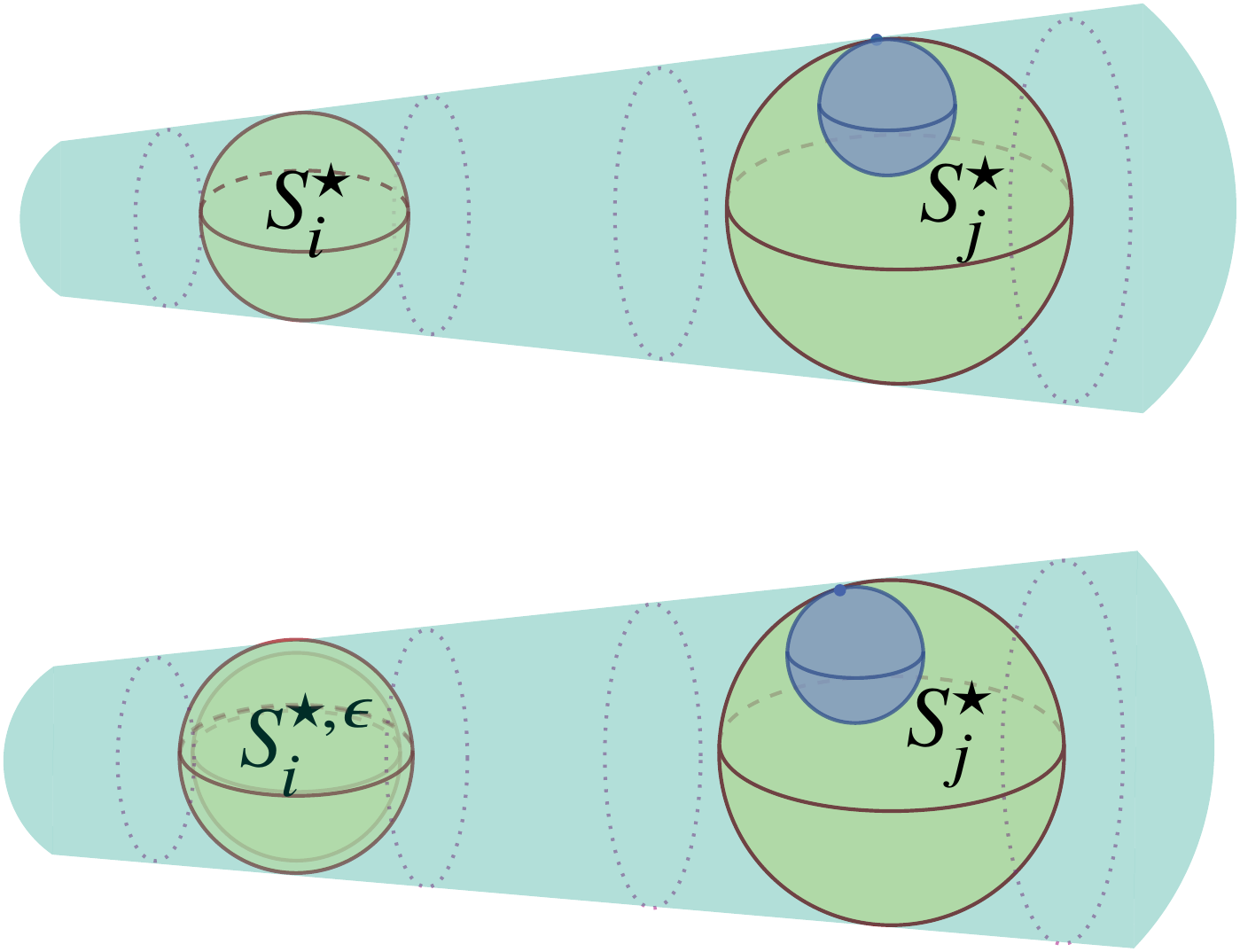}
  \caption[\shadow degeneracy when $t_j=t_a$ and $\sh{S_a}=(\chi,\phi)$]
  {In Subcase 1, if $\omega_a=\omega_j$ then $\inv{S_a}$ lies inside and is
  tangent to $\inv{S_j}$. After $\inv{S_i}$ is perturbed, the cone defined 
  by the perturbed sphere $\invp{S_i}$ and $\inv{S_j}$ will contain
  $\inv{S_a}$ and therefore \shp{$S_a$} will be $\emptyset$.}
  \label{fig:20}
  \end{figure}

\begin{figure}[tbp]
  \centering
  \includegraphics[width=0.6\textwidth]{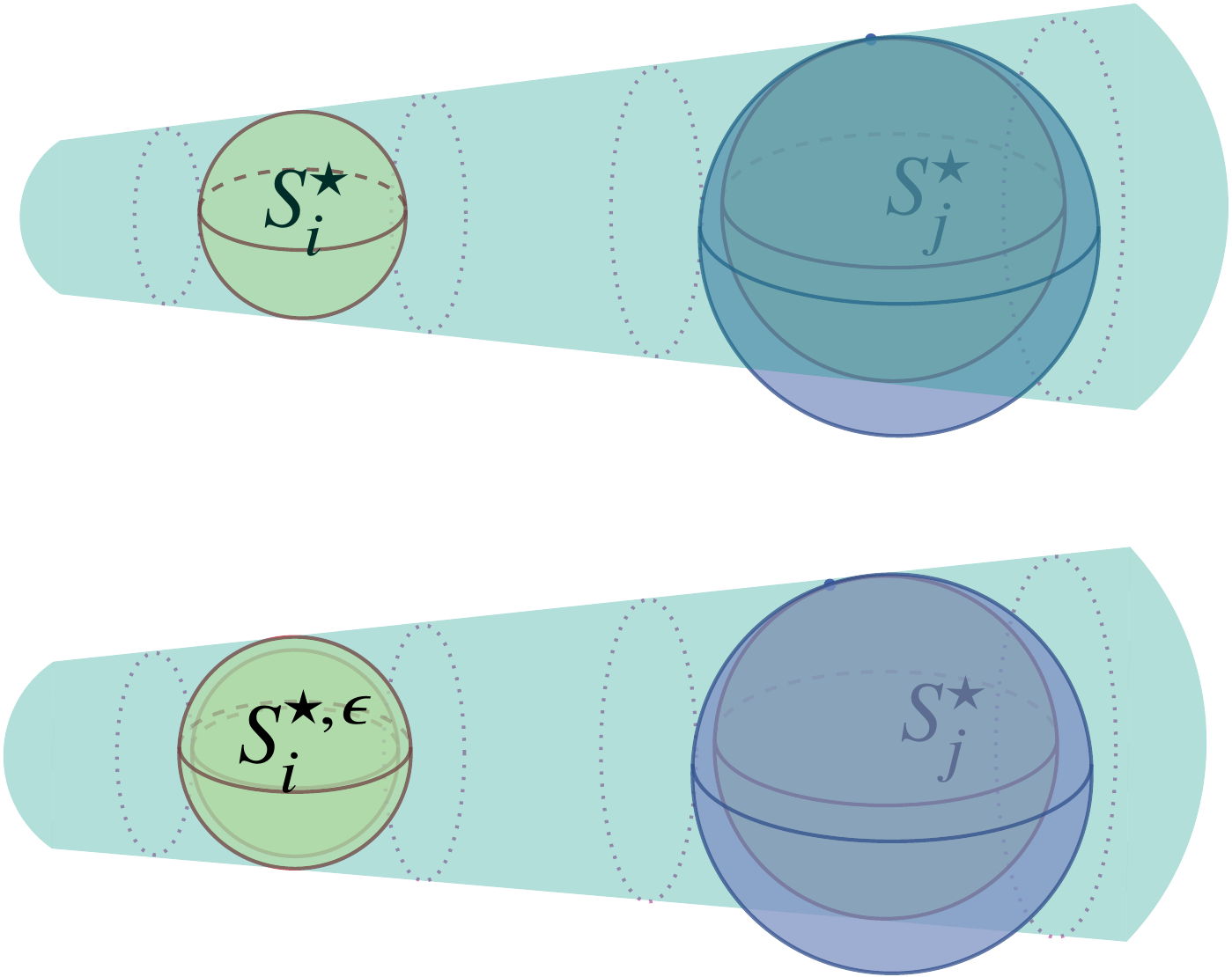}
  \caption[\shadow degeneracy when $t_j=t_a$ and 
  $\sh{S_a}=(-\infty,\phi)\cup(\chi,+\infty)$]
  {In Subcase 2, if $\omega_a=\omega_j$ then $\inv{S_j}$ lies inside and is
  tangent to $\inv{S_i}$. After $\inv{S_i}$ is perturbed, all planes
  tangent to the perturbed sphere $\invp{S_i}$ and $\inv{S_j}$ will 
  intersect $\inv{S_a}$ and therefore \shp{$S_a$} will be $\RR$.}
  \label{fig:21}
  \end{figure}

  Otherwise, let us consider the case where $i>j,k,a$; the case
  where $j>i,k,a$ is treated in the same way if we name exchange
  $S_i$ and $S_j$. 
  If $\omega_a=\omega_i$ or equivalently $t_a=t_i$, the inflation of 
  $\inv{S_i}$ will affect the cone $\wcone$. In Subcase 1, all 
  planes tangent to $\wcone$ will not intersect $\inv{S_a}$, 
  yielding $\text{\shp{$S_a$}}=\emptyset$ ((see Figure~\ref{fig:18})). If we are in a Subcase 2 
  configuration, some of these planes will no longer intersect
  $\inv{S_a}$,
  hence $\text{\shp{$S_a$}}=(-\infty,\phi)\cup(\chi,+\infty)$ 
  (see Figure~\ref{fig:19}). 
  In a respective way, if $\omega_a=\omega_j$ or equivalently $t_a=t_j$, 
  the perturbed shadow region $\text{\shp{$S_a$}}$ is 
  $\emptyset$ or $\RR$ in Subcase 1 or 2 respectively
  (see Figures~\ref{fig:20} and \ref{fig:21}). 

\begin{figure}[tbp]
  \centering
  \includegraphics[width=0.6\textwidth]{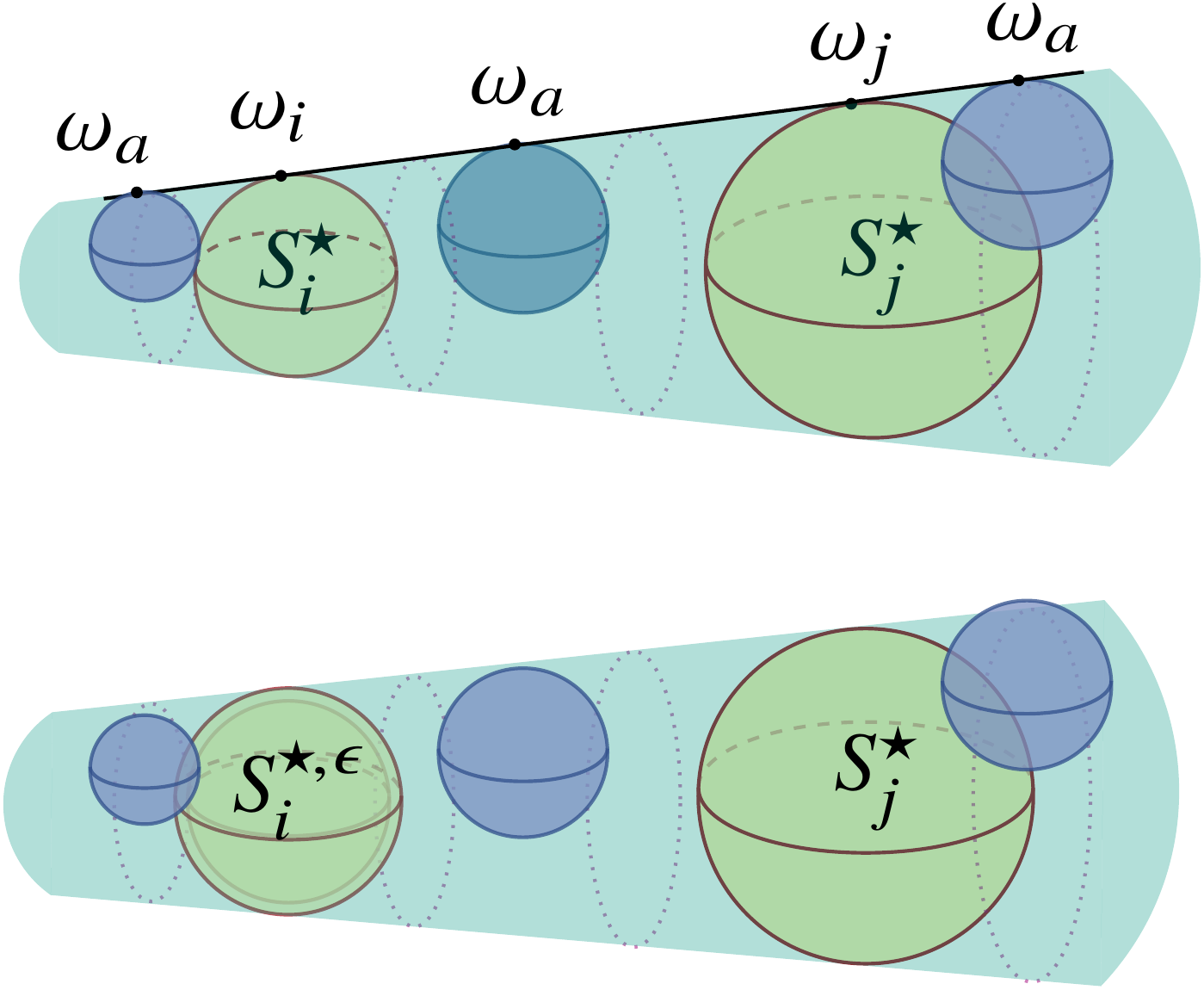}
  \caption[\shadow degeneracy when $\omega_i,\omega_j,\omega_a$ 
  are distinct and $\sh{S_a}=(\chi,\phi)$]
  {In Subcase 1, $\omega_a$ and 
  $\omega_i$ are on the same side with respect to $\omega_j$ if 
  $\inv{S_a}$ lie as the blue spheres on the left or the 
  center. In this case, after $\inv{S_i}$ is perturbed to $\invp{S_i}$,
  all planes commonly tangent to $\invp{S_i}$ and $\inv{S_j}$ will 
  not intersect $\inv{S_a}$ and therefore \shp{$S_a$} is $\emptyset$. 
  If $\inv{S_a}$ lies as the blue sphere on the right, the respective 
  \shp{$S_a$} is $(\chi,\phi)$, as there exist some planes tangent to 
  the perturbed cone that intersect $\inv{S_a}$.}
  \label{fig:22}
  \end{figure}

\begin{figure}[tbp]
  \centering
  \includegraphics[width=0.6\textwidth]{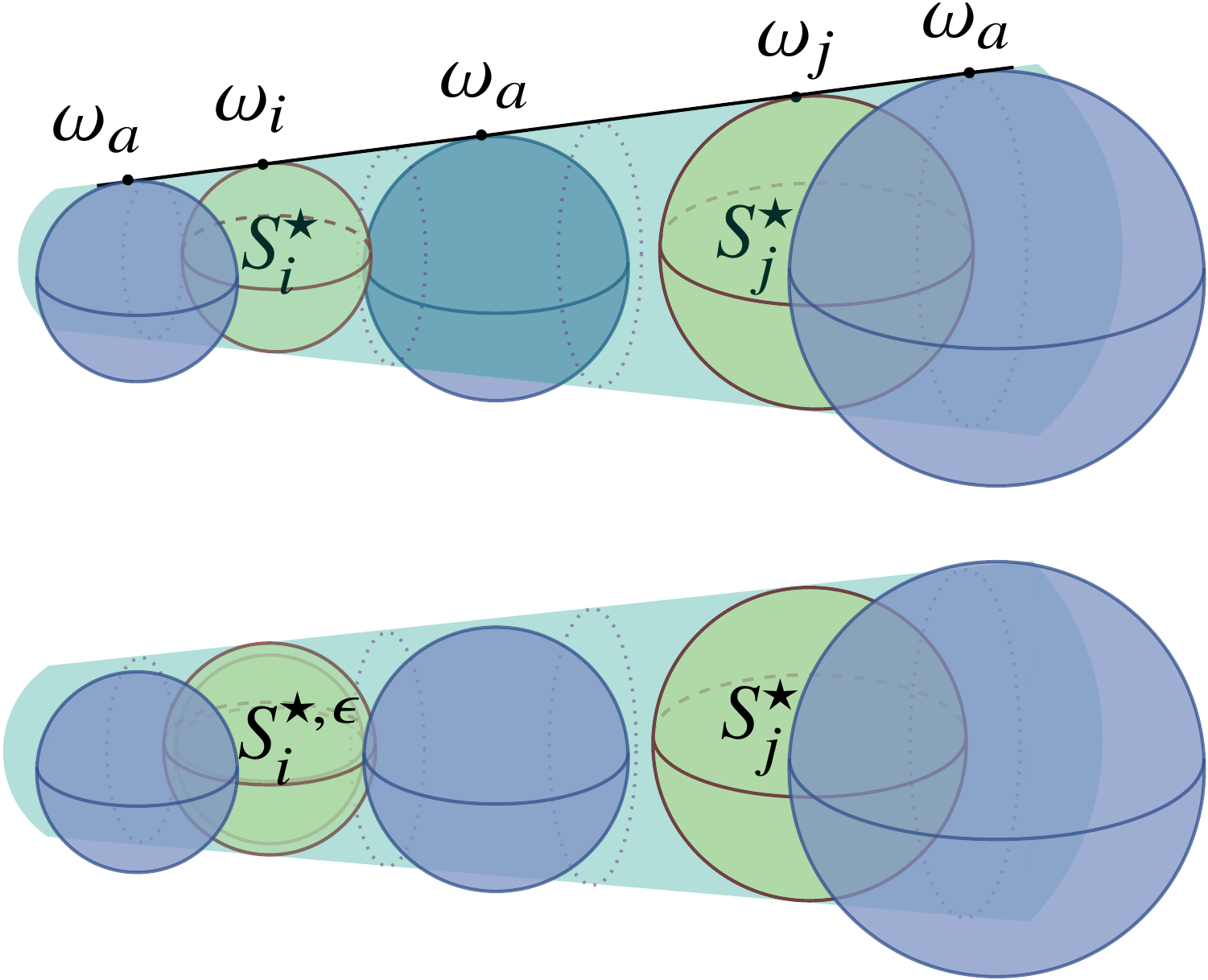}
  \caption[\shadow degeneracy when $\omega_i,\omega_j,\omega_a$ 
  are distinct and $\sh{S_a}=(-\infty,\phi)\cup(\chi,+\infty)$]
  {In Subcase 2, $\omega_a$ and 
  $\omega_i$ are on the same side with respect to $\omega_j$ if 
  $\inv{S_a}$ lie as the blue spheres on the left or the 
  center. In this case, after $\inv{S_i}$ is perturbed to $\invp{S_i}$,
  some commonly tangent to $\invp{S_i}$ and $\inv{S_j}$ will 
  not intersect $\inv{S_a}$ and therefore \shp{$S_a$} is 
  $(-\infty,\phi)\cup(\chi,+\infty)$. $\inv{S_a}$ lies as the blue 
  sphere on the right, the respective 
  \shp{$S_a$} is $\RR$ since all planes tangent to 
  the perturbed cone will intersect $\inv{S_a}$.}
  \label{fig:23}
  \end{figure}

  Lastly, we consider the scenario where $\omega_i,\omega_j$ and 
  $\omega_a$ are all distinct points; these tangency points must 
  lie on the same line due to the degenerate configuration we are studying.
  Using the same arguements as in the case where $\omega_a=\omega_i$,
  we conclude that, if $\omega_i$ and $\omega_a$ lie on the same side with
  respect to $\omega_j$ then \shp{$S_a$} becomes $\emptyset$ or
  $(-\infty,\phi)\cup(\chi,+\infty)$ in the Subcase 1 or 2 
  respectively. Otherwise, 
  if $\omega_i$ and $\omega_a$ lie on different sides with 
  respect to $\omega_j$ then \shp{$S_a$} becomes $\emptyset$ or
  $\RR$ in the Subcase 1 or 2 respectively (see Figure~\ref{fig:22} and
  \ref{fig:23}).

  For clarity we summarize the analysis for Case 2 into the following
  algorithm that returns the perturbed shadow region \shp{$S_a$}. 
  \begin{description}
  \item[Step 1.] 
  If $a>i,j,k$ return \shp{$S_a$} return $(\chi,\phi)$ if we 
  are in Subcase 1 or $\RR$ if we are in Subcase 2. Otherwise, 
  go to Step 2.
  \item[Step 2.] 
  If $j>i,k,a$, name exchange $S_i$ and $S_j$. Otherwise, 
  since it already holds that $i>j,k,a$, go to Step 3.
  \item[Step 3.] 
  If $t_a=t_i$ then return $\emptyset$ if we 
  are in Subcase 1, or $(-\infty,\phi)\cup(\chi,+\infty)$ if we 
  are in Subcase 2. If $t_a=t_j$ then return $\emptyset$ if we 
  are in Subcase 1, or $\RR$ if we 
  are in Subcase 2.  Otherwise go to Step 4.
  \item[Step 4.] Let $\omega_n$ denote the tangency points 
  of $\wcone$ and the inverted spheres $S_n$ for $n\in\{i,j,a\}$. 
  In the studied geometric configuration, $\omega_i$, 
  $\omega_j$ and $\omega_a$ are distinct and collinear. 
  If $\omega_i$ and $\omega_a$ lie on the same side with 
  respect to $\omega_j$ go to Step 4i, otherwise go to Step 4ii.
  \begin{enumerate}[i.]
  \item 
  Return $\emptyset$ if we 
  are in Subcase 1 or $(-\infty,\phi)\cup(\chi,+\infty)$ if we 
  are in Subcase 2.
  \item 
  Return $(\chi,\phi)$  if we are in Subcase 1 or $\RR$ if we 
  are in Subcase 2.
  \end{enumerate}
  \end{description}

  Recall that $t_a=t_n$ holds, for $n\in\{i,j\}$, if and only if
  $(x_a-x_n)^2+(y_a-y_n)^2+(z_a-z_n)^2=(r_a-r_n)^2$, which is a 2-degree
  demanding operation. The most difficult and degree-demanding 
  operation is to evaluate whether $\omega_i$ and $\omega_a$ lie on 
  the same side with respect to $\omega_j$, in Step 4. 

  Since the tangency points $\omega_i,\omega_j$ and $\omega_a$ lie 
  on the same line $\inv{\ell}$ in \wspace, 
  their preimages $t_i,t_j$ and $t_a$ 
  must be co-circular with $t_k$ in \zspace; this is a direct result
  of the properties of the inversion transformation 
  (see Section~\ref{sub:inversion}). Moreover, if $C_{ijk}$ denotes 
  the circle that contains the points $t_n$, for $n\in\{i,j,k,a\}$, 
  then $C_{ijk}$ must lie on the ``double'' Apollonius sphere 
  $\mathcal{A}$ of
  $S_i,S_j,S_k$ and $S_a$, centered at $V$. 

\begin{figure}[tbp]
  \centering
  \includegraphics[width=0.8\textwidth]{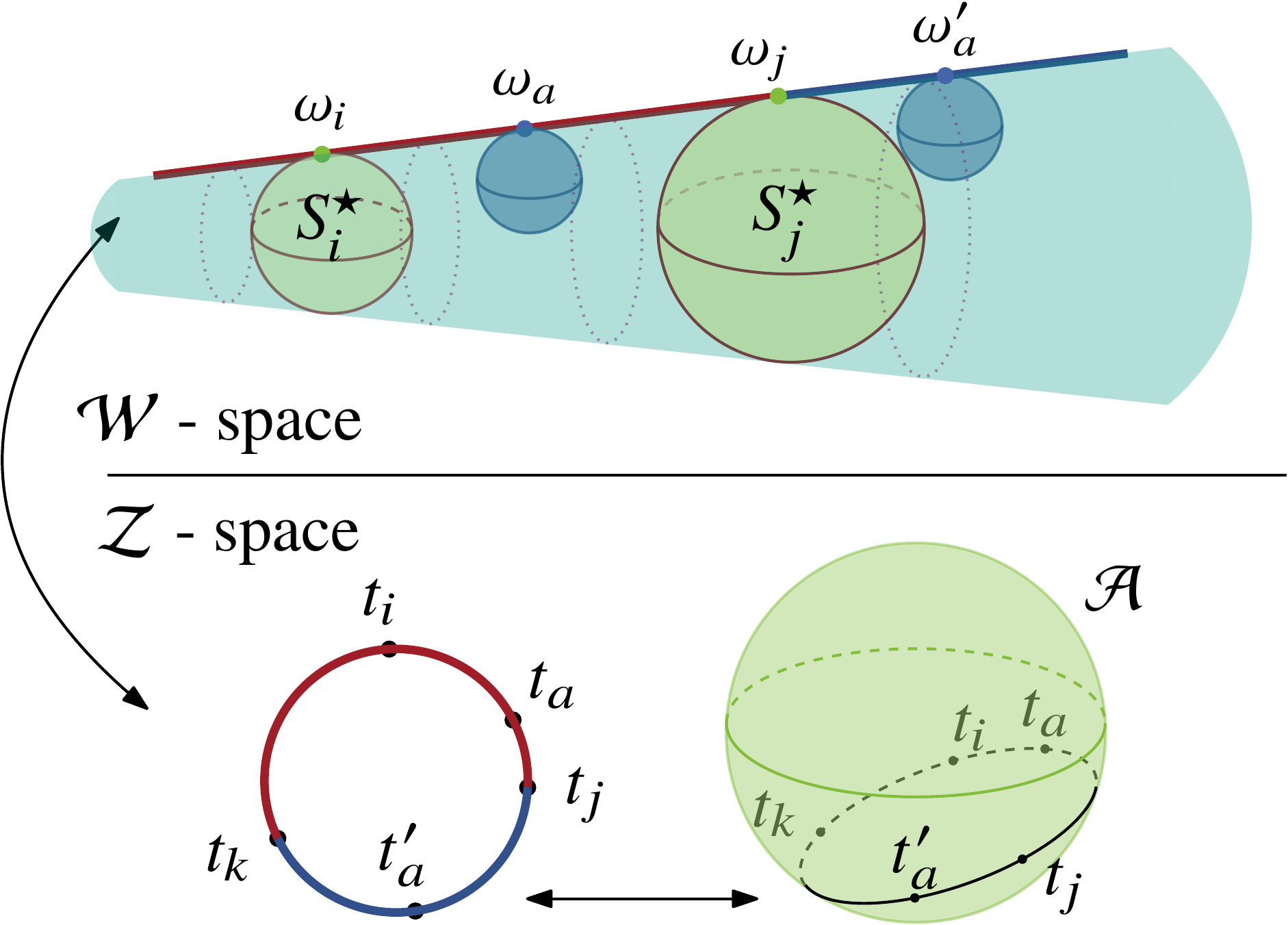}
  \caption[The arc A of \zspace and its equivalent in \wspace]
  {The line that contains $\omega_i$ and $\omega_j$ in \wspace, 
  corresponds in \yspace to a circle that contains the 
  tangency points $t_i,t_j$ and $t_k$, due to the properties of the 
  inversion transformation. Since the image of $t_k$ in \wspace is 
  the point at infinity, there is an equvalency between the arc A 
  (red arc) that is bounded by $t_j$ and $t_k$ and contains $t_i$ and 
  the (red) ray that starts from $\omega_j$, \emph{ends at the point 
  at infinity which is the image of $t_k$} and contains $\omega_i$. 
  The arc A lies on the circle defined by the three points $t_i,t_j$
  and $t_k$, on the surface of the Apollonius sphere $\mathcal{A}$.}
  \label{fig:24}
  \end{figure}

  Since $\inv{\ell}$ is the image of $C_{ijk}$ in \yspace, it  
  holds that $\omega_i$ and $\omega_a$ lie on 
  the same side with respect to $\omega_j$ on $\inv{\ell}$ if and 
  only if $t_a$ lies on the arc $A$ of $C_{ijk}$ that is bounded 
  by $t_j$ and $t_k$ and contains $t_i$ (see Figure~\ref{fig:24}). 

\begin{figure}[tbp]
  \centering
  \includegraphics[width=0.8\textwidth]{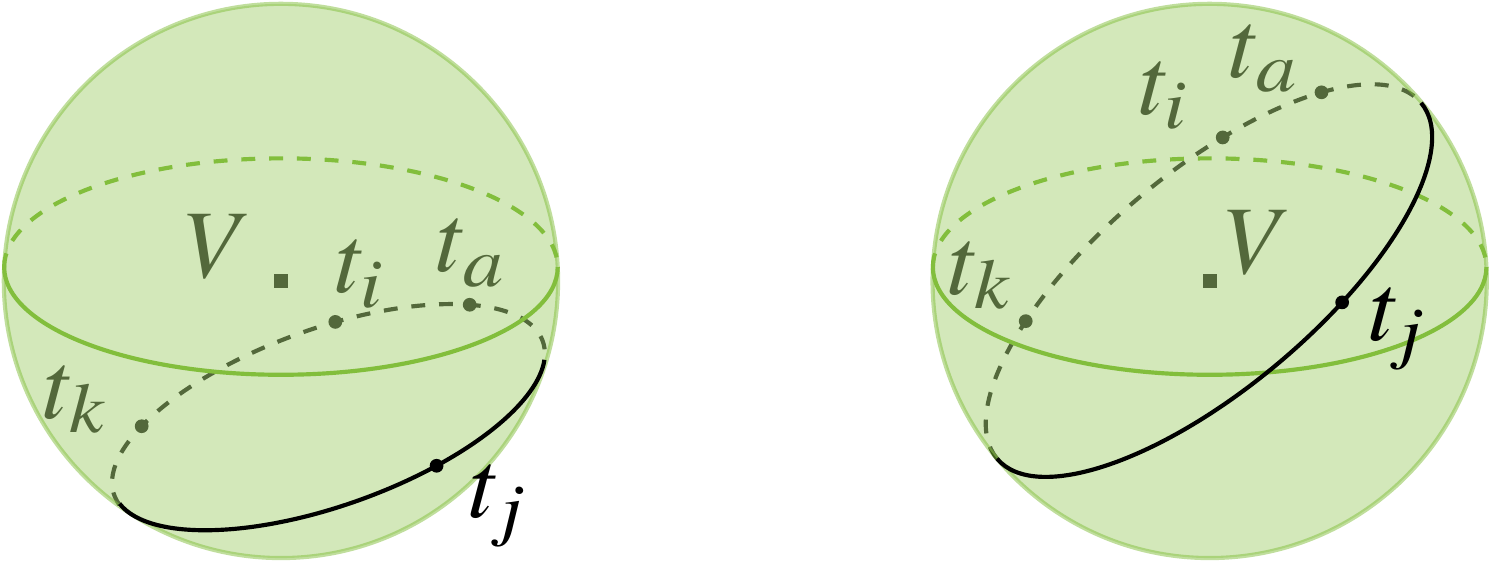}
  \caption[$V$ with respect to the plane of $t_i,t_j$ and $t_k$]
  {The Apollonius vertex $V$ can either be coplanar with the 
  tangency points $t_i,t_j$ and $t_k$ (right) or not (left). 
  The way the degeneracy is resolved is severely affected depending 
  on which of these cases applies.}
  \label{fig:25}
  \end{figure}

  In order to decide if $t_a$ lies on $A$ or not, we  consider 
  the relative position of the Apollonius vertex $V$ and the circle
  $C_{ijk}$. We distinguish two different cases depending on whether
  $V$ is coplanar with $C_{ijk}$ or not, which is equivalent to 
  all points $C_i,C_j,C_k,C_a$ and $V$ being coplanar or not 
  respectively (see Figure~\ref{fig:25}). Therefore, we evaluate 
  $\text{\orient}(C_i,C_j,C_k,C_a)=D^{xyz}_{ijka}$ and if it equals 0, 
  we are in the former case, otherwise we are in the latter. 

\begin{figure}[tbp]
  \centering
  \includegraphics[width=0.4\textwidth]{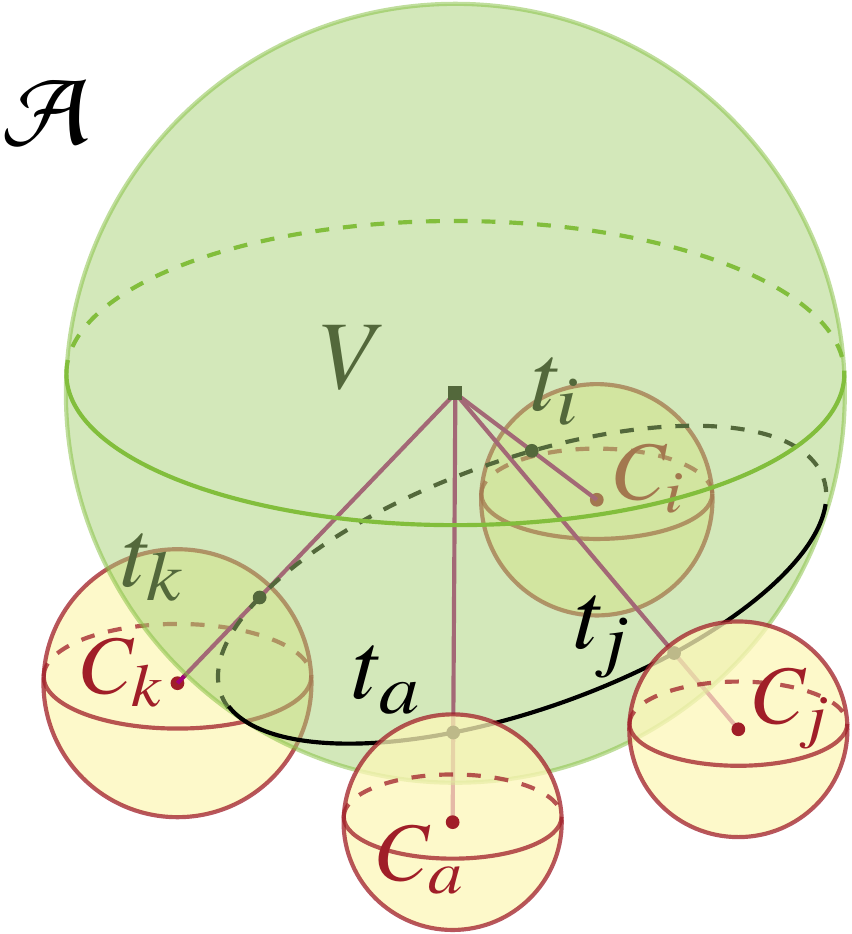}
  \caption[$V$ is not coplanar with $t_i,t_j$ and $t_k$]
  {When the Apollonius vertex $V$ is not coplanar with the 
  tangency points $t_i,t_j$ and $t_k$, we can determine if 
  $t_a$ and $t_i$ lie on the same arcs bounded by $t_k$ and 
  $t_j$ by checking various \orient predicates that involve 
  $V$ and three of the tangency points. The outcome of these 
  predicates remain the same if the tangency point $t_n$ is 
  replaced with the respective center $C_n$, as shown in the figure.}
  \label{fig:26}
  \end{figure}

  In the no-coplanar case, we can deduce that $t_a$ lies on the arc $A$
  if and only if the orientation predicates 
  $O_i=\text{\orient}(t_i,t_k,t_j,V)$ and 
  $O_a=\text{\orient}(t_a,t_k,t_j,V)$ have the same sign. Note that,
  since it holds that $O_n=\text{\orient}(C_n,C_k,C_j,V)$, for 
  $n\in\{i,a\}$, we do not need to evaluate the tangency 
  points $t_a,t_j$ and $t_k$ explicitly (see Figure~\ref{fig:26}). 

\begin{figure}[tbp]
  \centering
  \includegraphics[width=0.30\textwidth]{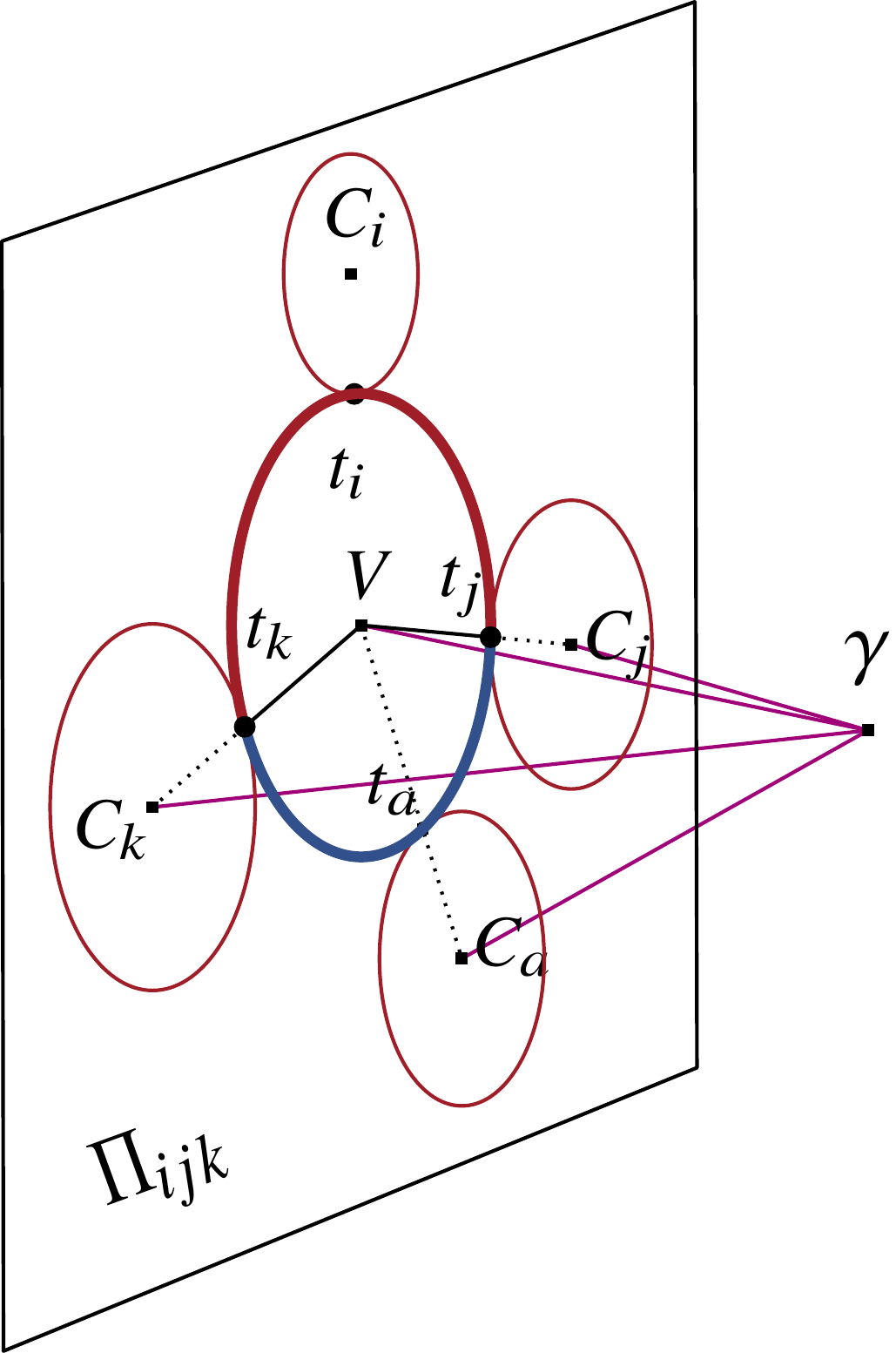}
  \caption[$V$ is coplanar with $t_i,t_j$ and $t_k$]
  {When the Apollonius vertex $V$ is coplanar with the 
  tangency points $t_i,t_j$ and $t_k$, an auxiliary point $\gamma$
  outside the plane is used to determine if $t_a$ lies on the 
  (red) arc A or not. This is determined by evaluating 
  various \orient predicates that involve 
  $V, \gamma$ and two of the tangency points. The outcome of these 
  predicates remain the same if the tangency point $t_n$ is 
  replaced with the respective center $C_n$, as shown in the figure.}
  \label{fig:27}
  \end{figure}

  In the coplanar case, the Apollonius vertex $V$ along with all 
  points $C_n$ and $t_n$, for $n\in\{i,j,k,a\}$, 
  lie on the same plane. In this scenario,
  we select a point $\gamma$ from 
  the set $\{(0,0,0),(1,0,0),(0,1,0),(0,0,1)\}$ such that 
  $\gamma\not\in\Pi(C_i,C_j,C_k)$ or equivalently 
  $O_\gamma=\text{\orient}(C_i,C_j,C_k,\gamma)\neq 0$ and we 
  denote $O_{\mu\nu}=\text{\orient}(C_\mu,\gamma,V,C_\nu)$, where 
  $\mu,\nu\in\{a,k,j\}$ (see Figure~\ref{fig:27}).

  Firstly, we evaluate $O_{kj}$; if it equals 0 then
  $C_k,C_j$ and $V$ are collinear. In this case, 
  $t_a$ does not lie on the arc $A$ if and only if
  $\text{\orient}(C_i,C_j,C_k,\gamma)\cdot 
  \text{\orient}(C_a,C_j,C_k,\gamma)<0$. 

  If $O_{kj}\neq 0$ then $t_a$ does not lie on the arc $A$ if and 
  only if $C_a$ and $C_i$ lie on different sides of the angle 
  $(C_k, V, C_j)$. This is equivalent to either
  \begin{itemize}
  \item 
  $O_{jk}\cdot O_{ak}, O_{kj}\cdot O_{aj}>0$ 
  and at least one of 
  $O_{jk}\cdot O_{ik}$,   $O_{kj}\cdot O_{ij}$ is non positive, Or
  \item 
  $O_{jk}\cdot O_{ik}, O_{kj}\cdot O_{ij}>0$ and at least one of 
  $O_{jk}\cdot O_{ak}$, $O_{kj}\cdot O_{aj}$ is non positive. 
  \end{itemize}

  Recollect that, if $t_a$ lies or not on the arc $A$, then we can 
  resolve resolve the degeneracy by following Step 4i or 4ii respectively.
  The most degree-demanding operation is the evaluation of the 
  signs of $O_n=\text{\orient}(C_n,C_k,C_j,V)$ and 
  $O_{\mu\nu}=\text{\orient}(C_\mu,\gamma,V,C_\nu)$, for $n\in\{i,a\}$ and
  $\mu,\nu\in\{a,k,j\}$ respectively. These \orient predicates
  can be expressed as 
  \begin{align}
  O_n &= \frac{1}{w_V}(-x_V D_{nkj}^{yz}+y_V D_{nkj}^{xz}
  -z_V D_{nkj}^{xy}+ w_V D_{nkj}^{xyz}) \\
  O_{\mu\nu} &= \frac{1}{w_V}(x_V D_{\mu \gamma\nu}^{yz}
  -y_V D_{\mu \gamma\nu}^{xz}+z_V D_{\mu \gamma\nu}^{xy}+w_V D_{nkj}^{xyz})
  \end{align}

  where $V=\frac{1}{w_V}(x_V,y_V,z_V)$. 

  The coordinates of the double Apollonius vertex $V$ can be evaluated
  from the analysis of the $\text{\exist}(S_i,S_j,S_k,S_a)$ predicate 
  (see Section~\ref{sub:the_existence_predicate_analysis}), if we 
  take into consideration three major facts. 
  \begin{enumerate}
  \item 
  The sphere that was used to apply the inversion was $S_k$ and 
  not $S_a$. Therefore, for all expressions of 
  Section~\ref{sub:the_existence_predicate_analysis}) that we will 
  use below, we exchange $k$ with $a$ in all subscripts appearing. 
  \item 
  Since we have a finite ``double'' Apollonius vertex, it holds that 
  $\Delta_M=0$ and $d\neq 0$. Moreover, the centers $\inv{C_i},\inv{C_j}$
  and $\inv{C_a}$ cannot be collinear since that would result in 
  infinite number of planes being tangent the cone $\wcone$, yielding
  a contradiction. Consequently, the quantities 
  $E^{xyp}_{ija}, E^{xzp}_{ija}$ and $E^{yzp}_{ija}$ are not all zero
  and therefore $M_2 > 0$. 
  \item 
  As shown in \cite{Iordanov}, if $au+bv+cw+d=0$ is the image 
  of the Apollonius sphere of $S_i,S_j,S_k,S_l$ in \wspace, then 
  the coordinates of the Apollonius vertex in \zspace is
  $(x_l-\frac{a}{2d},y_l-\frac{b}{2d},z_l-\frac{c}{2d})$. 
  \end{enumerate}

  Taking all the above facts into consideration, we can
  explicitly evaluate $a,b,c$ and $d$ in case $D^{uvw}_{ija}\neq 0$
  and $\Delta_M=0$. Indeed, $(a,b,c,d)=\frac{1}{\Phi}(A_a,A_b,A_c,A_d)$ 
  where
  \begin{align}
  A_a &= -E^{xzp}_{ija}E^{zrp}_{ija}-E^{xyp}_{ija}E^{yrp}_{ija},\\
  A_b &= E^{xyp}_{ija}E^{xrp}_{ija}-E^{yzp}_{ija}E^{zrp}_{ija},\\
  A_c &= E^{yzp}_{ija}E^{yrp}_{ija}+E^{xzp}_{ija}E^{xrp}_{ija},\\
  A_d &= E^{yzr}_{ija}E^{yzp}_{ija}+E^{xzr}_{ija}E^{xzp}_{ija}
  +E^{xyr}_{ija}E^{xyp}_{ija}\neq 0,\\
  \Phi &= (E^{xyp}_{ija})^2+(E^{xzp}_{ija})^2+(E^{yzp}_{ija})^2.
  \end{align}

  The following equations can also be obtained if one exchanges 
  $k$ with $a$ and substitutes $\Gamma=0$ in the expressions of 
  Iordanov in \cite{Iordanov}. Lastly, we can determine 
  \begin{align}
  V &=(x_k-\frac{a}{2d},y_k-\frac{b}{2d},z_k-\frac{c}{2d}) 
  = \frac{1}{2d}(2dx_k-a, 2dy_k-b,2 dz_k-c)\\
  & = \frac{1}{2A_d}(2A_d x_k-A_a, 2A_d y_k-A_b, 2A_d z_k-A_c)
  \end{align}
  and therefore $V=\frac{1}{w_V}(x_V,y_V,z_V)$, where 
  $x_V = 2A_d x_k-A_a, y_V = 2A_d y_k-A_b, z_V = 2A_d z_k-A_c$ and 
  $ w_V = 2A_d$. Since $x_V,y_V,z_V$ and $w_V$ are quantities of 
  algebraic degree
  8,8,8 and 7 respectively, it follows that the signs $O_n$ and 
  $O_{\mu\nu}$ both require algebraic operations of degree 10 on the
  input quantities to be evaluated. Since this is the most degree
  demanding operation in the evaluation of \shadowp, we have proven 
  the following lemma.

  \begin{lemma}
  The \shadowp predicate for hyperbolic trisectors can be evaluated 
  by determining the sign of quantities of algebraic degree at most 10
  (in the input quantities).
  \end{lemma}


  \section{%
  The \texorpdfstring{\existp}{perturbed Existence}
  Predicate for hyperbolic trisectors} 
  \label{sub:the_perturbed_existence_predicate}
  
  The \existp denotes the number of the ``finite'' endpoints of 
  \shp{$S_a$}, after applying the perturbation scheme. Therefore, 
  to determine the outcome of the $\text{\existp}(S_i,S_j,S_k,S_a)$ 
  predicate, we need to call the \shadowp predicate with the same 
  input. It is assumed that the trisector of $S_i, S_j$ and $S_k$ is 
  hyperbolic and therefore the outcome of the \shadowp predicate 
  is determined as described in Section~\ref{sub:the_perturbed_shadowregion_predicate}.
  \begin{itemize}
   \item 
   If \shadowp returns $\emptyset$ or $\RR$ then \existp returns 0.
   \item
   If \shadowp returns $(-\infty,\phi)$ or 
   $(\chi,+\infty)$ then \existp returns 1. Finally,
   \item
   If \shadowp returns $(-\infty,\phi)\cup (\chi,+\infty)$ or 
   $(\chi,\phi)$ then \existp returns 2. 
  \end{itemize}
  
  Since the evaluation of the \shadowp predicate requires 10-fold 
  operations with respect to the input quantities, we have proven 
  the following lemma.

  \begin{lemma}
  The \existp predicate for hyperbolic trisectors can be 
  evaluated by determining the sign of quantities of algebraic 
  degree at most 10 (in the input quantities).
  \end{lemma}


  \section{%
  The \texorpdfstring{\existp}{perturbed Existence}
  Predicate for elliptic trisectors} 
  \label{sub:the_perturbed_existence_predicate_elliptic}

  In the case where the trisector of the sites $S_i, S_j$ and $S_k$ 
  is elliptic, it is clear based on the previous analysis 
  that a non-degenerate shadow region type of a 
  site $S_a$ on $\tri{ijk}$ is either $\emptyset, \RR$ or $(\chi,\phi)$, 
  where $\RR$ denotes the entire trisector. The first two shadow region 
  types correspond to $\text{\exist}(S_i,S_j,S_k,S_a)=0$ as they do 
  not have ``finite''boundary points. On the other hand, $(\chi,\phi)$ 
  is the only option for $\mathcal{SR}(S_a)$  if 
  $\text{\exist}(S_i,S_j,S_k,S_a)=2$ as the last relation 
  denotes a two-finite-boundary shadow region and the option 
  $(-\infty,\phi)\cup(\chi,+\infty)$ in not feasible in elliptic 
  trisectors (in contrast with hyperbolic ones). 

  Therefore a clear 
  indication of a degerenate \exist, and equivalently \shadow, 
  outcome is the case where $\text{\exist}(S_i,S_j,S_k,S_a)=1$. In 
  this scenario, the shadow region of $S_a$ on the trisector $\tri{ijk}$ 
  is of the form $(\chi,\phi)$ (Subcase 1) or $\RR\backslash (\chi,\phi)$ 
  ((Subcase 2) with 
  $\chi\equiv\phi$. If the last shadow region form is viewed as 
  $(-\infty,\phi)\cup(\chi,+\infty)$ (in the sense that the shadow region 
  contains all points of $\tri{ijk}$ except one or simply put, 
  we may consider that $-\infty\equiv+\infty$) then we can determine 
  the \shp{$S_a$} based on the analysis of 
  Section~\ref{sub:the_perturbed_shadowregion_predicate}, 
  Degeneracies of Type B, Case 2.

  In order to apply the analysis of this section, we have 
  determine if we are in Subcase 1 or 2. This is easily accomplished as 
  the shadow region $\sh{S_a}$ in Subcase 1 denotes that all points 
  of the trisector $\tri{ijk}$ except one do not belong in $\sh{S_a}$ 
  and therefore all but one Apollonius sphere of the sites $S_i,S_j$ and
  $S_k$ do not intersect $S_a$. Since $v_{ijkl}$ and $v_{ikjm}$ are 
  considered to exist on $\tri{ijk}$ due to the fact that $e_{ijklm}$ is
  a valid Voronoi edge on the trisector we only need to evaluate 
  $I_1=\text{\insphere}(S_i,S_j,S_k,S_l,S_a)$ and 
  $I_2=\text{\insphere}(S_i,S_j,S_j,S_m,S_a)$. If at least one 
  of $\{I_1,I_2\}$ is positive (hence the other is necessarily 
  either positive or zero) the corresponding vertex 
  $\{v_{ijkl},v_{ikjm}\}$ does not belong to $\sh{S_a}$, indicating 
  we are in Subcase 1. On the other hand, if at least one 
  of $\{I_1,I_2\}$ is negative (hence the other is necessarily 
  either negative or zero) the corresponding vertex 
  $\{v_{ijkl},v_{ikjm}\}$ belongs to $\sh{S_a}$, indicating 
  we are in Subcase 2.

  Since we can distinguish if we are in Subcase 1 or 2, the analysis of 
  Section~\ref{sub:the_perturbed_shadowregion_predicate} can be applied,
  yielding one of the possible \shp{$S_a$}: $\emptyset, \RR, (\chi,\phi)$ 
  or $(-\infty,\phi)\cup(\chi,+\infty)$. The later result can be viewed 
  as $(\chi,\phi)$ as it denotes that all points of the elliptic trisector 
  $\tri{ijk}$ except for a segment belong to \shp{$S_a$} 
  (or simply put, since we considered $-\infty\equiv+\infty$). 

  Similar with Section~\ref{sub:the_perturbed_existence_predicate} for 
  hyperbolic trisectors, if \shp{$S_a$} returned $\emptyset$ or $\RR$, 
  the corresponding \existp predicate returns 0. Otherwise, if 
  \shp{$S_a$} returned $(\chi,\phi)$, the corresponding \existp 
  predicate returns 2. Since the evaluation of the \shadowp predicate 
  and the signs of $I_1$ and $I_2$ 
  demand, in the worst cast, operations of degree 10 in the input 
  quantities, we have proven the following lemma.

  \begin{lemma}
  The \existp predicate for elliptic trisectors can be 
  evaluated by determining the sign of quantities of algebraic 
  degree at most 10 (in the input quantities).
  \end{lemma}


  \section{Perturbation for the remaining predicates} 
  \label{sec:the_conflictp_predicate}
  

  Using the analysis of Section~\ref{sub:the_order_predicate_analysis}, 
  we can evaluate the \order predicate for hyperbolic trisectors and 
  non-degenerate input. Since the predicate requires the call of 
  \insphere, \exist, \shadow and \distance subpredicates, we can 
  also answer it for degenerate input if we could determine 
  the outcome of the perturbed \inspherep, \existp, \shadowp and 
  \distancep subpredicates. However the current known 
  algebraic degree of \inspherep, also known as \vconflictp predicate, 
  is known to be 28 (cf. \cite{devillers2017qualitative}, result 
  provided by the author). It is therefore apparent that the 
  algebraic degree of the \orderp predicate is subject to the following 
  lemma. 

  \begin{lemma}
  The \orderp predicate for hyperbolic trisectors can be 
  evaluated by determining the sign of quantities of algebraic 
  degree at $\max\{8,d\}$ (in the input quantities), where $d$ is 
  the algebraic degree of the \vconflictp predicate.
  \end{lemma}

  Using the algorithms presented in Section~\ref{sub:the_main_algorithm},
  we can answer the \conflict, \inflconflict and \infrconflict predicates
  for hyperbolic trisectors and non-degenerate inputs using 
  the \insphere, \order, \exist, \shadow and \distance subpredicates. 
  For degenerate inputs, we can use the outcomes of the respective 
  perturbed versions of these subpredicates and the following lemma is 
  proven. 

  \begin{lemma}
  The predicates \inflconflictp, \infrconflictp and \conflictp for hyperbolic trisectors can be 
  evaluated by determining the sign of quantities of algebraic 
  degree at $\max\{8,d\}$ (in the input quantities), where $d$ is 
  the algebraic degree of the \vconflictp predicate.
  \end{lemma} 

  Finally, the algorithm presented in Section~\ref{sec:non_degenerate_elliptic} can answer the \conflict predicate
  for elliptic trisectors and non-degenerate inputs using 
  the \insphere, \order and \exist subpredicates. In case of degenerate
  inputs, we can answer the \conflictp predicates using the respective 
  outcomes of the \vconflictp, \orderp and \existp subpredicates and 
  the following lemma is proven.

  \begin{lemma}
  The \conflictp predicate for elliptic trisectors can be 
  evaluated by determining the sign of quantities of algebraic 
  degree at $\max\{8,d\}$ (in the input quantities), where $d$ is 
  the algebraic degree of the \vconflictp predicate.
  \end{lemma}


\chapter{Analysis of parabolic trisectors} 
\label{cha:analysis_of_parabolic_trisectors}

In previous chapters, we analysed all arising predicates and subpredicates
under the assumption that the trisector of the first three input spheres
where either of hyperbolic or elliptic type. We will now consider in 
detail the case of a parabolic trisector, \ie, when the respective 
trisector is a parabola. 

This type of trisector is considered to be degenerate; it only arises 
when one of the spheres that define it point lies inside and it's 
boundary point-touches the convex hull of the other two. 
Equivalently, if the
the trisector $\tri{ijk}$ is parabolic and we properly exchange the names of the sites 
$S_i,S_j$ and $S_k$ such that the latter has the smaller radius,  
then the outcome of the $\text{\incone}(S_i,S_j;S_k)$ is necessarily 
\ptouch. 

This type of trisector can be viewed in two ways. We can either consider
it to be an elliptic trisector that contains the point at infinity or 
we can view it as a hyperbolic trisector where the points at $\pm\infty$ 
coincide. Although both of these considerations seem to contradict with our
common sense and the way we perceive a trisector in \zspace, these remarks
become clear when viewed in \wspace and \yspace. When the trisector 
$\tri{ijk}$ is a parabola, the point $\inv{O}$ of \wspace lies on the semi-cone defined by $\inv{S_i}$ and $\inv{S_j}$ and equivalently, 
the point $\yinv{O}$ of \yspace lies on the circle $\ycone$. Therefore, the 
image of $\tri{ijk}$ in \yspace is the circle  $\ycone$ that contains 
the image of the point of infinity of \zspace. 

If a perturbation scheme is applied, the outcome of 
$\text{\inconep}(S_i,S_j;S_k)$ will be either \outside or \inside; in this
case the pertubed spheres will define a hyperbolic or an elliptic trisector, 
respectively. In \yspace, the perturbation of the spheres will 
cause the point $\yinv{O}$ to move infinitesimally \outside or \inside 
the circle $\ycone$ respectively. This result is indicative of the fact that
a parabolic trisector can be viewed as an in-between state of elliptic 
and hyperbolic trisectors, as mentioned above. 

Due to the close relation of the parabolic trisectors with the other 
two types, we can determine the outcome of predicates that involve a 
parabolic trisector using variations of the algorithms that 
evaluate them for the respective hyperbolic or elliptic type. In the 
following section, we present a list of these predicates and 
the corresponding modifications that have to be applied in each case.

\section{Predicates for parabolic trisectors} 
\label{sec:predicates_for_parabolic_trisectors}

The type of the trisector $\tri{ijk}$ can be easily determined via 
the call of the $\text{\tritype}(S_i,\allowbreak S_j, \allowbreak S_k)$. If the predicate 
returns \emph{parabolic} then the following modifications have to 
be considered during the evaluation of the respective predicate. Notice 
that in the following list of predicates, only non-degenerate intermediate
results are considered and that the orientation of a parabolic trisector 
is defined in the same way as an elliptic trisector (see Section~\ref{ssub:elliptic_trisector_orientation}). 

\begin{description}
\item[$\text{\distance}(S_i,S_j,S_k,S_a)$ :] This predicate can be evaluated
as described in Section~\ref{sub:the_distance_predicate_analysis}, after 
taking into consideration that the centers $C_i,C_j$ and $C_k$ are 
not collinear and that there exist only one plane commonly tagent to the 
first three spheres. As a consequence, if $D_{ijka}^{xyz}\neq 0$, the quadratic polynomial $\Lambda(\epsilon)$, in terms of $\epsilon$, has a double root and therefore 
its only root is $\epsilon=-\frac{\Lambda_1}{2\Lambda_2}$,
where $\Lambda_2\neq 0$. It follows that the \distance predicate has 
to return a single (double) sign, $\sgn(\epsilon)=-\sgn(\Lambda_1)
=\sgn(D^{yzr}_{ijka}D^{yz}_{ijk}+D^{xzr}_{ijka}D^{xz}_{ijk}
              -D^{xyr}_{ijka}D^{xy}_{ijk} )$, 
which corresponds to the signed distance of $S_a$ from the plane 
cotagent to the sites $S_i,S_j$ and $S_k$. Alternatively, the 
predicate could return this \emph{double} sign in a tuple, \ie, 
$(-\sgn(\Lambda_1),-\sgn(\Lambda_1))$, for consistency with the 
hyperbolic case. Lastly, if $D_{ijka}^{xyz}=0$, we return 
$(\sgn(\epsilon), \sgn(\epsilon)) $, where  $\sgn(\epsilon)=-\sgn(D^{xyzr}_{ijka})\cdot \sgn(D^{xyz}_{ijk})$. 
\item[$\text{\exist}(S_i,S_j,S_k,S_a)$ :] This predicate can be evaluated 
as described in Section~\ref{tab:exist_distance_non_degenerate}, without 
modifications. Remember that the \exist predicate only returns the number 
of \emph{finite} Apollonius vertices that exist among 
$\{v_{ikja},v_{ijka}\}$ on the trisector $\tri{ijk}$.
\item[$\text{\shadow}(S_i,S_j,S_k,S_a)$ :] The outcomes of the \shadow 
predicate are the same with the hyperbolic case, with the only difference
that we can consider that $\pm\infty$ coincide. Therefore, the only possible 
topological forms of the shadow region are 
$\emptyset, \RR=(-\infty,+\infty), (-\infty, \phi)\cup(\chi,+\infty)$ and 
$(\chi,\phi)$. Notice that the latter two cases 
indicate that the image of the shadow region of $S_a$ on $\tri{ijk}$ in 
\yspace is an arc of $\ycone$ that either contains $\yinv{O}$ or 
not, respectively. We can answer the \shadow predicate using the analysis
presented in Section~\ref{sub:the_shadowregion_predicate_analysis}, 
although the cases $(-\infty,\phi)$ and $(\chi,+\infty)$ will never 
arise. 
\item[$\text{\conflict}(S_i,S_j,S_k,S_l,S_m,S_q)$ :] This predicate 
can be determined using the same algorithm presented for the respective 
elliptic case in Section~\ref{sub:conflict_elliptic}. 
\end{description}

Via the use of these modifications, we have proven the following lemma. 

\begin{lemma}
The \distance, \exist, \shadow and \conflict predicate for 
parabolic trisectors can be 
evaluated by determining the sign of quantities of algebraic 
degree at most 6, 8, 8 and 10 (in the input quantities), respectively.
\end{lemma}

In the case where an intermediate degenerate outcome arises during 
the evaluation of one of these predicates, we can resolve the 
degeneracy by applying a perturbation scheme. Below we present a list
of perturbed predicates and how we can determine their outcome, in the 
case where $\tri{ijk}$ is a parabola.

\begin{description}
\item[$\text{\distancep}(S_i,S_j,S_k,S_a)$ :] The sphere $S_a$ is 
tangent to the plane commonly tagent to the first three input spheres, 
\ie, the predicate $\text{\distance}(S_i,S_j,S_k,S_a)$ returned 0. Let 
$\Delta$ denote the outcome of the respective perturbed predicate, we 
determine the sign of $\Delta$, using similar arguements and notation with the ones presented in Section~\ref{ssub:the_perturbed_distance_predicate}.
\begin{description}
\item[\bf Step 1.] If $a >i,j,k$, return $\Delta=-$. Otherwise, name exchange the sites $S_i,S_j$ and $S_k$ such that $i>j,k,a$ and go to Step 2.
\item[\bf Step 2.] If $t_a\equiv t_i$ return $\Delta=+$, otherwise go to Step 3. 
\item[\bf Step 3.] If $t_a\equiv t_j$ return $\Delta=-$ if 
$r_a>r_j$ or $\Delta=-$ if $r_a<r_j$. Otherwise go to Step 4. 
\item[\bf Step 4.] If $t_a\equiv t_k$ return $\Delta=-$ if 
$r_a>r_k$ or $\Delta=-$ if $r_a<r_k$. Otherwise go to Step 5. 
\item[\bf Step 5.] If $t_j,t_k$ and $t_a$ are not collinear (equivalently 
$D_{ijka}^{xyz}\neq 0$) go to step 6. Otherwise we should return as 
$\Delta$ the sign of $-{D^{xyzr}_{ijka}}/{D^{xyz}_{ijk}}$ (see case $D^{xyz}_{ijka}=0$ in Section~\ref{sub:the_distance_predicate_analysis}) 
when the radius $r_a$ of the sphere $S_a$ infinitesimally becomes 
$r_a+\epsilon_a$ for some $\epsilon_a>0$. In this case,  
$\sgn(\Delta)=\sgn(D^{xyz}_{ijk})\cdot\sgn(\epsilon_a D^{xyz}_{jka})=
\sgn(D^{xyz}_{ijk})\cdot\sgn(D^{xyz}_{jka})$, 
since it originally holded that $D^{xyzr}_{ijka}=0$ (the quantity 
$-{D^{xyzr}_{ijka}}/{D^{xyz}_{ijk}}$ was zero as we are studying a degenerate case).
\item[\bf Step 6.] Since $D_{ijka}^{xyz}\neq 0$, we should return 
as $\Delta$ the $\sgn(-\Lambda_1)=\sgn(D^{yzr}_{ijka}D^{yz}_{ijk}+D^{xzr}_{ijka}D^{xz}_{ijk} -D^{xyr}_{ijka}D^{xy}_{ijk} )$ when the radius $r_a$ of the sphere $S_a$ infinitesimally becomes $r_a+\epsilon_a$ for some $\epsilon_a>0$. The last expression, initially equaled zero as we were in a degenerate 
case, but after substituting $r_a$ with $r_a+\epsilon_a$, it becomes
$\sgn(\epsilon_a)\sgn( D^{yz}_{jka}D^{yz}_{ijk}+D^{xz}_{jka}D^{xz}_{ijk} -D^{xy}_{jka}D^{xy}_{ijk} )=\sgn( D^{yz}_{jka}D^{yz}_{ijk}+D^{xz}_{jka}D^{xz}_{ijk} -D^{xy}_{jka}D^{xy}_{ijk} ).$
\end{description}

\item[$\text{\shadowp}(S_i,S_j,S_k,S_a)$ :] The non-degenerate outcomes
of the \shadow predicate for parabolic trisectors are $\emptyset, \RR=(-\infty,+\infty), (-\infty, \phi)\cup(\chi,+\infty)$ and 
$(\chi,\phi)$ (the last two cases, can be considered, for parabolic trisectors, as the set $(\chi,\phi)$ that either contains the point at infinity or not, respectively). Degenerate shadow regions occur when $\phi$ and/or $\chi$ coincide with each other and/or
the point at infinity. 

The non-degenerate outcomes of the $S=\text{\shadow}(S_i,S_j,S_k,S_a)$ predicate are identified 
from the combination of the outcomes $E=\text{\exist}(S_i,S_j,S_k,S_a) $
and $D=\text{\distance}(S_i,S_j,S_k,S_a)$. 
\begin{enumerate}
\item If $E=2$ and $D=(+,+)$ then $S=(\chi,\phi)$.
\item If $E=2$ and $D=(-,-)$ then $S=(-\infty, \phi)\cup(\chi,+\infty)$.
\item If $E=0$ and $D=(+,+)$ then $S=\emptyset$.
\item If $E=0$ and $D=(-,-)$ then $S=\RR=(-\infty,+\infty)$.
\end{enumerate} 

All other combinations of $S$ and $E$ outcomes are either infeasible or
yield a degenerate $S$ outcome. In order to determine, if such a combination
results in a degenerate shadow region and what the corresponding 
pertrubed shadow region would be, we follow an analysis similar with the
one presented in Section~\ref{sub:the_perturbed_shadowregion_predicate}. 
The feasible combinations are the following
\begin{description}
\item[Case 1.] If $E=0$ and $D=(0,0)$ then either $S=(\chi,\phi)$ where $\chi\equiv \phi \pm\infty$.
\item[Case 2.] If $E=1$ and $D=(0,0)$ then $S=(\chi,\phi)$ where either $\chi\equiv -\infty$ or $\phi\equiv +\infty$.
\item[Case 3.] If $E=1$ and $D=(+,+)$ or $(-,-)$ then $S=(-\infty, \phi)\cup(\chi,+\infty)$ or $(\chi,\phi)$, where $\chi\equiv \phi \not\equiv \pm\infty$.
\end{description}

To determine the perturbed shadow region $S^\epsilon$, we evaluate 
$D^\epsilon$ if $D=(0,0)$ and a non-zero $I$ that corresponds to either
$\text{\insphere}(S_i,S_j,S_k,S_n,S_a)$ or 
$\text{\insphere}(S_i,S_k,S_j,S_n,S_a)$ for some $S_n\not\equiv S_a$. 
Note that the \shadow predicate 
is assumed to be called during the evaluation of the \conflict predicate
and therefore the Apollonius vertices $v_{ijkl}$ and $v_{ikjm}$ and 
either one or both of $\{v_{ijkq},v_{ikjq}\}$ exist on
the trisector $\tri{ijk}$. The existence of these vertices guarantee that
the corresponding \insphere predicates are well defined for the respective 
inputs.  

To resolve degeneracies of Case 1, using the same arguements presented 
in Section~\ref{sub:the_perturbed_shadowregion_predicate}, we consider
$D^\epsilon$. If $D^\epsilon=(+,+)$ then $S^\epsilon=\emptyset$ if $I=+$ 
or $S^\epsilon=(-\infty, \phi)\cup(\chi,+\infty)$ if $I=-$. Otherwise, 
if $D^\epsilon=(-,-)$ then $S^\epsilon=(\chi,\phi)$ if $I=+$ 
or $S^\epsilon=\RR=(-\infty,+\infty)$ if $I=-$.

For degeneracies of Case 2, if $D^\epsilon=(+,+)$ 
or $(-,-)$ then $S^\epsilon=(\chi,\phi)$ or 
$(-\infty, \phi)\cup(\chi,+\infty)$, respectively. 

Finally, degeneracies of Case 3 are handled exactly as the degeneracies
of Type B, Case 2, Subcase 2, presented in Section~\ref{sub:the_perturbed_shadowregion_predicate}. This is also the most difficult 
degenerate case to handle in both hyperbolic and elliptic trisectors.

\item[$\text{\existp}(S_i,S_j,S_k,S_a)$ :] The outcome of the respective
\exist predicate is non-degenerate if and only if it does not equal 1 and 
the corresponding \distance predicate did not return $0$ (or $(0,0)$).
The outcome of the \existp predicate can be obtained by the 
evaluation of $\mathcal{SR}^\epsilon=\text{\shadowp}(S_i,S_j,S_k,S_a)$, as it expresses the 
number of finite boundary of the perturbed shadow region of $S_a$ on the
trisector $\tri{ijk}$:
\begin{enumerate}
\item If $\mathcal{SR}^\epsilon=\emptyset$ or $\RR=(-\infty,+\infty)$, 
then the \existp predicate returns 0. Otherwise,
\item If $\mathcal{SR}^\epsilon=(\chi,\phi)$ or 
$\RR=(-\infty,\phi)\cup(\chi+\infty)$, then the \existp predicate returns 2.
\end{enumerate}
\item[$\text{\conflictp}(S_i,S_j,S_k,S_l,S_m,S_q)$ :] The predicate makes 
use of the outcomes of the perturbed predicates \distancep, \existp, 
\shadowp, which can be evaluated as described above. 
\end{description}

The following remarks prove the following theorem. 

\begin{theorem}
Let $d$ denote the algebraic degree of the \vconflictp predicate. 
The \distancep, \existp, \shadowp and \conflictp predicate for 
parabolic trisectors can be evaluated by determining the sign 
of quantities of algebraic degree at 5, 10, 10 
and $\max\{10,d\}$ (in the input quantities), respectively.
\end{theorem} 



\chapter{Conclusion and Future Work} 
\label{sec:conclusion_and_future_progress}
  In this thesis, we presented a clever way of combining 
  various subpredicates in order to answer the  
  \conflict predicate. 
  The design of all predicates and primitives was made in such a way 
  that the maximum algebraic cost of answering them would be as 
  low as possible, on the input quantities. Based on current 
  bibliography, the resulting degree 10 of the main predicate 
  is quite small when compared with the respective 2D 
  version of the \conflict predicate (16 as shown in 
  \cite{Emiris2006Predicates} and 6 as shown in 
  \cite{millman2007degeneracy}). 
  Moreover, the fact that the inversion
  and perturbation techniques can be applied in the 2D Apollonius 
  diagram suggests that a similar analysis with the 
  respective 3D case will yield similar algebraic degrees with 
  \cite{millman2007degeneracy}, for both degenerate and 
  non-degenerate inputs. 
  It is also remarkable that both the \insphere 
  (equivalent to \vconflict) and the \conflict
  predicates share the same algebraic degree. 
  
  Through our attempt to answer the master predicate, various 
  useful primitives were also developed. These tools can also 
  be used in the context of an incremental algorithm that evaluates 
  the Apollonius diagram of a set of spheres.


  Ultimately, we would like to use the remarks that connect \zspace with
  \yspace to resolve the degenerate \vconflictp
  predicate described in \cite{devillers2017qualitative} with the 
  lowest algebraic degree, using a qualitative perturbation scheme.
  The resulting degree will inevitably be an upper bound for the 
  cost of the \orderp and \conflictp predicates for degenerate inputs.

  Lastly, it is our intent to implement the algorithms presented in this
  thesis. This would consist a great step towards the exact 
  construction of the 3D Apollonius diagram via an incremental algorithm 
  and would provide information of the viability of such algorithms 
  opposite existing implementations based on floating-point arithmetic.







\printbibliography[heading=bibintoc]


\end{document}